\newif\ifdraft\draftfalse

\ifdraft
\documentclass[a4paper,DIV=15,11pt]{scrartcl}
\else
\documentclass[compsoc,conference]{IEEEtran}
\IEEEoverridecommandlockouts
\usepackage{cite}
\usepackage{amsmath,amssymb,amsfonts}
\usepackage{algorithmic}
\usepackage{graphicx}
\usepackage{textcomp}
\usepackage{xcolor}
\usepackage{fontspec}

\def\BibTeX{{\rm B\kern-.05em{\sc i\kern-.025em b}\kern-.08em
    T\kern-.1667em\lower.7ex\hbox{E}\kern-.125emX}}
\fi

\usepackage{mysettings}
\usepackage{docmute}

\begin{document}

\ifdraft
\title{
Type-based information flow analysis for $\pi$-calculus with a dynamically extensible security lattice
}
\author{
Yukihiro Oda\thanks{yukihiro.oda.e5 [at] tohoku.ac.jp} 
\quad
Eijiro Sumii\thanks{sumii [at] tohoku.ac.jp}
\\
Graduate School of Information Sciences\\Tohoku University
}
\else
\title{
Type-based information flow analysis for $\pi$-calculus with a dynamically extensible security lattice
}

 \author{\IEEEauthorblockN{Yukihiro Oda}
\IEEEauthorblockA{\textit{Graduate School of Information Sciences} \\
\textit{Tohoku University}\\
Tohoku, Japan \\
yukihiro.oda.e5 [at] tohoku.ac.jp \\
ORCID:0000-0002-6216-3193}
\and
 \IEEEauthorblockN{Eijiro Sumii}
\IEEEauthorblockA{\textit{Graduate School of Information Sciences} \\
\textit{Tohoku University}\\
Tohoku, Japan \\
sumii [at] tohoku.ac.jp}
}
\fi

\maketitle
\ifdraft\else
\thispagestyle{plain}
\pagestyle{plain}
\fi

\begin{abstract}
  We develop a type system for secure information flow where new
  security levels can be created and inserted into the security
  lattice \emph{dynamically}, i.e., even in the middle of an execution
  of a system.  Our system is formalized by extending Kobayashi's
  type-based secure information flow analysis for Milner's
  pi-calculus, which is one of the most expressive models (or
  ``languages'') supporting both sequential and concurrent
  computations, with concise syntax, reduction-based semantics, and
  bisimulation equivalence as a robust formalization of secrecy as
  non-interference.  The development required careful treatment of
  extensions of lattices themselves as well as deliberate generalization from the
  simple 2-element lattice (consisting of only High and Low) in the
  original system.
\end{abstract}

\ifdraft\else
\begin{IEEEkeywords}
language-based security, pi-calculus, barbed bisimulation, non-interference, dynamic extension of security lattice, runtime creation and insertion of new security levels
\end{IEEEkeywords}
\fi

\section{Introduction}

Lattice-based secure information
flow~\cite{DBLP:journals/cacm/Denning76} classifies data into security
levels---such as H(igh) and L(ow)---that are elements of the security
lattice, and aims to prevent leakage of higher-level information into
lower-level actors.  Unlike mere access control, secure information
flow also addresses indirect information leakage---like
$\mathtt{if}~b^{\mathtt{H}}~\mathtt{then}~1^{\mathtt{L}}~\mathtt{else}~0^{\mathtt{L}}$
where $b^{\mathtt{H}}$ is a high-level Boolean value and
$1^{\mathtt{L}}$, $0^{\mathtt{L}}$ are low-level integers---and even
$\mathtt{if}~b^{\mathtt{H}}~\mathtt{then}~1^{\mathtt{L}}~\mathtt{else}~\mathit{diverge}()$
when termination-sensitive.  Absence of information leakage is
formalized as non-interference~\cite{DBLP:conf/sp/GoguenM82a}, which
asserts the equivalence of two systems with different high-level
information when observed from a low level.

To account for such indirect information flows, language-based---and,
more specifically, type-based---information flow
analysis~\cite{DBLP:journals/jcs/VolpanoIS96,DBLP:journals/jsac/SabelfeldM03}
models systems as programs and adopts a static type system that
imposes the classification by annotating types with security levels.
Though called ``language-based'' with ``programs,'' this approach is
not limited to programs or programming languages in the narrow sense
(such as C and Java), but is applicable to various models of systems
as well.  Traditionally, the ``languages'' (or models) have been
imperative~\cite{DBLP:journals/jcs/VolpanoIS96} or
functional~\cite{DBLP:conf/popl/HeintzeR98}, and sequential.  They
have also been extended with (mainly thread-based) concurrency, where
determinism has often been essential and non-interference has been
proved in probabilistic settings or under a strong restriction on the
concurrent computations---for instance, no low-level communication is
allowed after high-level synchronization, or observable
non-determinism is forbidden at all; see~\cite[Section
  IV-B]{DBLP:journals/jsac/SabelfeldM03} for a survey.

Kobayashi~\cite{Kobayashi2005} removed such strong restrictions by
adopting a type system for
lock-freedom~\cite{DBLP:journals/iandc/Kobayashi02} in
$\pi$-calculus~\cite{DBLP:books/daglib/0098267,DBLP:journals/iandc/MilnerPW92a,DBLP:journals/iandc/MilnerPW92b},
a rather general and expressive model of both sequential and
concurrent computations.  In short, even high-level synchronization is
allowed before low-level communications as long as the former is
lock-free.  Non-interference is proved as
(barbed~\cite{DBLP:conf/icalp/MilnerS92}) bisimulation-based
congruence, which is also a general and robust notion of equivalence
(and can be extended with probability; see, for instance,
\cite{DBLP:journals/iandc/LarsenS91,DBLP:conf/csfw/Smith03,DBLP:journals/jlap/CastiglioniLT24,DBLP:conf/concur/SporkBKPQ24},
among many others).\footnote{Equivalence (and therefore non-interference) of concurrent processes with
  general interactions (as opposed to ``threads'' with limited
  concurrency primitives) has by itself been a significant research
  challenge, e.g.~classically
  \cite{DBLP:books/el/01/Glabbeek01,DBLP:conf/concur/Glabbeek93} and
  more recently \cite{DBLP:conf/birthday/FinkbeinerO23}, just to name
  a few.  While no single definition of equivalence may be considered
  satisfactory in every respect, barbed bisimulation is ``fine'' (as opposed to coarse) and
  \emph{sufficient} as a proof of \emph{soundness} of a security
  type system.} However, Kobayashi~\cite{Kobayashi2005} only
considered the security lattice with 2 elements (high and low).

In the present work, we extend Kobayashi~\cite{Kobayashi2005} with
general security lattices.  Furthermore, we allow
\emph{dynamic}
extension of the
lattice, that is, new security levels can be added \emph{while} the
system is running, which is a mandatory functionality for many modern
systems that allow registration/creation of new users/accounts.  Even
the former generaliztion by itself is non-trivial---all the definitions, statements,
and proofs need to be carefully parameterized with the level $l$ of the
attacker (namely, observer for the equivalence) being considered
(\cref{def:secure-channel-type}, \cref{def:secure-env}, \cref{def:k-contexts}, and so forth).
The latter extension is, to the best of our knowledge, new, even for sequential
languages.  It also requires careful treatment of the ``current''
security lattice of the system, throughout our technical developments such as reduction (\cref{def:structural-preorder}) and typing rules (\cref{fig:typing-rules}), as
well as the definition of a ``safe'' extension of a lattice itself (\cref{def:newlev-lattice}) and all the proofs (in the Appendices).

The rest of this paper is as follows: \cref{sec:piL-calculus}
introduces the syntax and reduction semantics of our language along
with definitions concerning the security lattices.
\cref{sec:type-system} defines our type system for information flow
analysis.  \cref{sec:soundness} proves the non-interference theorem
and \cref{sec:conclusion} concludes.  Further details of our technical
developments and proofs
are given in the Appendices.

\section{$\pi^{L}$-calculus}
\label{sec:piL-calculus}
This section introduces our process calculus $\pi^{L}$ for
type-based information flow analysis, which extends the
$\pi$-calculus~\cite{DBLP:books/daglib/0098267,DBLP:journals/iandc/MilnerPW92a,DBLP:journals/iandc/MilnerPW92b}
with a lattice of secrecy levels.

\subsection{Syntax}
Our language is $\pi$-calculus~\cite{DBLP:books/daglib/0098267,DBLP:journals/iandc/MilnerPW92a,DBLP:journals/iandc/MilnerPW92b,Kobayashi2005} augmented with a
$\mathpinewsecrecylevelsy$ operation for extending the lattice of
secrecy levels, as in the definition below.  Intuitively, the new process form
${\mathpinewsecrecylevel{l}{l_1,\dots,l_m}{l_1',\dots,l_n'}{P}}$
creates a new level $l$ above (resp.~below) existing levels
$l_1,\dots,l_m$ (resp.~$l_1',\dots,l_n'$), and then execute $P$.
(Its formal semantics will be defined in the next sections.)

In the rest of this paper, we often write $\mathvect{l}$,
$\mathvect{x}$, etc.~to abbreviate sequences like $l_1,\dots,l_m$ and
$x_1,\dots,x_n$ when their lengths $m,n\ge 0$ are arbitrary or clear
from the context.  For a sequence $\mathvect{t}$ where $t$ is a
meta-variable of any kind, we write ${a}\in{\mathvect{t}}$ if $a$ is
an element of $\mathvect{t}$.

We write $\maththesetofnames$ and $\maththesetofsecrecylevels$ for the
distinct sets of \emph{channel names} and \emph{secrecy level names},
respectively.  Also, we assume at least two distinct secrecy level names
$\top,\bot\in\maththesetofsecrecylevels$.

\begin{definition}[Syntax of $\pi^{L}$-calculus]
 We define \emph{processes} as follows
 \begin{align*}
 x,y,\dots&\in \maththesetofnames &\text{(channel name)} \\
 \lk,l,\lm&\in \maththesetofsecrecylevels &\text{(secrecy level name)} \\
 c&\mathcoloneqq {\mathobtrue^{l}} \mathrelbar 
 {\mathobfalse^{l}} \mathrelbar 
 {\mathobunit}
 &\text{(constant value)} \\
 v&\mathcoloneqq 
 c \mathrelbar 
 x 
 &\text{(value)} \\
 P&\mathcoloneqq
\ifdraft\else\lefteqn{\fi
 {\mathnil} \mathrelbar 
 {\mleft(\mathpiparallel{P}{P}\mright)} \mathrelbar 
 {\mathpireplication{P}} \mathrelbar 
 {\mathpinew{\mathistype{x}{\xi}}{P}} \mathrelbar
\ifdraft\else}\fi
 &\text{(process)} \\ 
 &
\lefteqn{
 \phantom{{}\mathcoloneqq{}}
 {\mathpioutput{x}{\mathvect{v}}.P} \mathrelbar 
 {\mathpiinput{x}{\mathvect{x}}.P} \mathrelbar
 {\mathpinewsecrecylevel{l}{\mathvect{l}}{\mathvect{l}}{P}} \mathrelbar
}
 \\
 &
 \phantom{{}\mathcoloneqq{}} 
 {\mathobif{v}{P}{P}}
 \end{align*}
 where $\xi$ ranges over the core channel types (defined
 later in \cref{def:types-and-usages}).
\end{definition}
\noindent
 By convention, we give a lower
 precedence to $|$ than to other operators, so
 $\mathpiparallel{\mathpinew{\mathistype{x}{\xi}}{P_1}}{P_2}$ means
 $\mathpiparallel{\mleft(\mathpinew{\mathistype{x}{\xi}}{P_1}\mright)}{P_2}$
 for example.
 We also assume $|$ is left-associative.

As usual, every $y_i\in\mathvect{y}$ in
${\mathpiinput{x}{\mathvect{y}}.P}$, $x$ in
${\mathpinew{\mathistype{x}{\xi}}{P}}$, and, in particular, $l$ in
${\mathpinewsecrecylevel{l}{\mathvect{l_{1}}}{\mathvect{l_{2}}}{P}}$
are \emph{bound} in $P$ and subject to implicit $\alpha$-conversion
such that each of them is different from other (bound or unbound)
names.  We write $\mathFNof{P}$, $\mathFCNof{P}$,
and $\mathFSNof{P}$, respectively, for the set of free (i.e.,
unbound) names, free channel names, and free secrecy level names of
process $P$.  A \emph{sub-process} of a process $P$ is defined as a
subexpression of $P$ that is also a process.  We write
$\maththesetofprocesses$ for the set of processes.

We write ${\mathsubstbox{P}{\mathsubst{y_{0}}{v_{0}}, \dots,
    \mathsubst{y_{n}}{v_{n}}}}$ for the process obtained by
respectively replacing all the free occurrences of $y_{0}, \ldots,
y_{n}$ in $P$ with $v_{0}, \ldots, v_{n}$.  We often abbreviate
${\mathsubstbox{P}{\mathsubst{y_{0}}{v_{0}}, \dots,
    \mathsubst{y_{n}}{v_{n}}}}$ as
${\mathsubstbox{P}{\mathsubst{\mathvect{y}}{\mathvect{v}}}}$ and
$\mathpinew{\mathistype{x_{0}}{\xi_{0}}}{{}\dots
  \mathpinew{\mathistype{x_{n}}{\xi_{n}}}{P}}$ as
$\mathpinew{\mathistype{\mathvect{x}}{\mathvect{\xi}}}{P}$.

\subsection{Lattice of secrecy levels}
In this section, we give definitions and prove lemmas for the lattice
of secrecy levels.

\begin{definition}[Lattice]
 We define a lattice as a poset $(L, {\leq_{L}})$ where, 
 for any finite $S\subseteq L$,
 there exist the supremum and infimum of $S$.
\end{definition}
Note that the supremum of $\emptyset$ is the minimum of $(L, {\leq_{L}})$,
and the infimum of $\emptyset$ is the maximum of $(L, {\leq_{L}})$.
Thus, for each lattice, its maximum and minimum exist.%
\footnote{We adopt the present definition as in \cite{nLab:lattice}
  and \cite[p.~3, Remark]{Johnstone86}; the latter explains why it is
  more ``natural.''  An alternative term for this definition is a
  \emph{bounded lattice}~\cite[Section~2]{nLab:lattice}, which we
  avoid for brevity in this paper.}
We sometimes write ${\mathlatticemax}_{L}$ and ${\mathlatticemin}_{L}$ for
the maximum and minimum of $(L, {\leq_{L}})$, respectively.
For simplicity, we often write just $L$ for a lattice $(L, {\leq_{L}})$.
For $(L, {\leq_{L}})$ and ${a, b}\in {L}$, we write ${a} <_{L} {b}$ 
if ${a}\leq_{L}{b}$ and ${a}\neq{b}$.
We also write ${a} \not\leq_{L} {b}$ (resp.~${a} \not <_{L} {b}$)
when ${a}\leq_{L}{b}$ (resp.~${a} <_{L} {b}$) does not hold.
For $S\subseteq L$,
we write $\mathof{\sup_{L}}{S}$ and $\mathof{\inf_{L}}{S}$
for the supremum and infimum of $S$ in $L$, respectively.
Note also
$
 {a} \leq_{L} {b} 
 \iff \mathof{{\sup}_{L}}{\mathsetextension{a, b}}=b 
 \iff \mathof{{\inf}_{L}}{\mathsetextension{a, b}}=a
$.

The following definition is crucial for ``safe'' extension of a
lattice, as we will adopt in the rest of this paper.  In short, an
extension $L$ of a lattice $L'$ must be a ``superlattice'' of $L'$, that
is, $L'$ must be a sublattice of $L$.
\begin{definition}[Sublattice]
 \label[definition]{def:sublattice}
 For a lattice $(L, {\leq_{L}})$,
 we define a sublattice of $(L, {\leq_{L}})$ as a lattice $(L', \leq_{L'})$ satisfying
 the following conditions:
 \begin{enumerate}
  \item $L'\subseteq L$
  \item ${{\mathlatticemax}_{L}}, {{\mathlatticemin}_{L}}\in{L'}$
  \item $\mathof{\sup_{L'}}{S}=\mathof{\sup_{L}}{S}$ and $\mathof{\inf_{L'}}{S}=\mathof{\inf_{L}}{S}$ for every finite $S\subseteq L'$.
 \end{enumerate}
 We write $\mathissublattice{L'}{L}$ if $L'$ is a sublattice of $L$.
\end{definition}

\begin{lemma}
 Let $L$ be a lattice and $\mathissublattice{L'}{L}$.  For any $a,b\in L'$,
 \begin{enumerate}
  \item ${a} \leq_{L'} {b}$ if and only if ${a} \leq_{L} {b}$, and
  \item ${a} <_{L'} {b}$ if and only if ${a} <_{L} {b}$.
 \end{enumerate}
\end{lemma}
\noindent
We omit proofs when they are straightforward.

A \emph{lattice of secrecy levels} is a lattice such that its
underlying set is a finite subset of $\maththesetofsecrecylevels$, with
$\top$ and $\bot$ being the maximum and minimum, respectively.  We write
$\maththesetoflatticeforsec$ for the set of lattices of secrecy
levels.  We also write
${\mathvect{l}}\subseteq{L}$ if every ${l_i}\in{\mathvect{l}}$ belongs to $L$.

We then give a notation
${\mathpinewsecrecylevel{l}{\mathvect{l_{0}}}{\mathvect{l_{1}}}{L}}$
for extension of lattices:\footnote{In fact, \emph{any} extension of a
  lattice would do for the rest of our development as long as it
  yields a superlattice and parallel extensions are
  commutative. We adopt the present definition only for
  the sake of concreteness.}
\begin{definition} 
 \label[definition]{def:newlev-lattice}
 Let $(L, {\leq_{L}})$ be a lattice of secrecy levels with
 $\mathvect{l_{0}}, \mathvect{l_{1}}\subseteq L$, and $l\notin L$ for
 an $l\in\maththesetofsecrecylevels$.  Let furthermore
 $L'={L\cup{\mathsetextension{l}}}$ and $\leq_{L'}$ be the
 reflexive and transitive closure of
 ${\leq_{L}}\cup\mathsetintension{(l', l)}{l'\in \mathvect{l_{0}}}\cup
 \mathsetintension{(l, l')}{l'\in \mathvect{{l}_{1}}}$.  If $(L',
 \leq_{L'})$ is a lattice, then we write
 ${\mathpinewsecrecylevel{l}{\mathvect{l_{0}}}{\mathvect{l_{1}}}{L}}$
 for $L'$.
\end{definition}
\noindent
Note that $L'$ may not always be a lattice, e.g., if $L
=\mathsetextension{\bot, \top}$,
$\mathvect{l_{0}}=\top$, and
$\mathvect{l_{1}}=\bot$.  We reject such extensions by
definition.
Note also that $L'$ may not always be a superlattice of $L$, that is,
$L$ may not be a sublattice of $L'$, even if $L'$ is a lattice.
Later, we impose $\mathissublattice{L}{L'}$ by typing.

\subsection{Reduction}
This section defines reduction in $\pi^{L}$-calculus via structural
preorder, which is a variant of standard structural congruence in
$\pi$-calculus but is asymmetric for the sake of specifying
canonical forms.

\begin{definition}[Structural preorder]
 \label[definition]{def:structural-preorder}
 The binary relation $\mathpistructuralpo$ on processes, called 
 \emph{structural preorder}, is defined as
 the least reflexive and transitive relation satisfying the following rules,
 where ${P_{0}}\mathpistructuraleq {P_{1}}$ denotes that 
 both ${P_{0}}\mathpistructuralpo{P_{1}}$ and ${P_{1}}\mathpistructuralpo{P_{0}}$ hold.
 \begin{description}
  \item[\rulename{SP-Zero1}] ${P}\mathpistructuraleq{\mathpiparallel{P}{\mathnil}}$
  \item[\rulename{SP-Zero2}] ${\mathnil}\mathpistructuraleq{\mathpinew{\mathistype{x}{\xi}}{\mathnil}}$
  \item[\rulename{SP-Commut}] ${\mathpiparallel{P_{0}}{P_{1}}}\mathpistructuralpo{\mathpiparallel{P_{1}}{P_{0}}}$
  \item[\rulename{SP-Assoc}] ${\mathpiparallel{\mleft(\mathpiparallel{P_{0}}{P_{1}}\mright)}{P_{2}}}\mathpistructuralpo{\mathpiparallel{P_{0}}{\mleft(\mathpiparallel{P_{1}}{P_{2}}\mright)}}$~\footnote{The other direction can be derived from \rulename{SP-Commut} and \rulename{SP-Par}.}
  \item[\rulename{SP-New}] ${\mathpiparallel{\mathpinew{\mathistype{x}{\xi}}{P_{0}}}{P_{1}}}\mathpistructuraleq{\mathpinew{\mathistype{x}{\xi}}{\mleft(\mathpiparallel{P_{0}}{P_{1}}\mright)}}$ if $x\notin\mathFCNof{P_{1}}$
  \item[\rulename{SP-IfT}] ${\mathobif{\mathobtrue^{l}}{P_{0}}{P_{1}}}\mathpistructuralpo{P_{0}}$
  \item[\rulename{SP-IfF}] ${\mathobif{\mathobfalse^{l}}{P_{0}}{P_{1}}}\mathpistructuralpo{P_{1}}$
  \item[\rulename{SP-Rep}] ${\mathpireplication{P}}\mathpistructuralpo{\mathpiparallel{\mathpireplication{P}}{P}}$
  \item[\rulename{SP-Par}] If ${P_{0}}\mathpistructuralpo{P_{1}}$, then ${\mathpiparallel{P_{0}}{Q}}\mathpistructuralpo{\mathpiparallel{P_{1}}{Q}}$.
  \item[\rulename{SP-CNew}] If ${P_{0}}\mathpistructuralpo{P_{1}}$, then ${\mathpinew{\mathistype{x}{\xi}}{P_{0}}}\mathpistructuralpo{\mathpinew{\mathistype{x}{\xi}}{P_{1}}}$.
 \end{description}
\end{definition}

\begin{definition}[Reduction]
 \label[definition]{def:reduction}
 The binary relation $\mathpireduction$ on 
 ${\maththesetofprocesses}\times{\maththesetoflatticeforsec}$, called \emph{reduction},
 is defined as
 the least relation satisfying the following rules:
 \begin{description}
  \item[\rulename{R-Com}] ${\mathtuple{{\mathpiparallel{\mathpioutput{x}{\mathvect{v}}. P_{0}}{\mathpiinput{x}{\mathvect{y}}.P_{1}}}, L}}\mathpireduction{\mathtuple{\mathpiparallel{P_{0}}{\mathsubstbox{P_{1}}{\mathsubst{\mathvect{y}}{\mathvect{v}}}}, L}}$
  \item[\rulename{R-NewLev}] If ${\mathvect{l_{0}}},{\mathvect{l_{1}}}\subseteq{L}$ 
	     and 
	     ${\mathpinewsecrecylevel{l}{\mathvect{l_{0}}}{\mathvect{l_{1}}}{L}}$ 
	     is defined,\\ then
	     ${\mathtuple{{\mathpinewsecrecylevel{l}{\mathvect{l_{0}}}{\mathvect{l_{1}}}{P}}, L}}\mathpireduction{\mathtuple{P, {\mathpinewsecrecylevel{l}{\mathvect{l_{0}}}{\mathvect{l_{1}}}{L}}}}$.
  \item[\rulename{R-Par}] If ${\mathtuple{{P_{0}}, L}}\mathpireduction{\mathtuple{{P'_{0}}, L'}}$, 
	       then ${\mathtuple{{\mathpiparallel{P_{0}}{P_{1}}}, L}}\mathpireduction{\mathtuple{{\mathpiparallel{P'_{0}}{P_{1}}}, L'}}$.
  \item[\rulename{R-New}] If ${\mathtuple{{P}, L}}\mathpireduction{\mathtuple{{P'}, L'}}$, 
	       then ${\mathtuple{{\mathpinew{\mathistype{x}{\xi}}{P}}, L}}\mathpireduction{\mathtuple{{\mathpinew{\mathistype{x}{\xi}}{P'}}, L'}}$.
  \item[\rulename{R-SP}] If ${P_{0}}\mathpistructuralpo{P'_{0}}$, ${\mathtuple{P'_{0}, L_{0}}}\mathpireduction{\mathtuple{P'_{1}, L_{1}}}$, and ${P'_{1}}\mathpistructuralpo{P_{1}}$,\\
	then ${\mathtuple{P_{0}, L_{0}}}\mathpireduction{\mathtuple{P_{1}, L_{1}}}$.
 \end{description}
 We write $\mathpireductionkc$ 
 for the reflexive and transitive closure of $\mathpireduction$.
 We also write $\mathpireductiontc$ for the transitive closure of $\mathpireduction$.
\end{definition}

\section{Type System}
\label{sec:type-system}
Our type system is an extension of Kobayashi~\cite{Kobayashi2005}'s,
with a general lattice of secrecy levels and dynamic creation of new
levels.

\subsection{Types and Usages}
Before our extension, we basically repeat
Kobayashi~\cite{Kobayashi2005}'s definition of
\emph{usages}---originally proposed
in~\cite{DBLP:journals/entcs/SumiiK98}---albeit omitting recursion for
the sake of technical simplicity
(while keeping usage variables for substitutions as in
\cref{def:subusage} \cref{item:sub-in-def:subusage}).
Informally, a usage is part of a
channel type and expresses how (when and in what order) the channel is
used for input and output, so as to ensure lock-freedom by checking
the correctness---called \emph{reliability}---of the usage.  The
intuitive meanings of usage expressions are briefly summarized in
\cref{table:meaning-of-usages}.

\begin{definition}
 \label[definition]{def:types-and-usages}
 We define \emph{types} and \emph{usages} as follows:
 \begin{align*}
 \tau&\mathcoloneqq \ifdraft\else\lefteqn{\fi\mathbooltypewithsec{l} \mathrelbar 
 \mathunittype \mathrelbar 
 \xi/U\ifdraft\else}\fi
 & \text{(type)} \\
 \xi &\mathcoloneqq \mathprogramtypetuple{\mathvect{\tau}}^{l}
 & \text{(core channel type)} \\
 \rho&\in\maththesetofusagevar
 & \text{(usage variable)} \\
 t_{o}, t_{c}&\in{\mathnat \cup \mathsetextension{\infty}}
 & \text{(obligation and capability levels)} \\
 U &\mathcoloneqq \lefteqn{\mathusagenil \mathrelbar 
 \rho \mathrelbar
 \alpha^{t_{o}}_{t_{c}}.U \mathrelbar 
  {\mleft(\mathpiparallel{U}{U}\mright)} \mathrelbar\ifdraft\else}
 & \text{(usage)}  \\
 \phantom{U} &\phantom{{}\mathcoloneqq{}} 
 \lefteqn{\fi{\mathpireplication{U}} \mathrelbar 
 {\mathlevelraise{t_{o}}{t_{c}}{U}} \mathrelbar 
 {\mathusageor{U}{U}}}
 & \ifdraft\text{(usage)}\fi  \\
 \alpha &\mathcoloneqq I \mathrelbar O
 & \text{(input and output)} 
 \end{align*}
\end{definition}

\begin{table*}[t]
 \centering
 \begin{tabular}[t]{|c|p{0.85\textwidth}|}
  \hline
  Usages & Intuitive meaning (how the channel should be used) \\
  \hline \hline
  $\mathusagenil$  & cannot be used at all \\ \hline
  $\alpha^{t_{o}}_{t_{c}}.U$  & used once for input ($\alpha=I$) or output ($\alpha=O$), and then used according to $U$, where
  $t_{o}$ and $t_{c}$ are natural numbers (or $\infty$) respectively called \emph{obligation} and \emph{capability} levels (not to be confused with secrecy levels) and represent when (i.e., in what order) the input or output \emph{must} and \emph{can} be performed, so as to prevent deadlocks and livelocks caused by self or cyclic dependencies \\ \hline
  $\mathpiparallel{U_{0}}{U_{1}}$ 
  & used according to $U_{0}$ and $U_{1}$, possibly in parallel (the symbol $|$ here represents parallel composition as in $\pi$-calculus) \\ \hline
  ${\mathpireplication{U}}$ 
  & used according to $U$ infinitely many times in parallel \\ \hline
  ${\mathlevelraise{t_{I}}{t_{O}}{U}}$
  & used according to $U$, but input and output obligation levels are raised (at least) to $t_{I}$ and $t_{O}$, respectively \\ 
  \hline
  ${\mathusageor{U_{0}}{U_{1}}}$ & used according to either $U_{0}$ or $U_{1}$, as chosen by the ``user'' (the symbol $\&$ here represents additive conjunction as in linear logic~\cite{DBLP:journals/tcs/Girard87}) \\
  \hline
 \end{tabular}
 \medskip

 \caption{Intuitive meaning of usage expressions (similar to~\cite[Table~1]{Kobayashi2005})}
 \label{table:meaning-of-usages}
\end{table*}

We write $\mathFVof{U}$ for the set of usage variables occurring in a
usage $U$.  A usage $U$ is \emph{closed} if no usage variable occurs
in $U$.  Also, we write $\mathcoaction{\alpha}$ for the
\emph{co-action} of $\alpha$, defined as $\mathcoaction{I}=O$ and
$\mathcoaction{O}=I$.  We define \emph{the type of ${\mathobtrue^{l}}$
  and ${\mathobfalse^{l}}$} as $\mathbooltypewithsec{l}$ and \emph{the
  type of ${\mathobunit}$} as $\mathunittype$.  We assume
$|$ for usages is left-associative as well. We
call $\mathbooltypewithsec{l}$ and ${\mathunittype}$ \emph{base
  types}.  A type $\tau$ is called a \emph{channel type} if it is not
a base type.  \emph{The secrecy level of a type $\tau$} is defined as
$l$ if $\tau$ is either $\mathbooltypewithsec{l}$ or of the form
$\mathprogramtypetuple{\mathvect{\tau}}^{l}/U$.

For each occurrence of $\alpha^{t_{o}}_{t_{c}}$ in a usage, $t_{o}$
and $t_{c}$ are respectively called the \emph{obligation} and
\emph{capability level annotation} of the occurrence.  Intuitively,
they mean that the input or output $\alpha$ must be performed---though
may not succeed---by the ``time'' (relative ordering in terms of natural numbers and $\infty$) specified by
$t_{o}$, and will succeed by time $t_{c}$ if---though need not
be---performed.

 \begin{definition}[Capability]
  For ${\alpha}\in{\mathsetextension{I, O}}$,
  $\mathcapability{\alpha}{U}$ is defined by:
  \begin{gather*}
   \mathcapability{\alpha}{\mathusagenil}=
   \mathcapability{\alpha}{\rho}=
   \mathcapability{\alpha}{\mathcoaction{\alpha}^{t_{o}}_{t_{c}}.U}=
   \infty \\
   \mathcapability{\alpha}{\alpha^{t_{o}}_{t_{c}}.U}=t_{c} \\
   \mathcapability{\alpha}{\mathpiparallel{U_{0}}{U_{1}}}=
   \mathcapability{\alpha}{\mathusageor{U_{0}}{U_{1}}}=
   \mathof{\min}{\mathcapability{\alpha}{U_{0}}, \mathcapability{\alpha}{U_{1}}} \\
   \mathcapability{\alpha}{\mathpireplication{U}}=
   \mathcapability{\alpha}{\mathlevelraise{t_{I}}{t_{O}}{U}}=
   \mathcapability{\alpha}{U}
  \end{gather*}

  We call $\mathcapability{I}{U}$ and $\mathcapability{O}{U}$,
  respectively, the \emph{input} and \emph{output capability level}
  of a usage $U$.
 \end{definition}

 \begin{definition}[Obligation]
  For ${\alpha}\in{\mathsetextension{I, O}}$,
  $\mathobligation{\alpha}{U}$ is defined as:
  \begin{gather*}
   \mathobligation{\alpha}{\mathusagenil}=
   \mathobligation{\alpha}{\rho}=
   \mathobligation{\alpha}{\mathcoaction{\alpha}^{t_{o}}_{t_{c}}.U}=
   \infty, \\
   \mathobligation{\alpha}{\alpha^{t_{o}}_{t_{c}}.U}= t_{o}, \\
   \mathobligation{\alpha}{\mathpiparallel{U_{0}}{U_{1}}}=
   \mathof{\min}{\mathobligation{\alpha}{U_{0}}, \mathobligation{\alpha}{U_{1}}}, \\
   \mathobligation{\alpha}{\mathpireplication{U}}=\mathobligation{\alpha}{U}, \\
   \mathobligation{\alpha}{\mathlevelraise{t_{I}}{t_{O}}{U}}=
   \mathof{\max}{t_{\alpha}, \mathobligation{\alpha}{U}},\text{ and } \\
   \mathobligation{\alpha}{\mathusageor{U_{0}}{U_{1}}}=
   \mathof{\max}{\mathobligation{\alpha}{U_{0}}, \mathobligation{\alpha}{U_{1}}}.
  \end{gather*}

  We call, respectively, $\mathobligation{I}{U}$ and $\mathobligation{O}{U}$
  the \emph{input} and \emph{output obligation level of a usage $U$}.
  We define $\mathobligation{}{U}=\mathof{\min}{\mathobligation{I}{U}, \mathobligation{O}{U}}$,
  which is used as $\mathobligation{}{U}=\infty$ to mean $U$ has no obligation,
  and write $\mathlevelraiseuni{U}$ for $\mathlevelraise{t_{I}+1}{t_{O}+1}{U}$
  where $t_{I}=\mathobligation{I}{U}$ and 
  $t_{O}=\mathobligation{O}{U}$.
 \end{definition}

 We then define reduction of usages (with structural preorder, like we
 did for processes), which is required for defining their reliability
 and for the statement of type preservation (subject reduction), as is
 usual for behavioral type systems, where static types capture dynamic
 behavior of processes.

 \begin{definition}[Structural preorder for usages]
  \label[definition]{def:usagestructuralpo}
  The binary relation $\mathusagestructuralpo$ on usages is 
  the least reflexive and transitive relation satisfying the following rules:
  \begin{description}
   \item[\rulename{UP-Zero}] 
   ${\mathpiparallel{\mathusagenil}{U}}\mathusagestructuralpo{U}$
   \item[\rulename{UP-Commut}] 
   ${\mathpiparallel{U_{0}}{U_{1}}}\mathusagestructuralpo{\mathpiparallel{U_{1}}{U_{0}}}$
   \item[\rulename{UP-Assoc}] 
   ${\mathpiparallel{\mleft(\mathpiparallel{U_{0}}{U_{1}}\mright)}{U_{2}}}\mathusagestructuralpo{\mathpiparallel{U_{0}}{\mleft(\mathpiparallel{U_{1}}{U_{2}}\mright)}}$
   \item[\rulename{UP-CongP}] 
   If ${U_{0}}\mathusagestructuralpo{U'_{0}}$,
   then ${\mathpiparallel{U_{0}}{U_{1}}}\mathusagestructuralpo{\mathpiparallel{U'_{0}}{U_{1}}}$.
   \item[\rulename{UP-Rep}] 
   ${\mathpireplication{U}}\mathusagestructuralpo{\mathpiparallel{\mathpireplication{U}}{U}}$
   \item[\rulename{UP-$\mathlevelraiseunifunc$}]
   ${\mathlevelraise{t_{I}}{t_{O}}{\alpha^{t_{o}}_{t_{c}}. U}}\mathusagestructuralpo{\alpha^{\mathof{\max}{t_{o}, t_{\alpha}}}_{t_{c}}. U}$
   \item[\rulename{UP-Dist}] ${{\mathlevelraise{t_{I}}{t_{O}}{(\mathpiparallel{U_{0}}{U_{1}})}}\mathusagestructuralpo{\mathpiparallel{\mathlevelraise{t_{I}}{t_{O}}{U_{0}}}{\mathlevelraise{t_{I}}{t_{O}}{U_{1}}}}}$
   \item[\rulename{UP-Or}] 
   ${\mathusageor{U_{0}}{U_{1}}}\mathusagestructuralpo{U_{i}}$ for $i\in \mathsetextension{0, 1}$
   \item[\rulename{UP-Cong$\mathlevelraiseunifunc$}]
   If ${U}\mathusagestructuralpo{U'}$, 
   then ${\mathlevelraise{t_{I}}{t_{O}}{U}}\mathusagestructuralpo{\mathlevelraise{t_{I}}{t_{O}}{U'}}$.
   \item[\rulename{UP-Commut$\mathlevelraiseunifunc$}]
   ${\mathlevelraise{t_{I}}{t_{O}}{\mathlevelraise{t'_{I}}{t'_{O}}{U}}}\mathusagestructuralpo{\mathlevelraise{t'_{I}}{t'_{O}}{\mathlevelraise{t_{I}}{t_{O}}{U}}}$~\footnote{This rule was not present in \cite{Kobayashi2005} but is added for a part of our soundness proof.}
  \end{description}
 \end{definition}

 \begin{definition}[Usage reduction]
  The binary relation $\mathusagereduction$ on usages, called \emph{usage reduction},
  is defined as the least relation satisfying the following rules:
  \begin{enumerate}
   \item ${\mathpiparallel{O^{t_{o}}_{t_{c}}.U_{0}}{I^{t'_{o}}_{t'_{c}}.U_{1}}}\mathusagereduction{\mathpiparallel{U_{0}}{U_{1}}}$
   \item If ${U_{0}}\mathusagereduction {U'_{0}}$, 
             then ${\mathpiparallel{U_{0}}{U_{1}}}\mathusagereduction {\mathpiparallel{U'_{0}}{U_{1}}}$.
   \item If ${U_{0}}\mathusagestructuralpo{U'_{0}}$, ${U'_{0}}\mathusagereduction {U'_{1}}$ and 
	 ${U'_{1}}\mathusagestructuralpo{U_{1}}$, 
	 then ${U_{0}}\mathusagereduction {U_{1}}$.
  \end{enumerate}
   We write $\mathusagereductionkc$ 
   for the reflexive and transitive closure of $\mathusagereduction$ for usages as well.
 \end{definition}

 \begin{definition}[Reliability]
  \label[definition]{def:reliability}
  We write $\mathsuitable{\alpha}{U}$ 
  if ${\mathobligation{\mathcoaction{\alpha}}{U}}\leq{\mathcapability{\alpha}{U}}$,
  and $\mathsuitable{}{U}$ 
  if both $\mathsuitable{I}{U}$ and $\mathsuitable{O}{U}$.
  Then we say that a usage $U$ is \emph{reliable}, written $\mathreliable{U}$,
  if $\mathsuitable{}{U'}$ for any $U'$ 
  such that ${U}\mathusagereductionkc{U'}$.
 \end{definition}

 Next, we define the following \emph{subusage} and subtyping relations.

 \renewcommand{\theenumi}{\textup{(}\alph{enumi}\textup{)}}
 \renewcommand{\labelenumi}{\textup{(}\alph{enumi}\textup{)}}
 \begin{definition}[Subusage] 
  \label[definition]{def:subusage}
  The \emph{subusage relation} $\mathissubusagesy$ on closed usages is the largest binary relation 
  such that, whenever $\mathissubusage{U_{0}}{U_{1}}$, the following conditions hold:
  \begin{enumerate}
   \item $\mathissubusage{\mathsubstbox{U}{\mathsubst{\rho}{U_{0}}}}{\mathsubstbox{U}{\mathsubst{\rho}{U_{1}}}}$
	 for any usage $U$ with $\mathFVof{U}=\mathsetextension{\rho}$.
	 \label{item:sub-in-def:subusage}
   \item If ${U_{1}}\mathusagereduction{U'_{1}}$,
	 then there exists $U'_{0}$ such that 
	 ${U_{0}}\mathusagereduction{U'_{0}}$ and $\mathissubusage{U'_{0}}{U'_{1}}$.
	 \label{item:reduction-in-def:subusage}
   \item ${\mathcapability{\alpha}{U_{0}}}\leq{\mathcapability{\alpha}{U_{1}}}$
	 for each ${\alpha}\in{\mathsetextension{I, O}}$.
	 \label{item:capability-in-def:subusage}
   \item For each ${\alpha}\in{\mathsetextension{I, O}}$, 
	 if $\mathsuitable{\mathcoaction{\alpha}}{U_{0}}$,
	 then ${\mathobligation{\alpha}{U_{0}}}\geq{\mathobligation{\alpha}{U_{1}}}$.
	 \label{item:obligation-in-def:subusage}
  \end{enumerate}
 \end{definition}
 \renewcommand{\theenumi}{\textup{(}\arabic{enumi}\textup{)}}
 \renewcommand{\labelenumi}{\textup{(}\arabic{enumi}\textup{)}}

 \begin{proposition} 
  \label[proposition]{prop:property-of-subusages-and-reliability}
  \begin{enumerate}
   \item If $\mathissubusage{U_{0}}{U_{1}}$ and $\mathsuitable{\alpha}{U_{0}}$,
	 then $\mathsuitable{\alpha}{U_{1}}$.
	 \label{item:subusage-and-suitability-in-prop:posar}
   \item If $\mathissubusage{U_{0}}{U_{1}}$ and $\mathreliable{U_{0}}$,
	 then $\mathreliable{U_{1}}$.
	 \label{item:subusage-and-reliability-in-prop:posar}
   \item The subusage relation is reflexive and transitive.
	 \label{item:subusage-refl-tans-in-prop:posar}
  \end{enumerate}
 \end{proposition}

 \begin{definition}[Subtyping]
  \label[definition]{def:subtyping}
  The \emph{subtyping relation} $\mathissubtypesy$ on types is 
  the least reflexive relation such that
  $\mathissubtype{\xi/U}{\xi/U'}$ if $\mathissubusage{U}{U'}$.
 \end{definition}
 \noindent
 Note that
 if $\tau$ is a base type, then
 $\mathissubtype{\tau}{\tau'}$ implies
 ${\tau'}={\tau}$.
 Note also that the subtyping relation is transitive, i.e.,
 $\mathissubtype{\tau}{\tau'}$ and $\mathissubtype{\tau'}{\tau''}$
 imply $\mathissubtype{\tau}{\tau''}$.

 \begin{definition}\memo{[Obligation level of a type]}
  The \emph{obligation level of a type $\tau$}, written $\mathobligation{\alpha}{\tau}$, 
  is defined by
  \[
  \mathobligation{\alpha}{\tau}=
  \begin{cases}
  \infty & \text{if $\tau$ is a base type} \\
  \mathobligation{\alpha}{U} & \text{if $\tau = \xi/U$}
  \end{cases}
  \]
  along with $\mathobligation{}{\tau}=\mathof{\min}{\mathobligation{I}{\tau}, \mathobligation{O}{\tau}}$.
 We also define $\mathlevelraise{t_{I}}{t_{O}}{\tau}$, $\mathlevelraiseuni{\tau}$, 
 $\mathpireplication{\tau}$, and $\mathpiparallel{\tau_{0}}{\tau_{1}}$ as:
 \begin{align*}
  \mathlevelraise{t_{I}}{t_{O}}{\tau}&=
  \begin{cases}
   \tau & \text{if $\tau$ is a base type} \\
   \xi/\mathlevelraise{t_{I}}{t_{O}}{U} & \text{if $\tau = \xi/U$}
  \end{cases} \\
  \mathlevelraiseuni{\tau}&=
  \begin{cases}
   \tau & \text{if $\tau$ is a base type} \\
   \xi/\mathlevelraiseuni{U} & \text{if $\tau = \xi/U$}
  \end{cases} \\
  \mathpireplication{\tau}&=
  \begin{cases}
   \tau & \text{if $\tau$ is a base type} \\
   \xi/\mathpireplication{U} & \text{if $\tau = \xi/U$}
  \end{cases} \\
  \mathpiparallel{\tau_{0}}{\tau_{1}} &=
  \begin{cases}
   \tau_{0} & \text{if $\tau_{0}=\tau_{1}$ and is a base type} \\
   \xi/\mleft(\mathpiparallel{U_{0}}{U_{1}}\mright) & \text{if $\tau_{0} = \xi/U_{0}$ and $\tau_{1} = \xi/U_{1}$} \\
   \text{undefined} & \text{otherwise}
  \end{cases}
 \end{align*}
 \end{definition}
 
 We assume $\mathpiparallel{}{}$ for types is also left-associative, that is,
 $\mathpiparallel{\mathpiparallel{\tau_{0}}{\tau_{1}}}{\tau_{2}}$ 
 stands for $\mathpiparallel{(\mathpiparallel{\tau_{0}}{\tau_{1}})}{\tau_{2}}$.

 \begin{definition}[Equivalence except usages]
  The relation ${\tau}\mathequivexceptusagessy{\tau'}$ on types
  is the least equivalence relation satisfying
  ${\mathprogramtypetuple{\mathvect{\tau}}^{l}/U}\mathequivexceptusagessy{\mathprogramtypetuple{\mathvect{\tau}}^{l}/U'}$
  for ${l}\in{\maththesetofsecrecylevels}$.
 \end{definition}

 Then $\mathpiparallel{\tau_{0}}{\tau_{1}}$ is defined
 if and only if ${\tau_{0}}\mathequivexceptusagessy{\tau_{1}}$.
 Note also that
 $\mathissubtype{\tau_{0}}{\tau_{1}}$ implies
 ${\tau_{0}}\mathequivexceptusagessy{\tau_{1}}$.

\subsection{Type Environments} 
Our \emph{type environments} $\Gamma$ and $\Delta$ are defined as
functions from a finite set of values $v$ (consisting of channel names
$x$ and constants $c$) to types $\tau$, with constant values mapped to
their respective types.
We write $\emptyset$ for the empty type environment.
For a value ${v}\notin{\mathof{\mathdom}{\Gamma}}$, we write $\Gamma,
\mathistype{v}{\tau}$ for type environment $\Gamma'$ such that
$\mathof{\mathdom}{\Gamma'}=\mathof{\mathdom}{\Gamma}\cup
\mathsetextension{v}$, $\mathof{\Gamma'}{v}=\tau$, and
$\mathof{\Gamma'}{y}=\mathof{\Gamma}{y}$ for
$y\in\mathof{\mathdom}{\Gamma}$.  We write $\mathistype{v}{\tau}$ for
the type environment $\Gamma$ with $\mathof{\Gamma}{v}={\tau}$ and
$\mathof{\mathdom}{\Gamma}={\mathsetextension{v}}$.

  We extend the subtyping relation to a relation on type environments.

  \renewcommand{\theenumi}{\textup{(}\alph{enumi}\textup{)}}
  \renewcommand{\labelenumi}{\textup{(}\alph{enumi}\textup{)}}
  \begin{definition}[Subtyping relation on type environments]
   \label[definition]{def:subtyping-on-env}
   For type environments $\Gamma$ and $\Delta$,
   we write $\mathissubtype{\Gamma}{\Delta}$ if
   \begin{enumerate}
    \item ${\mathof{\mathdom}{\Gamma}}\supseteq{\mathof{\mathdom}{\Delta}}$,
	  \label{item:dom-def-subtyping-on-env}
    \item $\mathissubtype{\mathof{\Gamma}{x}}{\mathof{\Delta}{x}}$ 
	  for each ${x}\in{\mathof{\mathdom}{\Delta}}$, and
	  \label{item:subtyping-def-subtyping-on-env}
    \item $\mathobligation{}{\mathof{\Gamma}{x}}=\infty$
	  for each 
	  ${x}\in{{\mathof{\mathdom}{\Gamma}}\setminus{\mathof{\mathdom}{\Delta}}}$.
	  \label{item:obligation-def-subtyping-on-env}
   \end{enumerate}
  \end{definition}
  \renewcommand{\theenumi}{\textup{(}\arabic{enumi}\textup{)}}
  \renewcommand{\labelenumi}{\textup{(}\arabic{enumi}\textup{)}}

  We say that a type environment $\Gamma$ is \emph{closed}~\cite[p.~316]{Kobayashi2005}\cite[p.~238,~Definition~6.1.2]{DBLP:books/daglib/0004377}
  if $\mathof{\Gamma}{x}$ is a channel type for each channel name
  ${x}\in{\mathof{\mathdom}{\Gamma}}$.
  We also say that $\Gamma$ is \emph{reliable}, written $\mathreliable{\Gamma}$,
  if, for any ${x}\in{\mathof{\mathdom}{\Gamma}}$,
  $\mathof{\Gamma}{x}$ is a channel type ${\xi}/{U}$ with
  $\mathreliable{U}$.
  Note that the subtyping relation on type environments is also transitive, i.e.,
  if $\mathissubtype{\Gamma}{\Gamma'}$ and $\mathissubtype{\Gamma'}{\Gamma''}$,
  then $\mathissubtype{\Gamma}{\Gamma''}$.
  
  For type environments $\Gamma$ and $\Delta$,
  we define :
  \[
  \mathof{\mleft(\mathpiparallel{\Gamma}{\Delta}\mright)}{x} =
  \begin{cases}
   \mathpiparallel{\mathof{\Gamma}{x}}{\mathof{\Delta}{x}} & \text{if $x\in{{\mathof{\mathdom}{\Gamma}}\cap{\mathof{\mathdom}{\Delta}}}$} \\
   \mathof{\Gamma}{x} & \text{if $x\in{{\mathof{\mathdom}{\Gamma}}\setminus{\mathof{\mathdom}{\Delta}}}$} \\
   \mathof{\Delta}{x} & \text{if $x\in{{\mathof{\mathdom}{\Delta}}\setminus{\mathof{\mathdom}{\Gamma}}}$}
  \end{cases}
  \]
  For a type environment $\Gamma$,
  we write $\mathpireplication{\Gamma}$, $\mathlevelraise{t_{I}}{t_{O}}{\Gamma}$, and
  $\mathlevelraiseuni{\Gamma}$ for the type environments satisfying
  $\mathof{(\mathpireplication{\Gamma})}{x}=\mathpireplication{(\mathof{\Gamma}{x})}$,
  ${\mathof{(\mathlevelraise{t_{I}}{t_{O}}{\Gamma})}{x}}={\mathlevelraise{t_{I}}{t_{O}}{(\mathof{\Gamma}{x})}}$,
  and ${\mathof{(\mathlevelraiseuni{\Gamma})}{x}}={\mathlevelraiseuni{(\mathof{\Gamma}{x})}}$,
  respectively.
  Note
  ${\mathof{\mathdom}{\mathpiparallel{\Gamma}{\Delta}}}={{\mathof{\mathdom}{\Gamma}}\cup{\mathof{\mathdom}{\Delta}}}$
  and
  ${\mathof{\mathdom}{\mathpireplication{\Gamma}}}=\mathof{\mathdom}{\mathlevelraise{t_{I}}{t_{O}}{\Gamma}}=\mathof{\mathdom}{\mathlevelraiseuni{\Gamma}}={\mathof{\mathdom}{\Gamma}}$.
  Again, we assume $|$ is left-associative, and abbreviate $\mathpiparallel{\mathistype{v_{1}}{\tau_{1}}}{\mathpiparallel{{}\cdots}{\mathistype{v_{n}}{\tau_{n}}}}$
  as $\mathistype{\mathvect{v}}{\mathvect{\tau}}$.
   Furthermore,
   we write ${\Gamma}\mathequivexceptusagessy{\Gamma'}$ 
   if ${\mathof{\mathdom}{\Gamma}}={\mathof{\mathdom}{\Gamma'}}$ and 
   ${\mathof{\Gamma}{x}}\mathequivexceptusagessy{\mathof{\Gamma'}{x}}$
   for every ${x}\in{\mathof{\mathdom}{\Gamma}}$.

   We now start to extend the type system with general lattices.  The
   first definition below generalizes well-formed channel
   types~\cite[Definition~14]{Kobayashi2005}.  Informally, it means
   the channel type is well-formed when the level of the ``attacker''
   is $l$ (which was just $\bot$, written $\mathbf{L}$ in
   \cite{Kobayashi2005}, in the 2-element lattice).

   \begin{definition}[$l$-secure channel type]
    \label[definition]{def:secure-channel-type}
    We say that a channel type ${\mathprogramtypetuple{\mathvect{\tau}}^{l'}/U}$ is \emph{$l$-secure in $L$}
    when
    \begin{enumerate}
     \item if
	   $l' \leq_{L} l$, then
	   all the capability level annotations in $U$ are $\infty$, and
     \item $l' \leq_{L} l''$
	   for any secrecy level $l''$ occurring in $\mathvect{\tau}$.
    \end{enumerate}
   \end{definition}

   We write $\mathtypeenvandsecrecylatice{\Gamma}{L}$ 
   for the pair $\mathtuple{\Gamma, L}$ and call it an \emph{environment}.
   
   \begin{definition}[$l$-secure environment]
    \label[definition]{def:secure-env}
    For a type environment $\Gamma$ and a lattice of secrecy levels $L$,
    we say that $\mathtypeenvandsecrecylatice{\Gamma}{L}$ is \emph{$l$-secure} 
    if every channel type in the range of $\Gamma$ is $l$-secure in $L$.
   \end{definition}

\subsection{Typing Rules}

 \begin{figure*}[t]
    \centering
    \begin{tabular}{cc}
     \begin{inlineprooftree}
      \AxiomC{$\lm\in L$}
      \RightLabel{\rulename{T-Zero}}
      \UnaryInfC{$\mathtypejudgementwithsecrecy{\emptyset}{L}{\lm}{\mathnil}$}
     \end{inlineprooftree}
     & 
     \begin{inlineprooftree}
      \AxiomC{$\mathtypejudgementwithsecrecy{\Gamma, {\mathistype{x}{\xi/U}}}{L}{\lm}{P}$}
      \AxiomC{$\mathreliable{U}$}
      \RightLabel{\rulename{T-New}}
      \BinaryInfC{$\mathtypejudgementwithsecrecy{\Gamma}{L}{\lm}{\mathpinew{\mathistype{x}{\xi}}{P}}$}
     \end{inlineprooftree}
	 \\
     \begin{inlineprooftree}
      \AxiomC{$\mathtypejudgementwithsecrecy{\Gamma}{L}{\lm}{P}$}
      \RightLabel{\rulename{T-Rep}}
      \UnaryInfC{$\mathtypejudgementwithsecrecy{\mathpireplication{\Gamma}}{L}{\lm}{\mathpireplication{P}}$}
     \end{inlineprooftree}
     & 
     \begin{inlineprooftree}
      \AxiomC{$\mathtypejudgementwithsecrecy{\Gamma_{0}}{L}{\lm}{P_{0}}$}
      \AxiomC{$\mathtypejudgementwithsecrecy{\Gamma_{1}}{L}{\lm}{P_{1}}$}
      \RightLabel{\rulename{T-Par}}
      \BinaryInfC{$\mathtypejudgementwithsecrecy{\mathpiparallel{\Gamma_{0}}{\Gamma_{1}}}{L}{\lm}{\mathpiparallel{P_{0}}{P_{1}}}$}
     \end{inlineprooftree}
    \end{tabular}

    \begin{inlineprooftree}
     \AxiomC{$\mathtypejudgementwithsecrecy{\Gamma}{L}{\lm}{P}$}
     \AxiomC{$\mathtypejudgementwithsecrecy{\Gamma}{L}{\lm}{Q}$}
     \RightLabel{\rulename{T-If}}
     \BinaryInfC{$\mathtypejudgementwithsecrecy{\mathpiparallel{\Gamma}{\mathistype{v}{\mathbooltypewithsec{\lm}}}}{L}{\lm}{\mathobif{v}{P}{Q}}$}
    \end{inlineprooftree}

    \begin{inlineprooftree} 
      \AxiomC{$\mathtypejudgementwithsecrecy{\Gamma, {\mathistype{x}{\mathprogramtypetuple{\mathvect{\tau}}^{l_{0}}/U}}}{L}{l_{1}}{P}$}
      \AxiomC{${\lm}\leq_{L}{l_{0}}$}
      \AxiomC{${\lm}\leq_{L}{l_{1}}$}
      \AxiomC{$t_{c} = \infty \Longrightarrow {l_{0}}\leq_{L}{l_{1}}$}
      \RightLabel{\rulename{T-Out}}
      \QuaternaryInfC{$\mathtypejudgementwithsecrecy{\mathpiparallel{\mathlevelraise{t_{c}+1}{t_{c}+1}{\mleft(\mathpiparallel{\Gamma}{\mathistype{\mathvect{v}}{\mathlevelraiseuni{\mathvect{\tau}}}}\mright)}}{\mathistype{x}{\mathprogramtypetuple{\mathvect{\tau}}^{l_{0}}/O^{0}_{t_{c}}.U}}}{L}{\lm}{\mathpioutput{x}{\mathvect{v}}.P}$}
    \end{inlineprooftree}

    \begin{inlineprooftree} 
      \AxiomC{$\mathtypejudgementwithsecrecy{\Gamma, {\mathistype{x}{\mathprogramtypetuple{\mathvect{\tau}}^{l_{0}}/U}}, {\mathistype{\mathvect{y}}{\mathvect{\tau}}}}{L}{l_{1}}{P}$}
      \AxiomC{${\lm}\leq_{L}{l_{0}}$}
      \AxiomC{${\lm}\leq_{L}{l_{1}}$}
      \AxiomC{$t_{c} = \infty \Longrightarrow {l_{0}}\leq_{L}{l_{1}}$}
      \RightLabel{\rulename{T-In}}
      \QuaternaryInfC{$\mathtypejudgementwithsecrecy{\mathlevelraise{t_{c}+1}{t_{c}+1}{(\Gamma)}, \mathistype{x}{\mathprogramtypetuple{\mathvect{\tau}}^{l_{0}}/I^{0}_{t_{c}}.U}}{L}{\lm}{\mathpiinput{x}{\mathvect{y}}.P}$}
    \end{inlineprooftree}

     \begin{inlineprooftree}
      \AxiomC{$\mathtypejudgementwithsecrecy{\Gamma}{\mathpinewsecrecylevel{l_{0}}{\mathvect{l_{1}}}{\mathvect{l_{2}}}{L}}{\lm}{P}$}
      \AxiomC{$\mathissublattice{L}{\mathpinewsecrecylevel{l_{0}}{\mathvect{l_{1}}}{\mathvect{l_{2}}}{L}}$}
      \AxiomC{$\lm \leq_{L} \mathvect{l_{1}}, \mathvect{l_{2}}$}
      \RightLabel{\rulename{T-NewSec}}
      \TrinaryInfC{$\mathtypejudgementwithsecrecy{\Gamma}{L}{\lm}{\mathpinewsecrecylevel{l_{0}}{\mathvect{l_{1}}}{\mathvect{l_{2}}}{P}}$}
     \end{inlineprooftree}

     \begin{inlineprooftree} 
      \AxiomC{$\mathtypejudgementwithsecrecy{\Gamma'}{L}{\lm'}{P}$}
      \AxiomC{$\mathissubtype{\Gamma}{\Gamma'}$}

      \AxiomC{$\lm\leq_{L}\lm'$}
      \RightLabel{\rulename{T-Weak}}
      \TrinaryInfC{$\mathtypejudgementwithsecrecy{\Gamma}{L}{\lm}{P}$}
     \end{inlineprooftree}

    \caption{Typing rules}
    \label{fig:typing-rules}
 \end{figure*}

   Our \emph{type judgement}
   $\mathtypejudgementwithsecrecy{\Gamma}{L}{\lm}{P}$ is a tuple
   $\mathtuple{\mathtypeenvandsecrecylatice{\Gamma}{L}, \lm, P}$
   of an environment $\mathtypeenvandsecrecylatice{\Gamma}{L}$,
   a secrecy level ${\lm}\in {L}$, and a process $P$, where
   every secrecy level occurring in $P$ and $\Gamma$ is in $L$.
   Intuitively, it means that the process $P$ is secure (i.e., does
   not leak information about high-secrecy values to low-level
   ``attackers'' or contexts) under the type environment $\Gamma$ and
   secrecy lattice $L$, where $\lm$ is a lower bound of the levels that
   $P$ may interact with (as in \cite{Kobayashi2005}). 

   The typing rules other than \rulename{T-NewSec} are similar to
   previous ones \cite{Kobayashi2005} except that they are all
   parameterized by the general lattice of secrecy levels $L$.
   The second premise $\mathissublattice{L}{\mathpinewsecrecylevel{l_{0}}{\mathvect{l_{1}}}{\mathvect{l_{2}}}{L}}$ in \rulename{T-NewSec} ensures safe
   extension thanks to our previous definitions on lattices (\cref{def:sublattice} and \cref{def:newlev-lattice}).
   The ``side'' condition $\lm \leq_{L} \mathvect{l_{1}}, \mathvect{l_{2}}$, which is more subtle, guarantees that high-secrecy operations can be erased in the proof of non-interference (the third last case in \cref{def:Er}).

   We say a type judgement $\mathtypejudgementwithsecrecy{\Gamma}{L}{\lm}{P}$
   is \emph{$l$-secure} 
   if $\mathtypeenvandsecrecylatice{\Gamma}{L}$ is $l$-secure.
   An \emph{$l$-secure derivation tree} is defined as a tree of $l$-secure type judgements
   constructed by instances of the rules in \cref{fig:typing-rules}.
   An $l$-secure derivation tree with root $\mathtypejudgementwithsecrecy{\Gamma}{L}{\lm}{P}$
   is called an \emph{$l$-secure derivation tree of 
   $\mathtypejudgementwithsecrecy{\Gamma}{L}{\lm}{P}$}.
   We say that $\mathtypejudgementwithsecrecy{\Gamma}{L}{\lm}{P}$ is
   \emph{$l$-securely derivable
   from $\mathtypejudgementwithsecrecy{\Delta}{L'}{\lm'}{P'}$}
   if there exists an $l$-secure derivation tree of 
   $\mathtypejudgementwithsecrecy{\Gamma}{L}{\lm}{P}$
   whose leaves are either $\mathtypejudgementwithsecrecy{\Delta}{L'}{\lm'}{P'}$
   or constructed by \rulename{T-Zero}.
   We also say that $\mathtypejudgementwithsecrecy{\Gamma}{L}{\lm}{P}$ is
   \emph{$l$-securely derivable}
   if there exists a derivation tree of
   $\mathtypejudgementwithsecrecy{\Gamma}{L}{\lm}{P}$
   whose leaves are constructed by \rulename{T-Zero},
   and that $\mathtypejudgementwithsecrecy{\Gamma}{L}{\lm}{P}$ is
   \emph{derivable} (from $\mathtypejudgementwithsecrecy{\Delta}{L'}{\lm'}{P'}$)
   if it is $l$-securely derivable
   (from $\mathtypejudgementwithsecrecy{\Delta}{L'}{\lm'}{P'}$) for some $l$.

   All the careful definitions above are needed for soundness and
   related proofs of our type system with a generalized (let alone
   extensible) security lattice (that is, lattice of secrecy levels).

   \begin{figure*}[t]
    \centering
    \begin{inlineprooftree}
     \AxiomC{}
     \UnaryInfC{$\mathtypejudgementwithsecrecy{\emptyset}{L'}{\bot}{\mathnil}$}
     \UnaryInfC{$\mathtypejudgementwithsecrecy{\mathistype{x}{\mathprogramtypetuple{\mathbooltypewithsec{l}}^{l}/{\mathusagenil}}, \mathistype{y}{\mathprogramtypetuple{}^{\bot}/{\mathusagenil}}}{L'}{\bot}{\mathnil}$}
     \UnaryInfC{$\mathtypejudgementwithsecrecy{\mathistype{x}{\mathprogramtypetuple{\mathbooltypewithsec{l}}^{l}/{\mathusagenil}}, \mathistype{y}{\mathprogramtypetuple{}^{\bot}/{I^{0}_{\infty}}}}{L'}{\bot}{\mathpiinput{y}{}}$}
     \UnaryInfC{$\mathtypejudgementwithsecrecy{\mathistype{x}{\mathprogramtypetuple{\mathbooltypewithsec{l}}^{l}/{O^{0}_{0}}}, \mathistype{y}{\mathprogramtypetuple{}^{\bot}/{\mathlevelraise{1}{1}{I^{0}_{\infty}}}}}{L'}{\bot}{\mathpioutput{x}{\mathobtrue^{l}}.\mathpiinput{y}{}}$}

     \AxiomC{}
     \UnaryInfC{$\mathtypejudgementwithsecrecy{\emptyset}{L'}{\bot}{\mathnil}$}
     \UnaryInfC{$\mathtypejudgementwithsecrecy{\mathistype{x}{\mathprogramtypetuple{\mathbooltypewithsec{l}}^{l}/{\mathusagenil}}, \mathistype{y}{\mathprogramtypetuple{}^{\bot}/\mathusagenil}, \mathistype{b}{\mathbooltypewithsec{l}}}{L'}{\bot}{\mathnil}$}
     \UnaryInfC{$\mathtypejudgementwithsecrecy{\mathistype{x}{\mathprogramtypetuple{\mathbooltypewithsec{l}}^{l}/{I^{0}_{0}}}, \mathistype{y}{\mathprogramtypetuple{}^{\bot}/\mathusagenil}}{L'}{\bot}{\mathpiinput{x}{b}}$}

     \BinaryInfC{$\mathtypejudgementwithsecrecy{\mathistype{x}{\mathprogramtypetuple{\mathbooltypewithsec{l}}^{l}/{\mathpiparallel{O^{0}_{0}}{I^{0}_{0}}}}, \mathistype{y}{\mathprogramtypetuple{}^{\bot}/{\mathpiparallel{\mathlevelraise{1}{1}{I^{0}_{\infty}}}{\mathusagenil}}}}{L'}{\bot}{\mathpiparallel{\mathpioutput{x}{\mathobtrue^{l}}.\mathpiinput{y}{}}{\mathpiinput{x}{b}}}$}
     \UnaryInfC{$\mathtypejudgementwithsecrecy{\mathistype{y}{\mathprogramtypetuple{}^{\bot}/{\mathpiparallel{\mathlevelraise{1}{1}{I^{0}_{\infty}}}{\mathusagenil}}}}{L'}{\bot}{\mathpinew{\mathistype{x}{\mathprogramtypetuple{\mathbooltypewithsec{l}}^{l}}}{(\mathpiparallel{\mathpioutput{x}{\mathobtrue^{l}}.\mathpiinput{y}{}}{\mathpiinput{x}{b}})}}$}
     \UnaryInfC{$\mathtypejudgementwithsecrecy{\mathistype{y}{\mathprogramtypetuple{}^{\bot}/{\mathpiparallel{\mathlevelraise{1}{1}{I^{0}_{\infty}}}{\mathusagenil}}}}{L}{\bot}{\mathpinewsecrecylevel{l}{\bot}{\top}{\mathpinew{\mathistype{x}{\mathprogramtypetuple{\mathbooltypewithsec{l}}^{l}}}{(\mathpiparallel{\mathpioutput{x}{\mathobtrue^{l}}.\mathpiinput{y}{}}{\mathpiinput{x}{b}})}}}$}
    \end{inlineprooftree}
    \caption{An example of typing (rule names and trailing $\mathnil$ are omitted), where $L'=\mathpinewsecrecylevel{l}{(\bot)}{(\top)}{L}$}
    \label{fig:typing-example}
   \end{figure*}

   \begin{example}
     \cref{fig:typing-example} shows a $\top$-secure derivation tree for
     $\mathtypejudgementwithsecrecy{\Gamma}{L}{\bot}{P}$ where the process $P$ is
     \[\mathpinewsecrecylevel{l}{\top}{\bot}{\mathpinew{\mathistype{x}{\mathprogramtypetuple{\mathistype{v}{\mathbooltypewithsec{l}}}^{l}}}{(\mathpiparallel{\mathpioutput{x}{\mathobtrue^{l}}.\mathpiinput{y}{}.\mathnil}{\mathpiinput{x}{b}. \mathnil})}},\]
     $\Gamma =
     \mathistype{y}{\mathprogramtypetuple{}^{\bot}/{\mathpiparallel{\mathlevelraise{1}{1}{(I^{0}_{\infty}.\mathusagenil)}}{\mathusagenil}}}$,
     and $L=\mathsetextension{\bot, \top}$.  Intuitively,
     $P$ creates a new secrecy level $l$ and internally communicates a
     Boolean value $\mathobtrue^{l}$ of that level.  Non-interferece
     means that replacing it with $\mathobfalse^{l}$ makes no
     difference to low-level observers.
   \end{example}

   The following proposition allows a weakening, namely, extension of the lattice $L$ in an environment $\mathtypeenvandsecrecylatice{\Gamma}{L}$.
\begin{proposition}
 \label[proposition]{lemma:weakening-lattice}
 If $\mathtypejudgementwithsecrecy{\Gamma}{L}{\lm}{P}$ 
 is $l$-securely derivable
 (resp.~from $\mathtypejudgementwithsecrecy{\Delta}{L'}{\lm'}{P'}$),
 then\linebreak[4] $\mathtypejudgementwithsecrecy{\Gamma}{\mathpinewsecrecylevel{l'}{\mathvect{l_{0}}}{\mathvect{l_{1}}}{L}}{\lm}{P}$ is also $l$-securely derivable
 (resp.~from $\mathtypejudgementwithsecrecy{\Delta}{\mathpinewsecrecylevel{l'}{\mathvect{l_{0}}}{\mathvect{l_{1}}}{L'}}{\lm'}{P'}$), 
 where $l'$ is fresh.
\end{proposition}

\section{Soundness}
\label{sec:soundness}
In this section, we will prove our main theorem: non-interference.
To that goal, we first show two important properties of well-typed processes:
subject reduction (type preservation) and lock-freedom.

\subsection{Subject reduction}
We define ${\Gamma}\mathtypeenvreduction{\Gamma'}$
as ${\Gamma}={(\Gamma_{0}, \mathistype{x}{{\xi}/{U}})}$
for some ${x}$ with ${U}\mathusagereduction{U'}$ and
${\Gamma'}={(\Gamma_{0}, \mathistype{x}{{\xi}/{U'}})}$.
Again, we write $\mathtypeenvreductionkc$ 
for the reflexive and transitive closure of $\mathtypeenvreduction$.

\begin{lemma}
 If $\mathtypeenvandsecrecylatice{\Gamma}{L}$ is $l$-secure and 
 ${\Gamma}\mathtypeenvreduction{\Gamma'}$,
 then $\mathtypeenvandsecrecylatice{\Gamma'}{L}$ is $l$-secure.
\end{lemma}

\begin{lemma}[Structural preorder preserves typing] \label[lemma]{lemma:spo-preservation}
 If $\mathtypejudgementwithsecrecy{\Gamma}{L}{\lm}{P}$
 is $l$-securely derivable and 
 ${P}\mathpistructuralpo{P'}$,
 then $\mathtypejudgementwithsecrecy{\Gamma}{L}{\lm}{P'}$ 
 is also $l$-securely derivable.
\end{lemma}

\begin{proof}
 See \cref{sec:proof-of-lemma-spo-preservation}.
\end{proof}

\begin{figure*}[t]
  \[
  \mathof{\mathsubstbox{\Gamma}{\mathsubst{\mathvect{x}}{\mathvect{v}}}}{w}=
  \begin{cases}
   \mathof{\Gamma}{w} & \text{if 
   ${w}\notin{\mathvect{v}}$ and ${w}\notin{\mathvect{x}}$} \medskip \\  
   \mathpiparallel{\mathpiparallel{\mathof{\Gamma}{x_{j_{0}}}}{\cdots}}{\mathof{\Gamma}{x_{j_{k}}}} & 
   \begin{aligned}

    &\text{if } {w}\in{\mathvect{v}}\text{ and }{w}\notin{\mathof{\mathdom}{\Gamma}} \text{ with} \\
    &{\mathsetintension{{x_{i}}\in{\mathvect{x}}}{{w}={v_{i}} \text{ and } {x_{i}}\in{\mathof{\mathdom}{\Gamma}}}}={}\\
    &{\mathsetextension{x_{j_{0}}, \dots, x_{j_{k}}}}
    \text{ for } {0} \leq {j_{0}} < \dots < {j_{k}} \leq {n}
   \end{aligned}
   \medskip \\ 
   \mathpiparallel{\mathpiparallel{\mathpiparallel{\mathof{\Gamma}{x_{j_{0}}}}{\cdots}}{\mathof{\Gamma}{x_{j_{k}}}}}{\mathof{\Gamma}{w}} & 
   \begin{aligned}
    &\text{if }{w}\in{\mathvect{v}}\text{ and }{w}\in{\mathof{\mathdom}{\Gamma}} \text{ with} \\
    &{\mathsetintension{{x_{i}}\in{\mathvect{x}}}{{w}={v_{i}} \text{ and } {x_{i}}\in{\mathof{\mathdom}{\Gamma}}}}={}\\
    &{\mathsetextension{x_{j_{0}}, \dots, x_{j_{k}}}}
    \text{ for } {0} \leq {j_{0}} < \dots < {j_{k}} \leq {n}
   \end{aligned}
  \end{cases}
  \]

 \caption{Substitution on type environment}
 \label{fig:Substitution-in-type-environment}
\end{figure*}

\begin{definition}[Substitution on type environment]
 For a type environment $\Gamma$, variables $\mathvect{x}$, and values
 $\mathvect{v}$ with $\{\mathvect{x}\}\cap\{\mathvect{v}\}=\emptyset$,
 we define a type environment
 $\mathsubstbox{\Gamma}{\mathsubst{\mathvect{x}}{\mathvect{v}}}$ as in
 \cref{fig:Substitution-in-type-environment}.
 Furthermore, we say that
 \emph{$\mathsubstbox{\Gamma}{\mathsubst{\mathvect{x}}{\mathvect{v}}}$ is well-defined}
 if 
 ${\mathof{\mathdom}{\mathsubstbox{\Gamma}{\mathsubst{\mathvect{x}}{\mathvect{v}}}}}=
 {\mleft({\mathof{\mathdom}{\Gamma}}\setminus{\mathsetextension{\mathvect{x}}}\mright)}\cup
 {\mathsetintension{v_{i}}{{x_{i}}\in{\mathof{\mathdom}{\Gamma}}}}$.
\end{definition}

  \begin{lemma}[Substitution lemma] \label[lemma]{lemma:substitution-lemma}
   If $\mathtypejudgementwithsecrecy{\Gamma}{L}{\lm}{P}$ is
   $l$-securely derivable
   and ${\mathsubstbox{\Gamma}{\mathsubst{\mathvect{x}}{\mathvect{v}}}}$ is well-defined,
   then
   $\mathtypejudgementwithsecrecy{\mathsubstbox{\Gamma}{\mathsubst{\mathvect{x}}{\mathvect{v}}}}{L}{\lm}{\mathsubstbox{P}{\mathsubst{\mathvect{x}}{\mathvect{v}}}}$
   is also $l$-securely derivable.
  \end{lemma}

\begin{proof}
 See \cref{sec:proof-of-substitution-lemma}.
\end{proof}

\begin{proposition}[Subject reduction]
 \label[proposition]{prop:subject-reduction}
 If $\mathtypejudgementwithsecrecy{\Gamma}{L}{\lm}{P}$
 is $l$-securely derivable and
 ${\mathtuple{P, L}}\mathpireduction{\mathtuple{P', L'}}$,
 then there exists $\Gamma'$ such that 
 either ${\Gamma}={\Gamma'}$ or ${\Gamma}\mathtypeenvreduction{\Gamma'}$,
 and that
 $\mathtypejudgementwithsecrecy{\Gamma'}{L'}{\lm}{P'}$
 is $l$-securely derivable.
\end{proposition}

\begin{proof}
  By induction on the derivation of
  ${\mathtuple{P, L}}\mathpireduction{\mathtuple{P', L'}}$.
  See \cref{sec:proof-of-subject-reduction} for details.
\end{proof}
Note that the $L'$ above is always a superlattice of $L$---that is,
$L$ is a sublattice of $L'$---as can be proved by simple induction
according to \cref{def:reduction}.

\subsection{Lock-freedom}
Next, we prove the lock-freedom property, which, in short, guarantees
that every input or output with finite (namely, not $\infty$)
capability will eventually succeed and, as a result, such high-level
synchronization can securely be performed even before low-level
communication.\footnote{Although we do not consider timing in the
  present paper, it can also be
  incorporated~\cite{DBLP:conf/ifipTCS/Kobayashi00}.}

  \begin{definition}[Strong barbs]
   \label[definition]{def:Strong-barbs}
   Let $P$ be a process.
   We define $\maththestrongbarbsof{O}{P}$, $\maththestrongbarbsof{I}{P}$, and
   $\maththestrongbarbsof{}{P}$ as:
   \begin{align*}
    {\maththestrongbarbsof{O}{P}}&={\mathsetintension{x}{
    \ifdraft\else\begin{gathered}\fi
     {P}\mathpistructuralpo{\mathpinew{\mathvect{y}}{(\mathpiparallel{\mathpioutput{x}{\mathvect{v}}. P_{0}}{P_{1}})}} \text{ with } \ifdraft\else\\\fi
     {{x}\notin{\mathvect{y}}} \text{ for some } P_{0}, P_{1} 
    \ifdraft\else\end{gathered}\fi
    }
    } \\
    {\maththestrongbarbsof{I}{P}}&={\mathsetintension{x}{
    \ifdraft\else\begin{gathered}\fi
     {P}\mathpistructuralpo{\mathpinew{\mathvect{y}}{(\mathpiparallel{\mathpiinput{x}{\mathvect{z}}. P_{0}}{P_{1}})}} \text{ with } \ifdraft\else\\\fi
     {{x}\notin{\mathvect{y}}} \text{ for some } P_{0}, P_{1}
    \ifdraft\else\end{gathered}\fi
    }} \\
    {\maththestrongbarbsof{}{P}}&={\maththestrongbarbsof{O}{P}\cup\maththestrongbarbsof{I}{P}}
   \end{align*}
  \end{definition}

  The following definition specifies reduction by internal communication on a secret channel (of level $l'$) that the attacker (of level $l$) cannot access ($l'\not\leq_{L} l$).
  \begin{definition}[Secret reduction]
   \label[definition]{def:reduction-under-Gamma}
   For a type environment $\Gamma$,
   the binary relation $\mathpisecreduction{\Gamma}{l}$  is defined as
   the least relation satisfying the following rules:
   \begin{enumerate}
    \item ${\mathtuple{{\mathpiparallel{\mathpioutput{x}{(v_{0}, \dots, v_{n})}. P_{0}}{\mathpiinput{x}{(y_{0}, \dots, y_{n})}.P_{1}}}, L}}\mathpisecreduction{\Gamma}{l}{\mathtuple{{\mathpiparallel{P_{0}}{\mathsubstbox{P_{1}}{\mathsubst{y_{0}}{v_{0}}, \dots, \mathsubst{y_{n}}{v_{n}}}}}, L}}$,
	  where the secrecy level of $\mathof{\Gamma}{x}$ is $l'$ with $l'\not\leq_{L} l$
	    \label{item:r-com-def-reduction-under-Gamma}
  \item If ${\mathvect{l_{0}}},{\mathvect{l_{1}}}\subseteq{L}$ 
	and 
	${\mathpinewsecrecylevel{l}{\mathvect{l_{0}}}{\mathvect{l_{1}}}{L}}$ 
	is defined,
	then \linebreak[3]
	${\mathtuple{{\mathpinewsecrecylevel{l}{\mathvect{l_{0}}}{\mathvect{l_{1}}}{P}}, L}}\mathpisecreduction{\Gamma}{l}{\mathtuple{P, {\mathpinewsecrecylevel{l}{\mathvect{l_{0}}}{\mathvect{l_{1}}}{L}}}}$.
	\label{item:r-newlev-reduction-under-Gamma}
  \item If ${\mathtuple{{P_{0}}, L}}\mathpisecreduction{\Gamma}{l}{\mathtuple{{P'_{0}}, L'}}$, 
	       then ${\mathtuple{{\mathpiparallel{P_{0}}{P_{1}}}, L}}\mathpisecreduction{\Gamma}{l}{\mathtuple{{\mathpiparallel{P'_{0}}{P_{1}}}, L'}}$,
\label{item:r-par-reduction-under-Gamma}
  \item If ${\mathtuple{{P}, L}}\mathpisecreduction{\Gamma, \mathistype{x}{\xi/U}}{l}{\mathtuple{{P'}, L'}}$,
	       then ${\mathtuple{{\mathpinew{\mathistype{x}{\xi}}{P}}, L}}\mathpisecreduction{\Gamma}{l}{\mathtuple{{\mathpinew{\mathistype{x}{\xi}}{P'}}, L'}}$,
	\label{item:r-new-reduction-under-Gamma}
  \item If ${P_{0}}\mathpistructuralpo{P'_{0}}$, ${\mathtuple{P'_{0}, L_{0}}}\mathpisecreduction{\Gamma}{l}{\mathtuple{P'_{1}, L_{1}}}$, and ${P'_{1}}\mathpistructuralpo{P_{1}}$,
	then ${\mathtuple{P_{0}, L_{0}}}\mathpisecreduction{\Gamma}{l}{\mathtuple{P_{1}, L_{1}}}$.
	\label{item:r-sp-reduction-under-Gamma}
   \end{enumerate}
   We write $\mathpisecreductionkc{\Gamma}{l}$ 
   for the reflexive and transitive closure of $\mathpisecreduction{\Gamma}{l}$.
  \end{definition}

  Note that the secret reduction ${\mathtuple{P_{0},
      L_{0}}}\mathpisecreduction{\Gamma}{l}{\mathtuple{P_{1}, L_{1}}}$
  implies usual reduction ${\mathtuple{P_{0},
      L_{0}}}\mathpireduction{\mathtuple{P_{1}, L_{1}}}$.

  \begin{lemma}
   \label[lemma]{lemma:lock-freedom}
   Suppose that 
   $\mathtypejudgementwithsecrecy{\Gamma}{L}{m}{P}$ and
   $\mathtypejudgementwithsecrecy{\Delta}{L}{m}{Q}$ 
   are $k$-securely derivable,
   $\mathpiparallel{\Gamma}{\Delta}$ is reliable,
   $\mathtypeenvandsecrecylatice{\mathpiparallel{\Gamma}{\Delta}}{L}$ is $k$-secure, 
   and $\mathobligation{\alpha}{\mathof{\Gamma}{x}}$ is finite,
   where $\alpha={I}$ or $\alpha={O}$.
   Then, there exist a process $R$ and a lattice for secrecy levels $\hat{L}$ such that
   $\mathtuple{\mathpiparallel{P}{Q}, L}\mathpisecreductionkc{\mathpiparallel{\Gamma}{\Delta}}{k}{\mathtuple{R, \hat{L}}}$ and
   ${x}\in{\maththestrongbarbsof{\alpha}{R}}$.
  \end{lemma}

  \begin{proof}
   See \cref{sec:proof-of-lock-freedom}.
  \end{proof}
  
  \begin{definition}[Context]
   \label[definition]{def:context}
   A \emph{context} is defined as an expression obtained from a process 
   by replacing a sub-process
   with $\mathcontext{}{\;}$. We write $\mathcontext{C}{P}$ for the process obtained
   by replacing $\mathcontext{}{\;}$ in $C$ with $P$.

   A \emph{context with two holes} is defined as an expression obtained from a process 
   by replacing just two sub-processes
   with $\firstholemathcontexttwoholes{\;}$ and $\secondholemathcontexttwoholes{\;}$. 
   We write $\mathcontexttwoholes{C}{P_{0}}{P_{1}}$ for the process obtained
   by replacing $\firstholemathcontexttwoholes{\;}$ and 
   $\secondholemathcontexttwoholes{\;}$ in $C$ 
   with $P_{0}$ and $P_{1}$, respectively.
  \end{definition}

  \begin{figure}[t]
   \begin{align*}
    \mathdepth{\firstholemathcontexttwoholes{\;}}&= \mathdepth{\secondholemathcontexttwoholes{\;}} =0 \\
    \mathdepth{P}&=0\begin{gathered}\text{if } \firstholemathcontexttwoholes{\;} \text{ and } \secondholemathcontexttwoholes{\;} \\\qquad\text{do not occur in } P\end{gathered} \\
    \mathdepth{\mathpiparallel{C_{0}}{C_{1}}}&=\mathdepth{C_{0}} + \mathdepth{C_{1}} \\
    \mathdepth{\mathpioutput{x}{\mathvect{v}}. C_{0}}&=\mathdepth{C_{0}}+1 \\
    \mathdepth{\mathpiinput{x}{\mathvect{y}}. C_{0}}&=\mathdepth{C_{0}}+1 \\
    \mathdepth{\mathpireplication{C_{0}}}&=\mathdepth{C_{0}} \\
    \mathdepth{\mathpinew{\mathistype{x}{\xi}}{C_{0}}}&=\mathdepth{C_{0}} \\
    \mathdepth{\mathpinewsecrecylevel{l}{\mathvect{l_{0}}}{\mathvect{l_{1}}}{C_{0}}}&=\mathdepth{C_{0}}+1 \\
    \mathdepth{\mathobif{v}{C_{0}}{C_{1}}}&=\mathdepth{C_{0}} + \mathdepth{C_{1}}
   \end{align*}

 \caption{Depth of context}
 \label{fig:depth-of-context}
  \end{figure}

 \begin{definition}[Depth of context]
  \label[definition]{def:depth-of-context}
  For a context with two holes $C$,
  we inductively define the depth of $C$, written $\mathdepth{C}$,
  as in \cref{fig:depth-of-context}.
  We also define the depth of a context with one hole in the same manner.
 \end{definition}

  \begin{definition}[$k$-constrained derivation tree, $k$-finite level context, $k$-evaluation context]
   \label[definition]{def:k-contexts}
   For a context $C$ (resp.~with two holes),
   we define a \emph{$k$-constrained derivation tree}
   of $\mathtypejudgementwithsecrecy{\Gamma}{L}{l}{C}$
   as a $k$-secure derivation tree of 
   $\mathtypejudgementwithsecrecy{\Gamma}{L}{l}{C}$
   from 
   $\mathtypejudgementwithsecrecy{\Delta}{L'}{l'}{\mathcontext{}{\;}}$ 
   (resp.~$\mathtypejudgementwithsecrecy{\Delta_{1}}{L'_{1}}{l'_{1}}{\firstholemathcontexttwoholes{\;}}$
   and
   $\mathtypejudgementwithsecrecy{\Delta_{2}}{L'_{2}}{l'_{2}}{\secondholemathcontexttwoholes{\;}}$)
   where $l_{0}\leq_{L}k$ or $t_{c}$ is finite in
   every instance of \rulename{T-Out} or \rulename{T-In} if
   $\mathcontext{}{\;}$ 
   (resp.~$\firstholemathcontexttwoholes{\;}$ or $\secondholemathcontexttwoholes{\;}$)
   occurs in $P$.

   We then define an \emph{$k$-finite level context in $L$} as a context $F$ 
   satisfying the following conditions:
   \begin{enumerate}
    \item $F$ is of the following forms:
	  \begin{align*}
	  {F}&\mathcoloneqq \mathcontext{}{\;} \mathrelbar (\mathpiparallel{P}{F}) \mathrelbar (\mathpiparallel{F}{P}) \mathrelbar \mathpioutput{x}{\mathvect{v}}. F \mathrelbar \mathpiinput{x}{\mathvect{y}}. F \mathrelbar \ifdraft\else{}\\&\phantom{{}\mathcoloneqq{}}\fi
	   \mathpinew{\mathistype{x}{\xi}}{F}
	   \mathrelbar {\mathpinewsecrecylevel{l}{\mathvect{l_{0}}}{\mathvect{l_{1}}}{F}}
	  \end{align*}
    \item There exists a $k$-constrained derivation tree
	  of $\mathtypejudgementwithsecrecy{\Gamma}{L}{l}{F}$.
   \end{enumerate}

   We also define an \emph{$k$-finite level context in $L$ with two holes} 
   as a context with two holes $F$ satisfying the following conditions:
   \begin{enumerate}
    \item $F$ is of the following forms:
	  \begin{align*}
	   {F}&\mathcoloneqq (\mathpiparallel{F^{(1)}}{F^{(2)}}) \mathrelbar (\mathpiparallel{F^{(2)}}{F^{(1)}}) \mathrelbar (\mathpiparallel{P}{F}) \mathrelbar (\mathpiparallel{F}{P}) \mathrelbar \ifdraft\else{}\\&\phantom{{}\mathcoloneqq{}}\fi \mathpioutput{x}{\mathvect{v}}. F \mathrelbar \mathpiinput{x}{\mathvect{y}}. F \mathrelbar 
	   \mathpinew{\mathistype{x}{\xi}}{F}
	   \mathrelbar {\mathpinewsecrecylevel{l}{\mathvect{l_{0}}}{\mathvect{l_{1}}}{F}}
	   \\
	   {F^{(1)}}&\mathcoloneqq  \firstholemathcontexttwoholes{\;} \mathrelbar (\mathpiparallel{P}{F^{(1)}}) \mathrelbar (\mathpiparallel{F^{(1)}}{P}) \mathrelbar \mathpioutput{x}{\mathvect{v}}. F^{(1)} \mathrelbar \ifdraft\else{}\\&\phantom{{}\mathcoloneqq{}}\fi \mathpiinput{x}{\mathvect{y}}. F^{(1)} \mathrelbar 
	   \mathpinew{\mathistype{x}{\xi}}{F^{(1)}}
	   \mathrelbar {\mathpinewsecrecylevel{l}{\mathvect{l_{0}}}{\mathvect{l_{1}}}{F^{(1)}}}
	   \\
	   {F^{(2)}}&\mathcoloneqq  \secondholemathcontexttwoholes{\;} \mathrelbar (\mathpiparallel{P}{F^{(2)}}) \mathrelbar (\mathpiparallel{F^{(2)}}{P}) \mathrelbar \mathpioutput{x}{\mathvect{v}}. F^{(2)} \mathrelbar \ifdraft\else{}\\&\phantom{{}\mathcoloneqq{}}\fi \mathpiinput{x}{\mathvect{y}}. F^{(2)} \mathrelbar 
	   \mathpinew{\mathistype{x}{\xi}}{F^{(2)}}
	   \mathrelbar {\mathpinewsecrecylevel{l}{\mathvect{l_{0}}}{\mathvect{l_{1}}}{F^{(2)}}}
	  \end{align*}
    \item There exists a $k$-constrained derivation tree
	  of $\mathtypejudgementwithsecrecy{\Gamma}{L}{l}{F}$.
   \end{enumerate}

   Finally, we define an \emph{$k$-evaluation context} (resp.~\emph{with two holes})
   as a $k$-finite level context (resp.~with two holes) of depth $0$.
  \end{definition}

  \begin{lemma}
   \label[lemma]{lemma:lock-freedom-context}
   Let $\mathpiparallel{D_{0}}{D_{1}}$ be a context with two holes
   (hence, $D_{i}$ is a process, a context, or a context with two holes for $i=0, 1$). 
   Suppose that 
   $\mathtypejudgementwithsecrecy{\Gamma}{L}{l}{D_{0}}$ and
   $\mathtypejudgementwithsecrecy{\Delta}{L}{l}{D_{1}}$ 
   are $k$-securely derivable 
   from $\mathtypejudgementwithsecrecy{\Gamma_{0}}{L_{0}}{l_{0}}{\firstholemathcontexttwoholes{\;}}$ and
   $\mathtypejudgementwithsecrecy{\Gamma_{1}}{L_{1}}{l_{1}}{\secondholemathcontexttwoholes{\;}}$,
   $\mathpiparallel{\Gamma}{\Delta}$ is reliable,
   $\mathtypeenvandsecrecylatice{\mathpiparallel{\Gamma}{\Delta}}{L}$ is $k$-secure, 
   and $\mathobligation{\alpha}{\mathof{\Gamma}{x}}$ is finite,
   where $\alpha={I}$ or $\alpha={O}$.
   Then, there exist a context with two holes $C$ and a lattice for secrecy levels $\hat{L}$ such that
   $\mathtuple{\mathpiparallel{D_{0}}{D_{1}}, L}\mathpisecreductionkc{\mathpiparallel{\Gamma}{\Delta}}{k}{\mathtuple{C, \hat{L}}}$ and
   ${x}\in{\maththestrongbarbsof{\alpha}{C}}$.
  \end{lemma}

  \begin{proof}
   Similar to \cref{lemma:lock-freedom}.
  \end{proof}

  \begin{lemma}
   \label[lemma]{lemma:lock-freedom-new-free-finite-level-to-eval}
   Let $F$ be a $k$-finite level context with two holes,
   and $\mathtypejudgementwithsecrecy{\Gamma}{L}{l}{F}$ be
   the root of a $k$-constrained derivation tree.
   If $\Gamma$ is reliable, 
   then there exist an evaluation context $E$ and a lattice for secrecy levels $\hat{L}$
   such that
   $\mathtuple{F, L}\mathpisecreductionkc{\Gamma}{k}\mathtuple{E, \hat{L}}$.
  \end{lemma}

  \begin{proof}
   By induction on the depth of $F$.

   In case the depth of $F$ is $0$, we have the claimed result immediately.
   
   Assume that $F$ is of the form $\mathcontext{E_{0}}{\mathpioutput{x}{\mathvect{v}}. C_{0}}$
   for an evaluation context $E_{0}$.
   Since $\Gamma$ is reliable,
   $\mathobligation{I}{\mathof{\Gamma}{x}}\leq{\mathcapability{O}{\mathof{\Gamma}{x}}}$. 

   Assume $x$ is free.
   Since $\mathtypejudgementwithsecrecy{\Gamma}{L}{l}{F}$ is
   the root of the $k$-constrained derivation tree of $F$, 
   $\mathcapability{O}{\mathof{\Gamma}{x}}$ is finite.
   Hence, $\mathobligation{I}{\mathof{\Gamma}{x}}$ is finite.
   By \cref{lemma:lock-freedom-context},
   there exist $R$ and $\hat{L}$ such that
   $\mathtuple{\mathcontext{E_{0}}{\mathpioutput{x}{\mathvect{v}}. C_{0}}, L}\mathpisecreductionkc{\Gamma}{k}{\mathtuple{R, \hat{L}}}$
   and ${x}\in{\maththestrongbarbsof{I}{R}}$.
   Then
   $\mathtuple{\mathcontext{E_{0}}{\mathpioutput{x}{\mathvect{v}}. F_{0}}, L}\mathpisecreductionkc{\Gamma}{k}{\mathtuple{\mathcontext{E'_{0}}{F_{0}}, \hat{L}_{0}}}$
   for an evaluation context $E'_{0}$.
   By the induction hypothesis,
   there exists an evaluation context $E$ and a lattice for secrecy levels $\hat{L}$
   such that
   ${\mathtuple{\mathcontext{E'_{0}}{C_{0}}, \hat{L}_{0}}}\mathpisecreductionkc{\Gamma}{k}\mathtuple{E, \hat{L}}$.
   Thus,
   ${\mathtuple{F, L}}\mathpisecreductionkc{\Gamma}{k}\mathtuple{E, \hat{L}}$
   
   In the case that $x$ is not free, $\mathcontext{E_{0}}{\mathpioutput{x}{\mathvect{v}}. F_{0}}\mathpistructuralpo\mathpinew{x}{\mathcontext{E'_{0}}{\mathpioutput{x}{\mathvect{v}}. F_{0}}}$ for a evaluation context $E'_{0}$.
  Since there is a $k$-constrained derivation tree of 
   $\mathtypejudgementwithsecrecy{\Gamma}{L}{l}{\mathcontext{E_{0}}{\mathpioutput{x}{\mathvect{v}}. F_{0}}}$,
   there is a $k$-constrained derivation tree of 
   $\mathtypejudgementwithsecrecy{\Gamma}{L}{l}{\mathpinew{x}{\mathcontext{E'_{0}}{\mathpioutput{x}{\mathvect{v}}. F_{0}}}}$.
   Hence, 
   there is a $k$-constrained derivation tree of 
   $\mathtypejudgementwithsecrecy{\Gamma', {\mathistype{x}{\xi/U}}}{L}{l}{\mathcontext{E'_{0}}{\mathpioutput{x}{\mathvect{v}}. F_{0}}}$ with $\mathreliable{U}$.
   Therefore, we can show the claim in the same way to the case that $x$ is free.

   In case $F$ is of the form $\mathcontext{E}{\mathpiinput{x}{\mathvect{y}}. F_{0}}$,
   we can show the claim in the similar way to the case 
   $F\equiv\mathcontext{E}{\mathpioutput{x}{\mathvect{v}}. F_{0}}$.

   In case $F$ is of the form $\mathcontext{E}{\mathpinewsecrecylevel{l}{\mathvect{l_{0}}}{\mathvect{l_{1}}}{F_{0}}}$,
   we can show the claim easily.
  \end{proof}

\subsection{Non-interference}
We will now show our main theorem: the non-interference property of well-typed processes.

\begin{definition}[Barbs]
 \label[definition]{def:Barbs}
 Let $P$ be a process, and $L$ be a lattice of secrecy levels.
 We define the \emph{barbs of $\mathtuple{P, L}$}, written $\maththebarbsof{P, L}$, \ifdraft as follows:\else as:\fi
\ifdraft\begin{align*}\else\[\fi
  \ifdraft{\maththebarbsof{P, L}}=&\fi
  {\mathsetintension{x}{
  \begin{gathered}
   {\mathtuple{P, L}}\mathpireductionkc\mathtuple{P', L'} \text{ and } \\
   {{P'}={\mathpinew{\mathvect{y}}{\mathpiparallel{\mathpioutput{x}{\mathvect{v}}. P_{0}}{P_{1}}}}} \text{ or }
   {{P'}={\mathpinew{\mathvect{y}}{\mathpiparallel{\mathpiinput{x}{\mathvect{z}}. P_{0}}{P_{1}}}}}\\ \text{ with }
   {{x}\notin{\mathvect{y}}} \text{ for some } P_{0}, P_{1} 
  \end{gathered}
  }
  }
\ifdraft\end{align*}\else\]\fi
\end{definition}

\begin{definition}[Barbed bisimulation]
  \label[definition]{def:Barbed-bisimulation}
 A \emph{barbed bisimulation} is defined as
 a binary relation $\mathbarbedbisim{R}$ on 
 ${\maththesetofprocesses}\times{\maththesetoflatticeforsec}$ 
 satisfying the following conditions for every
 ${\mathtuple{\mathtuple{P_{0}, L_{0}}, \mathtuple{P_{1}, L_{1}}}}\in{\mathbarbedbisim{R}}$.
 \begin{enumerate}
  \item If ${\mathtuple{P_{0}, L_{0}}}\mathpireduction{\mathtuple{P'_{0}, L'_{0}}}$,
	then there exists ${\mathtuple{P'_{1}, L'_{1}}}$ such that
	${\mathtuple{P_{1}, L_{1}}}\mathpireductionkc{\mathtuple{P'_{1}, L'_{1}}}$
	with
	${\mathtuple{\mathtuple{P'_{0}, L'_{0}}, \mathtuple{P'_{1}, L'_{1}}}}\in{\mathbarbedbisim{R}}$.
	\label{item:left-def-Barbed-bisimulation}
  \item If ${\mathtuple{P_{1}, L_{1}}}\mathpireduction{\mathtuple{P'_{1}, L'_{1}}}$,
	then there exists ${\mathtuple{P'_{0}, L'_{0}}}$ such that
	${\mathtuple{P_{0}, L_{0}}}\mathpireductionkc{\mathtuple{P'_{0}, L'_{0}}}$
	with
	${\mathtuple{\mathtuple{P'_{0}, L'_{0}}, \mathtuple{P'_{1}, L'_{1}}}}\in{\mathbarbedbisim{R}}$.
	\label{item:right-def-Barbed-bisimulation}
  \item ${\maththebarbsof{P_{0}, L_{0}}}={\maththebarbsof{P_{1}, L_{1}}}$
	\label{item:barbs-def-Barbed-bisimulation}
 \end{enumerate}
\end{definition}

  We say that $\mathtuple{P_{0}, L_{0}}$ and $\mathtuple{P_{1},
    L_{1}}$ are \emph{barbed bisimilar}, written
  $\mathisbisimilar{\mathtuple{P_{0}, L_{0}}}{\mathtuple{P_{1},
      L_{1}}}$, if there exists a barbed bisimulation
  ${\mathbarbedbisim{R}}$ such that ${\mathtuple{\mathtuple{P_{0},
        L_{0}}, \mathtuple{P_{1}, L_{1}}}}\in{\mathbarbedbisim{R}}$.
  
  \begin{definition}\memo{[$\mathenvandlevel{\Gamma}{L}{\lm}$-$\mathenvandlevel{\Delta}{L'}{\lm'}$-Context]}
   \label[definition]{def:typed-context}
   A context $C$ is called an 
   \emph{%
   $\mathenvandlevel{\Gamma}{L}{\lm}$-$\mathenvandlevel{\Delta}{L'}{\lm'}$-context%
   }
   if $\mathtypejudgementwithsecrecy{\Delta}{L'}{\lm'}{C}$ is derivable from
   $\mathtypejudgementwithsecrecy{\Gamma}{L}{\lm}{\mathcontext{}{\;}}$.
  \end{definition}

  \noindent Note that $\mathissublattice{L'}{L}$ if
  $\mathtypejudgementwithsecrecy{\Delta}{L'}{\lm'}{C}$ is derivable
  from
  $\mathtypejudgementwithsecrecy{\Gamma}{L}{\lm}{\mathcontext{}{\;}}$.

  \begin{definition}[Barbed congruence]
   \label[definition]{def:barbed-congruence}
   For processes $P_{0}$ and $P_{1}$, we say that $P_{0}$ and $P_{1}$ are barbed
   $\mathenvandlevel{\Gamma}{L}{\lm}$-congruent,
   written $\mathiscongruentwithenvandlevel{\Gamma}{L}{\lm}{P_{0}}{P_{1}}$,
   if
   \begin{enumerate}
    \item $\mathtypejudgementwithsecrecy{\Gamma}{L}{\lm}{P_{i}}$ is derivable 
	  for ${i}={0, 1}$, and
	  \label{item:type-in-def-barbed-congruence}
    \item for any closed $\Delta$, lattice of secrecy levels $L'$,
	  secrecy level $\lm'$, and any $\mathenvandlevel{\Gamma}{L}{\lm}$-$\mathenvandlevel{\Delta}{L'}{\lm'}$-context 
	  $C$,
	  $\mathisbisimilar{\mathtuple{\mathcontext{C}{P_{0}}, L'}}{\mathtuple{\mathcontext{C}{P_{1}}, L'}}$.
	  \label{item:bisimilar-in-def-barbed-congruence}
   \end{enumerate}
  \end{definition}

  We say that \emph{the secrecy level of
    $\mathtypeenvandsecrecylatice{\Gamma}{L}$ is $l$} if $l$ is the
  supremum in $L$ of all the secrecy levels
  that appear in $\Gamma(x)$ for every channel name
  ${x}\in{\mathof{\mathdom}{\Gamma}}$.

  Now, we state the non-interference theorems.  Intuitively, they
  guarantee secrets---values of level $l'$, or behavior of processes
  of level $m'$---do not leak to attackers of level $l$ or $k$.

  \begin{theorem}[Non-interference (1)]
   \label[theorem]{thm:non-interference-value}
   For any type environment $\Gamma$, lattice of secrecy levels $L$,
   process $P$, and secrecy levels $l$, $l'$,
   we have
   \[\mathiscongruentwithenvandlevel{\Gamma}{L}{\lm}{{\mathsubstbox{P}{\mathsubst{x}{\mathobtrue^{l'}}}}}{{\mathsubstbox{P}{\mathsubst{x}{\mathobfalse^{l'}}}}}\]
   if $\mathtypejudgementwithsecrecy{\Gamma}{L}{\lm}{\mathsubstbox{P}{\mathsubst{x}{\mathobtrue^{l'}}}}$ is $\lk$-securely derivable,
   the secrecy level of $\mathtypeenvandsecrecylatice{\Gamma}{L}$ is $l$,
   ${l'}\not \leq_{L}{l}$, and ${l'}\not\leq_{L} {k}$.
  \end{theorem}

\begin{proof}
It suffices to show that, 
for any closed $\Delta$, a lattice for secrecy levels $L'$,
 and a secrecy level $m'$, 
 $\mathisbisimilar{\mathtuple{\mathcontext{C}{\mathsubstbox{P}{\mathsubst{x}{\mathobtrue^{l'}}}}, L'}}{\mathtuple{\mathcontext{C}{{\mathsubstbox{P}{\mathsubst{x}{\mathobfalse^{l'}}}}, L'}}}$ with any 
 $\mathenvandlevel{\Gamma}{L}{m}$-$\mathenvandlevel{\Delta}{L'}{m'}$-context $C$.
 We can construct a process $Q$, where
 \[
  \mathisbisimilar{\mathtuple{\mathcontext{C}{\mathsubstbox{P}{\mathsubst{x}{\mathobtrue^{l'}}}}, L'}}{\mathisbisimilar{\mathtuple{Q, L'}}{\mathtuple{\mathcontext{C}{{\mathsubstbox{P}{\mathsubst{x}{\mathobfalse^{l'}}}}, L'}}}}.
 \]
 Hence, we have the claimed result.
 $Q$ is obtained by eliminating channels with secrecy level higher than $\lk$
 from $\mathcontext{C}{\mathsubstbox{P}{\mathsubst{x}{\mathobtrue^{l'}}}}$.
 See \cref{subsec:proof-thm-non-interference-value} for details. \ifdraft\qed\fi
\end{proof}

  \begin{theorem}[Non-interference (2)]
   \label[theorem]{thm:non-interference-process}
   For any type environments $\Gamma$,  $\Delta$,
   lattices for secrecy levels $L$, $L'$,
   processes $P_{0}$, $P_{1}$, and
   any $\mathenvandlevel{\Delta}{L'}{\lm'}$-$\mathenvandlevel{\Gamma}{L}{\lm}$-context $C$, we have
   \[\mathiscongruentwithenvandlevel{\Gamma}{L}{\lm}{\mathcontext{C}{P_{0}}}{\mathcontext{C}{P_{1}}}\]
   if $\mathtypejudgementwithsecrecy{\Delta}{L'}{\lm'}{P_{i}}$
   is $\lk$-securely derivable for ${i}={0, 1}$,
   the secrecy level of $\mathtypeenvandsecrecylatice{\Gamma}{L}$ is $l$,
   ${\lm'}\not \leq_{L}{l}$, and ${\lm'}\not \leq_{L}{k}$.
  \end{theorem}

  \begin{proof}
      Let $\Pi$ be a closed type environment, and
   $C$ be 
   $\mathenvandlevel{\Gamma}{L}{m}$-$\mathenvandlevel{\Pi}{L'}{m''}$-context.
 We can construct a context $D$, where
   \ifdraft\[\else$\fi
 \mathisbisimilar{\mathtuple{\mathcontext{C}{\mathcontext{\hat{C}}{P_{0}}}, L}}{\mathisbisimilar{\mathtuple{\mathcontext{D}{\mathcontext{\hat{C}}{P_{0}}}, L}}{\mathisbisimilar{\mathtuple{\mathcontext{D}{\mathcontext{\hat{C}}{P_{1}}}, L}}{\mathtuple{\mathcontext{C}{\mathcontext{\hat{C}}{P_{1}}}, L}}}}.
   \ifdraft\]\else$ \fi
   $D$ is obtained by eliminating channels with secrecy level higher than $\lk$ from $C$.
   See \cref{subsec:proof-thm-non-interference-process} for details. \qed
  \end{proof}

\section{Conclusion}
\label{sec:conclusion}
We have defined $\pi^{L}$-calculus,
an extension of $\pi$-calculus with secrecy types and an operation
to extend the lattice of secrecy levels.
Then, we have given a type system for secure information flow
and shown the lock-freedom and non-interference properties.
.
Our system has extended previous work~\cite{Kobayashi2005} with
general lattices and its dynamic extensions, requiring deliberate
definitions and proofs for sound generalization and safe extension of
the security lattices.  Future work would include further extending
the calculus with polymorphism for secrecy levels so that processes
can communicate and share the new levels they create, as well as
considering other operations---such as deletion---to dynamically
change the security lattice.

\ifdraft
\section*{Acknowledgments}

This work was supported by JST, CREST Grant Number JPMJCR22M3, Japan
and partially by JSPS KAKENHI Grant Number 22K19766 and 23K20379
(20H04161).

\fi

\bibliography{refs}

\ifdraft
\appendix

\section*{Appendices}
\else
\onecolumn

\appendices
\fi

\section{Basic properties of $\pi^{L}$-calculus}

\begin{lemma}
 \label[lemma]{lemma:sublatice-double-NewLev}
 For a lattice of secrecy levels $L$,
 assume that
 $\mathpinewsecrecylevel{l}{\mathvect{l_{0}}}{\mathvect{l_{1}}}{L}$ is defined, and
 $\mathissublattice{L}{\mathpinewsecrecylevel{l}{\mathvect{l_{0}}}{\mathvect{l_{1}}}{L}}$.
 Let $L'=\mathpinewsecrecylevel{l}{\mathvect{l_{0}}}{\mathvect{l_{1}}}{L}$.
 Then 
 $\mathpinewsecrecylevel{l'}{\mathvect{l_{0}}}{\mathvect{l_{1}}}{L'}$ is defined and
 $\mathissublattice{L'}{\mathpinewsecrecylevel{l'}{\mathvect{l_{0}}}{\mathvect{l_{1}}}{L'}}$.
\end{lemma}

\begin{proof}
 Straightforward.
\end{proof}

\begin{lemma}
 \label[lemma]{lemma:sublatice-swap}
 For a lattice of secrecy levels $L$ and 
$\mathvect{l_{0}}, \mathvect{l_{1}}, \mathvect{l'_{0}}, \mathvect{l'_{1}}\subseteq L$,
 assume that
 $\mathpinewsecrecylevel{l}{\mathvect{l_{0}}}{\mathvect{l_{1}}}{L}$ and
 $\mathpinewsecrecylevel{l'}{\mathvect{l'_{0}}}{\mathvect{l'_{1}}}{\mathpinewsecrecylevel{l}{\mathvect{l_{0}}}{\mathvect{l_{1}}}{L}}$ are defined, and
 $\mathissublattice{L}{\mathpinewsecrecylevel{l}{\mathvect{l_{0}}}{\mathvect{l_{1}}}{L}}$
 and
 $\mathissublattice{\mathpinewsecrecylevel{l}{\mathvect{l_{0}}}{\mathvect{l_{1}}}{L}}{\mathpinewsecrecylevel{l'}{\mathvect{l'_{0}}}{\mathvect{l'_{1}}}{\mathpinewsecrecylevel{l}{\mathvect{l_{0}}}{\mathvect{l_{1}}}{L}}}$.
 Then 
 $\mathpinewsecrecylevel{l'}{\mathvect{l'_{0}}}{\mathvect{l'_{1}}}{L}$ and
 $\mathpinewsecrecylevel{l}{\mathvect{l_{0}}}{\mathvect{l_{1}}}{\mathpinewsecrecylevel{l'}{\mathvect{l'_{0}}}{\mathvect{l'_{1}}}{L}}$ are defined and
 $\mathissublattice{L}{\mathpinewsecrecylevel{l'}{\mathvect{l'_{0}}}{\mathvect{l'_{1}}}{L}}$,
 $\mathissublattice{\mathpinewsecrecylevel{l'}{\mathvect{l'_{0}}}{\mathvect{l'_{1}}}{L}}{\mathpinewsecrecylevel{l}{\mathvect{l_{0}}}{\mathvect{l_{1}}}{\mathpinewsecrecylevel{l'}{\mathvect{l'_{0}}}{\mathvect{l'_{1}}}{L}}}$,
 and \linebreak[4]
 $\mathpinewsecrecylevel{l'}{\mathvect{l'_{0}}}{\mathvect{l'_{1}}}{\mathpinewsecrecylevel{l}{\mathvect{l_{0}}}{\mathvect{l_{1}}}{L}} = \mathpinewsecrecylevel{l}{\mathvect{l_{0}}}{\mathvect{l_{1}}}{\mathpinewsecrecylevel{l'}{\mathvect{l'_{0}}}{\mathvect{l'_{1}}}{L}}$.
\end{lemma}

\begin{proof}
 Straightforward.
\end{proof}

\begin{proposition}
 \label[proposition]{lemma:structural-po-extra-rules-cong}
 If 
 ${P_{0}}\mathpistructuralpo{P'_{0}}$ and 
 ${P_{1}}\mathpistructuralpo{P'_{1}}$,
 then 
 ${\mathpiparallel{P_{0}}{P_{1}}}\mathpistructuralpo{\mathpiparallel{P'_{0}}{P'_{1}}}$.
\end{proposition}
 
 \begin{proof}
  Assume ${P_{0}}\mathpistructuralpo{P'_{0}}$ and 
  ${P_{1}}\mathpistructuralpo{P'_{1}}$.
  By \rulename{SP-Commut} and \rulename{SP-Par}, we see
  \begin{align*}
   {\mathpiparallel{P_{0}}{P_{1}}}
   &\mathpistructuralpo{\mathpiparallel{P'_{0}}{P_{1}}}
   &\text{\rulename{SP-Par}} \\
   &\mathpistructuralpo{\mathpiparallel{P_{1}}{P'_{0}}}
   &\text{\rulename{SP-Commut}} \\
   &\mathpistructuralpo{\mathpiparallel{P'_{1}}{P'_{0}}}
   &\text{\rulename{SP-Par}} \\
   &\mathpistructuralpo{\mathpiparallel{P'_{0}}{P'_{1}}}
   &\text{\rulename{SP-Commut}\lefteqn{.}}
  \end{align*}
 \end{proof}

 \begin{lemma}
  \label{lemma:structural-po-extra-rules-assoc}
  ${\mathpiparallel{P_{0}}{\mleft(\mathpiparallel{P_{1}}{P_{2}}\mright)}}\mathpistructuralpo{\mathpiparallel{\mleft(\mathpiparallel{P_{0}}{P_{1}}\mright)}{P_{2}}}$.
 \end{lemma}
 
 \begin{proof} 
  By \rulename{SP-Commut}, \rulename{SP-Assoc} and \rulename{SP-Par}, we see
  \begin{align*}
   {\mathpiparallel{P_{0}}{\mleft(\mathpiparallel{P_{1}}{P_{2}}\mright)}}
   &\mathpistructuralpo{\mathpiparallel{\mleft(\mathpiparallel{P_{1}}{P_{2}}\mright)}{P_{0}}} &\text{\rulename{SP-Commut}} \\
   &\mathpistructuralpo{\mathpiparallel{\mleft(\mathpiparallel{P_{2}}{P_{1}}\mright)}{P_{0}}} &\text{\rulename{SP-Commut} and \rulename{SP-Par}} \\
   &\mathpistructuralpo{\mathpiparallel{P_{2}}{\mleft(\mathpiparallel{P_{1}}{P_{0}}\mright)}} &\text{\rulename{SP-Assoc}}\\
   &\mathpistructuralpo{\mathpiparallel{P_{2}}{\mleft(\mathpiparallel{P_{0}}{P_{1}}\mright)}} &\text{\rulename{SP-Commut} and \rulename{SP-Par}} \\
   &\mathpistructuralpo{\mathpiparallel{\mleft(\mathpiparallel{P_{0}}{P_{1}}\mright)}{P_{2}}} &\text{\rulename{SP-Commut}\lefteqn{.}} 
  \end{align*}
 \end{proof}

 \begin{lemma}
  \label[lemma]{lemma:structural-po-and-set-of-free-names}
  \begin{enumerate}
   \item If ${P_{0}}\mathpistructuralpo{P_{1}}$, 
	 then ${\mathFNof{P_{0}}}\supseteq{\mathFNof{P_{1}}}$. 
	 \label{item:po-lemma-structural-po-and-set-of-free-names}
   \item If ${P_{0}}\mathpistructuraleq{P_{1}}$, 
	 then ${\mathFNof{P_{0}}}={\mathFNof{P_{1}}}$. 
	 \label{item:eq-lemma-structural-po-and-set-of-free-names}
  \end{enumerate}
 \end{lemma}

 \begin{proof}
  It suffices to show \cref{item:po-lemma-structural-po-and-set-of-free-names}.
  We see \cref{item:po-lemma-structural-po-and-set-of-free-names} by induction on 
  the construction of ${P_{0}}\mathpistructuralpo{P_{1}}$.
 \end{proof}

\begin{proposition}
 \label{prop:reduction-basic}
 If ${\mathtuple{P, L}}\mathpireduction{\mathtuple{\hat{P}, \hat{L}}}$,
 then either 
 \begin{enumerate}
 \item ${P}\mathpistructuralpo{\mathpinew{\mathistype{\mathvect{x}}{\mathvect{\xi}}}{\mathpiparallel{\mathpiparallel{\mathpioutput{z}{\mathvect{v}}. P_{0}}{\mathpiinput{z}{\mathvect{y}}.P_{1}}}{P_{2}}}}$, 
       ${\mathpinew{\mathistype{\mathvect{x}}{\mathvect{\xi}}}{\mathpiparallel{\mathpiparallel{P_{0}}{\mathsubstbox{P_{1}}{\mathsubst{\mathvect{y}}{\mathvect{v}}}}}{P_{2}}}}\mathpistructuralpo{\hat{P}}$ and 
       $\hat{L}=L$, or
 \item ${P}\mathpistructuralpo{\mathpinew{\mathistype{\mathvect{x}}{\mathvect{\xi}}}{\mathpiparallel{\mathpinewsecrecylevel{l}{\mathvect{l_{0}}}{\mathvect{l_{1}}}{P_{0}}}{P_{1}}}}$,
       ${\mathpinew{\mathistype{\mathvect{x}}{\mathvect{\xi}}}{\mathpiparallel{P_{0}}{P_{1}}}}\mathpistructuralpo{\hat{P}}$ and
       $\hat{L}=\mathpinewsecrecylevel{l}{\mathvect{l_{0}}}{\mathvect{l_{1}}}{L}$.
 \end{enumerate}
\end{proposition}

\begin{proof}
 By induction on the construction of
 ${\mathtuple{P, L}}\mathpireduction{\mathtuple{\hat{P}, \hat{L}}}$.
\end{proof}

\section{Basic properties of usages}

\subsection{Propositions for usages}

\begin{lemma}
 \label[lemma]{lemma:cap_and_ob_subst}
 Let $U$, $U_{0}$, and $U_{1}$ be usages. 
 Let ${\alpha}\in{\mathsetextension{I, O}}$.
 Let $F$ be a partial mapping from usage variables to obligation levels.
 \begin{enumerate}
  \item If ${\mathcapability{\alpha}{U_{0}}}\leq{\mathcapability{\alpha}{U_{1}}}$,
	then 
	${\mathcapability{\alpha}{\mathsubstbox{U}{\mathsubst{\rho}{U_{0}}}}}\leq{\mathcapability{\alpha}{\mathsubstbox{U}{\mathsubst{\rho}{U_{1}}}}}$.
	\label{item:cap_leq_lemma-cap_and_ob_subst}
  \item If ${\mathcapability{\alpha}{U_{0}}}={\mathcapability{\alpha}{U_{1}}}$,
	then 
	${\mathcapability{\alpha}{\mathsubstbox{U}{\mathsubst{\rho}{U_{0}}}}}={\mathcapability{\alpha}{\mathsubstbox{U}{\mathsubst{\rho}{U_{1}}}}}$.
	\label{item:cap_eq_lemma-cap_and_ob_subst}
  \item If ${\mathobligation{\alpha}{U_{0}}}\geq{\mathobligation{\alpha}{U_{1}}}$, then
	${\mathobligation{\alpha}{\mathsubstbox{U}{\mathsubst{\rho}{U_{0}}}}}\geq{\mathobligation{\alpha}{\mathsubstbox{U}{\mathsubst{\rho}{U_{1}}}}}$.
	 \label{item:ob_geq_lemma-cap_and_ob_subst}
  \item If ${\mathobligation{\alpha}{U_{0}}}={\mathobligation{\alpha}{U_{1}}}$, then
	${\mathobligation{\alpha}{\mathsubstbox{U}{\mathsubst{\rho}{U_{0}}}}}={\mathobligation{\alpha}{\mathsubstbox{U}{\mathsubst{\rho}{U_{1}}}}}$.
	\label{item:ob_eq_lemma-cap_and_ob_subst}
 \end{enumerate}
 \end{lemma}

 \begin{proof}
  \noindent \cref{item:cap_leq_lemma-cap_and_ob_subst} By induction on the construction of $U$.

  \noindent \cref{item:cap_eq_lemma-cap_and_ob_subst} By induction on the construction of $U$.

  \noindent \cref{item:ob_geq_lemma-cap_and_ob_subst} By induction on the construction of $U$.

  \noindent \cref{item:ob_eq_lemma-cap_and_ob_subst} By induction on the construction of $U$.
 \end{proof}

 \begin{lemma}
  \label[lemma]{lemma:usage-extra-rules}
  \begin{enumerate}
   \item If 
	 ${U_{0}}\mathusagestructuralpo{U'_{0}}$ and 
	 ${U_{1}}\mathusagestructuralpo{U'_{1}}$,
	 then 
	 ${\mathpiparallel{U_{0}}{U_{1}}}\mathusagestructuralpo{\mathpiparallel{U'_{0}}{U'_{1}}}$.
	 \label{item:cong-in-lemma-usage-extra-rules}
   \item ${\mathpiparallel{U_{0}}{\mleft(\mathpiparallel{U_{1}}{U_{2}}\mright)}}\mathusagestructuralpo{\mathpiparallel{\mleft(\mathpiparallel{U_{0}}{U_{1}}\mright)}{U_{2}}}$.
	 \label{item:assoc-in-lemma-usage-extra-rules}
  \end{enumerate}
 \end{lemma}
 
 \begin{proof} We show each clause.
  
  \noindent \cref{item:cong-in-lemma-usage-extra-rules}
  Assume ${U_{0}}\mathusagestructuralpo{U'_{0}}$ and 
  ${U_{1}}\mathusagestructuralpo{U'_{1}}$.
  By \rulename{UP-Commut} and \rulename{UP-CongP}, we see
  \begin{align*}
   {\mathpiparallel{U_{0}}{U_{1}}}
   &\mathusagestructuralpo{\mathpiparallel{U'_{0}}{U_{1}}}
   &\text{\rulename{UP-CongP}} \\
   &\mathusagestructuralpo{\mathpiparallel{U_{1}}{U'_{0}}}
   &\text{\rulename{UP-Commut}} \\
   &\mathusagestructuralpo{\mathpiparallel{U'_{1}}{U'_{0}}}
   &\text{\rulename{UP-CongP}} \\
   &\mathusagestructuralpo{\mathpiparallel{U'_{0}}{U'_{1}}}
   &\text{\rulename{UP-Commut}}\lefteqn{.}
  \end{align*}

  \noindent \cref{item:assoc-in-lemma-usage-extra-rules}
  By \rulename{UP-Commut}, \rulename{UP-Assoc} and \rulename{UP-CongP}, we see
  \begin{align*}
   {\mathpiparallel{U_{0}}{\mleft(\mathpiparallel{U_{1}}{U_{2}}\mright)}}
   &\mathusagestructuralpo{\mathpiparallel{\mleft(\mathpiparallel{U_{1}}{U_{2}}\mright)}{U_{0}}} &\text{\rulename{UP-Commut}} \\
   &\mathusagestructuralpo{\mathpiparallel{\mleft(\mathpiparallel{U_{2}}{U_{1}}\mright)}{U_{0}}} &\text{\rulename{UP-Commut} and \rulename{UP-CongP}} \\
   &\mathusagestructuralpo{\mathpiparallel{U_{2}}{\mleft(\mathpiparallel{U_{1}}{U_{0}}\mright)}} &\text{\rulename{UP-Assoc}}\\
   &\mathusagestructuralpo{\mathpiparallel{U_{2}}{\mleft(\mathpiparallel{U_{0}}{U_{1}}\mright)}} &\text{\rulename{UP-Commut} and \rulename{UP-CongP}} \\
   &\mathusagestructuralpo{\mathpiparallel{\mleft(\mathpiparallel{U_{0}}{U_{1}}\mright)}{U_{2}}} &\text{\rulename{UP-Commut}}\lefteqn{.} 
  \end{align*}
 \end{proof}

 \begin{lemma} \label[lemma]{lemma:usagestrucuralpo-cap-ob}
  If ${U}\mathusagestructuralpo{U'}$,
  then ${\mathcapability{\alpha}{U}}\leq{\mathcapability{\alpha}{U'}}$ and
  ${\mathobligation{\alpha}{U}}\geq{\mathobligation{\alpha}{U'}}$
  for ${\alpha}\in{\mathsetextension{I, O}}$ and 
  a partial mapping $F$ from usage variables to obligation levels.
 \end{lemma}

 \begin{proof}
  We show the claim
  by induction on the construction of ${U}\mathusagestructuralpo{U'}$.
  We consider cases according to the clauses of the definition.

 \noindent Case 1. If ${U}={U'}$, then we have the claimed result obviously.

 \noindent Case 2. Assume ${U}\mathusagestructuralpo{U''}$ and ${U''}\mathusagestructuralpo{U'}$.
  By the induction hypothesis, 
  ${\mathcapability{\alpha}{U}}\leq{\mathcapability{\alpha}{U''}}$,
  ${\mathobligation{\alpha}{U}}\geq{\mathobligation{\alpha}{U''}}$,
  ${\mathcapability{\alpha}{U''}}\leq{\mathcapability{\alpha}{U'}}$, and
  ${\mathobligation{\alpha}{U''}}\geq{\mathobligation{\alpha}{U'}}$.
  Then, we have
  ${\mathcapability{\alpha}{U}}\leq{\mathcapability{\alpha}{U'}}$ and
  ${\mathobligation{\alpha}{U}}\geq{\mathobligation{\alpha}{U'}}$.

  \noindent Case 3. \rulename{UP-Zero}.
  Assume ${U}={\mathpiparallel{\mathusagenil}{U_{1}}}$ and ${U'}={U_{1}}$.
  Then
  \begin{align*}
   {\mathcapability{\alpha}{U}}
   &={\mathcapability{\alpha}{\mathpiparallel{\mathusagenil}{U_{1}}}}
   ={\mathof{\min}{\mathcapability{\alpha}{\mathusagenil}, \mathcapability{\alpha}{U_{1}}}}
   ={\mathof{\min}{\infty, \mathcapability{\alpha}{U_{1}}}}
   ={\mathcapability{\alpha}{U_{1}}}
   ={\mathcapability{\alpha}{U'}}
   \intertext{and}
   {\mathobligation{\alpha}{U}}
   &={\mathobligation{\alpha}{\mathpiparallel{\mathusagenil}{U_{1}}}}
   ={\mathof{\min}{\mathobligation{\alpha}{\mathusagenil}, \mathobligation{\alpha}{U_{1}}}}
   ={\mathof{\min}{\infty, \mathobligation{\alpha}{U_{1}}}}
   ={\mathobligation{\alpha}{U_{1}}}
   ={\mathobligation{\alpha}{U'}}.
  \end{align*}

 \noindent Case 4. \rulename{UP-Commut}.
  Assume ${U}={\mathpiparallel{U_{1}}{U_{2}}}$ and
  ${U'}={\mathpiparallel{U_{2}}{U_{1}}}$.
  Then
  \begin{align*}
   {\mathcapability{\alpha}{U}}
   &={\mathcapability{\alpha}{\mathpiparallel{U_{1}}{U_{2}}}}
   ={\mathof{\min}{\mathcapability{\alpha}{U_{1}}, \mathcapability{\alpha}{U_{2}}}}
   ={\mathcapability{\alpha}{\mathpiparallel{U_{2}}{U_{1}}}}
   ={\mathcapability{\alpha}{U'}}
   \intertext{and}
   {\mathobligation{\alpha}{U}}
   &={\mathobligation{\alpha}{\mathpiparallel{U_{1}}{U_{2}}}}
   ={\mathof{\min}{\mathobligation{\alpha}{U_{1}}, \mathobligation{\alpha}{U_{2}}}}
   ={\mathobligation{\alpha}{\mathpiparallel{U_{2}}{U_{1}}}}
   ={\mathobligation{\alpha}{U'}}.
  \end{align*}

 \noindent Case 5. \rulename{UP-Assoc}.
  Assume ${U}={\mathpiparallel{\mleft(\mathpiparallel{U_{1}}{U_{2}}\mright)}{U_{3}}}$ 
  and
  ${U'}={\mathpiparallel{U_{1}}{\mleft(\mathpiparallel{U_{2}}{U_{3}}\mright)}}$.
  Then
  \begin{align*}
   {\mathcapability{\alpha}{U}}
   &={\mathcapability{\alpha}{\mathpiparallel{\mleft(\mathpiparallel{U_{1}}{U_{2}}\mright)}{U_{3}}}} \\
   &={\mathof{\min}{\mathcapability{\alpha}{\mathpiparallel{U_{1}}{U_{2}}}, \mathcapability{\alpha}{U_{3}}}} \\
   &={\mathof{\min}{{\mathof{\min}{\mathcapability{\alpha}{U_{1}}, \mathcapability{\alpha}{U_{2}}}}, \mathcapability{\alpha}{U_{3}}}} \\
   &={\mathof{\min}{\mathcapability{\alpha}{U_{1}}, \mathcapability{\alpha}{U_{2}}, \mathcapability{\alpha}{U_{3}}}} \\
   &={\mathof{\min}{\mathcapability{\alpha}{U_{1}}, {\mathof{\min}{\mathcapability{\alpha}{U_{2}}, \mathcapability{\alpha}{U_{3}}}} }} \\
   &={\mathof{\min}{\mathcapability{\alpha}{\mathcapability{\alpha}{U_{1}}, \mathpiparallel{U_{2}}{U_{3}}}}} \\
   &={\mathcapability{\alpha}{\mathpiparallel{U_{1}}{\mleft(\mathpiparallel{U_{2}}{U_{3}}\mright)}}} \\
   &={\mathcapability{\alpha}{U'}}
   \intertext{and}
   {\mathobligation{\alpha}{U}}&=
   {\mathobligation{\alpha}{{\mathpiparallel{\mleft(\mathpiparallel{U_{1}}{U_{2}}\mright)}{U_{3}}}}} \\
   &={\mathof{\min}{\mathobligation{\alpha}{\mathpiparallel{U_{1}}{U_{2}}}, \mathobligation{\alpha}{U_{3}}}} \\
   &={\mathof{\min}{\mathobligation{\alpha}{U_{1}}, \mathobligation{\alpha}{U_{2}}, \mathobligation{\alpha}{U_{3}}}} \\
   &={\mathof{\min}{\mathobligation{\alpha}{U_{1}}, \mathobligation{\alpha}{\mathpiparallel{U_{2}}{U_{3}}}}} \\
   &={\mathobligation{\alpha}{\mathpiparallel{U_{1}}{\mleft(\mathpiparallel{U_{2}}{U_{3}}\mright)}}} \\
   &={\mathobligation{\alpha}{U'}}.
  \end{align*}

 \noindent Case 6. \rulename{UP-CongP}.
  Assume ${U_{1}}\mathusagestructuralpo {U'_{1}}$.
  We also assume ${U}={\mathpiparallel{U_{1}}{U_{2}}}$ and 
  ${U'}={\mathpiparallel{U'_{1}}{U_{2}}}$.
  By the induction hypothesis, we have
  ${\mathcapability{\alpha}{U_{1}}}\leq{\mathcapability{\alpha}{U'_{1}}}$ and
  ${\mathobligation{\alpha}{U_{1}}}\geq{\mathobligation{\alpha}{U'_{1}}}$.
  Then 
  \begin{align*}
   {\mathcapability{\alpha}{U}}
   &={\mathcapability{\alpha}{\mathpiparallel{U_{1}}{U_{2}}}} \\
   &={\mathof{\min}{\mathcapability{\alpha}{U_{1}}, \mathcapability{\alpha}{U_{2}}}} \\
   &\leq{\mathof{\min}{\mathcapability{\alpha}{U'_{1}}, \mathcapability{\alpha}{U_{2}}}} \\
   &={\mathcapability{\alpha}{\mathpiparallel{U'_{1}}{U_{2}}}} \\
   &={\mathcapability{\alpha}{U'}}
   \intertext{and}
   {\mathobligation{\alpha}{U}}&={\mathobligation{\alpha}{\mathpiparallel{U_{1}}{U_{2}}}} \\
   &={\mathof{\min}{\mathobligation{\alpha}{U_{1}}, \mathobligation{\alpha}{U_{2}}}} \\
   &\geq{\mathof{\min}{\mathobligation{\alpha}{U'_{1}}, \mathobligation{\alpha}{U_{2}}}} \\
   &={\mathobligation{\alpha}{\mathpiparallel{U'_{1}}{U_{2}}}} \\
   &={\mathobligation{\alpha}{U'}}.
  \end{align*}

  \noindent Case 7. \rulename{UP-Rep}.
  Assume ${U}={\mathpireplication{U_{0}}}$ and 
  ${U'}={\mathpiparallel{\mathpireplication{U_{0}}}{U_{0}}}$.
  Then 
  \begin{align*}
   {\mathcapability{\alpha}{U'}}
   &={\mathcapability{\alpha}{\mathpiparallel{\mathpireplication{U_{0}}}{U_{0}}}} \\
   &={\mathof{\min}{\mathcapability{\alpha}{\mathpireplication{U_{0}}}, \mathcapability{\alpha}{U_{0}}}} \\
   &={\mathcapability{\alpha}{\mathpireplication{U_{0}}}} \\
   &={\mathcapability{\alpha}{U}}
   \intertext{and}
   {\mathobligation{\alpha}{U'}}
   &={\mathobligation{\alpha}{\mathpiparallel{\mathpireplication{U_{0}}}{U_{0}}}} \\
   &={\mathof{\min}{\mathobligation{\alpha}{\mathpireplication{U_{0}}}, \mathobligation{\alpha}{U_{0}}}} \\
   &={\mathobligation{\alpha}{\mathpireplication{U_{0}}}} \\
   &={\mathobligation{\alpha}{U}}.
  \end{align*}
  
  \noindent Case 8. \rulename{UP-$\mathlevelraisefunc{\ast}{\ast}$}.
  Assume ${U}={\mathlevelraise{t_{I}}{t_{O}}{{\beta}^{t_{1}}_{t_{2}}. U_{0}}}$ and 
  ${U'}={{\beta}^{\mathof{\max}{t_{1}, t_{\alpha}}}_{t_{2}}. U_{0}}$.

  Assume ${\beta}={\alpha}$.
  \begin{align*}
   {\mathcapability{\alpha}{U}}
   &={\mathcapability{\alpha}{\mathlevelraise{t_{I}}{t_{O}}{{\beta}^{t_{1}}_{t_{2}}. U_{0}}}} \\
   &={\mathcapability{\alpha}{{\beta}^{t_{1}}_{t_{2}}. U_{0}}} \\
   &={t_{2}} \\
   &={\mathcapability{\alpha}{{\beta}^{\mathof{\max}{t_{1}, t_{\alpha}}}_{t_{2}}. U_{0}}} \\
   &={\mathcapability{\alpha}{U'}}
   \intertext{and}
   {\mathobligation{\alpha}{U}}
   &={\mathobligation{\alpha}{\mathlevelraise{t_{I}}{t_{O}}{{\beta}^{t_{1}}_{t_{2}}. U_{0}}}} \\
   &={\mathof{\max}{t_{\alpha}, \mathobligation{\alpha}{{\beta}^{t_{1}}_{t_{2}}. U_{0}}}} \\
   &={\mathof{\max}{t_{\alpha}, t_{1}}} \\
   &={\mathobligation{\alpha}{{\beta}^{\mathof{\max}{t_{1}, t_{\alpha}}}_{t_{2}}. U_{0}}} \\
   &={\mathobligation{\alpha}{U'}}.
  \end{align*}

  Assume ${\beta}={\mathcoaction{\alpha}}$.
  \begin{align*}
   {\mathcapability{\alpha}{U}}
   &={\mathcapability{\alpha}{\mathlevelraise{t_{I}}{t_{O}}{{\beta}^{t_{1}}_{t_{2}}. U_{0}}}} \\
   &={\mathcapability{\alpha}{{\beta}^{t_{1}}_{t_{2}}. U_{0}}} \\
   &={\infty} \\
   &={\mathcapability{\alpha}{{\beta}^{\mathof{\max}{t_{1}, t_{\alpha}}}_{t_{2}}. U_{0}}} \\
   &={\mathcapability{\alpha}{U'}}
   \intertext{and}
   {\mathobligation{\alpha}{U}}
   &={\mathobligation{\alpha}{\mathlevelraise{t_{I}}{t_{O}}{{\beta}^{t_{1}}_{t_{2}}. U_{0}}}} \\
   &={\mathof{\max}{t_{\alpha}, \mathobligation{\alpha}{{\beta}^{t_{1}}_{t_{2}}. U_{0}}}} \\
   &={\mathof{\max}{t_{\alpha}, \infty}} \\
   &={\mathobligation{\alpha}{{\beta}^{\mathof{\max}{t_{1}, t_{\alpha}}}_{t_{2}}. U_{0}}} \\
   &={\mathobligation{\alpha}{U'}}.
  \end{align*}

  \noindent Case 9. \rulename{UP-Dist}.
  Assume ${U}={\mathlevelraise{t_{I}}{t_{O}}{\mathpiparallel{U_{1}}{U_{2}}}}$ and 
  ${U'}={\mathpiparallel{\mathlevelraise{t_{I}}{t_{O}}{U_{1}}}{\mathlevelraise{t_{I}}{t_{O}}{U_{2}}}}$.
  Then
  \begin{align*}
   {\mathcapability{\alpha}{U}}
   &={\mathcapability{\alpha}{\mathlevelraise{t_{I}}{t_{O}}{\mathpiparallel{U_{1}}{U_{2}}}}} \\
   &={\mathcapability{\alpha}{\mathpiparallel{U_{1}}{U_{2}}}} \\
   &={\mathof{\min}{\mathcapability{\alpha}{U_{1}}, \mathcapability{\alpha}{U_{2}}}} \\
   &={\mathof{\min}{\mathcapability{\alpha}{\mathlevelraise{t_{I}}{t_{O}}{U_{1}}}, \mathcapability{\alpha}{\mathlevelraise{t_{I}}{t_{O}}{U_{2}}}}} \\
   &={\mathcapability{\alpha}{\mathpiparallel{\mathlevelraise{t_{I}}{t_{O}}{U_{1}}}{\mathlevelraise{t_{I}}{t_{O}}{U_{2}}}}} \\
   &={\mathcapability{\alpha}{U'}}
   \intertext{and}
   {\mathobligation{\alpha}{U}}
   &={\mathobligation{\alpha}{\mathlevelraise{t_{I}}{t_{O}}{\mathpiparallel{U_{1}}{U_{2}}}}} \\
   &={\mathof{\max}{t_{\alpha}, \mathobligation{\alpha}{\mathpiparallel{U_{1}}{U_{2}}}}} \\
   &={\mathof{\max}{t_{\alpha}, \mathof{\min}{\mathobligation{\alpha}{U_{1}}, \mathobligation{\alpha}{U_{2}}}}} \\
   &\geq{\mathof{\min}{{\mathof{\max}{t_{\alpha}, \mathobligation{\alpha}{U_{1}}}}, {\mathof{\max}{t_{\alpha}, \mathobligation{\alpha}{U_{2}}}}}} \\
   &={\mathof{\min}{{\mathobligation{\alpha}{\mathlevelraise{t_{I}}{t_{O}}{U_{1}}}}, {\mathobligation{\alpha}{\mathlevelraise{t_{I}}{t_{O}}{U_{2}}}}}} \\
   &={\mathobligation{\alpha}{\mathpiparallel{\mathlevelraise{t_{I}}{t_{O}}{U_{1}}}{\mathlevelraise{t_{I}}{t_{O}}{U_{2}}}}} \\
   &={\mathobligation{\alpha}{U'}}.
  \end{align*}

  \noindent Case 10. \rulename{UP-Or}.
  Fix $i\in \mathsetextension{1, 2}$.
  Assume ${U}={\mathusageor{U_{1}}{U_{2}}}$ and ${U'}={U_{i}}$.
  Then
  \begin{align*}
   {\mathcapability{\alpha}{U}}
   &={\mathcapability{\alpha}{\mathusageor{U_{1}}{U_{2}}}} \\
   &={\mathof{\min}{\mathcapability{\alpha}{U_{1}}, \mathcapability{\alpha}{U_{2}}}} \\
   &\leq{\mathcapability{\alpha}{U_{i}}} \\
   &={\mathcapability{\alpha}{U'}}
   \intertext{and}
   {\mathobligation{\alpha}{U}}
   &={\mathobligation{\alpha}{\mathusageor{U_{1}}{U_{2}}}} \\
   &={\mathof{\max}{{\mathcapability{\alpha}{U_{1}}}, {\mathcapability{\alpha}{U_{2}}}}} \\
   &\geq{\mathobligation{\alpha}{U_{i}}} \\
   &={\mathobligation{\alpha}{U'}}.
  \end{align*}

  \noindent Case 11. \rulename{UP-Cong $\mathlevelraisefunc{\ast}{\ast}$}.
  Assume ${U_{0}}\mathusagestructuralpo{U'_{0}}$, 
  ${U}={\mathlevelraise{t_{I}}{t_{O}}{U_{0}}}$, and
  ${U'}={\mathlevelraise{t_{I}}{t_{O}}{U'_{0}}}$.
  By the induction hypothesis, we have
  ${\mathcapability{\alpha}{U_{0}}}\leq{\mathcapability{\alpha}{U'_{0}}}$ and
  ${\mathobligation{\alpha}{U_{0}}}\geq{\mathobligation{\alpha}{U'_{0}}}$.
  Then
  \begin{align*}
   {\mathcapability{\alpha}{U}}
   &={\mathcapability{\alpha}{\mathlevelraise{t_{I}}{t_{O}}{U_{0}}}} \\
   &={\mathcapability{\alpha}{U_{0}}} \\
   &\leq {\mathcapability{\alpha}{U'_{0}}}\\
   &={\mathcapability{\alpha}{\mathlevelraise{t_{I}}{t_{O}}{U'_{0}}}} \\
   &={\mathcapability{\alpha}{U'}}
   \intertext{and}
   {\mathobligation{\alpha}{U}}
   &={\mathobligation{\alpha}{\mathlevelraise{t_{I}}{t_{O}}{U_{0}}}} \\
   &={\mathof{\max}{t_{\alpha}, \mathobligation{\alpha}{U_{0}}}} \\
   &\geq {\mathof{\max}{t_{\alpha}, \mathobligation{\alpha}{U'_{0}}}} \\
   &={\mathobligation{\alpha}{\mathlevelraise{t_{I}}{t_{O}}{U'_{0}}}} \\
   &={\mathobligation{\alpha}{U'}}.
  \end{align*}

  \noindent Case 12. \rulename{UP-Commut $\mathlevelraisefunc{\ast}{\ast}$}.
  Assume ${U}={\mathlevelraise{t_{I}}{t_{O}}{\mathlevelraise{t'_{I}}{t'_{O}}{U_{0}}}}$
  and ${U'}={\mathlevelraise{t'_{I}}{t'_{O}}{\mathlevelraise{t_{I}}{t_{O}}{U_{0}}}}$.
  Then
  \begin{align*}
   {\mathcapability{\alpha}{U}}
   &={\mathcapability{\alpha}{\mathlevelraise{t_{I}}{t_{O}}{\mathlevelraise{t'_{I}}{t'_{O}}{U_{0}}}}} \\
   &={\mathcapability{\alpha}{\mathlevelraise{t'_{I}}{t'_{O}}{U_{0}}}} \\
   &={\mathcapability{\alpha}{U_{0}}} \\
   &={\mathcapability{\alpha}{\mathlevelraise{t_{I}}{t_{O}}{U_{0}}}} \\
   &={\mathcapability{\alpha}{\mathlevelraise{t'_{I}}{t'_{O}}{\mathlevelraise{t_{I}}{t_{O}}{U_{0}}}}} \\
   &={\mathcapability{\alpha}{U'}}
   \intertext{and}
   {\mathobligation{\alpha}{U}}
   &={\mathobligation{\alpha}{\mathlevelraise{t_{I}}{t_{O}}{\mathlevelraise{t'_{I}}{t'_{O}}{U_{0}}}}} \\
   &={\mathof{\max}{t_{\alpha}, {\mathobligation{\alpha}{\mathlevelraise{t'_{I}}{t'_{O}}{U_{0}}}}}} \\
   &={\mathof{\max}{t_{\alpha}, \mathof{\max}{t'_{\alpha}, {\mathobligation{\alpha}{U_{0}}}}}} \\
   &={\mathof{\max}{t'_{\alpha}, \mathof{\max}{t_{\alpha}, {\mathobligation{\alpha}{U_{0}}}}}} \\
   &={\mathof{\max}{t'_{\alpha}, {\mathobligation{\alpha}{\mathlevelraise{t_{I}}{t_{O}}{U_{0}}}}}} \\
   &={\mathobligation{\alpha}{\mathlevelraise{t'_{I}}{t'_{O}}{\mathlevelraise{t_{I}}{t_{O}}{U_{0}}}}} \\
   &={\mathobligation{\alpha}{U'}}.
  \end{align*}
 \end{proof}

 \subsection{Proof of \cref{prop:property-of-subusages-and-reliability}} % \label[appendix]{subsec:proof-of-prop-property-of-subusage}

  We show each claim.
  
  \noindent \cref{item:subusage-and-suitability-in-prop:posar}
  Let $U_{0}$ and $U_{1}$ usages.
  Assume $\mathissubusage{U_{0}}{U_{1}}$ and $\mathsuitable{\alpha}{U_{0}}$.
  By \cref{def:subusage} \cref{item:capability-in-def:subusage}, we have
  ${\mathcapability{\alpha}{U_{0}}}\leq{\mathcapability{\alpha}{U_{1}}}$.
  By \cref{def:subusage} \cref{item:obligation-in-def:subusage}, we have
  ${\mathobligation{\mathcoaction{\alpha}}{U_{0}}}\geq{\mathobligation{\mathcoaction{\alpha}}{U_{1}}}$.
  By \cref{def:reliability}, we have 
  ${\mathobligation{\mathcoaction{\alpha}}{U_{0}}}\leq{\mathcapability{\alpha}{U_{0}}}$.
  Then 
  ${\mathobligation{\mathcoaction{\alpha}}{U_{1}}}\leq{\mathobligation{\mathcoaction{\alpha}}{U_{0}}}\leq{\mathcapability{\alpha}{U_{0}}}\leq{\mathcapability{\alpha}{U_{1}}}$.
  Thus, $\mathsuitable{\alpha}{U_{1}}$.

  \noindent \cref{item:subusage-and-reliability-in-prop:posar}
  Let $U_{0}$ and $U_{1}$ usages.
  Assume $\mathissubusage{U_{0}}{U_{1}}$ and $\mathreliable{U_{0}}$.
  Fix a usage $U_{1}'$, where ${U_{1}}\mathusagereductionkc{U_{1}'}$.
  By induction and \cref{def:subusage} \cref{item:reduction-in-def:subusage},
  we see that there exists a usage $U_{0}'$ such that
  ${U_{0}}\mathusagereductionkc{U_{0}'}$ and $\mathissubusage{U_{0}'}{U_{1}'}$.
  Since $\mathreliable{U_{0}}$, we have $\mathsuitable{}{U_{0}'}$.
  By \cref{item:subusage-and-suitability-in-prop:posar} in this proposition,
  we have $\mathsuitable{}{U_{1}'}$.
  Thus, $\mathreliable{U_{1}}$.

  \noindent \cref{item:subusage-refl-tans-in-prop:posar}
  Since it is obvious that the identity relation on closed usages satisfies
  all the conditions of \cref{def:subusage}, 
  we have the reflexivity of the subusage relation.

  We show the transitivity of the subusage relation.
  Let 
  \[
   R=\mathsetintension{\mathtuple{U_{0}, U_{1}}}{
  \begin{gathered}
   \mathissubusage{U_{0}}{U_{1}}, \text{ or} \\
   \text{there exits } U_{2} \text{ such that } \\
   \mathissubusage{U_{0}}{U_{2}} \text{ and } \mathissubusage{U_{2}}{U_{1}}
  \end{gathered}
}. 
  \]
  It suffices to show that $R$ satisfies all the conditions of \cref{def:subusage}.
  
  Assume ${\mathtuple{U_{0}, U_{1}}}\in{R}$. 
  Then, either $\mathissubusage{U_{0}}{U_{1}}$, or there exits $U_{2}$ such that
  $\mathissubusage{U_{0}}{U_{2}}$ and $\mathissubusage{U_{2}}{U_{1}}$.
  If $\mathissubusage{U_{0}}{U_{1}}$, 
  then all the conditions of \cref{def:subusage} hold by \cref{def:subusage}.
  Assume $\mathissubusage{U_{0}}{U_{2}}$ and $\mathissubusage{U_{2}}{U_{1}}$.

 \noindent \cref{item:sub-in-def:subusage}
  Fix a usage $U$, where $\mathFVof{U}=\mathsetextension{\rho}$.
  By \cref{def:subusage} \cref{item:sub-in-def:subusage},
  we have $\mathissubusage{\mathsubstbox{U}{\mathsubst{\rho}{U_{0}}}}{\mathsubstbox{U}{\mathsubst{\rho}{U_{2}}}}$
  and 
  $\mathissubusage{\mathsubstbox{U}{\mathsubst{\rho}{U_{2}}}}{\mathsubstbox{U}{\mathsubst{\rho}{U_{1}}}}$.
  Then, we have
  ${\mathtuple{{\mathsubstbox{U}{\mathsubst{\rho}{U_{0}}}}, {\mathsubstbox{U}{\mathsubst{\rho}{U_{1}}}}}}\in{R}$.

  \noindent \cref{item:reduction-in-def:subusage}
  Assume ${U_{1}}\mathusagereduction{U'_{1}}$.
  By \cref{def:subusage} \cref{item:reduction-in-def:subusage},
  there exists $U'_{2}$ such that 
  ${U_{2}}\mathusagereduction{U'_{2}}$ and $\mathissubusage{U'_{2}}{U'_{1}}$.
  By \cref{def:subusage} \cref{item:reduction-in-def:subusage},
  there exists $U'_{0}$ such that 
  ${U_{0}}\mathusagereduction{U'_{0}}$ and $\mathissubusage{U'_{0}}{U'_{2}}$.
  Then, we see that
  there exists $U'_{0}$ such that 
  ${U_{0}}\mathusagereduction{U'_{0}}$ and ${\mathtuple{U'_{0}, U'_{1}}}\in{R}$.

  \noindent \cref{item:capability-in-def:subusage}
  By \cref{def:subusage} \cref{item:capability-in-def:subusage},
  we have
  ${\mathcapability{\alpha}{U_{0}}}\leq{\mathcapability{\alpha}{U_{2}}}$
  and ${\mathcapability{\alpha}{U_{2}}}\leq{\mathcapability{\alpha}{U_{1}}}$,
  for each ${\alpha}\in{\mathsetextension{I, O}}$.
  Then, we have
  ${\mathcapability{\alpha}{U_{0}}}\leq{\mathcapability{\alpha}{U_{1}}}$,
  for each ${\alpha}\in{\mathsetextension{I, O}}$.

  \noindent \cref{item:obligation-in-def:subusage}
  Fix ${\alpha}\in{\mathsetextension{I, O}}$.
  Assume $\mathsuitable{\mathcoaction{\alpha}}{U_{0}}$.
  By \cref{prop:property-of-subusages-and-reliability} \cref{item:subusage-and-suitability-in-prop:posar},
  we have $\mathsuitable{\mathcoaction{\alpha}}{U_{2}}$.
  Then, by \cref{def:subusage} \cref{item:obligation-in-def:subusage},
  we have
  ${\mathobligation{\alpha}{U_{0}}}\geq{\mathobligation{\alpha}{U_{2}}}$ and
  ${\mathobligation{\alpha}{U_{2}}}\geq{\mathobligation{\alpha}{U_{1}}}$.
  Thus,
  ${\mathobligation{\alpha}{U_{0}}}\geq{\mathobligation{\alpha}{U_{1}}}$.
  \qed

\subsection{Property of subusages}

 \begin{proposition} \label[proposition]{prop:property-of-subusage}
  \begin{enumerate}
   \item For closed usages $U$ and $U'$,
	 if ${U}\mathusagestructuralpo{U'}$, then $\mathissubusage{U}{U'}$.
	 \label{item:usagestructuralpo-in-prop:pos}
   \item For closed usages $U$,
	 if $\mathobligation{}{U}=\infty$,
	 $\mathissubusage{U}{\mathusagenil}$.
	 \label{item:ob-infty-and-zero-in-prop:pos}
   \item For closed usages $U$ and $U'$,
	 if $\mathobligation{}{U}=\infty$, then
	 $\mathissubusage{\mathpiparallel{U}{U'}}{U'}$.
	 \label{item:ob-infty-and-parallel-in-prop:pos}
   \item Let $U_{0}$ and $U_{1}$ be closed usages.
	 Then $\mathissubusage{(\mathpiparallel{\mathpireplication{U_{0}}}{\mathpireplication{U_{1}}})}{\mathpireplication{\mleft(\mathpiparallel{U_{0}}{U_{1}}\mright)}}$.
	 \label{item:2-rep-and-parallel-in-prop:pos}
   \item Let $U_{0}, \dots, U_{n}$ be closed usages.
	 Then $\mathissubusage{(\mathpiparallel{\mathpireplication{U_{0}}}{\mathpiparallel{\cdots}{\mathpireplication{U_{n}}}})}{\mathpireplication{\mleft(\mathpiparallel{U_{0}}{\mathpiparallel{\cdots}{U_{n}}}\mright)}}$.
	 \label{item:rep-and-parallel-in-prop:pos}
   \item For usages $U_{0}$, $U_{1}$, $U'_{0}$, and $U'_{1}$, 
	 if $\mathissubusage{U_{0}}{U'_{0}}$ and $\mathissubusage{U_{1}}{U'_{1}}$,
	 then 
	 $\mathissubusage{\mathpiparallel{U_{0}}{U_{1}}}{\mathpiparallel{U'_{0}}{U'_{1}}}$.
	 \label{item:cong-parallel-in-prop:pos}
   \item $\mathissubusage{\mathlevelraise{t_{o}}{t_{c}}{U}}{U}$
	 for a usage $U$ and
	 $t_{o}, t_{c}\in{\mathnat \cup \mathsetextension{\infty}}$.
	 \label{item:gen-raise-op-in-prop:pos}
   \item $\mathissubusage{\mathlevelraiseuni{U}}{U}$ for a usage $U$.
	 \label{item:uni-raise-op-in-prop:pos}
  \end{enumerate}
 \end{proposition}

\begin{proof}
  We show each claim of \cref{prop:property-of-subusage}.

  \noindent \cref{item:usagestructuralpo-in-prop:pos}
  For closed usages $U$ and $U'$ with ${U}\mathusagestructuralpo{U'}$, let
  \[
  {R_{1}^{(U, U')}}={\mathsetintension{\mathtuple{{\mathsubstbox{U_{0}}{\mathsubst{\rho}{U}}}, {\mathsubstbox{U_{0}}{\mathsubst{\rho}{U'}}}}}{
  U_{0} \text{ is a usage with } {\mathFVof{U_{0}}\subseteq{\mathsetextension{\rho}}}
   } }.
  \]
  
  It suffices to show that $R_{1}^{(U, U')}$ satisfies all the conditions of
  \cref{def:subusage}.

  Fix closed usages $U$ and $U'$ with ${U}\mathusagestructuralpo{U'}$.
  Assume \linebreak[3]
  ${\mathtuple{{\mathsubstbox{U_{0}}{\mathsubst{\rho}{U}}}, {\mathsubstbox{U_{0}}{\mathsubst{\rho}{U'}}}}}\in{R_{1}^{(U, U')}}$.

  \noindent \cref{item:sub-in-def:subusage}
  Let $U'_{0}$ be a usage with $\mathFVof{U'_{0}}=\mathsetextension{\rho'}$.
  Since ${\mathFVof{U_{0}}\subseteq{\mathsetextension{\rho}}}$ and 
  $\mathFVof{U'_{0}}=\mathsetextension{\rho'}$, we have \linebreak[3] 
  ${\mathsubstbox{U'_{0}}{\mathsubst{\rho'}{(\mathsubstbox{U_{0}}{\mathsubst{\rho}{U}})}}}={\mathsubstbox{(\mathsubstbox{U'_{0}}{\mathsubst{\rho'}{U_{0}}})}{\mathsubst{\rho}{U}}}$
 and \linebreak[3]
  ${\mathsubstbox{U'_{0}}{\mathsubst{\rho'}{(\mathsubstbox{U_{0}}{\mathsubst{\rho}{U'}})}}}={\mathsubstbox{(\mathsubstbox{U'_{0}}{\mathsubst{\rho'}{U_{0}}})}{\mathsubst{\rho}{U'}}}$.
  \linebreak[3] Hence, \linebreak[3] 
  ${\mathtuple{{\mathsubstbox{U'_{0}}{\mathsubst{\rho'}{(\mathsubstbox{U_{0}}{\mathsubst{\rho}{U}})}}}, {\mathsubstbox{U'_{0}}{\mathsubst{\rho'}{(\mathsubstbox{U_{0}}{\mathsubst{\rho}{U'}})}}}}}\in{R_{1}^{(U, U')}}$.

  \noindent \cref{item:reduction-in-def:subusage}
  To show that $R_{1}^{(U, U')}$ satisfies 
  \cref{def:subusage} \cref{item:reduction-in-def:subusage},
  we show that \linebreak[3] 
  if ${\mathtuple{{\mathsubstbox{U_{0}}{\mathsubst{\rho}{U}}}, {\mathsubstbox{U_{0}}{\mathsubst{\rho}{U'}}}}}\in{R_{1}^{(U, U')}}$, 
  and
  ${\mathsubstbox{U_{0}}{\mathsubst{\rho}{U'}}}\mathusagestructuralpo{\hat{V}}$, then
  there exists a closed usage $\check{V}$ such that
  ${\mathsubstbox{U_{0}}{\mathsubst{\rho}{U}}}\mathusagestructuralpo{\check{V}}$ and
  ${\mathtuple{\check{V}, \hat{V}}}\in{R_{1}^{(U, U')}}$.

  Assume ${\mathtuple{{\mathsubstbox{U_{0}}{\mathsubst{\rho}{U}}}, {\mathsubstbox{U_{0}}{\mathsubst{\rho}{U'}}}}}\in{R_{1}^{(U, U')}}$
  and
  ${\mathsubstbox{U_{0}}{\mathsubst{\rho}{U'}}}\mathusagestructuralpo{\hat{V}}$.
  We show that there exists a closed usage $\check{V}$ such that
  ${\mathsubstbox{U_{0}}{\mathsubst{\rho}{U}}}\mathusagestructuralpo{\check{V}}$ and
  ${\mathtuple{\check{V}, \hat{V}}}\in{R_{1}^{(U, U')}}$.
  The proof is by induction on the construction of 
  ${\mathsubstbox{U_{0}}{\mathsubst{\rho}{U}}}\mathusagestructuralpo{\hat{V}}$.
  
  Assume ${U_{0}}={\rho}$. 
  Then, we have ${\mathsubstbox{U_{0}}{\mathsubst{\rho}{U'}}}={U'}$.
  Since ${U}\mathusagestructuralpo{U'}$ and ${U'}\mathusagestructuralpo{\hat{V}}$,
  we have ${U}\mathusagestructuralpo{\hat{V}}$.
  Let ${\check{V}}={\hat{V}}$.
  Then, we see that
  ${\mathsubstbox{U_{0}}{\mathsubst{\rho}{U}}}\mathusagestructuralpo{\check{V}}$ and
  ${\mathtuple{\check{V}, \hat{V}}}\in{R_{1}^{(U, U')}}$
  if ${\check{V}}={\hat{V}}$.

  We consider other cases according to the last rule of the construction of 
  ${\mathsubstbox{U_{0}}{\mathsubst{\rho}{U}}}\mathusagestructuralpo{\hat{V}}$.
  
  \noindent Case 1. 
  Assume ${\hat{V}}={\mathsubstbox{U_{0}}{\mathsubst{\rho}{U'}}}$.
  Let ${\check{V}}={\mathsubstbox{U_{0}}{\mathsubst{\rho}{U}}}$.
  Then, we have 
  ${\mathsubstbox{U_{0}}{\mathsubst{\rho}{U}}}\mathusagestructuralpo{\check{V}}$ and
  ${\mathtuple{\check{V}, \hat{V}}}\in{R_{1}^{(U, U')}}$.

  \noindent Case 2. 
  Assume that there exists $\hat{V}'$ such that
  ${\mathsubstbox{U_{0}}{\mathsubst{\rho}{U'}}}\mathusagestructuralpo{\hat{V}'}$ and
  ${\hat{V}'}\mathusagestructuralpo{\hat{V}}$.
  By the induction hypothesis, there exists a closed usage $\check{V}'$ such that
  ${\mathsubstbox{U_{0}}{\mathsubst{\rho}{U}}}\mathusagestructuralpo{\check{V}'}$ and
  ${\mathtuple{\check{V}', \hat{V}'}}\in{R_{1}^{(U, U')}}$.
  Since ${\mathtuple{\check{V}', \hat{V}'}}\in{R_{1}^{(U, U')}}$ and 
  ${\hat{V}'}\mathusagestructuralpo{\hat{V}}$,
  the induction hypothesis implies that there exists a closed usage $\check{V}$ such that
  ${\check{V}'}\mathusagestructuralpo{\check{V}}$ and
  ${\mathtuple{\check{V}, \hat{V}}}\in{R_{1}^{(U, U')}}$.
  Since 
  ${{\mathsubstbox{U_{0}}{\mathsubst{\rho}{U}}}}\mathusagestructuralpo{\check{V}'}$, 
  we have ${\mathsubstbox{U_{0}}{\mathsubst{\rho}{U}}}\mathusagestructuralpo{\check{V}}$.

  \noindent Case 3. \rulename{UP-Zero}.
  Assume ${U_{0}}={\mathpiparallel{\mathusagenil}{U_{1}}}$ and
  ${\hat{V}}={\mathsubstbox{U_{1}}{\mathsubst{\rho}{U'}}}$.
  Let ${\check{V}}={\mathsubstbox{U_{1}}{\mathsubst{\rho}{U}}}$.
  Then, we have 
  ${\mathsubstbox{U_{0}}{\mathsubst{\rho}{U}}}\mathusagestructuralpo{\check{V}}$ and
  ${\mathtuple{\check{V}, \hat{V}}}\in{R_{1}^{(U, U')}}$.

  \noindent Case 4. \rulename{UP-Commut}.
  Assume ${U_{0}}={\mathpiparallel{U_{1}}{U_{2}}}$ and
  ${\hat{V}}={\mathpiparallel{\mathsubstbox{U_{2}}{\mathsubst{\rho}{U'}}}{\mathsubstbox{U_{1}}{\mathsubst{\rho}{U'}}}}$.
  Let ${\check{V}}={\mathpiparallel{\mathsubstbox{U_{2}}{\mathsubst{\rho}{U}}}{\mathsubstbox{U_{1}}{\mathsubst{\rho}{U}}}}$.
  Then, we have 
  ${\mathsubstbox{U_{0}}{\mathsubst{\rho}{U}}}\mathusagestructuralpo{\check{V}}$ and
  ${\mathtuple{\check{V}, \hat{V}}}\in{R_{1}^{(U, U')}}$.

  \noindent Case 5. \rulename{UP-Assoc}.
  Assume ${U_{0}}={\mathpiparallel{\rho}{U_{3}}}$, 
  ${U'}={\mathpiparallel{U_{1}}{U_{2}}}$ and
  ${\hat{V}}={\mathpiparallel{U_{1}}{\mleft( \mathpiparallel{U_{2}}{\mathsubstbox{U_{3}}{\mathsubst{\rho}{U'}}} \mright)}}$.
  Then ${\mathsubstbox{U_{0}}{\mathsubst{\rho}{U}}}={\mathpiparallel{U}{\mathsubstbox{U_{3}}{\mathsubst{\rho}{U}}}}$.
  By \rulename{UP-CongP} and transitivity, we have
  \[
   {\mathpiparallel{U}{\mathsubstbox{U_{3}}{\mathsubst{\rho}{U}}}}
  \mathusagestructuralpo{\mathpiparallel{\mleft( \mathpiparallel{U_{1}}{U_{2}} \mright)}{\mathsubstbox{U_{3}}{\mathsubst{\rho}{U}}}}
  \mathusagestructuralpo{\mathpiparallel{U_{1}}{\mleft( \mathpiparallel{U_{2}}{\mathsubstbox{U_{3}}{\mathsubst{\rho}{U}}} \mright)}}.
  \]
  Let \linebreak[3]
  ${\check{V}}={\mathpiparallel{U_{1}}{\mleft( \mathpiparallel{U_{2}}{\mathsubstbox{U_{3}}{\mathsubst{\rho}{U'}}} \mright)}}$.
  Then, we have 
  ${\mathsubstbox{U_{0}}{\mathsubst{\rho}{U}}}\mathusagestructuralpo{\check{V}}$ and
  ${\mathtuple{\check{V}, \hat{V}}}\in{R_{1}^{(U, U')}}$.

  Assume ${U_{0}}={\mathpiparallel{\mleft(\mathpiparallel{U_{1}}{U_{2}}\mright)}{U_{3}}}$ and
  ${\hat{V}}={\mathpiparallel{\mathsubstbox{U_{1}}{\mathsubst{\rho}{U'}}}{\mleft(\mathpiparallel{\mathsubstbox{U_{2}}{\mathsubst{\rho}{U'}}}{\mathsubstbox{U_{3}}{\mathsubst{\rho}{U'}}}\mright)}}$.
  Let \linebreak[3] \\ 
  ${\check{V}}={\mathpiparallel{\mathsubstbox{U_{1}}{\mathsubst{\rho}{U}}}{\mleft( \mathpiparallel{\mathsubstbox{U_{2}}{\mathsubst{\rho}{U}}}{\mathsubstbox{U_{3}}{\mathsubst{\rho}{U}}} \mright)}}$.
  Then, we have 
  ${\mathsubstbox{U_{0}}{\mathsubst{\rho}{U}}}\mathusagestructuralpo{\check{V}}$ and
  ${\mathtuple{\check{V}, \hat{V}}}\in{R_{1}^{(U, U')}}$.

  \noindent Case 6. \rulename{UP-CongP}.
  Assume ${U_{0}}={\mathpiparallel{U_{1}}{U_{2}}}$,
  ${\mathsubstbox{U_{1}}{\mathsubst{\rho}{U'}}}\mathusagestructuralpo{\hat{V}_{1}}$,
  and
  ${\hat{V}}={\mathpiparallel{\hat{V}_{1}}{\mathsubstbox{U_{2}}{\mathsubst{\rho}{U'}}}}$.
  Then, we have \linebreak[3] 
  ${\mathtuple{{\mathsubstbox{U_{1}}{\mathsubst{\rho}{U}}}, {\mathsubstbox{U_{1}}{\mathsubst{\rho}{U'}}}}}\in{R_{1}^{(U, U')}}$.
  By the induction hypothesis, there exists a usage $\check{V}_{1}$ such that 
  ${\mathsubstbox{U_{1}}{\mathsubst{\rho}{U}}}\mathusagestructuralpo{\check{V}_{1}}$ and
  ${\mathtuple{\check{V}_{1}, \hat{V}_{1}}}\in{R_{1}^{(U, U')}}$.
  Since ${\mathtuple{\check{V}_{1}, \hat{V}_{1}}}\in{R_{1}^{(U, U')}}$,
  there exists a usage $U'_{1}$ such that
  ${\mathFVof{U'_{1}}}\subseteq{\mathsetextension{\rho}}$,
  ${\check{V}_{1}}={\mathsubstbox{U'_{1}}{\mathsubst{\rho}{U}}}$, and
  ${\hat{V}_{1}}={\mathsubstbox{U'_{1}}{\mathsubst{\rho}{U'}}}$.
  Let ${\check{V}}={\mathpiparallel{\mathsubstbox{U'_{1}}{\mathsubst{\rho}{U}}}{\mathsubstbox{U_{2}}{\mathsubst{\rho}{U}}}}$.
  Then, we have ${\mathtuple{\check{V}, \hat{V}}}\in{R_{1}^{(U, U')}}$.
  By \rulename{UP-CongP} and transitivity, we see
  \[
   {\mathpiparallel{\mathsubstbox{U_{1}}{\mathsubst{\rho}{U}}}{\mathsubstbox{U_{2}}{\mathsubst{\rho}{U}}}}
  \mathusagestructuralpo
  {\mathpiparallel{\mathsubstbox{U'_{1}}{\mathsubst{\rho}{U}}}{\mathsubstbox{U_{2}}{\mathsubst{\rho}{U}}}}.
  \]

  \noindent Case 7. \rulename{UP-Rep}.
  Assume ${U_{0}}={\mathpireplication{U_{1}}}$ and
  ${\hat{V}}={\mathpiparallel{\mathpireplication{\mathsubstbox{U_{1}}{\mathsubst{\rho}{U'}}}}{\mathsubstbox{U_{1}}{\mathsubst{\rho}{U'}}}}$.
  Let ${\check{V}}= \mathpiparallel{\mathpireplication{\mathsubstbox{U_{1}}{\mathsubst{\rho}{U}}}}{\mathsubstbox{U_{1}}{\mathsubst{\rho}{U'}}}$.
  Then, we have
  ${\mathsubstbox{U_{0}}{\mathsubst{\rho}{U}}}\mathusagestructuralpo{\check{V}}$ and
  ${\mathtuple{\check{V}, \hat{V}}}\in{R_{1}^{(U, U')}}$.

  \noindent Case 8. \rulename{UP-$\mathlevelraisefunc{\ast}{\ast}$}.
  Assume ${U_{0}}={\mathlevelraise{t_{I}}{t_{O}}{\rho}}$,
  ${U'}={\alpha^{t_{1}}_{t_{2}}. U_{1}}$, and
  ${\hat{V}}={\alpha^{\mathof{\max}{t_{1}, t_{\alpha}}}_{t_{2}}. U_{1}}$.
  Since ${U}\mathusagestructuralpo{U'}$, we have
  either ${U}={U'}$ or 
  ${U}={\mathlevelraise{t'_{I}}{t'_{O}}{\alpha^{t'_{1}}_{t_{2}}. U_{1}}}$ 
  with ${t_{1}}={\mathof{\max}{t'_{1}, t'_{\alpha}}}$.

  Assume ${U}={U'}$. Let ${\check{V}}={\hat{V}}$. Then, we have
  ${\mathsubstbox{U_{0}}{\mathsubst{\rho}{U}}}\mathusagestructuralpo{\check{V}}$ and
  ${\mathtuple{\check{V}, \hat{V}}}\in{R_{1}^{(U, U')}}$.
  
  Assume ${U}={\mathlevelraise{t'_{I}}{t'_{O}}{\alpha^{t'_{1}}_{t_{2}}. U_{1}}}$ 
  with ${t_{1}}={\mathof{\max}{t'_{1}, t'_{\alpha}}}$.
  By \rulename{UP-Cong $\mathlevelraisefunc{\ast}{\ast}$} and transitivity,
  we have
  \[
  {\mathlevelraise{t_{I}}{t_{O}}{\mathlevelraise{t'_{I}}{t'_{O}}{\alpha^{t'_{1}}_{t_{2}}. U_{1}}}}
  \mathusagestructuralpo
  {\mathlevelraise{t_{I}}{t_{O}}{\alpha^{t_{1}}_{t_{2}}. U_{1}}}
  \mathusagestructuralpo
  {\alpha^{\mathof{\max}{t_{1}, t_{\alpha}}}_{t_{2}}. U_{1}}.
  \]
  Let ${\check{V}}={\hat{V}}$. Then, we have
  ${\mathsubstbox{U_{0}}{\mathsubst{\rho}{U}}}\mathusagestructuralpo{\check{V}}$ and
  ${\mathtuple{\check{V}, \hat{V}}}\in{R_{1}^{(U, U')}}$.

  Assume ${U_{0}}={\mathlevelraise{t_{I}}{t_{O}}{\alpha^{t_{1}}_{t_{2}}. U_{1}}}$ and
  ${\hat{V}}={\alpha^{\mathof{\max}{t_{1}, t_{\alpha}}}_{t_{2}}. \mathsubstbox{U_{1}}{\mathsubst{\rho}{U'}}}$.
  Let ${\check{V}}={\alpha^{\mathof{\max}{t_{1}, t_{\alpha}}}_{t_{2}}. \mathsubstbox{U_{1}}{\mathsubst{\rho}{U}}}$.
  Then, we have
  ${\mathsubstbox{U_{0}}{\mathsubst{\rho}{U}}}\mathusagestructuralpo{\check{V}}$ and
  ${\mathtuple{\check{V}, \hat{V}}}\in{R_{1}^{(U, U')}}$.

 \noindent Case 9. \rulename{UP-Dist}.
  Assume ${U_{0}}={\mathlevelraise{t_{I}}{t_{O}}{\rho}}$, 
  ${U'}={\mathpiparallel{U_{1}}{U_{2}}}$, and
  ${\hat{V}}={\mathpiparallel{\mathlevelraise{t_{I}}{t_{O}}{U_{1}}}{\mathlevelraise{t_{I}}{t_{O}}{U_{2}}}}$.
  By \rulename{UP-Cong $\mathlevelraisefunc{\ast}{\ast}$} and transitivity,
  \[
  {\mathlevelraise{t_{I}}{t_{O}}{U}}
  \mathusagestructuralpo
  {\mathlevelraise{t_{I}}{t_{O}}{U'}}
  \mathusagestructuralpo
  {\mathpiparallel{\mathlevelraise{t_{I}}{t_{O}}{U_{1}}}{\mathlevelraise{t_{I}}{t_{O}}{U_{2}}}}.
  \]
  Let ${\check{V}}={\hat{V}}$. Then, we have
  ${\mathsubstbox{U_{0}}{\mathsubst{\rho}{U}}}\mathusagestructuralpo{\check{V}}$ and
  ${\mathtuple{\check{V}, \hat{V}}}\in{R_{1}^{(U, U')}}$.

  Assume 
  ${U_{0}}={\mathlevelraise{t_{I}}{t_{O}}{\mathpiparallel{U_{1}}{U_{2}}}}$
  and
  ${\hat{V}}={\mathpiparallel{\mathlevelraise{t_{I}}{t_{O}}{\mathsubstbox{U_{1}}{\mathsubst{\rho}{U'}}}}{\mathlevelraise{t_{I}}{t_{O}}{\mathsubstbox{U_{2}}{\mathsubst{\rho}{U'}}}}}$.
  Let
  ${\check{V}}= \mathpiparallel{\mathlevelraise{t_{I}}{t_{O}}{\mathsubstbox{U_{1}}{\mathsubst{\rho}{U}}}}{\mathlevelraise{t_{I}}{t_{O}}{\mathsubstbox{U_{2}}{\mathsubst{\rho}{U}}}}$.
  Then, we have
  ${\mathsubstbox{U_{0}}{\mathsubst{\rho}{U}}}\mathusagestructuralpo{\check{V}}$ and
  ${\mathtuple{\check{V}, \hat{V}}}\in{R_{1}^{(U, U')}}$.

  \noindent Case 10. \rulename{UP-Or}.
  Fix $i\in \mathsetextension{1, 2}$.
  Assume 
  ${U_{0}}={\mathusageor{U_{1}}{U_{2}}}$
  and
  ${\hat{V}}={\mathsubstbox{U_{i}}{\mathsubst{\rho}{U'}}}$.
  Let ${\check{V}}={\mathsubstbox{U_{i}}{\mathsubst{\rho}{U}}}$.
  Then, we have
  ${\mathsubstbox{U_{0}}{\mathsubst{\rho}{U}}}\mathusagestructuralpo{\check{V}}$ and
  ${\mathtuple{\check{V}, \hat{V}}}\in{R_{1}^{(U, U')}}$.

  \noindent Case 11. \rulename{UP-Cong $\mathlevelraisefunc{\ast}{\ast}$}.
  Assume ${U_{0}}={\mathlevelraise{t_{I}}{t_{O}}{\rho}}$,
  ${U'}\mathusagestructuralpo{U''}$, and 
  ${\hat{V}}={\mathlevelraise{t_{I}}{t_{O}}{U''}}$.
  By \rulename{UP-Cong $\mathlevelraisefunc{\ast}{\ast}$} and transitivity, we have
  \[
  {\mathlevelraise{t_{I}}{t_{O}}{U}}
  \mathusagestructuralpo
  {\mathlevelraise{t_{I}}{t_{O}}{U'}}
  \mathusagestructuralpo
  {\mathlevelraise{t_{I}}{t_{O}}{U''}}.
  \]

  Assume ${U_{0}}={\mathlevelraise{t_{I}}{t_{O}}{U_{1}}}$,
  ${\mathsubstbox{U_{1}}{\mathsubst{\rho}{U'}}}\mathusagestructuralpo{\hat{V}_{1}}$, and 
  ${\hat{V}}={\mathlevelraise{t_{I}}{t_{O}}{\hat{V}_{1}}}$.
  By the induction hypothesis,
  there exists a usage $\check{V}_{1}$ such that 
  ${\mathsubstbox{U_{1}}{\mathsubst{\rho}{U}}}\mathusagestructuralpo{\check{V}_{1}}$ and
  ${\mathtuple{\check{V}_{1}, \hat{V}_{1}}}\in{R_{1}^{(U, U')}}$.
  By \rulename{UP-Cong $\mathlevelraisefunc{\ast}{\ast}$}, we have
  \[
  {\mathlevelraise{t_{I}}{t_{O}}{\mathsubstbox{U_{1}}{\mathsubst{\rho}{U}}}}
  \mathusagestructuralpo
  {\mathlevelraise{t_{I}}{t_{O}}{\check{V}_{1}}}.
  \]
  Let ${\check{V}}={\mathlevelraise{t_{I}}{t_{O}}{\check{V}_{1}}}$.
  Then, we have
  ${\mathsubstbox{U_{0}}{\mathsubst{\rho}{U}}}\mathusagestructuralpo{\check{V}}$ and
  ${\mathtuple{\check{V}, \hat{V}}}\in{R_{1}^{(U, U')}}$.

  \noindent Case 12. \rulename{UP-Commut $\mathlevelraisefunc{\ast}{\ast}$}.
  Assume ${U_{0}}={\mathlevelraise{t_{I}}{t_{O}}{\rho}}$,
  ${U'}={\mathlevelraise{t'_{I}}{t'_{O}}{U_{1}}}$, and
  ${\hat{V}}={\mathlevelraise{t'_{I}}{t'_{O}}{\mathlevelraise{t_{I}}{t_{O}}{U_{1}}}}$.
  Let ${\check{V}}={\hat{V}}$.
  Then ${\mathtuple{\check{V}, \hat{V}}}\in{R_{1}^{(U, U')}}$.
  By \rulename{UP-Cong $\mathlevelraisefunc{\ast}{\ast}$}, we have
  \[
  {\mathsubstbox{U_{0}}{\mathsubst{\rho}{U}}}={\mathlevelraise{t_{I}}{t_{O}}{U}}\mathusagestructuralpo{{\mathlevelraise{t_{I}}{t_{O}}{U'}}}\mathusagestructuralpo{\mathlevelraise{t'_{I}}{t'_{O}}{\mathlevelraise{t_{I}}{t_{O}}{U_{1}}}}={\check{V}}. 
  \]
  Hence, 
  ${\mathsubstbox{U_{0}}{\mathsubst{\rho}{U}}}\mathusagestructuralpo{\check{V}}$.

  Assume ${U_{0}}={\mathlevelraise{t_{I}}{t_{O}}{\mathlevelraise{t'_{I}}{t'_{O}}{U_{1}}}}$ and 
  ${\hat{V}}={\mathlevelraise{t'_{I}}{t'_{O}}{\mathlevelraise{t_{I}}{t_{O}}{\mathsubstbox{U_{1}}{\mathsubst{\rho}{U'}}}}}$.
  Let ${\check{V}}= \linebreak[3] \mathlevelraise{t'_{I}}{t'_{O}}{\mathlevelraise{t_{I}}{t_{O}}{\mathsubstbox{U_{1}}{\mathsubst{\rho}{U}}}}$.
  Then, we have
  ${\mathsubstbox{U_{0}}{\mathsubst{\rho}{U}}}\mathusagestructuralpo{\check{V}}$ and
  ${\mathtuple{\check{V}, \hat{V}}}\in{R_{1}^{(U, U')}}$.

  Now, we show that 
  if ${\mathtuple{{\mathsubstbox{U_{0}}{\mathsubst{\rho}{U}}}, {\mathsubstbox{U_{0}}{\mathsubst{\rho}{U'}}}}}\in{R_{1}^{(U, U')}}$ and
  there exists a closed usage ${\hat{V}}$ 
  such that ${\mathsubstbox{U_{0}}{\mathsubst{\rho}{U'}}}\mathusagereduction{\hat{V}}$, 
  then there exists a usage $\check{V}$ such that
  ${\mathsubstbox{U_{0}}{\mathsubst{\rho}{U}}}\mathusagereduction{\check{V}}$ and
  ${\mathtuple{\check{V}, \hat{V}}}\in{R_{1}^{(U, U')}}$.

  Assume ${\mathtuple{{\mathsubstbox{U_{0}}{\mathsubst{\rho}{U}}}, {\mathsubstbox{U_{0}}{\mathsubst{\rho}{U'}}}}}\in{R_{1}^{(U, U')}}$ 
  and ${\mathsubstbox{U_{0}}{\mathsubst{\rho}{U'}}}\mathusagereduction{\hat{V}}$.
  We show that there exists a usage $\check{V}$ such that
  ${\mathsubstbox{U_{0}}{\mathsubst{\rho}{U}}}\mathusagereduction{\check{V}}$ and
  ${\mathtuple{\check{V}, \hat{V}}}\in{R_{1}^{(U, U')}}$.
  We show the claim by induction on the construction of
  ${\mathsubstbox{U_{0}}{\mathsubst{\rho}{U'}}}\mathusagereduction{\hat{V}}$.

  Assume ${U_{0}}={\rho}$. 
  Then, we have ${\mathsubstbox{U_{0}}{\mathsubst{\rho}{U'}}}={U'}$
  and ${\mathsubstbox{U_{0}}{\mathsubst{\rho}{U}}}={U}$.
  Then, we see
  \[
  {U}\mathusagestructuralpo{U'}\mathusagereduction{\hat{V}}
  \]
  Let ${\check{V}}={\hat{V}}$.
  Then, we see that
  ${\mathsubstbox{U_{0}}{\mathsubst{\rho}{U}}}\mathusagereduction{\check{V}}$ and
  ${\mathtuple{\check{V}, \hat{V}}}\in{R_{1}^{(U, U')}}$
  if ${\check{V}}={\hat{V}}$.

  We consider other cases according to the last rule of the construction of 
  ${\mathsubstbox{U_{0}}{\mathsubst{\rho}{U'}}}\mathusagereduction{\hat{V}}$.

  \noindent Case 1. 
  Assume ${U_{0}}={\mathpiparallel{\rho}{O^{t'_{o}}_{t'_{c}}.U_{2}}}$,
  ${U'}={I^{t_{o}}_{t_{c}}.U_{1}}$, and
  ${\hat{V}}={\mathpiparallel{U_{1}}{\mathsubstbox{U_{2}}{\mathsubst{\rho}{U'}}}}$.

  Since ${U}\mathusagestructuralpo{U'}$, we have
  either ${U}={U'}$ or 
  ${U}={\mathlevelraise{t''_{I}}{t''_{O}}{I^{t_{1}}_{t_{c}}. U_{1}}}$ 
  with ${t_{o}}={\mathof{\max}{t_{1}, t''_{I}}}$.

  Assume ${U}={U'}$. Let ${\check{V}}={\hat{V}}$. Then, we have
  ${\mathsubstbox{U_{0}}{\mathsubst{\rho}{U}}}\mathusagereduction{\check{V}}$ and
  ${\mathtuple{\check{V}, \hat{V}}}\in{R_{1}^{(U, U')}}$.
  
  Assume ${U}={\mathlevelraise{t''_{I}}{t''_{O}}{I^{t_{1}}_{t_{c}}. U_{1}}}$ 
  with ${t_{o}}={\mathof{\max}{t_{1}, t''_{I}}}$.
  By \rulename{UP-$\mathlevelraisefunc{\ast}{\ast}$},
  we have
  \[
  {\mathpiparallel{\mathlevelraise{t''_{I}}{t''_{O}}{I^{t_{1}}_{t_{c}}. U_{1}}}{O^{t'_{o}}_{t'_{c}}. {\mathsubstbox{U_{2}}{\mathsubst{\rho}{U}}}}}
  \mathusagestructuralpo
  {\mathpiparallel{I^{t_{o}}_{t_{c}}. U_{1}}{O^{t'_{o}}_{t'_{c}}. {\mathsubstbox{U_{2}}{\mathsubst{\rho}{U}}}}}
  \mathusagereduction
  {\mathpiparallel{U_{1}}{\mathsubstbox{U_{2}}{\mathsubst{\rho}{U}}}}.
  \]
  Let ${\check{V}}={\mathpiparallel{U_{1}}{\mathsubstbox{U_{2}}{\mathsubst{\rho}{U}}}}$.
  Then, we have
  ${\mathsubstbox{U_{0}}{\mathsubst{\rho}{U}}}\mathusagestructuralpo{\check{V}}$ and
  ${\mathtuple{\check{V}, \hat{V}}}\in{R_{1}^{(U, U')}}$.

  Assume ${U_{0}}={\mathpiparallel{I^{t_{o}}_{t_{c}}.U_{1}}{\rho}}$,
  ${U'}={O^{t'_{o}}_{t'_{c}}.U_{2}}$, and
  ${\hat{V}}={\mathpiparallel{\mathsubstbox{U_{1}}{\mathsubst{\rho}{U'}}}{U_{2}}}$.

  Since ${U}\mathusagestructuralpo{U'}$, we have
  either ${U}={U'}$ or 
  ${U}={\mathlevelraise{t''_{I}}{t''_{O}}{O^{t_{1}}_{t'_{c}}. U_{1}}}$ 
  with ${t'_{o}}={\mathof{\max}{t_{1}, t''_{O}}}$.

  Assume ${U}={U'}$. Let ${\check{V}}={\hat{V}}$. Then, we have
  ${\mathsubstbox{U_{0}}{\mathsubst{\rho}{U}}}\mathusagereduction{\check{V}}$ and
  ${\mathtuple{\check{V}, \hat{V}}}\in{R_{1}^{(U, U')}}$.
  
  Assume ${U}={\mathlevelraise{t''_{I}}{t''_{O}}{O^{t_{1}}_{t'_{c}}. U_{1}}}$ 
  with ${t'_{o}}={\mathof{\max}{t_{1}, t''_{O}}}$.
  By \rulename{UP-Commut} and \rulename{UP-$\mathlevelraisefunc{\ast}{\ast}$},
  we have
  \begin{align*}
   {\mathpiparallel{I^{t_{o}}_{t_{c}}. {\mathsubstbox{U_{1}}{\mathsubst{\rho}{U}}}}{\mathlevelraise{t''_{I}}{t''_{O}}{O^{t_{1}}_{t'_{c}}. U_{2}}}}
  &\mathusagestructuralpo
  {\mathpiparallel{\mathlevelraise{t''_{I}}{t''_{O}}{O^{t_{1}}_{t'_{c}}. U_{2}}}{I^{t_{o}}_{t_{c}}. {\mathsubstbox{U_{1}}{\mathsubst{\rho}{U}}}}} \\
  &\mathusagestructuralpo
  {\mathpiparallel{O^{t'_{o}}_{t'_{c}}. U_{2}}{I^{t_{o}}_{t_{c}}. {\mathsubstbox{U_{1}}{\mathsubst{\rho}{U}}}}} \\
  &\mathusagestructuralpo
  {\mathpiparallel{I^{t_{o}}_{t_{c}}. {\mathsubstbox{U_{1}}{\mathsubst{\rho}{U}}}}{O^{t'_{o}}_{t'_{c}}. U_{2}}} \\
  &\mathusagereduction
  {\mathpiparallel{\mathsubstbox{U_{1}}{\mathsubst{\rho}{U}}}{U_{2}}}.
  \end{align*}
  Let ${\check{V}}={\mathpiparallel{\mathsubstbox{U_{1}}{\mathsubst{\rho}{U}}}{U_{2}}}$.
  Then, we have
  ${\mathsubstbox{U_{0}}{\mathsubst{\rho}{U}}}\mathusagestructuralpo{\check{V}}$ and
  ${\mathtuple{\check{V}, \hat{V}}}\in{R_{1}^{(U, U')}}$.

  Assume 
  ${U_{0}}={\mathpiparallel{I^{t_{o}}_{t_{c}}.U_{1}}{O^{t'_{o}}_{t'_{c}}.U_{2}}}$
  and
  ${\hat{V}}={\mathpiparallel{\mathsubstbox{U_{1}}{\mathsubst{\rho}{U'}}}{\mathsubstbox{U_{2}}{\mathsubst{\rho}{U'}}}}$.
  Let ${\check{V}}={\mathpiparallel{\mathsubstbox{U_{1}}{\mathsubst{\rho}{U}}}{\mathsubstbox{U_{2}}{\mathsubst{\rho}{U}}}}$.
  Then, we have
  ${\mathsubstbox{U_{0}}{\mathsubst{\rho}{U}}}\mathusagereduction{\check{V}}$ and
  ${\mathtuple{\check{V}, \hat{V}}}\in{R_{1}^{(U, U')}}$.

  \noindent Case 2.
  Assume ${U_{0}}={\mathpiparallel{U_{1}}{U_{2}}}$,
  ${\mathsubstbox{U_{1}}{\mathsubst{\rho}{U'}}}\mathusagereduction{\hat{V}_{1}}$, and
  ${\hat{V}}={\mathpiparallel{\hat{V}_{1}}{\mathsubstbox{U_{2}}{\mathsubst{\rho}{U'}}}}$.
  By the induction hypothesis,
  there exists a usage $\check{V}_{1}$ such that
  ${\mathsubstbox{U_{1}}{\mathsubst{\rho}{U}}}\mathusagereduction{\check{V}_{1}}$ and
  ${\mathtuple{\check{V}_{1}, \hat{V}_{1}}}\in{R_{1}^{(U, U')}}$.
  Let ${\check{V}}={\mathpiparallel{\check{V}_{1}}{\mathsubstbox{U_{2}}{\mathsubst{\rho}{U}}}}$.
  Since ${\mathtuple{\check{V}_{1}, \hat{V}_{1}}}\in{R_{1}^{(U, U')}}$,
  there exists a usage $U'_{1}$ such that
  ${\mathFVof{U'_{1}}}\subseteq{\mathsetextension{\rho}}$,
  ${\check{V}_{1}}={\mathsubstbox{U'_{1}}{\mathsubst{\rho}{U}}}$, and
  ${\hat{V}_{1}}={\mathsubstbox{U'_{1}}{\mathsubst{\rho}{U'}}}$.
  Hence, we see
  ${\check{V}}={\mathpiparallel{\mathsubstbox{U'_{1}}{\mathsubst{\rho}{U}}}{\mathsubstbox{U_{2}}{\mathsubst{\rho}{U}}}}$
  and
  ${\hat{V}}={\mathpiparallel{\mathsubstbox{U'_{1}}{\mathsubst{\rho}{U'}}}{\mathsubstbox{U_{2}}{\mathsubst{\rho}{U'}}}}$.
  Then, we have
  ${\mathsubstbox{U_{0}}{\mathsubst{\rho}{U}}}\mathusagereduction{\check{V}}$ and
  ${\mathtuple{\check{V}, \hat{V}}}\in{R_{1}^{(U, U')}}$.

  \noindent Case 3.
  Assume that there exist usages $\hat{V}_{1}$ and $\hat{V}_{2}$ such that
  ${{\mathsubstbox{U_{0}}{\mathsubst{\rho}{U'}}}}\mathusagestructuralpo{\hat{V}_{1}}$,
  ${\hat{V}_{1}}\mathusagereduction{\hat{V}_{2}}$, and ${\hat{V}_{2}}\mathusagestructuralpo{\hat{V}}$.
  Since 
  ${\mathtuple{{\mathsubstbox{U_{0}}{\mathsubst{\rho}{U}}}, {\mathsubstbox{U_{0}}{\mathsubst{\rho}{U'}}}}}\in{R_{1}^{(U, U')}}$
  and
  ${\mathsubstbox{U_{0}}{\mathsubst{\rho}{U'}}}\mathusagestructuralpo{\hat{V}_{1}}$,
  there exists a closed usage $\check{V}_{1}$ such that
  ${\mathsubstbox{U_{0}}{\mathsubst{\rho}{U}}}\mathusagestructuralpo{\check{V}_{1}}$ and
  ${\mathtuple{\check{V}_{1}, \hat{V}_{1}}}\in{R_{1}^{(U, U')}}$.
  By the induction hypothesis,
  there exists a closed usage $\check{V}_{2}$ such that
  ${\check{V}_{1}}\mathusagereduction{\check{V}_{2}}$ and
  ${\mathtuple{\check{V}_{2}, \hat{V}_{2}}}\in{R_{1}^{(U, U')}}$.
  Since 
  ${\mathtuple{\check{V}_{2}, \hat{V}_{2}}}\in{R_{1}^{(U, U')}}$
  and ${\hat{V}_{2}}\mathusagestructuralpo{\hat{V}}$,
  there exists a closed usage $\check{V}$ such that
  ${\check{V}_{2}}\mathusagestructuralpo{\check{V}}$ and
  ${\mathtuple{\check{V}, \hat{V}}}\in{R_{1}^{(U, U')}}$.
  Since
  ${{\mathsubstbox{U_{0}}{\mathsubst{\rho}{U}}}}\mathusagestructuralpo{\check{V}_{1}}$,
  ${\check{V}_{1}}\mathusagereduction{\check{V}_{2}}$, and ${\check{V}_{2}}\mathusagestructuralpo{\check{V}}$,
  we have ${{\mathsubstbox{U_{0}}{\mathsubst{\rho}{U}}}}\mathusagereduction{\check{V}}$.
    
  \noindent \cref{item:capability-in-def:subusage}
  By \cref{lemma:cap_and_ob_subst} \cref{item:cap_leq_lemma-cap_and_ob_subst} and \cref{lemma:usagestrucuralpo-cap-ob}.

  \noindent \cref{item:obligation-in-def:subusage}
  By \cref{lemma:cap_and_ob_subst} \cref{item:ob_geq_lemma-cap_and_ob_subst} and \cref{lemma:usagestrucuralpo-cap-ob}.

  %%%%%%%%%%%%%%%%%%%%%%%%%%%%%%%%%%%%%%%%%%%%%%%%%%%%%%%%%%%%%

  \noindent \cref{item:ob-infty-and-zero-in-prop:pos}
  For a closed usage $U$ with ${\mathobligation{}{U}=\infty}$, let
  \[
  {R_{2}^{(U)}}={\mathsetintension{\mathtuple{{\mathsubstbox{U_{0}}{\mathsubst{\rho}{U}}}, {\mathsubstbox{U_{0}}{\mathsubst{\rho}{\mathusagenil}}}}}{ 
  U_{0} \text{ is a usage with } {\mathFVof{U_{0}}={\mathsetextension{\rho}}} \\
  }}.
  \]
  
  It suffices to show that $R_{2}^{(U)}$ satisfies all the conditions of \cref{def:subusage}.

  Fix a usage $U_{0}$ with ${\mathFVof{U_{0}}\subseteq{\mathsetextension{\rho}}}$.
  Fix a closed usage $U$ with ${\mathobligation{}{U}=\infty}$.
  Assume \linebreak[4] ${\mathtuple{{\mathsubstbox{U_{0}}{\mathsubst{\rho}{U}}}, {\mathsubstbox{U_{0}}{\mathsubst{\rho}{\mathusagenil}}}}}\in{R_{2}^{(U)}}$.

  \noindent \cref{item:sub-in-def:subusage}
  Let $U'$ be a usage with $\mathFVof{U'}=\mathsetextension{\rho'}$.
  Since ${\mathFVof{U_{0}}\subseteq{\mathsetextension{\rho}}}$ and 
  $\mathFVof{U'}=\mathsetextension{\rho'}$, we have
  ${\mathsubstbox{U'}{\mathsubst{\rho'}{(\mathsubstbox{U_{0}}{\mathsubst{\rho}{U}})}}}={\mathsubstbox{(\mathsubstbox{U'}{\mathsubst{\rho'}{U_{0}}})}{\mathsubst{\rho}{U}}}$
 and 
  ${\mathsubstbox{U'}{\mathsubst{\rho'}{(\mathsubstbox{U_{0}}{\mathsubst{\rho}{\mathnil}})}}}={\mathsubstbox{(\mathsubstbox{U'}{\mathsubst{\rho'}{U_{0}}})}{\mathsubst{\rho}{\mathnil}}}$. \linebreak[3]
  Hence, \linebreak[3] 
  ${\mathtuple{{\mathsubstbox{U'}{\mathsubst{\rho'}{(\mathsubstbox{U_{0}}{\mathsubst{\rho}{U}})}}}, {\mathsubstbox{U'}{\mathsubst{\rho'}{(\mathsubstbox{U_{0}}{\mathsubst{\rho}{\mathnil}})}}}}}\in{R_{2}^{(U)}}$.

  \noindent \cref{item:reduction-in-def:subusage}
  To show that $R_{2}^{(U)}$ satisfies
  \cref{def:subusage} \cref{item:reduction-in-def:subusage},
  if ${\mathtuple{{\mathsubstbox{U_{0}}{\mathsubst{\rho}{U}}}, {\mathsubstbox{U_{0}}{\mathsubst{\rho}{\mathusagenil}}}}}\in{R_{2}^{(U)}}$
  and
  ${\mathsubstbox{U_{0}}{\mathsubst{\rho}{\mathusagenil}}}\mathusagestructuralpo{\hat{V}}$, 
  then there exists $\check{V}$ such that 
  ${\mathsubstbox{U_{0}}{\mathsubst{\rho}{U}}}\mathusagestructuralpo{\check{V}}$
  and 
  ${\mathtuple{\check{V}, \hat{V}}}\in{R_{2}^{(U)}}$.

  Assume 
  ${\mathtuple{{\mathsubstbox{U_{0}}{\mathsubst{\rho}{U}}}, {\mathsubstbox{U_{0}}{\mathsubst{\rho}{\mathusagenil}}}}}\in{R_{2}^{(U)}}$
  and
  ${\mathsubstbox{U_{0}}{\mathsubst{\rho}{\mathusagenil}}}\mathusagestructuralpo{\hat{V}}$.
  We show that there exists a closed usage $\check{V}$ such that 
  ${\mathsubstbox{U_{0}}{\mathsubst{\rho}{U}}}\mathusagestructuralpo{\check{V}}$
  and 
  ${\mathtuple{\check{V}, \hat{V}}}\in{R_{2}^{(U)}}$.
  The proof is by the induction on the construction of 
  ${\mathsubstbox{U_{0}}{\mathsubst{\rho}{\mathusagenil}}}\mathusagestructuralpo{\hat{V}}$.
  We consider cases according to the last rule of the construction.

  \noindent Case 1. 
  Assume ${\hat{V}}={\mathsubstbox{U_{0}}{\mathsubst{\rho}{\mathusagenil}}}$. 
  Let ${\check{V}}={\mathsubstbox{U_{0}}{\mathsubst{\rho}{U}}}$.
  Then, we have 
  ${\mathsubstbox{U_{0}}{\mathsubst{\rho}{U}}}\mathusagestructuralpo{\check{V}}$
  and 
  ${\mathtuple{\check{V}, \hat{V}}}\in{R_{2}^{(U)}}$.

  \noindent Case 2. Assume ${\mathsubstbox{U_{0}}{\mathsubst{\rho}{\mathusagenil}}}\mathusagestructuralpo{V'}$ and ${V'}\mathusagestructuralpo{\hat{V}}$.
  By the induction hypothesis,
  there exists a closed usage $\check{V}'$ such that 
  ${\mathsubstbox{U_{0}}{\mathsubst{\rho}{U}}}\mathusagestructuralpo{\check{V}'}$ and 
  ${\mathtuple{\check{V}', V'}}\in{R_{2}^{(U)}}$.
  Since ${\mathtuple{\check{V}', V'}}\in{R_{2}^{(U)}}$ and 
  ${V'}\mathusagestructuralpo{\hat{V}}$,
  the induction hypothesis implies that
  there exists a closed usage $\check{V}$ such that 
  ${\check{V}'}\mathusagestructuralpo{\check{V}}$ and 
  ${\mathtuple{\check{V}, \hat{V}}}\in{R_{2}^{(U)}}$.
  Since ${\mathsubstbox{U_{0}}{\mathsubst{\rho}{U}}}\mathusagestructuralpo{\check{V}'}$
  and ${\check{V}'}\mathusagestructuralpo{\check{V}}$,
  we have ${\mathsubstbox{U_{0}}{\mathsubst{\rho}{U}}}\mathusagestructuralpo{\check{V}}$.

  \noindent Case 3. \rulename{UP-Zero}.
  Assume ${U_{0}}={\mathpiparallel{\mathusagenil}{U_{1}}}$ and
  ${\hat{V}}={\mathsubstbox{U_{1}}{\mathsubst{\rho}{\mathusagenil}}}$.
  Let ${\check{V}}={\mathsubstbox{U_{1}}{\mathsubst{\rho}{U}}}$.
  Then, we have 
  ${\mathsubstbox{U_{0}}{\mathsubst{\rho}{U}}}\mathusagestructuralpo{\check{V}}$ and
  ${\mathtuple{\check{V}, \hat{V}}}\in{R_{2}^{(U)}}$.

  \noindent Case 4. \rulename{UP-Commut}.
  Assume ${U_{0}}={\mathpiparallel{U_{1}}{U_{2}}}$ and
  ${\hat{V}}={\mathpiparallel{\mathsubstbox{U_{2}}{\mathsubst{\rho}{\mathusagenil}} }{ \mathsubstbox{U_{1}}{\mathsubst{\rho}{\mathusagenil}}}}$.
  Let $\check{V}=\mathpiparallel{\mathsubstbox{U_{2}}{\mathsubst{\rho}{U}}}{\mathsubstbox{U_{1}}{\mathsubst{\rho}{U}}}$.
  Then, we have 
  ${\mathsubstbox{U_{0}}{\mathsubst{\rho}{U}}}\mathusagestructuralpo{\check{V}}$ and 
  ${\mathtuple{\check{V}, \hat{V}}}\in{R_{2}^{(U)}}$.

  \noindent Case 5. \rulename{UP-Assoc}.
  Assume ${U_{0}}={\mathpiparallel{\mleft(\mathpiparallel{U_{1}}{U_{2}}\mright)}{U_{3}}}$ 
  and
  ${\hat{V}}={\mathpiparallel{\mathsubstbox{U_{1}}{\mathsubst{\rho}{\mathusagenil}}}{\mleft(\mathpiparallel{\mathsubstbox{U_{2}}{\mathsubst{\rho}{\mathusagenil}}}{\mathsubstbox{U_{3}}{\mathsubst{\rho}{\mathusagenil}}}\mright)}}$.
  Let ${\check{V}}={\mathpiparallel{\mathsubstbox{U_{1}}{\mathsubst{\rho}{U}}}{\mleft(\mathpiparallel{\mathsubstbox{U_{2}}{\mathsubst{\rho}{U}}}{\mathsubstbox{U_{3}}{\mathsubst{\rho}{U}}}\mright)}}$.
  Then, we have 
  ${\mathsubstbox{U_{0}}{\mathsubst{\rho}{U}}}\mathusagestructuralpo{\check{V}}$ and 
  ${\mathtuple{\check{V}, \hat{V}}}\in{R_{2}^{(U)}}$.

  \noindent Case 6. \rulename{UP-CongP}.
  Assume ${U_{0}}={\mathpiparallel{U_{1}}{U_{2}}}$, 
  ${\mathsubstbox{U_{1}}{\mathsubst{\rho}{\mathusagenil}}}\mathusagestructuralpo{\hat{V}_{1}}$, and
  ${\hat{V}}={\mathpiparallel{\hat{V}_{1}}{\mathsubstbox{U_{2}}{\mathsubst{\rho}{\mathusagenil}}}}$.
  By the induction hypothesis, there exists a closed usage $\check{V}_{1}$ such that 
  ${\mathsubstbox{U_{1}}{\mathsubst{\rho}{U}}}\mathusagestructuralpo{\check{V}_{1}}$ and 
  ${\mathtuple{\check{V}_{1}, \hat{V}_{1}}}\in{R_{2}^{(U)}}$.
  Since ${\mathtuple{\check{V}_{1}, \hat{V}_{1}}}\in{R_{2}^{(U)}}$,
  there exists $\check{U}_{0}$ such that
  ${\check{V}_{1}}={\mathsubstbox{\check{U}_{1}}{\mathsubst{\rho}{U}}}$ and
  ${\hat{V}_{1}}={\mathsubstbox{\check{U}_{1}}{\mathsubst{\rho}{\mathusagenil}}}$.
  Let ${\check{V}}={\mathpiparallel{\mathsubstbox{\check{U}_{1}}{\mathsubst{\rho}{U}}}{\mathsubstbox{U_{2}}{\mathsubst{\rho}{U}}}}$.
  Then, we have ${\mathtuple{\check{V}, \hat{V}}}\in{R_{2}^{(U)}}$.
  By \rulename{UP-CongP},
  we have 
  ${\mathpiparallel{\mathsubstbox{U_{1}}{\mathsubst{\rho}{U}}}{\mathsubstbox{U_{2}}{\mathsubst{\rho}{U}}}}\mathusagestructuralpo{\mathpiparallel{\mathsubstbox{\check{U}_{1}}{\mathsubst{\rho}{U}}}{\mathsubstbox{U_{2}}{\mathsubst{\rho}{U}}}}$.

  \noindent Case 7. \rulename{UP-Rep}.
  Assume ${U_{0}}={\mathpireplication{U_{1}}}$ and 
  ${\hat{V}}={\mathpiparallel{\mathpireplication{\mathsubstbox{U_{1}}{\mathsubst{\rho}{\mathusagenil}}}}{\mathsubstbox{U_{1}}{\mathsubst{\rho}{\mathusagenil}}}}$.
  Let
  ${\check{V}}={\mathpiparallel{\mathpireplication{\mathsubstbox{U_{1}}{\mathsubst{\rho}{U}}}}{\mathsubstbox{U_{1}}{\mathsubst{\rho}{U}}}}$.
  Then, we have 
  ${\mathsubstbox{U_{0}}{\mathsubst{\rho}{U}}}\mathusagestructuralpo{\check{V}}$ and 
  ${\mathtuple{\check{V}, \hat{V}}}\in{R_{2}^{(U)}}$.

  \noindent Case 8. \rulename{UP-$\mathlevelraisefunc{\ast}{\ast}$}.
  Assume ${U_{0}}={\mathlevelraise{t_{I}}{t_{O}}{{\alpha}^{t_{1}}_{t_{2}}. U_{1}}}$ 
  and 
  ${\hat{V}}={{{\alpha}^{\mathof{\max}{t_{1}, t_{\alpha}}}_{t_{2}}}. \mathsubstbox{U_{1}}{\mathsubst{\rho}{\mathusagenil}}}$.
  Let ${\check{V}}={{{\alpha}^{\mathof{\max}{t_{1}, t_{\alpha}}}_{t_{2}}}. \mathsubstbox{U_{1}}{\mathsubst{\rho}{U}}}$.
  Then, we have 
  ${\mathsubstbox{U_{0}}{\mathsubst{\rho}{U}}}\mathusagestructuralpo{\check{V}}$ and 
  ${\mathtuple{\check{V}, \hat{V}}}\in{R_{2}^{(U)}}$.

  \noindent Case 9. \rulename{UP-Dist}.
  Assume ${U_{0}}={\mathlevelraise{t_{I}}{t_{O}}{\mathpiparallel{U_{1}}{U_{2}}}}$ and 
  ${\hat{V}}={\mathpiparallel{\mathlevelraise{t_{I}}{t_{O}}{\mathsubstbox{U_{1}}{\mathsubst{\rho}{\mathusagenil}}}}{\mathlevelraise{t_{I}}{t_{O}}{\mathsubstbox{U_{2}}{\mathsubst{\rho}{\mathusagenil}}}}}$.
  Let ${\check{V}}={\mathpiparallel{\mathlevelraise{t_{I}}{t_{O}}{\mathsubstbox{U_{1}}{\mathsubst{\rho}{U}}}}{\mathlevelraise{t_{I}}{t_{O}}{\mathsubstbox{U_{2}}{\mathsubst{\rho}{U}}}}}$.
  Then, we have 
  ${\mathsubstbox{U_{0}}{\mathsubst{\rho}{U}}}\mathusagestructuralpo{\check{V}}$ and 
  ${\mathtuple{\check{V}, \hat{V}}}\in{R_{2}^{(U)}}$.

  \noindent Case 10. \rulename{UP-Or}.
  Fix $i\in \mathsetextension{1, 2}$.
  Assume ${U_{0}}={\mathusageor{U_{1}}{U_{2}}}$ and 
  ${\hat{V}}={\mathsubstbox{U_{i}}{\mathsubst{\rho}{\mathusagenil}}}$.
  Let ${\check{V}}={\mathsubstbox{U_{i}}{\mathsubst{\rho}{U}}}$.
  Then, we have 
  ${\mathsubstbox{U_{0}}{\mathsubst{\rho}{U}}}\mathusagestructuralpo{\check{V}}$ and 
  ${\mathtuple{\check{V}, \hat{V}}}\in{R_{2}^{(U)}}$.

  \noindent Case 11. \rulename{UP-Cong $\mathlevelraisefunc{\ast}{\ast}$}.
  Assume 
  ${U_{0}}={\mathlevelraise{t_{I}}{t_{O}}{U_{1}}}$,
  ${\mathsubstbox{U_{1}}{\mathsubst{\rho}{\mathusagenil}}}\mathusagestructuralpo{\hat{V}_{1}}$, and
  ${\hat{V}}={\mathlevelraise{t_{I}}{t_{O}}{\hat{V}_{1}}}$.
  By the induction hypothesis, there exists a closed usage $\check{V}_{1}$ such that 
  ${\mathsubstbox{U_{1}}{\mathsubst{\rho}{U}}}\mathusagestructuralpo{\check{V}_{1}}$ and 
  ${\mathtuple{\check{V}_{1}, \hat{V}_{1}}}\in{R_{2}^{(U)}}$.
  Since ${\mathtuple{\check{V}_{1}, \hat{V}_{1}}}\in{R_{2}^{(U)}}$,
  there exists $\check{U}_{0}$ such that
  ${\check{V}_{1}}={\mathsubstbox{\check{U}_{1}}{\mathsubst{\rho}{U}}}$ and
  ${\hat{V}_{1}}={\mathsubstbox{\check{U}_{1}}{\mathsubst{\rho}{\mathusagenil}}}$.
  Let ${\check{V}}={\mathlevelraise{t_{I}}{t_{O}}{\mathsubstbox{\check{U}_{1}}{\mathsubst{\rho}{U}}}}$.
  Then, we have ${\mathtuple{\check{V}, \hat{V}}}\in{R_{2}^{(U)}}$.
  By \rulename{UP-Cong $\mathlevelraisefunc{\ast}{\ast}$},
  we have
  ${\mathlevelraise{t_{I}}{t_{O}}{\mathsubstbox{U_{1}}{\mathsubst{\rho}{U}}}}\mathusagestructuralpo{\mathlevelraise{t_{I}}{t_{O}}{\mathsubstbox{\check{U}_{1}}{\mathsubst{\rho}{U}}}}$.

  \noindent Case 12. \rulename{UP-Commut $\mathlevelraisefunc{\ast}{\ast}$}.
  Assume ${U_{0}}={\mathlevelraise{t_{I}}{t_{O}}{\mathlevelraise{t'_{I}}{t'_{O}}{U_{1}}}}$ and 
  ${\hat{V}}={\mathlevelraise{t'_{I}}{t'_{O}}{\mathlevelraise{t_{I}}{t_{O}}{\mathsubstbox{U_{1}}{\mathsubst{\rho}{\mathusagenil}}}}}$.
  Let ${\check{V}}= \mathlevelraise{t'_{I}}{t'_{O}}{\mathlevelraise{t_{I}}{t_{O}}{\mathsubstbox{U_{1}}{\mathsubst{\rho}{U}}}}$.
  Then, we have
  ${\mathsubstbox{U_{0}}{\mathsubst{\rho}{U}}}\mathusagestructuralpo{\check{V}}$ and
  ${\mathtuple{\check{V}, \hat{V}}}\in{R_{2}^{(U)}}$.

  Now, we show that
  if ${\mathtuple{{\mathsubstbox{U_{0}}{\mathsubst{\rho}{U}}}, {\mathsubstbox{U_{0}}{\mathsubst{\rho}{\mathusagenil}}}}}\in{R_{2}^{(U)}}$
  and
  ${\mathsubstbox{U_{0}}{\mathsubst{\rho}{\mathusagenil}}}\mathusagereduction{\hat{V}}$, 
  then there exists $\check{V}$ such that 
  ${\mathsubstbox{U_{0}}{\mathsubst{\rho}{U}}}\mathusagereduction{\check{V}}$
  and 
  ${\mathtuple{\check{V}, \hat{V}}}\in{R_{2}^{(U)}}$.

  Assume 
  ${\mathtuple{{\mathsubstbox{U_{0}}{\mathsubst{\rho}{U}}}, {\mathsubstbox{U_{0}}{\mathsubst{\rho}{\mathusagenil}}}}}\in{R_{2}^{(U)}}$
  and
  ${\mathsubstbox{U_{0}}{\mathsubst{\rho}{\mathusagenil}}}\mathusagereduction{\hat{V}}$.
  We show that there exists a closed usage $\check{V}$ such that 
  ${\mathsubstbox{U_{0}}{\mathsubst{\rho}{U}}}\mathusagereduction{\check{V}}$
  and 
  ${\mathtuple{\check{V}, \hat{V}}}\in{R_{2}^{(U)}}$.
  The proof is by the induction on the construction of 
  ${\mathsubstbox{U_{0}}{\mathsubst{\rho}{\mathusagenil}}}\mathusagereduction{\hat{V}}$.
  We consider cases according to the last rule of the construction.
 
  \noindent Case 1. 
  Assume 
  ${U_{0}}={\mathpiparallel{I^{t_{o}}_{t_{c}}.U_{1}}{O^{t'_{o}}_{t'_{c}}.U_{2}}}$
  and
  ${\hat{V}}={\mathpiparallel{\mathsubstbox{U_{1}}{\mathsubst{\rho}{\mathusagenil}}}{\mathsubstbox{U_{2}}{\mathsubst{\rho}{\mathusagenil}}}}$.
  Let ${\check{V}}={\mathpiparallel{\mathsubstbox{U_{1}}{\mathsubst{\rho}{U}}}{\mathsubstbox{U_{2}}{\mathsubst{\rho}{U}}}}$.
  Then, we have
  ${\mathsubstbox{U_{0}}{\mathsubst{\rho}{U}}}\mathusagereduction{\check{V}}$ and 
  ${\mathtuple{\check{V}, \hat{V}}}\in{R_{2}^{(U)}}$.

  \noindent Case 2.
  Assume ${U_{0}}={\mathpiparallel{U_{1}}{U_{2}}}$,
  ${\mathsubstbox{U_{1}}{\mathsubst{\rho}{\mathusagenil}}}\mathusagereduction{\hat{V}_{1}}$, 
  and 
  ${\hat{V}}={\mathpiparallel{\hat{V}_{1}}{\mathsubstbox{U_{2}}{\mathsubst{\rho}{\mathusagenil}}}}$.
  By the induction hypothesis, there exists $\check{V}_{1}$ such that
  ${\mathsubstbox{U_{1}}{\mathsubst{\rho}{U}}}\mathusagereduction{\check{V}_{1}}$ and 
  ${\mathtuple{\check{V}_{1}, \hat{V}_{1}}}\in{R_{2}^{(U)}}$.
  Since ${\mathtuple{\check{V}_{1}, \hat{V}_{1}}}\in{R_{2}^{(U)}}$,
  there exists $\check{U}_{1}$ such that
  ${\check{V}_{1}}={\mathsubstbox{\check{U}_{1}}{\mathsubst{\rho}{U}}}$ and
  ${\hat{V}_{1}}={\mathsubstbox{\check{U}_{1}}{\mathsubst{\rho}{\mathusagenil}}}$.
  Let ${\check{V}}={\mathpiparallel{\mathsubstbox{\check{U}_{1}}{\mathsubst{\rho}{U}}}{\mathsubstbox{U_{2}}{\mathsubst{\rho}{U}}}}$.
  Then, we have
  ${\mathsubstbox{U_{0}}{\mathsubst{\rho}{U}}}\mathusagereduction{\check{V}}$ and 
  ${\mathtuple{\check{V}, \hat{V}}}\in{R_{2}^{(U)}}$.

  \noindent Case 3.
  Assume there exists usages $V_{1}$ and $V_{2}$ such that
  ${\mathsubstbox{U_{0}}{\mathsubst{\rho}{\mathusagenil}}}\mathusagestructuralpo{V_{1}}$,
  ${V_{1}}\mathusagereduction{V_{2}}$,
  and ${V_{2}}\mathusagestructuralpo{\hat{V}}$.
  Since  
  ${\mathsubstbox{U_{0}}{\mathsubst{\rho}{\mathusagenil}}}\mathusagestructuralpo{V_{1}}$,
  there exists $\check{V}_{1}$ such that
  ${\mathsubstbox{U_{1}}{\mathsubst{\rho}{U}}}\mathusagestructuralpo{\check{V}_{1}}$ and 
  ${\mathtuple{\check{V}_{1}, V_{1}}}\in{R_{2}^{(U)}}$.
  Since ${\mathtuple{\check{V}_{1}, V_{1}}}\in{R_{2}^{(U)}}$ and
  ${V_{1}}\mathusagereduction{V_{2}}$,
  the induction hypothesis implies that
  there exists $\check{V}_{2}$ such that
  ${{\check{V}_{1}}}\mathusagereduction{\check{V}_{2}}$ and 
  ${\mathtuple{\check{V}_{2}, V_{2}}}\in{R_{2}^{(U)}}$.
  Since  
  ${V_{2}}\mathusagestructuralpo{\hat{V}}$,
  there exists $\check{V}$ such that
  ${\check{V}_{2}}\mathusagestructuralpo{\check{V}}$ and 
  ${\mathtuple{\check{V}, \hat{V}}}\in{R_{2}^{(U)}}$.
  Since 
  ${\mathsubstbox{U_{1}}{\mathsubst{\rho}{U}}}\mathusagestructuralpo{\check{V}_{1}}$,
  ${{\check{V}_{1}}}\mathusagereduction{\check{V}_{2}}$, and
  ${\check{V}_{2}}\mathusagestructuralpo{\check{V}}$,
  we have ${\mathsubstbox{U_{1}}{\mathsubst{\rho}{U}}}\mathusagereduction{\check{V}}$.

  \noindent \cref{item:capability-in-def:subusage}
  Let ${\alpha}\in{\mathsetextension{I, O}}$. 
  Then, we have ${\mathcapability{\alpha}{U}}\leq{\infty}={\mathcapability{\alpha}{\mathnil}}$.
  By \cref{lemma:cap_and_ob_subst} \cref{item:cap_leq_lemma-cap_and_ob_subst},
  we see
  ${\mathcapability{\alpha}{\mathsubstbox{U_{0}}{\mathsubst{\rho}{U}}}}\leq{\mathcapability{\alpha}{\mathsubstbox{U_{0}}{\mathsubst{\rho}{\mathusagenil}}}}$.

  \noindent \cref{item:obligation-in-def:subusage}
  Let ${\alpha}\in{\mathsetextension{I, O}}$. 
  By assumption, ${\mathobligation{\alpha}{U}}={\infty}$.
  Then ${\mathobligation{\alpha}{U}}=\infty={\mathobligation{\alpha}{\mathusagenil}}$.
  By \cref{lemma:cap_and_ob_subst} \cref{item:ob_eq_lemma-cap_and_ob_subst}, we have
  ${\mathobligation{\alpha}{\mathsubstbox{U_{0}}{\mathsubst{\rho}{U}}}}={\mathobligation{\alpha}{\mathsubstbox{U_{0}}{\mathsubst{\rho}{\mathusagenil}}}}$.
  Thus,
  ${\mathobligation{\alpha}{\mathsubstbox{U_{0}}{\mathsubst{\rho}{U}}}}\geq{\mathobligation{\alpha}{\mathsubstbox{U_{0}}{\mathsubst{\rho}{\mathusagenil}}}}$.

  %%%%%%%%%%%%%%%%%%%%%%%%%%%%%%%%%%%%%%%%%%%%%%%%%%%%%%%%%%%%%

  \noindent \cref{item:ob-infty-and-parallel-in-prop:pos}
  For closed usages $U$ and $U'$, assume $\mathobligation{}{U}=\infty$.
  By this proposition \cref{item:ob-infty-and-zero-in-prop:pos}, 
  we have $\mathissubusage{U}{\mathusagenil}$.
  From \cref{def:subusage} \cref{item:sub-in-def:subusage}, we see
  $\mathissubusage{\mathpiparallel{U}{U'}}{\mathpiparallel{\mathusagenil}{U'}}$.
  By \rulename{UP-Zero}, ${\mathpiparallel{\mathusagenil}{U'}}\mathusagestructuralpo{U'}$.
  By this proposition \cref{item:usagestructuralpo-in-prop:pos},
  we have $\mathissubusage{\mathpiparallel{\mathusagenil}{U'}}{U'}$.
  By \cref{prop:property-of-subusages-and-reliability} \cref{item:subusage-refl-tans-in-prop:posar},
  we have $\mathissubusage{\mathpiparallel{U}{U'}}{U'}$.

  %%%%%%%%%%%%%%%%%%%%%%%%%%%%%%%%%%%%%%%%%%%%%%%%%%%%%%%%%%%%
  \noindent \cref{item:2-rep-and-parallel-in-prop:pos}
  For closed usages $U_{0}$, $U_{1}$, let
  \[
    {R_{4}^{(U_{0}, U_{1})}}={\mathsetintension{\mathtuple{{\mathsubstbox{U}{\mathsubst{\rho}{\mathpiparallel{\mathpireplication{U_{0}}}{\mathpireplication{U_{1}}}}}}, {\mathsubstbox{U}{\mathsubst{\rho}{\mathpireplication{\mleft(\mathpiparallel{U_{0}}{U_{1}}\mright)}}}}}}{ 
  U \text{ is a usage with } {\mathFVof{U}\subseteq{\mathsetextension{\rho}}}
  }}.
  \]

  It suffices to show that $R_{4}^{(U_{0}, U_{1})}$ satisfies all the conditions of \cref{def:subusage}.

  Fix closed usages $U_{0}$, $U_{1}$.
  Fix a usage $U$ with ${\mathFVof{U}\subseteq{\mathsetextension{\rho}}}$.
  Assume 
  \[
  {\mathtuple{{\mathsubstbox{U}{\mathsubst{\rho}{(\mathpiparallel{\mathpireplication{U_{0}}}{\mathpireplication{U_{1}}})}}}, {\mathsubstbox{U}{\mathsubst{\rho}{\mathpireplication{\mleft(\mathpiparallel{U_{0}}{U_{1}}\mright)}}}}}}\in{R_{4}^{(U_{0}, U_{1})}}. 
  \]

  Let 
  ${W_{0}}={\mathpiparallel{\mathpireplication{U_{0}}}{\mathpireplication{U_{1}}}}$
  and
  ${W_{1}}={\mathpireplication{\mleft(\mathpiparallel{U_{0}}{U_{1}}\mright)}}$.

  \noindent \cref{item:sub-in-def:subusage}
  Let $U'$, where $\mathFVof{U'}=\mathsetextension{\rho'}$.
  Since ${\mathFVof{U}={\mathsetextension{\rho}}}$ and 
  $\mathFVof{U'}=\mathsetextension{\rho'}$, \linebreak[3] we have \linebreak[4]
  $\mathsubstbox{U'}{\mathsubst{\rho'}{(\mathsubstbox{U}{\mathsubst{\rho}{W_{0}}})}}=\mathsubstbox{(\mathsubstbox{U'}{\mathsubst{\rho'}{U}})}{\mathsubst{\rho}{W_{0}}}$
  and
  ${\mathsubstbox{U'}{\mathsubst{\rho'}{(\mathsubstbox{U}{\mathsubst{\rho}{W_{1}}})}}}={\mathsubstbox{(\mathsubstbox{U'}{\mathsubst{\rho'}{U}})}{\mathsubst{\rho}{W_{1}}}}$.
  Hence, 
  \[
  {\mathtuple{{\mathsubstbox{U'}{\mathsubst{\rho'}{(\mathsubstbox{U}{\mathsubst{\rho}{\mleft(\mathpiparallel{\mathpireplication{U_{0}}}{\mathpireplication{U_{1}}}\mright)}})}}}, {\mathsubstbox{U'}{\mathsubst{\rho'}{(\mathsubstbox{U}{\mathsubst{\rho}{\mathpireplication{\mathpiparallel{U_{0}}{U_{1}}}}})}}}}}\in{R_{4}^{(U_{0}, U_{1})}}. 
  \]

  \noindent \cref{item:reduction-in-def:subusage}
  To show that $R_{4}^{(U_{0}, U_{1})}$ satisfies 
  the condition \cref{def:subusage} \cref{item:reduction-in-def:subusage},
  we show that if \linebreak[4] $\mathtuple{{\mathsubstbox{U}{\mathsubst{\rho}{W_{0}}}}, {\mathsubstbox{U}{\mathsubst{\rho}{W_{1}}}}}\in R_{4}^{(U_{0}, U_{1})}$ and
  there exists a usage $\hat{V}$ such that 
  ${\mathsubstbox{U}{\mathsubst{\rho}{W_{1}}}}\mathusagestructuralpo{\hat{V}}$,
  then there exists $\check{V}$ such that \linebreak[2]
  ${\mathsubstbox{U}{\mathsubst{\rho}{W_{0}}}}\mathusagestructuralpo{\check{V}}$
  and ${\mathtuple{\check{V}, \hat{V}}}\in{R_{4}^{(U_{0}, U_{1})}}$.

  Assume ${\mathsubstbox{U}{\mathsubst{\rho}{W_{1}}}}\mathusagestructuralpo{\hat{V}}$.
  The proof is by induction on the construction of \linebreak[3]
  ${{\mathsubstbox{V}{\mathsubst{\rho}{W_{1}}}}}\mathusagestructuralpo{\hat{V}}$.
  We consider cases according to the last rule of the construction.

  \noindent Case 1. Assume ${\hat{V}}={{\mathsubstbox{U}{\mathsubst{\rho}{W_{1}}}}}$.
  Let ${\check{V}}={\mathsubstbox{U}{\mathsubst{\rho}{W_{0}}}}$.
  Then, we have
  ${\mathsubstbox{U}{\mathsubst{\rho}{W_{0}}}}\mathusagestructuralpo{\check{V}}$
  and ${\mathtuple{\check{V}, \hat{V}}}\in{R_{4}^{(U_{0}, U_{1})}}$.  

  \noindent Case 2. Assume 
  ${\mathsubstbox{U}{\mathsubst{\rho}{W_{1}}}}\mathusagestructuralpo{\hat{V}'}$ and
  ${\hat{V}'}\mathusagestructuralpo{\hat{V}}$.
  By the induction hypothesis,
  then there exists $\check{V}'$ such that \linebreak[2]
  ${\mathsubstbox{U}{\mathsubst{\rho}{W_{0}}}}\mathusagestructuralpo{\check{V}'}$ and 
  ${\mathtuple{\check{V}', \hat{V}'}}\in{R_{4}^{(U_{0}, U_{1})}}$.
  Since ${\mathtuple{\check{V}', \hat{V}'}}\in{R_{4}^{(U_{0}, U_{1})}}$ and
  ${\hat{V}'}\mathusagestructuralpo{\hat{V}}$,
  induction implies that there exists $\check{V}$ such that \linebreak[2]
  ${\check{V}'}\mathusagestructuralpo{\check{V}}$ and 
  ${\mathtuple{\check{V}, \hat{V}}}\in{R_{4}^{(U_{0}, U_{1})}}$.
  Since ${\mathsubstbox{U}{\mathsubst{\rho}{W_{0}}}}\mathusagestructuralpo{\check{V}'}$
  and ${\check{V}'}\mathusagestructuralpo{\check{V}}$,
  we have ${\mathsubstbox{U}{\mathsubst{\rho}{W_{0}}}}\mathusagestructuralpo{\check{V}}$.

  \noindent Case 3. \rulename{UP-Zero}.
  Assume ${U}={\mathpiparallel{\mathusagenil}{U'}}$ and
  ${\hat{V}}={\mathsubstbox{U'}{\mathsubst{\rho}{W_{1}}}}$.
  Let ${\check{V}}={\mathsubstbox{U'}{\mathsubst{\rho}{W_{0}}}}$.
  Then, we have
  ${\mathsubstbox{U}{\mathsubst{\rho}{W_{0}}}}\mathusagestructuralpo{\check{V}}$
  and ${\mathtuple{\check{V}, \hat{V}}}\in{R_{4}^{(U_{0}, U_{1})}}$. 

  \noindent Case 4. \rulename{UP-Commut}.
  Assume ${U}={\mathpiparallel{U'_{0}}{U'_{1}}}$ and
  ${\hat{V}}={\mathpiparallel{\mathsubstbox{U'_{1}}{\mathsubst{\rho}{W_{1}}}}{\mathsubstbox{U'_{0}}{\mathsubst{\rho}{W_{1}}}}}$.
  Let ${\check{V}}={\mathpiparallel{\mathsubstbox{U'_{1}}{\mathsubst{\rho}{W_{0}}}}{\mathsubstbox{U'_{0}}{\mathsubst{\rho}{W_{0}}}}}$.
  Then, we have 
  ${\mathsubstbox{U}{\mathsubst{\rho}{W_{0}}}}\mathusagestructuralpo{\check{V}}$ and 
  ${\mathtuple{\check{V}, \hat{V}}}\in{R_{4}^{(U_{0}, U_{1})}}$. 

 \noindent Case 5. \rulename{UP-Assoc}.
  Assume ${U}={\mathpiparallel{\mleft(\mathpiparallel{U'_{0}}{U'_{1}}\mright)}{U'_{2}}}$ 
  and
  ${\hat{V}}={\mathpiparallel{\mathsubstbox{U'_{0}}{\mathsubst{\rho}{W_{1}}}}{\mleft(\mathpiparallel{\mathsubstbox{U'_{1}}{\mathsubst{\rho}{W_{1}}}}{\mathsubstbox{U'_{2}}{\mathsubst{\rho}{W_{1}}}}\mright)}}$.
  Let ${\hat{V}}={\mathpiparallel{\mathsubstbox{U'_{0}}{\mathsubst{\rho}{W_{0}}}}{\mleft(\mathpiparallel{\mathsubstbox{U'_{1}}{\mathsubst{\rho}{W_{0}}}}{\mathsubstbox{U'_{2}}{\mathsubst{\rho}{W_{0}}}}\mright)}}$.
  Then, we have 
  ${\mathsubstbox{U}{\mathsubst{\rho}{W_{0}}}}\mathusagestructuralpo{\check{V}}$ and
  ${\mathtuple{\check{V}, \hat{V}}}\in{R_{4}^{(U_{0}, U_{1})}}$. 

  \noindent Case 6. \rulename{UP-CongP}.
  Assume ${U}={\mathpiparallel{U'_{0}}{U'_{1}}}$, 
  ${\mathsubstbox{U'_{0}}{\mathsubst{\rho}{W_{1}}}}\mathusagestructuralpo{\hat{V}_{0}}$, 
  and
  ${\hat{V}}={\mathpiparallel{\hat{V}_{0}}{\mathsubstbox{U'_{1}}{\mathsubst{\rho}{W_{1}}}}}$.
  By the induction hypothesis, there exists a closed usage $\check{V}_{0}$ such that 
  ${\mathsubstbox{U'_{0}}{\mathsubst{\rho}{W_{0}}}}\mathusagestructuralpo{\check{V}_{0}}$ and
  ${\mathtuple{\check{V}_{0}, \hat{V}_{0}}}\in{R_{4}^{(U_{0}, U_{1})}}$. 
  Since ${\mathtuple{\check{V}_{0}, \hat{V}_{0}}}\in{R_{4}^{(U_{0}, U_{1})}}$,
  there exists $\check{U}_{0}$ such that
  ${\check{V}_{0}}={\mathsubstbox{\check{U}_{0}}{\mathsubst{\rho}{W_{0}}}}$ and
  ${\check{V}_{1}}={\mathsubstbox{\check{U}_{0}}{\mathsubst{\rho}{W_{1}}}}$.
  Let ${\check{V}}={\mathpiparallel{\mathsubstbox{\check{U}_{0}}{\mathsubst{\rho}{W_{0}}}}{\mathsubstbox{U'_{1}}{\mathsubst{\rho}{W_{0}}}}}$.
  Then, we have 
  ${\mathsubstbox{U}{\mathsubst{\rho}{W_{0}}}}\mathusagestructuralpo{\check{V}}$ and
  ${\mathtuple{\check{V}, \hat{V}}}\in{R_{4}^{(U_{0}, U_{1})}}$.   

  \noindent Case 7. \rulename{UP-Rep}.
  Assume ${U}={\rho}$ and
  ${\hat{V}}={\mathpiparallel{\mleft({\mathpireplication{\mleft(\mathpiparallel{U_{0}}{U_{1}}\mright)}}\mright)}{\mleft(\mathpiparallel{U_{0}}{U_{1}}\mright)}}$.
  Let ${U'}={\mathpiparallel{\rho}{\mleft(\mathpiparallel{U_{0}}{U_{1}}\mright)}}$
  and ${\check{V}}={\mathsubstbox{U'}{\mathsubst{\rho}{W_{0}}}}$.
  Then, we have ${\hat{V}}={\mathsubstbox{U'}{\mathsubst{\rho}{W_{1}}}}$.
  Hence, we have ${\mathtuple{\check{V}, \hat{V}}}\in{R_{4}^{(U_{0}, U_{1})}}$.
  By \rulename{UP-Commut}, \rulename{UP-Rep} and \cref{lemma:usage-extra-rules}, we have
  \begin{align*}
   {W_{0}}
   &={\mathpiparallel{\mathpireplication{U_{0}}}{\mathpireplication{U_{1}}}} \\
   &\mathusagestructuralpo{\mathpiparallel{\mleft(\mathpiparallel{\mathpireplication{U_{0}}}{U_{0}}\mright)}{\mleft(\mathpiparallel{\mathpireplication{U_{1}}}{U_{1}}\mright)}} \\
   &\mathusagestructuralpo{\mathpiparallel{\mathpiparallel{\mathpireplication{U_{0}}}{\mathpireplication{U_{1}}}}{\mleft(\mathpiparallel{U_{0}}{U_{1}}\mright)}} \\
   &={\check{V}}.
  \end{align*}

  Assume ${U}={\mathpireplication{U'}}$ and
  ${\hat{V}}={\mathpiparallel{\mathpireplication{\mathsubstbox{U'}{\mathsubst{\rho}{W_{1}}}}}{\mathsubstbox{U'}{\mathsubst{\rho}{W_{1}}}}}$.
  Let ${\check{V}}={\mathpiparallel{\mathpireplication{\mathsubstbox{U'}{\mathsubst{\rho}{W_{0}}}}}{\mathsubstbox{U'}{\mathsubst{\rho}{W_{0}}}}}$. 
  Then, we have 
  ${\mathsubstbox{U}{\mathsubst{\rho}{W_{0}}}}\mathusagestructuralpo{\check{V}}$ and
  ${\mathtuple{\check{V}, \hat{V}}}\in{R_{4}^{(U_{0}, U_{1})}}$. 

  \noindent Case 8. \rulename{UP-$\mathlevelraisefunc{\ast}{\ast}$}.
  Assume ${U}={\mathlevelraise{t_{I}}{t_{O}}{{\alpha}^{t_{1}}_{t_{2}}. U'}}$ 
  and 
  ${\hat{V}}={{{\alpha}^{\mathof{\max}{t_{1}, t_{\alpha}}}_{t_{2}}}. \mathsubstbox{U'}{\mathsubst{\rho}{W_{1}}}}$.
  Let ${\check{V}}={{{\alpha}^{\mathof{\max}{t_{1}, t_{\alpha}}}_{t_{2}}}. \mathsubstbox{U'}{\mathsubst{\rho}{W_{0}}}}$.
  Then, we have 
  ${\mathsubstbox{U}{\mathsubst{\rho}{W_{0}}}}\mathusagestructuralpo{\check{V}}$ and
  ${\mathtuple{\check{V}, \hat{V}}}\in{R_{4}^{(U_{0}, U_{1})}}$. 

  \noindent Case 9. \rulename{UP-Dist}.
  Assume ${U}={\mathlevelraise{t_{I}}{t_{O}}{\mathpiparallel{U'_{0}}{U'_{1}}}}$ and 
  ${\hat{V}}={\mathpiparallel{\mathlevelraise{t_{I}}{t_{O}}{\mathsubstbox{U'_{0}}{\mathsubst{\rho}{W_{1}}}}}{\mathlevelraise{t_{I}}{t_{O}}{\mathsubstbox{U'_{1}}{\mathsubst{\rho}{W_{1}}}}}}$.
  Let ${\hat{V}}={\mathpiparallel{\mathlevelraise{t_{I}}{t_{O}}{\mathsubstbox{U'_{0}}{\mathsubst{\rho}{W_{0}}}}}{\mathlevelraise{t_{I}}{t_{O}}{\mathsubstbox{U'_{1}}{\mathsubst{\rho}{W_{0}}}}}}$.
  Then, we have 
  ${\mathsubstbox{U}{\mathsubst{\rho}{W_{0}}}}\mathusagestructuralpo{\check{V}}$ and
  ${\mathtuple{\check{V}, \hat{V}}}\in{R_{4}^{(U_{0}, U_{1})}}$. 

  \noindent Case 10. \rulename{UP-Or}.
  Fix $i\in \mathsetextension{0, 1}$.
  Assume ${U}={\mathusageor{U'_{0}}{U'_{1}}}$ and 
  ${\hat{V}}={\mathsubstbox{U'_{i}}{\mathsubst{\rho}{W_{1}}}}$.
  Let ${\check{V}}={\mathsubstbox{U'_{i}}{\mathsubst{\rho}{W_{0}}}}$.
  Then, we have 
  ${\mathsubstbox{U}{\mathsubst{\rho}{W_{0}}}}\mathusagestructuralpo{\check{V}}$ and
  ${\mathtuple{\check{V}, \hat{V}}}\in{R_{4}^{(U_{0}, U_{1})}}$. 

  \noindent Case 11. \rulename{UP-Cong $\mathlevelraisefunc{\ast}{\ast}$}.
  Assume ${U}={\mathlevelraise{t_{I}}{t_{O}}{U'}}$,
  ${\mathsubstbox{U'}{\mathsubst{\rho}{W_{1}}}}\mathusagestructuralpo{\hat{V}'}$, and
  ${\hat{V}}={\mathlevelraise{t_{I}}{t_{O}}{\hat{V}'}}$.
  By the induction hypothesis, there exists a closed usage $\check{V}'$ such that 
  ${\mathsubstbox{U'}{\mathsubst{\rho}{W_{0}}}}\mathusagestructuralpo{\check{V}'}$ and
  ${\mathtuple{\check{V}', \hat{V}'}}\in{R_{4}^{(U_{0}, U_{1})}}$. 
  Let ${\check{V}}={\mathlevelraise{t_{I}}{t_{O}}{\check{V}'}}$.
  Then, we have ${\mathsubstbox{U}{\mathsubst{\rho}{W_{0}}}}\mathusagestructuralpo{\check{V}}$.
  By \cref{def:subusage} \cref{item:sub-in-def:subusage} we have shown,
  ${\mathtuple{{\mathlevelraise{t_{I}}{t_{O}}{\check{V}'}}, {\mathlevelraise{t_{I}}{t_{O}}{\hat{V}'}}}}\in{R_{4}^{(U_{0}, U_{1})}}$ i.\ e.\ 
  ${\mathtuple{\check{V}, \hat{V}}}\in{R_{4}^{(U_{0}, U_{1})}}$. 

  \noindent Case 12. \rulename{UP-Commut $\mathlevelraisefunc{\ast}{\ast}$}.
  Assume ${U_{0}}={\mathlevelraise{t_{I}}{t_{O}}{\mathlevelraise{t'_{I}}{t'_{O}}{U_{1}}}}$ and 
  ${\hat{V}}={\mathlevelraise{t'_{I}}{t'_{O}}{\mathlevelraise{t_{I}}{t_{O}}{\mathsubstbox{U_{1}}{\mathsubst{\rho}{W_{1}}}}}}$.
  Let ${\check{V}}= \mathlevelraise{t'_{I}}{t'_{O}}{\mathlevelraise{t_{I}}{t_{O}}{\mathsubstbox{U_{1}}{\mathsubst{\rho}{W_{0}}}}}$.
  Then, we have
  ${\mathsubstbox{U_{0}}{\mathsubst{\rho}{U}}}\mathusagestructuralpo{\check{V}}$ and
  ${\mathtuple{\check{V}, \hat{V}}}\in{R_{4}^{(U_{0}, U_{1})}}$.

  Now, we show that if ${\mathtuple{{\mathsubstbox{U}{\mathsubst{\rho}{W_{0}}}}, {\mathsubstbox{U}{\mathsubst{\rho}{W_{1}}}}}}\in{R_{4}^{(U_{0}, U_{1})}}$ and
  there exists a usage $\hat{V}$ such that 
  ${\mathsubstbox{U}{\mathsubst{\rho}{W_{1}}}}\mathusagereduction{\hat{V}}$,
  then there exists $\check{V}$ such that \linebreak[2]
  ${\mathsubstbox{U}{\mathsubst{\rho}{W_{0}}}}\mathusagereduction{\check{V}}$
  and ${\mathtuple{\check{V}, \hat{V}}}\in{R_{4}^{(U_{0}, U_{1})}}$.

  Assume ${\mathsubstbox{U}{\mathsubst{\rho}{W_{1}}}}\mathusagereduction{\hat{V}}$.
  The proof is by induction on the construction of \linebreak[3]
  ${{\mathsubstbox{V}{\mathsubst{\rho}{W_{1}}}}}\mathusagereduction{\hat{V}}$.
  We consider cases according to the last rule of the construction.

  \noindent Case 1. 
  Assume 
  ${U}={\mathpiparallel{I^{t_{o}}_{t_{c}}.U'_{0}}{O^{t'_{o}}_{t'_{c}}.U'_{1}}}$
  and
  ${\hat{V}}={\mathpiparallel{\mathsubstbox{U'_{0}}{\mathsubst{\rho}{W_{1}}}}{\mathsubstbox{U'_{1}}{\mathsubst{\rho}{W_{1}}}}}$.
  Let ${\check{V}}={\mathpiparallel{\mathsubstbox{U'_{0}}{\mathsubst{\rho}{W_{0}}}}{\mathsubstbox{U'_{1}}{\mathsubst{\rho}{W_{0}}}}}$.
  Then, we have
  ${\mathsubstbox{U}{\mathsubst{\rho}{W_{0}}}}\mathusagereduction{\check{V}}$
  and ${\mathtuple{\check{V}, \hat{V}}}\in{R_{4}^{(U_{0}, U_{1})}}$.

  \noindent Case 2.
  Assume ${U}={\mathpiparallel{U'_{0}}{U'_{1}}}$,
  ${\mathsubstbox{U'_{0}}{\mathsubst{\rho}{W_{1}}}}\mathusagereduction{\hat{V}_{0}}$, 
  and 
  ${\hat{V}}={\mathpiparallel{\hat{V}_{0}}{\mathsubstbox{U'_{1}}{\mathsubst{\rho}{W_{1}}}}}$.
  By the induction hypothesis, there exists $\check{V}_{0}$ such that
  ${\mathsubstbox{U'_{0}}{\mathsubst{\rho}{W_{0}}}}\mathusagereduction{\check{V}_{0}}$
  and ${\mathtuple{\check{V}_{0}, \hat{V}_{0}}}\in{R_{4}^{(U_{0}, U_{1})}}$.
  Since ${\mathtuple{\check{V}_{0}, \hat{V}_{0}}}\in{R_{4}^{(U_{0}, U_{1})}}$,
  there exists $\check{U}_{0}$ such that
  ${\check{V}_{0}}={\mathsubstbox{\check{U}_{0}}{\mathsubst{\rho}{W_{0}}}}$ and
  ${\check{V}_{1}}={\mathsubstbox{\check{U}_{0}}{\mathsubst{\rho}{W_{1}}}}$.
  Let ${\check{V}}={\mathpiparallel{\mathsubstbox{\check{U}_{0}}{\mathsubst{\rho}{W_{0}}}}{\mathsubstbox{U'_{1}}{\mathsubst{\rho}{W_{0}}}}}$.
  Then, we have 
  ${\mathsubstbox{U}{\mathsubst{\rho}{W_{0}}}}\mathusagereduction{\check{V}}$
  and ${\mathtuple{\check{V}, \hat{V}}}\in{R_{4}^{(U_{0}, U_{1})}}$.

  \noindent Case 3.
  Assume there exists usages $V_{0}$ and $V_{1}$ such that
  ${\mathsubstbox{U}{\mathsubst{\rho}{W_{1}}}}\mathusagestructuralpo{V_{0}}$,
  ${V_{0}}\mathusagereduction{V_{1}}$, and ${V_{1}}\mathusagestructuralpo{\hat{V}}$.
  Since  
  ${\mathsubstbox{U}{\mathsubst{\rho}{W_{1}}}}\mathusagestructuralpo{V_{0}}$,
  there exists $\check{V}_{0}$ such that
  ${\mathsubstbox{U}{\mathsubst{\rho}{W_{0}}}}\mathusagestructuralpo{\check{V}_{0}}$
  and ${\mathtuple{\check{V}_{0}, V_{0}}}\in{R_{4}^{(U_{0}, U_{1})}}$.
  Since ${\mathtuple{\check{V}_{0}, V_{0}}}\in{R_{4}^{(U_{0}, U_{1})}}$ and 
  ${V_{0}}\mathusagereduction{V_{1}}$,
  the induction hypothesis implies that there exists $\check{V}_{1}$ such that
  ${\check{V}_{0}}\mathusagereduction{\check{V}_{1}}$ and 
  ${\mathtuple{\check{V}_{1}, V_{1}}}\in{R_{4}^{(U_{0}, U_{1})}}$.
  Since ${\mathtuple{\check{V}_{1}, V_{1}}}\in{R_{4}^{(U_{0}, U_{1})}}$ and
  ${V_{1}}\mathusagestructuralpo{\hat{V}}$,
  there exists $\check{V}$ such that ${\check{V}_{1}}\mathusagestructuralpo{\check{V}}$ and 
  ${\mathtuple{\check{V}, \hat{V}}}\in{R_{4}^{(U_{0}, U_{1})}}$.
  Since
  ${\mathsubstbox{U}{\mathsubst{\rho}{W_{0}}}}\mathusagestructuralpo{\check{V}_{0}}$,
  ${\check{V}_{0}}\mathusagereduction{\check{V}_{1}}$, and 
  ${\check{V}_{1}}\mathusagestructuralpo{\check{V}}$,
  we have ${\mathsubstbox{U}{\mathsubst{\rho}{W_{0}}}}\mathusagereduction{\check{V}}$.

  \noindent \cref{item:capability-in-def:subusage}
  Let ${\alpha}\in{\mathsetextension{I, O}}$. Then
  \begin{align*}
   {\mathcapability{\alpha}{\mathpireplication{\mleft(\mathpiparallel{U_{0}}{U_{1}}\mright)}}}
   &={\mathcapability{\alpha}{\mathpiparallel{U_{0}}{U_{1}}}} \\
   &={\mathof{\min}{{\mathcapability{\alpha}{U_{0}}}, {\mathcapability{\alpha}{U_{1}}}}} \\
   &={\mathof{\min}{{\mathcapability{\alpha}{\mathpireplication{U_{0}}}}, {\mathcapability{\alpha}{\mathpireplication{U_{1}}}}}} \\
   &={\mathcapability{\alpha}{\mathpiparallel{\mathpireplication{U_{0}}}{\mathpireplication{U_{1}}}}}.
  \end{align*}
  By \cref{lemma:cap_and_ob_subst} \cref{item:cap_eq_lemma-cap_and_ob_subst},
  we have
  ${\mathcapability{\alpha}{\mathsubstbox{U}{\mathsubst{\rho}{\mathpireplication{\mleft(\mathpiparallel{U_{0}}{U_{1}}\mright)}}}}}={\mathcapability{\alpha}{\mathsubstbox{U}{\mathsubst{\rho}{(\mathpiparallel{\mathpireplication{U_{0}}}{\mathpireplication{U_{1}}})}}}}$.
  Thus, 
  \[
  {\mathcapability{\alpha}{\mathsubstbox{U}{\mathsubst{\rho}{\mathpireplication{\mleft(\mathpiparallel{U_{0}}{U_{1}}\mright)}}}}}\leq{\mathcapability{\alpha}{\mathsubstbox{U}{\mathsubst{\rho}{(\mathpiparallel{\mathpireplication{U_{0}}}{\mathpireplication{U_{1}}})}}}}. 
  \]

  \noindent \cref{item:obligation-in-def:subusage}
  Let ${\alpha}\in{\mathsetextension{I, O}}$. Then
  \begin{align*}
   {\mathobligation{\alpha}{\mathpireplication{\mleft(\mathpiparallel{U_{0}}{U_{1}}\mright)}}}
   &={\mathobligation{\alpha}{\mathpiparallel{U_{0}}{U_{1}}}} \\
   &={\mathof{\min}{{\mathobligation{\alpha}{U_{0}}}, {\mathobligation{\alpha}{U_{1}}}}} \\
   &={\mathof{\min}{{\mathobligation{\alpha}{\mathpireplication{U_{0}}}}, {\mathobligation{\alpha}{\mathpireplication{U_{1}}}}}} \\
   &={\mathobligation{\alpha}{\mathpiparallel{\mathpireplication{U_{0}}}{\mathpireplication{U_{1}}}}}.
  \end{align*}
  By \cref{lemma:cap_and_ob_subst} \cref{item:ob_eq_lemma-cap_and_ob_subst},
  we have
  ${\mathobligation{\alpha}{\mathsubstbox{U}{\mathsubst{\rho}{\mathpireplication{(\mathpiparallel{U_{0}}{U_{1}})}}}}}={\mathobligation{\alpha}{\mathsubstbox{U}{\mathsubst{\rho}{(\mathpiparallel{\mathpireplication{U_{0}}}{\mathpireplication{U_{1}}})}}}}$.
  Thus,
  \[
  {\mathobligation{\alpha}{\mathsubstbox{U}{\mathsubst{\rho}{\mathpireplication{(\mathpiparallel{U_{0}}{U_{1}})}}}}}\geq{\mathobligation{\alpha}{\mathsubstbox{U}{\mathsubst{\rho}{(\mathpiparallel{\mathpireplication{U_{0}}}{\mathpireplication{U_{1}}})}}}}.
  \]

  \noindent \cref{item:rep-and-parallel-in-prop:pos}
  By this proposition \cref{item:2-rep-and-parallel-in-prop:pos} and 
  \cref{prop:property-of-subusages-and-reliability} \cref{item:subusage-refl-tans-in-prop:posar},
  \begin{align*}
   {\mathpireplication{(\mathpiparallel{U_{0}}{\mathpiparallel{\cdots}{U_{n}}})}}
   &\mathissubusagesyrev{\mathpiparallel{\mathpireplication{(\mathpiparallel{U_{0}}{\mathpiparallel{\cdots}{U_{n-1}}})}}{\mathpireplication{U_{n}}}} \\
   &\mathissubusagesyrev\qquad \vdots               \\
   &\vdots\qquad \vdots               \\
   &\mathissubusagesyrev{(\mathpiparallel{\mathpireplication{U_{0}}}{\mathpiparallel{\cdots}{\mathpireplication{U_{n}}}})}.
  \end{align*}

  \noindent \cref{item:cong-parallel-in-prop:pos}
  Assume $\mathissubusage{U_{0}}{U'_{0}}$ and $\mathissubusage{U_{1}}{U'_{1}}$.
  By \cref{def:subusage} \cref{item:sub-in-def:subusage}, we have
  $\mathissubusage{\mathpiparallel{U_{0}}{U_{1}}}{\mathpiparallel{U'_{0}}{U_{1}}}$. 
  We also have
  $\mathissubusage{\mathpiparallel{U'_{0}}{U_{1}}}{\mathpiparallel{U'_{0}}{U'_{1}}}$.
  By \cref{prop:property-of-subusages-and-reliability} \cref{item:subusage-refl-tans-in-prop:posar},
  we see
  $\mathissubusage{\mathpiparallel{U_{0}}{U_{1}}}{\mathpiparallel{U'_{0}}{U'_{1}}}$. 

  %%%%%%%%%%%%%%%%%%%%%%%%%%%%%%%%%%%%%%%%%%%

  \noindent \cref{item:gen-raise-op-in-prop:pos}
  For a tuple of variables ${\mathvect{\rho}}={\mathtuple{\rho_{0}, \dots, \rho_{n}}}$ and
  a tuple of usages ${\mathvect{U}}={\mathtuple{U_{0}, \dots, U_{n}}}$,
  we abbreviate 
  ${\mathsubstbox{U}{\mathsubst{\rho_{0}}{U_{0}}, \dots, \mathsubst{\rho_{n}}{U_{n}}}}$ 
  for ${\mathsubstbox{U}{\mathsubst{\mathvect{\rho}}{\mathvect{U}}}}$.

  Let
  \[
  {R_{5}}={
  \bigcup_{n=0}^{\infty}
  \mathsetintension{\mathtuple{
  \begin{aligned}
   &{\mathsubstbox{V}{
   {\mathsubst{\rho_{0}}{\mathlevelraise{{t_{I}}_{0}}{{t_{O}}_{0}}{U_{0}}}}, \dots, 
   {\mathsubst{\rho_{n}}{\mathlevelraise{{t_{I}}_{n}}{{t_{O}}_{n}}{U_{n}}}}   
   }}, \\
   &{\mathsubstbox{V}{{\mathsubst{\rho_{0}}{U_{0}}}, \dots, {\mathsubst{\rho_{n}}{U_{n}}}}}
  \end{aligned}
  }}{ 
  \begin{aligned} 
   &U_{0}, \dots, U_{n} \text{ are closed usages, and } \\
   &V \text{ is a usage with } {\mathFVof{V}\subseteq{\mathsetextension{\rho_{0}, \dots, \rho_{n}}}} 
  \end{aligned}
  }}.
  \]
  
  It suffices to show that $R_{5}$ satisfies all the conditions of \cref{def:subusage}.

  Fix closed usages $U_{0}, \dots, U_{n}$,
  a usage $V$ with ${\mathFVof{V}\subseteq{\mathsetextension{\rho_{0}, \dots, \rho_{n}}}}$,
  and
  ${{{t_{O}}_{0}}, \dots, {{t_{O}}_{n}}, {{t_{I}}_{0}}, \dots {{t_{I}}_{n}}}\in{\mathnat \cup \mathsetextension{\infty}}$.
  Let ${\mathvect{\rho}}={\mathtuple{\rho_{0}, \dots, \rho_{n}}}$,
  ${\mathvect{U}}={\mathtuple{U_{0}, \dots, U_{n}}}$, and
  ${\mathvect{U'}}={\mathtuple{{\mathlevelraise{{t_{I}}_{0}}{{t_{O}}_{0}}{U_{0}}}, \dots, {\mathlevelraise{{t_{I}}_{n}}{{t_{O}}_{n}}{U_{n}}}}}$.
  Assume ${\mathtuple{{\mathsubstbox{V}{\mathsubst{\mathvect{\rho}}{\mathvect{U'}}}}, {\mathsubstbox{V}{\mathsubst{\mathvect{\rho}}{\mathvect{U}}}}}}\in{R_{5}}$.

  \noindent \cref{item:sub-in-def:subusage}
  Let $V'$ be a usage with $\mathFVof{V'}=\mathsetextension{\rho'}$.
  Since ${\mathFVof{V}\subseteq{\mathsetextension{\rho_{0}, \dots, \rho_{n}}}}$ and 
  $\mathFVof{V'}=\mathsetextension{\rho'}$, we have
  ${\mathsubstbox{V'}{\mathsubst{\rho'}{\mleft(\mathsubstbox{V}{\mathsubst{\mathvect{\rho}}{\mathvect{U}}}\mright)}}}={\mathsubstbox{(\mathsubstbox{V'}{\mathsubst{\rho'}{V}})}{\mathsubst{\mathvect{\rho}}{\mathvect{U}}}}$
  and 
  ${\mathsubstbox{V'}{\mathsubst{\rho'}{\mleft(\mathsubstbox{V}{\mathsubst{\mathvect{\rho}}{\mathvect{U'}}}\mright)}}}={\mathsubstbox{(\mathsubstbox{V'}{\mathsubst{\rho'}{V}})}{\mathsubst{\mathvect{\rho}}{\mathvect{U'}}}}$.
  Hence, \linebreak[3] 
  ${\mathtuple{{\mathsubstbox{V'}{\mathsubst{\rho'}{\mleft(\mathsubstbox{V}{\mathsubst{\mathvect{\rho}}{\mathvect{U'}}}\mright)}}}, {\mathsubstbox{V'}{\mathsubst{\rho'}{\mleft(\mathsubstbox{V}{\mathsubst{\mathvect{\rho}}{\mathvect{U}}}\mright)}}}}}\in{R_{5}}$.

  \noindent \cref{item:reduction-in-def:subusage}
  To show that $R_{5}$ satisfies
  \cref{def:subusage} \cref{item:reduction-in-def:subusage},
  if ${\mathtuple{{\mathsubstbox{V}{\mathsubst{\mathvect{\rho}}{\mathvect{U'}}}}, {\mathsubstbox{V}{\mathsubst{\mathvect{\rho}}{\mathvect{U}}}}}}\in{R_{5}}$
  and
  ${\mathsubstbox{V}{\mathsubst{\mathvect{\rho}}{\mathvect{U}}}}\mathusagestructuralpo{\hat{V}}$, 
  then there exists $\check{V}$ such that 
  ${\mathsubstbox{V}{\mathsubst{\mathvect{\rho}}{\mathvect{U'}}}}\mathusagestructuralpo{\check{V}}$
  and ${\mathtuple{\check{V}, \hat{V}}}\in{R_{5}}$.

  Assume 
  ${\mathtuple{{\mathsubstbox{V}{\mathsubst{\mathvect{\rho}}{\mathvect{U'}}}}, {\mathsubstbox{V}{\mathsubst{\mathvect{\rho}}{\mathvect{U}}}}}}\in{R_{5}}$
  and
  ${\mathsubstbox{V}{\mathsubst{\mathvect{\rho}}{\mathvect{U}}}}\mathusagestructuralpo{\hat{V}}$.
  We show that there exists a closed usage $\check{V}$ such that 
  ${\mathsubstbox{V}{\mathsubst{\mathvect{\rho}}{\mathvect{U'}}}}\mathusagestructuralpo{\check{V}}$
  and ${\mathtuple{\check{V}, \hat{V}}}\in{R_{5}}$.
  The proof is by the induction on the construction of 
  ${\mathsubstbox{V}{\mathsubst{\mathvect{\rho}}{\mathvect{U}}}}\mathusagestructuralpo{\hat{V}}$.

  Assume ${V}={\rho_{i}}$ with some $i=0, \dots, n$. 
  Then, we have ${\mathsubstbox{V}{\mathsubst{\mathvect{\rho}}{\mathvect{U}}}}={U_{i}}$ and
  ${\mathsubstbox{V}{\mathsubst{\mathvect{\rho}}{\mathvect{U'}}}}={\mathlevelraise{t_{I_{i}}}{t_{O_{i}}}{U_{i}}}$.
  Let ${\check{V}}={\mathlevelraise{t_{I_{i}}}{t_{O_{i}}}{\hat{V}}}$.
  Then, we have 
  ${\mathtuple{{\mathlevelraise{t_{I_{i}}}{t_{O_{i}}}{\hat{V}}}, \hat{V}}}\in{R_{5}}$.
  By \rulename{UP-Cong $\mathlevelraisefunc{\ast}{\ast}$},
  ${U_{i}}\mathusagestructuralpo{\hat{V}}$ implies
  ${\mathlevelraise{t_{I_{i}}}{t_{O_{i}}}{U_{i}}}\mathusagestructuralpo{\check{V}}$.

  We consider other cases according to the last rule of the construction of ${\mathsubstbox{V}{\mathsubst{\mathvect{\rho}}{\mathvect{U}}}}\mathusagestructuralpo{\hat{V}}$.

  \noindent Case 1. 
  Assume ${\hat{V}}={\mathsubstbox{V}{\mathsubst{\mathvect{\rho}}{\mathvect{U}}}}$. 
  Let ${\check{V}}={\mathsubstbox{V}{\mathsubst{\mathvect{\rho}}{\mathvect{U'}}}}$.
  Then, we have 
  ${\mathsubstbox{V}{\mathsubst{\mathvect{\rho}}{\mathvect{U'}}}}\mathusagestructuralpo{\check{V}}$
  and 
  ${\mathtuple{\check{V}, \hat{V}}}\in{R_{5}}$.

  \noindent Case 2. Assume ${\mathsubstbox{V}{\mathsubst{\mathvect{\rho}}{\mathvect{U}}}}\mathusagestructuralpo{V'}$ and ${V'}\mathusagestructuralpo{\hat{V}}$.
  By the induction hypothesis,
  there exists a closed usage $\check{V}'$ such that 
  ${\mathsubstbox{V}{\mathsubst{\mathvect{\rho}}{\mathvect{U'}}}}\mathusagestructuralpo{\check{V}'}$ and 
  ${\mathtuple{\check{V}', V'}}\in{R_{5}}$.
  Since ${\mathtuple{\check{V}', V'}}\in{R_{5}}$ and 
  ${V'}\mathusagestructuralpo{\hat{V}}$,
  the induction hypothesis implies that
  there exists a closed usage $\check{V}$ such that 
  ${\check{V}'}\mathusagestructuralpo{\check{V}}$ and 
  ${\mathtuple{\check{V}, \hat{V}}}\in{R_{5}}$.
  Since ${\mathsubstbox{V}{\mathsubst{\mathvect{\rho}}{\mathvect{U'}}}}\mathusagestructuralpo{\check{V}'}$
  and ${\check{V}'}\mathusagestructuralpo{\check{V}}$,
  we have ${\mathsubstbox{V}{\mathsubst{\mathvect{\rho}}{\mathvect{U'}}}}\mathusagestructuralpo{\check{V}}$.

  \noindent Case 3. \rulename{UP-Zero}.
  Assume ${V}={\mathpiparallel{\mathusagenil}{V_{0}}}$ and
  ${\hat{V}}={\mathsubstbox{V_{0}}{\mathsubst{\mathvect{\rho}}{\mathvect{U}}}}$.
  Let ${\check{V}}={\mathsubstbox{V_{0}}{\mathsubst{\mathvect{\rho}}{\mathvect{U'}}}}$.
  Then, we have 
  ${\mathsubstbox{V}{\mathsubst{\mathvect{\rho}}{\mathvect{U'}}}}\mathusagestructuralpo{\check{V}}$ and
  ${\mathtuple{\check{V}, \hat{V}}}\in{R_{5}}$.

  \noindent Case 4. \rulename{UP-Commut}.
  Assume ${V}={\mathpiparallel{V_{0}}{V_{1}}}$ and
  ${\hat{V}}={\mathpiparallel{\mathsubstbox{V_{1}}{\mathsubst{\mathvect{\rho}}{\mathvect{U}}}}{\mathsubstbox{V_{0}}{\mathsubst{\mathvect{\rho}}{\mathvect{U}}}}}$.
  Let ${\check{V}}=\mathpiparallel{\mathsubstbox{V_{1}}{\mathsubst{\mathvect{\rho}}{\mathvect{U'}}}}{\mathsubstbox{V_{0}}{\mathsubst{\mathvect{\rho}}{\mathvect{U'}}}}$.
  Then, we have 
  ${\mathsubstbox{V}{\mathsubst{\mathvect{\rho}}{\mathvect{U'}}}}\mathusagestructuralpo{\check{V}}$ and 
  ${\mathtuple{\check{V}, \hat{V}}}\in{R_{5}}$.

 \noindent Case 5. \rulename{UP-Assoc}.
  Assume ${V}={\mathpiparallel{\rho_{i}}{V_{2}}}$, 
  ${U_{i}}={\mathpiparallel{V_{0}}{V_{1}}}$ and
  ${\hat{V}}={\mathpiparallel{V_{0}}{\mleft( \mathpiparallel{V_{1}}{\mathsubstbox{V_{2}}{\mathsubst{\mathvect{\rho}}{\mathvect{U}}}} \mright)}}$
  for some $i=0, \dots, n$. 
  Then ${\mathsubstbox{V}{\mathsubst{\mathvect{\rho}}{\mathvect{U'}}}}={\mathpiparallel{\mleft(\mathlevelraise{t_{I_{i}}}{t_{O_{i}}}{\mleft(\mathpiparallel{V_{0}}{V_{1}}\mright)}\mright)}{\mathsubstbox{V_{2}}{\mathsubst{\mathvect{\rho}}{\mathvect{U'}}}}}$.
  By \rulename{UP-Dist} and transitivity, we have
  \begin{align*}
   {\mathpiparallel{\mathlevelraise{t_{I_{i}}}{t_{O_{i}}}{\mleft(\mathpiparallel{V_{0}}{V_{1}}\mright)}}{\mathsubstbox{V_{2}}{\mathsubst{\mathvect{\rho}}{\mathvect{U'}}}}}
  &\mathusagestructuralpo{\mathpiparallel{\mathpiparallel{\mathlevelraise{t_{I_{i}}}{t_{O_{i}}}{V_{0}}}{\mathlevelraise{t_{I_{i}}}{t_{O_{i}}}{V_{1}}}}{\mathsubstbox{V_{2}}{\mathsubst{\mathvect{\rho}}{\mathvect{U'}}}}} \\
  &\mathusagestructuralpo{\mathpiparallel{\mathlevelraise{t_{I_{i}}}{t_{O_{i}}}{V_{0}}}{\mleft( \mathpiparallel{\mathlevelraise{t_{I_{i}}}{t_{O_{i}}}{V_{1}}}{\mathsubstbox{V_{2}}{\mathsubst{\mathvect{\rho}}{\mathvect{U'}}}} \mright)}}.
  \end{align*}
  Let \linebreak[3]
  ${\check{V}}={\mathpiparallel{\mathlevelraise{t_{I_{i}}}{t_{O_{i}}}{V_{0}}}{\mleft( \mathpiparallel{\mathlevelraise{t_{I_{i}}}{t_{O_{i}}}{V_{1}}}{\mathsubstbox{V_{2}}{\mathsubst{\mathvect{\rho}}{\mathvect{U'}}}} \mright)}}$.
  Then, we have 
  ${\mathsubstbox{V}{\mathsubst{\mathvect{\rho}}{\mathvect{U'}}}}\mathusagestructuralpo{\check{V}}$ and
  ${\mathtuple{\check{V}, \hat{V}}}\in{R_{5}}$.

  Assume ${V}={\mathpiparallel{\mleft(\mathpiparallel{V_{0}}{V_{1}}\mright)}{V_{2}}}$ 
  and
  ${\hat{V}}={\mathpiparallel{\mathsubstbox{V_{0}}{\mathsubst{\mathvect{\rho}}{\mathvect{U}}}}{\mleft(\mathpiparallel{\mathsubstbox{V_{1}}{\mathsubst{\mathvect{\rho}}{\mathvect{U}}}}{\mathsubstbox{V_{2}}{\mathsubst{\mathvect{\rho}}{\mathvect{U}}}}\mright)}}$.
  Let ${\check{V}}=\mathpiparallel{\mathsubstbox{V_{0}}{\mathsubst{\mathvect{\rho}}{\mathvect{U'}}}}{\mleft(\mathpiparallel{\mathsubstbox{V_{1}}{\mathsubst{\mathvect{\rho}}{\mathvect{U'}}}}{\mathsubstbox{V_{2}}{\mathsubst{\mathvect{\rho}}{\mathvect{U'}}}}\mright)}$.
  Then, we have 
  ${\mathsubstbox{V}{\mathsubst{\mathvect{\rho}}{\mathvect{U'}}}}\mathusagestructuralpo{\check{V}}$ and 
  ${\mathtuple{\check{V}, \hat{V}}}\in{R_{5}}$.

  \noindent Case 6. \rulename{UP-CongP}.
  Assume ${V}={\mathpiparallel{V_{0}}{V_{1}}}$, 
  ${\mathsubstbox{V_{0}}{\mathsubst{\mathvect{\rho}}{\mathvect{U}}}}\mathusagestructuralpo{\hat{V}_{0}}$, and
  ${\hat{V}}={\mathpiparallel{\hat{V}_{0}}{\mathsubstbox{V_{1}}{\mathsubst{\mathvect{\rho}}{\mathvect{U}}}}}$.
  By the induction hypothesis, there exists a closed usage $\check{V}_{0}$ such that 
  ${\mathsubstbox{V_{0}}{\mathsubst{\mathvect{\rho}}{\mathvect{U'}}}}\mathusagestructuralpo{\check{V}_{0}}$ and 
  ${\mathtuple{\check{V}_{0}, \hat{V}_{0}}}\in{R_{5}}$.
  Since ${\mathtuple{\check{V}_{0}, \hat{V}_{0}}}\in{R_{5}}$,
  there exists closed usages $W_{0}, \dots, W_{m}$, 
  a usage $V'$ with $\mathFVof{V'}\subseteq{\mathsetextension{\rho'_{0}, \dots, \rho'_{m}}}$, 
  and
  ${{{t'_{O}}_{0}}, \dots, {{t'_{O}}_{m}}, {{t'_{I}}_{0}}, \dots {{t'_{I}}_{m}}}\in{\mathnat \cup \mathsetextension{\infty}}$
  such that
  \begin{align*}
  {\check{V}_{0}}&={\mathsubstbox{V'}{\mathsubst{\rho'_{0}}{\mathlevelraise{t_{I_{0}}}{t_{O_{0}}}{W_{0}}, \dots, \mathsubst{\rho'_{m}}{\mathlevelraise{t_{I_{m}}}{t_{O_{m}}}{W_{m}}}}}} \text{ and } \\
  {\hat{V}_{0}}&={\mathsubstbox{V'}{\mathsubst{\rho'_{0}}{W_{0}, \dots, \mathsubst{\rho'_{m}}{W_{m}}}}}. 
  \end{align*}
  Without loss of generality, we can assume
  ${\mathsetextension{\rho_{0}, \dots, \rho_{n}}}\cap{\mathsetextension{\rho'_{0}, \dots, \rho'_{m}}}=\emptyset$.
  Let ${\mathvect{\rho'}}={\mathtuple{\rho'_{0}, \dots, \rho'_{m}}}$,
  ${\mathvect{W}}={\mathtuple{W_{0}, \dots, W_{m}}}$, and
  ${\mathvect{W'}}={\mathtuple{{\mathlevelraise{{t'_{I}}_{0}}{{t'_{O}}_{0}}{W_{0}}}, \dots, {\mathlevelraise{{t'_{I}}_{m}}{{t'_{O}}_{m}}{W_{m}}}}}$.
  
  Let ${\check{V}}={\mathpiparallel{\mathsubstbox{V'}{\mathsubst{\mathvect{\rho'}}{\mathvect{W'}}}}{\mathsubstbox{V_{1}}{\mathsubst{\mathvect{\rho}}{\mathvect{U'}}}}}$.
  Then, we have ${\mathtuple{\check{V}, \hat{V}}}\in{R_{5}}$.
  By \rulename{UP-CongP}, we have 
  ${\mathpiparallel{\mathsubstbox{V_{0}}{\mathsubst{\mathvect{\rho}}{\mathvect{U'}}}}{\mathsubstbox{V_{1}}{\mathsubst{\mathvect{\rho}}{\mathvect{U'}}}}}\mathusagestructuralpo{\mathpiparallel{\mathsubstbox{V'}{\mathsubst{\mathvect{\rho'}}{\mathvect{W'}}}}{\mathsubstbox{V_{1}}{\mathsubst{\mathvect{\rho}}{\mathvect{U'}}}}}$.

 \noindent Case 7. \rulename{UP-Rep}.
  Assume ${V}={\mathpireplication{V_{0}}}$ and 
  ${\hat{V}}={\mathpiparallel{\mathpireplication{\mathsubstbox{V_{0}}{\mathsubst{\mathvect{\rho}}{\mathvect{U}}}}}{\mathsubstbox{V_{0}}{\mathsubst{\mathvect{\rho}}{\mathvect{U}}}}}$.
  Let
  ${\check{V}}={\mathpiparallel{\mathpireplication{\mathsubstbox{V_{0}}{\mathsubst{\mathvect{\rho}}{\mathvect{U'}}}}}{\mathsubstbox{V_{0}}{\mathsubst{\mathvect{\rho}}{\mathvect{U'}}}}}$.
  Then, we have 
  ${\mathsubstbox{V_{0}}{\mathsubst{\mathvect{\rho}}{\mathvect{U'}}}}\mathusagestructuralpo{\check{V}}$ 
  and ${\mathtuple{\check{V}, \hat{V}}}\in{R_{5}}$.

  \noindent Case 8. \rulename{UP-$\mathlevelraisefunc{\ast}{\ast}$}.
  Assume ${V}={\mathlevelraise{t_{I}}{t_{O}}{\rho_{i}}}$,
  ${U_{i}}={{\alpha}^{t_{1}}_{t_{2}}. V_{0}}$, and
  ${\hat{V}}={{{\alpha}^{\mathof{\max}{t_{1}, t_{\alpha}}}_{t_{2}}}. V_{0}}$.
  Let ${\check{V}}={\mathlevelraise{t_{I_{i}}}{t_{O_{i}}}{{{\alpha}^{\mathof{\max}{t_{1}, t_{\alpha}}}_{t_{2}}}. V_{0}}}$.
  Then, we have ${\mathtuple{\check{V}, \hat{V}}}\in{R_{5}}$.
  By \rulename{UP-Commut $\mathlevelraisefunc{\ast}{\ast}$} and \rulename{UP-Rep}, we have
  \[
  {\mathsubstbox{V}{\mathsubst{\mathvect{\rho}}{\mathvect{U'}}}}
  =
  {\mathlevelraise{t_{I}}{t_{O}}{ \mathlevelraise{t_{I_{i}}}{t_{O_{i}}}{{\alpha}^{t_{1}}_{t_{2}}. V_{0}} }}
  \mathusagestructuralpo
  {\mathlevelraise{t_{I_{i}}}{t_{O_{i}}}{ \mathlevelraise{t_{I}}{t_{O}}{{\alpha}^{t_{1}}_{t_{2}}. V_{0}} }}
  \mathusagestructuralpo
  {\mathlevelraise{t_{I_{i}}}{t_{O_{i}}}{{{\alpha}^{\mathof{\max}{t_{1}, t_{\alpha}}}_{t_{2}}}. V_{0}}}
  ={\check{V}}.
  \]
  Therefore, we have 
  ${\mathsubstbox{V}{\mathsubst{\mathvect{\rho}}{\mathvect{U'}}}}\mathusagestructuralpo{\check{V}}$.

  Assume ${V}={\mathlevelraise{t_{I}}{t_{O}}{{\alpha}^{t_{1}}_{t_{2}}. V_{0}}}$ 
  and 
  ${\hat{V}}={{{\alpha}^{\mathof{\max}{t_{1}, t_{\alpha}}}_{t_{2}}}. \mathsubstbox{V_{0}}{\mathsubst{\mathvect{\rho}}{\mathvect{U}}}}$.
  Let ${\check{V}}={{{\alpha}^{\mathof{\max}{t_{1}, t_{\alpha}}}_{t_{2}}}. \mathsubstbox{V_{0}}{\mathsubst{\mathvect{\rho}}{\mathvect{U'}}}}$.
  Then, we have 
  ${\mathsubstbox{V}{\mathsubst{\mathvect{\rho}}{\mathvect{U'}}}}\mathusagestructuralpo{\check{V}}$ and 
  ${\mathtuple{\check{V}, \hat{V}}}\in{R_{5}}$.

  \noindent Case 9. \rulename{UP-Dist}.
  Assume ${V}={\mathlevelraise{t_{I}}{t_{O}}{\mathpiparallel{V_{0}}{V_{1}}}}$ and 
  ${\hat{V}}={\mathpiparallel{\mathlevelraise{t_{I}}{t_{O}}{\mathsubstbox{V_{0}}{\mathsubst{\mathvect{\rho}}{\mathvect{U}}}}}{\mathlevelraise{t_{I}}{t_{O}}{\mathsubstbox{V_{1}}{\mathsubst{\mathvect{\rho}}{\mathvect{U}}}}}}$.
  Let ${\check{V}}={\mathpiparallel{\mathlevelraise{t_{I}}{t_{O}}{\mathsubstbox{V_{0}}{\mathsubst{\mathvect{\rho}}{\mathvect{U'}}}}}{\mathlevelraise{t_{I}}{t_{O}}{\mathsubstbox{V_{1}}{\mathsubst{\mathvect{\rho}}{\mathvect{U'}}}}}}$.
  Then, we have 
  ${\mathsubstbox{V}{\mathsubst{\mathvect{\rho}}{\mathvect{U'}}}}\mathusagestructuralpo{\check{V}}$ and 
  ${\mathtuple{\check{V}, \hat{V}}}\in{R_{5}}$.

  \noindent Case 10. \rulename{UP-Or}.
  Fix $j \in \mathsetextension{0, 1}$.
  Assume ${V}={\mathusageor{V_{0}}{V_{1}}}$ and 
  ${\hat{V}}={\mathsubstbox{V_{j}}{\mathsubst{\mathvect{\rho}}{\mathvect{U}}}}$.
  Let ${\check{V}}={\mathsubstbox{V_{j}}{\mathsubst{\mathvect{\rho}}{\mathvect{U'}}}}$.
  Then, we have 
  ${\mathsubstbox{V}{\mathsubst{\mathvect{\rho}}{\mathvect{U'}}}}\mathusagestructuralpo{\check{V}}$ and 
  ${\mathtuple{\check{V}, \hat{V}}}\in{R_{5}}$.

  \noindent Case 11. \rulename{UP-Cong $\mathlevelraisefunc{\ast}{\ast}$}.
  Assume 
  ${V}={\mathlevelraise{t_{I}}{t_{O}}{V_{0}}}$,
  ${\mathsubstbox{V_{0}}{\mathsubst{\mathvect{\rho}}{\mathvect{U}}}}\mathusagestructuralpo{\hat{V}_{0}}$, and
  ${\hat{V}}={\mathlevelraise{t_{I}}{t_{O}}{\hat{V}_{0}}}$.
  By the induction hypothesis, there exists a closed usage $\check{V}_{0}$ such that 
  ${\mathsubstbox{V_{0}}{\mathsubst{\mathvect{\rho}}{\mathvect{U'}}}}\mathusagestructuralpo{\check{V}_{0}}$ and 
  ${\mathtuple{\check{V}_{0}, \hat{V}_{0}}}\in{R_{5}}$.
  Since ${\mathtuple{\check{V}_{0}, \hat{V}_{0}}}\in{R_{5}}$,
  there exist closed usages $W_{0}, \dots, W_{m}$, 
  a usage $V'$ with $\mathFVof{V'}\subseteq{\mathsetextension{\rho'_{0}, \dots, \rho'_{m}}}$, and
  ${{{t'_{O}}_{0}}, \dots, {{t'_{O}}_{m}}, {{t'_{I}}_{0}}, \dots {{t'_{I}}_{m}}}\in{\mathnat \cup \mathsetextension{\infty}}$ 
  such that
  \begin{align*}
  {\check{V}_{0}}&={\mathsubstbox{V'}{\mathsubst{\rho'_{0}}{\mathlevelraise{t_{I_{0}}}{t_{O_{0}}}{W_{0}}, \dots, \mathsubst{\rho'_{m}}{\mathlevelraise{t_{I_{m}}}{t_{O_{m}}}{W_{m}}}}}} \text{ and } \\
  {\hat{V}_{0}}&={\mathsubstbox{V'}{\mathsubst{\rho'_{0}}{W_{0}, \dots, \mathsubst{\rho'_{m}}{W_{m}}}}}. 
  \end{align*}
  Let ${\mathvect{\rho'}}={\mathtuple{\rho'_{0}, \dots, \rho'_{m}}}$,
  ${\mathvect{W}}={\mathtuple{W_{0}, \dots, W_{m}}}$, and
  ${\mathvect{W'}}={\mathtuple{{\mathlevelraise{{t'_{I}}_{0}}{{t'_{O}}_{0}}{W_{0}}}, \dots, {\mathlevelraise{{t'_{I}}_{m}}{{t'_{O}}_{m}}{W_{m}}}}}$.
  Let ${\check{V}}={\mathlevelraise{t_{I}}{t_{O}}{\mathsubstbox{V'}{\mathsubst{\rho'_{0}}{W_{0}, \dots, \mathsubst{\rho'_{m}}{W_{m}}}}}}$.
  Then, we have ${\mathtuple{\check{V}, \hat{V}}}\in{R_{5}}$.
  By \rulename{UP-Cong $\mathlevelraisefunc{\ast}{\ast}$}, we have 
  ${\mathlevelraise{t_{I}}{t_{O}}{\mathsubstbox{V_{0}}{\mathsubst{\mathvect{\rho}}{\mathvect{U'}}}}}\mathusagestructuralpo{\mathlevelraise{t_{I}}{t_{O}}{\mathsubstbox{V'}{\mathsubst{\mathvect{\rho'}}{\mathvect{W'}}}}}$.

  \noindent Case 12. \rulename{UP-Commut $\mathlevelraisefunc{\ast}{\ast}$}.
  Assume ${V}={\mathlevelraise{t_{I}}{t_{O}}{\rho_{i}}}$,
  ${U_{i}}={\mathlevelraise{t'_{I}}{t'_{O}}{V_{0}}}$, and
  ${\hat{V}}={\mathlevelraise{t'_{I}}{t'_{O}}{\mathlevelraise{t_{I}}{t_{O}}{V_{0}}}}$.
  Let ${\check{V}}={\mathlevelraise{t'_{I}}{t'_{O}}{\mathlevelraise{t_{I}}{t_{O}}{\mathlevelraise{t_{I_{i}}}{t_{O_{i}}}{V_{0}}}}}$.
  Then, we have ${\mathtuple{\check{V}, \hat{V}}}\in{R_{5}}$.
  By \rulename{UP-Commut $\mathlevelraisefunc{\ast}{\ast}$} and
  \rulename{UP-Cong $\mathlevelraisefunc{\ast}{\ast}$}, we have 
  \begin{align*}
  {\mathsubstbox{V}{\mathsubst{\mathvect{\rho}}{\mathvect{U'}}}}
  &=
  {\mathlevelraise{t_{I}}{t_{O}}{\mathlevelraise{t_{I_{i}}}{t_{O_{i}}}{\mathlevelraise{t'_{I}}{t'_{O}}{V_{0}}}}} \\
  &\mathusagestructuralpo
  {\mathlevelraise{t_{I}}{t_{O}}{\mathlevelraise{t'_{I}}{t'_{O}}{\mathlevelraise{t_{I_{i}}}{t_{O_{i}}}{V_{0}}}}} \\
  &\mathusagestructuralpo
  {\mathlevelraise{t'_{I}}{t'_{O}}{\mathlevelraise{t_{I}}{t_{O}}{\mathlevelraise{t_{I_{i}}}{t_{O_{i}}}{V_{0}}}}} \\
   &={\check{V}}.
  \end{align*}
  Hence, we have
  ${\mathsubstbox{V}{\mathsubst{\mathvect{\rho}}{\mathvect{U'}}}}\mathusagestructuralpo{\check{V}}$.

  Assume ${V}={\mathlevelraise{t_{I}}{t_{O}}{\mathlevelraise{t'_{I}}{t'_{O}}{V_{0}}}}$
  and 
  ${\hat{V}}={\mathlevelraise{t'_{I}}{t'_{O}}{\mathlevelraise{t_{I}}{t_{O}}{\mathsubstbox{V_{0}}{\mathsubst{\mathvect{\rho}}{\mathvect{U}}}}}}$.
  Let ${\check{V}}= \mathlevelraise{t'_{I}}{t'_{O}}{\mathlevelraise{t_{I}}{t_{O}}{\mathsubstbox{V_{0}}{\mathsubst{\mathvect{\rho}}{\mathvect{U'}}}}}$.
  Then, we have
  ${\mathsubstbox{V}{\mathsubst{\mathvect{\rho}}{\mathvect{U'}}}}\mathusagestructuralpo{\check{V}}$ and 
  ${\mathtuple{\check{V}, \hat{V}}}\in{R_{5}}$.

  Now, we show that
  if ${\mathtuple{{\mathsubstbox{V}{\mathsubst{\mathvect{\rho}}{\mathvect{U'}}}}, {\mathsubstbox{V}{\mathsubst{\mathvect{\rho}}{\mathvect{U}}}}}}\in{R_{5}}$
  and
  ${\mathsubstbox{V}{\mathsubst{\mathvect{\rho}}{\mathvect{U}}}}\mathusagereduction{\hat{V}}$, 
  then there exists $\check{V}$ such that 
  ${\mathsubstbox{V}{\mathsubst{\mathvect{\rho}}{\mathvect{U'}}}}\mathusagereduction{\check{V}}$
  and ${\mathtuple{\check{V}, \hat{V}}}\in{R_{5}}$.

  Assume 
  ${\mathtuple{{\mathsubstbox{V}{\mathsubst{\mathvect{\rho}}{\mathvect{U'}}}}, {\mathsubstbox{V}{\mathsubst{\mathvect{\rho}}{\mathvect{U}}}}}}\in{R_{5}}$
  and
  ${\mathsubstbox{V}{\mathsubst{\mathvect{\rho}}{\mathvect{U}}}}\mathusagereduction{\hat{V}}$.
  We show that there exists a closed usage $\check{V}$ such that 
  ${\mathsubstbox{V}{\mathsubst{\mathvect{\rho}}{\mathvect{U'}}}}\mathusagereduction{\check{V}}$
  and ${\mathtuple{\check{V}, \hat{V}}}\in{R_{5}}$.
  The proof is by the induction on the construction of 
  ${\mathsubstbox{V}{\mathsubst{\mathvect{\rho}}{\mathvect{U}}}}\mathusagereduction{\hat{V}}$.

  \noindent Case 1. 
  Assume 
  ${V}={\mathpiparallel{I^{t_{o}}_{t_{c}}.V_{0}}{O^{t'_{o}}_{t'_{c}}.V_{1}}}$
  and
  ${\hat{V}}={\mathpiparallel{\mathsubstbox{V_{0}}{\mathsubst{\mathvect{\rho}}{\mathvect{U}}}}{\mathsubstbox{V_{1}}{\mathsubst{\mathvect{\rho}}{\mathvect{U}}}}}$.
  Let ${\hat{V}}= \mathpiparallel{\mathsubstbox{V_{0}}{\mathsubst{\mathvect{\rho}}{\mathvect{U'}}}}{\mathsubstbox{V_{1}}{\mathsubst{\mathvect{\rho}}{\mathvect{U'}}}}$.
  Then, we have
  ${\mathsubstbox{V}{\mathsubst{\mathvect{\rho}}{\mathvect{U'}}}}\mathusagereduction{\check{V}}$
  and ${\mathtuple{\check{V}, \hat{V}}}\in{R_{5}}$.

  \noindent Case 2.
  Assume ${V}={\mathpiparallel{V_{0}}{V_{1}}}$,
  ${\mathsubstbox{V_{0}}{\mathsubst{\mathvect{\rho}}{\mathvect{U}}}}\mathusagereduction{\hat{V}_{0}}$, 
  and 
  ${\hat{V}}={\mathpiparallel{\hat{V}_{0}}{\mathsubstbox{V_{1}}{\mathsubst{\mathvect{\rho}}{\mathvect{U}}}}}$.
  By the induction hypothesis, there exists a closed usage $\check{V}_{0}$ such that 
  ${\mathsubstbox{V_{0}}{\mathsubst{\mathvect{\rho}}{\mathvect{U'}}}}\mathusagereduction{\check{V}_{0}}$ and 
  ${\mathtuple{\check{V}_{0}, \hat{V}_{0}}}\in{R_{5}}$.
  Since ${\mathtuple{\check{V}_{0}, \hat{V}_{0}}}\in{R_{5}}$,
  there exists closed usages $W_{0}, \dots, W_{m}$, 
  a usage $V'$ with $\mathFVof{V'}\subseteq{\mathsetextension{\rho'_{0}, \dots, \rho'_{m}}}$, 
  and
  ${{{t'_{O}}_{0}}, \dots, {{t'_{O}}_{m}}, {{t'_{I}}_{0}}, \dots {{t'_{I}}_{m}}}\in{\mathnat \cup \mathsetextension{\infty}}$
  such that
  \begin{align*}
  {\check{V}_{0}}&={\mathsubstbox{V'}{\mathsubst{\rho'_{0}}{\mathlevelraise{t_{I_{0}}}{t_{O_{0}}}{W_{0}}, \dots, \mathsubst{\rho'_{m}}{\mathlevelraise{t_{I_{m}}}{t_{O_{m}}}{W_{m}}}}}} \text{ and } \\
  {\hat{V}_{0}}&={\mathsubstbox{V'}{\mathsubst{\rho'_{0}}{W_{0}, \dots, \mathsubst{\rho'_{m}}{W_{m}}}}}. 
  \end{align*}
  Without loss of generality, we can assume
  ${\mathsetextension{\rho_{0}, \dots, \rho_{n}}}\cap{\mathsetextension{\rho'_{0}, \dots, \rho'_{m}}}=\emptyset$.
  Let ${\mathvect{\rho'}}={\mathtuple{\rho'_{0}, \dots, \rho'_{m}}}$,
  ${\mathvect{W}}={\mathtuple{W_{0}, \dots, W_{m}}}$, and
  ${\mathvect{W'}}={\mathtuple{{\mathlevelraise{{t'_{I}}_{0}}{{t'_{O}}_{0}}{W_{0}}}, \dots, {\mathlevelraise{{t'_{I}}_{m}}{{t'_{O}}_{m}}{W_{m}}}}}$.
  
  Let ${\check{V}}={\mathpiparallel{\mathsubstbox{V'}{\mathsubst{\mathvect{\rho'}}{\mathvect{W'}}}}{\mathsubstbox{V_{1}}{\mathsubst{\mathvect{\rho}}{\mathvect{U'}}}}}$.
  Then, we have ${\mathtuple{\check{V}, \hat{V}}}\in{R_{5}}$.
  We also have 
  $\mathpiparallel{\mathsubstbox{V_{0}}{\mathsubst{\mathvect{\rho}}{\mathvect{U'}}}}{\mathsubstbox{V_{1}}{\mathsubst{\mathvect{\rho}}{\mathvect{U'}}}}\mathusagereduction \mathpiparallel{\mathsubstbox{V'}{\mathsubst{\mathvect{\rho'}}{\mathvect{W'}}}}{\mathsubstbox{V_{1}}{\mathsubst{\mathvect{\rho}}{\mathvect{U'}}}}$.

  \noindent Case 3.
  Assume there exists usages $V_{0}$ and $V_{1}$ such that
  ${\mathsubstbox{V}{\mathsubst{\mathvect{\rho}}{\mathvect{U}}}}\mathusagestructuralpo{V_{0}}$,
  ${V_{0}}\mathusagereduction{V_{1}}$,
  and ${V_{1}}\mathusagestructuralpo{\hat{V}}$.
  Since  
  ${\mathsubstbox{V}{\mathsubst{\mathvect{\rho}}{\mathvect{U}}}}\mathusagestructuralpo{V_{0}}$,
  there exists $\check{V}_{0}$ such that
  ${\mathsubstbox{V}{\mathsubst{\mathvect{\rho}}{\mathvect{U'}}}}\mathusagestructuralpo{\check{V}_{0}}$ and 
  ${\mathtuple{\check{V}_{0}, V_{0}}}\in{R_{5}}$.
  Since ${\mathtuple{\check{V}_{0}, V_{0}}}\in{R_{5}}$ and
  ${V_{0}}\mathusagereduction{V_{1}}$,
  the induction hypothesis implies that there exists $\check{V}_{1}$ such that
  ${{\check{V}_{0}}}\mathusagereduction{\check{V}_{1}}$ and 
  ${\mathtuple{\check{V}_{1}, V_{1}}}\in{R_{5}}$.
  Since  
  ${V_{1}}\mathusagestructuralpo{\hat{V}}$,
  there exists $\check{V}$ such that
  ${\check{V}_{1}}\mathusagestructuralpo{\check{V}}$ and 
  ${\mathtuple{\check{V}, \hat{V}}}\in{R_{5}}$.
  Since 
  ${\mathsubstbox{V}{\mathsubst{\mathvect{\rho}}{\mathvect{U'}}}}\mathusagestructuralpo{\check{V}_{0}}$,
  ${{\check{V}_{0}}}\mathusagereduction{\check{V}_{1}}$ and 
  ${\check{V}_{1}}\mathusagestructuralpo{\check{V}}$,
  we have ${\mathsubstbox{V}{\mathsubst{\mathvect{\rho}}{\mathvect{U'}}}}\mathusagereduction{\check{V}}$.

  \noindent \cref{item:capability-in-def:subusage}
  By \cref{lemma:cap_and_ob_subst} \cref{item:cap_eq_lemma-cap_and_ob_subst}.

  \noindent \cref{item:obligation-in-def:subusage}
  By \cref{lemma:cap_and_ob_subst} \cref{item:ob_eq_lemma-cap_and_ob_subst}.

  %%%%%%%%%%%%%%%%%%%%%%%%%%%%%%%%%%%%%%%%%%%%%%%%%%%%%%%%%%%%%

  \noindent \cref{item:uni-raise-op-in-prop:pos}
By this proposition \cref{item:gen-raise-op-in-prop:pos}. 
\end{proof}

\section{Basic properties of types and type environments}

\subsection{A basic property of types}

 \begin{proposition} 
  \label[proposition]{prop:property-of-subtyping}
  \begin{enumerate} 
   \item Let $\tau_{0}$ and $\tau_{1}$ be types.
	 Then $\mathissubtype{\mathpiparallel{\tau_{0}}{\tau_{1}}}{\mathpiparallel{\tau_{1}}{\tau_{0}}}$. 
	 \label{item:comm-parallel-in-prop:property-of-subtyping}
   \item Let $\tau_{0}$, $\tau_{1}$, and $\tau_{2}$ be types.
	 Then $\mathissubtype{\mathpiparallel{(\mathpiparallel{\tau_{0}}{\tau_{1}})}{\tau_{2}}}{\mathpiparallel{\tau_{0}}{(\mathpiparallel{\tau_{1}}{\tau_{2}})}}$. 
	 \label{item:assoc1-parallel-in-prop:property-of-subtyping}
   \item For types $\tau_{0}$, $\tau'_{0}$, $\tau_{1}$ and $\tau'_{1}$,
	 if $\mathissubtype{\tau_{i}}{\tau'_{i}}$ for each $i=0, 1$,
	 then $\mathissubtype{\mathpiparallel{\tau_{0}}{\tau_{1}}}{\mathpiparallel{\tau'_{0}}{\tau'_{1}}}$. 
	 \label{item:parallel-in-prop:property-of-subtyping}
   \item Let $\tau_{0}$, $\tau_{1}$, and $\tau_{2}$ be types.
	 Then $\mathissubtype{\mathpiparallel{\tau_{0}}{(\mathpiparallel{\tau_{1}}{\tau_{2}})}}{\mathpiparallel{(\mathpiparallel{\tau_{0}}{\tau_{1}})}{\tau_{2}}}$.
	 \label{item:assoc2-parallel-in-prop:property-of-subtyping}
   \item For types $\tau$ and $\tau'$,
	 $\mathobligation{}{\tau}=\infty$ implies
	 $\mathissubtype{\mathpiparallel{\tau}{\tau'}}{\tau'}$.
	 \label{item:ob-infty-and-parallel-in-prop:property-of-subtyping}
   \item For types $\tau$ and $\tau'$,
	 if $\mathissubtype{\tau}{\tau'}$,
	 then $\mathissubtype{\mathpireplication{\tau}}{\mathpireplication{\tau'}}$.
	 \label{item:replication-cong-in-prop:property-of-subtyping}
   \item Let $\tau$ be a type.
	 Then $\mathissubtype{\mathpireplication{\tau}}{\mathpiparallel{\mathpireplication{\tau}}{\tau}}$.
	 \label{item:replication-in-prop:property-of-subtyping}
   \item Let $\tau_{0}, \dots, \tau_{n}$ be types.
	 Then $\mathissubtype{(\mathpiparallel{\mathpireplication{\tau_{0}}}{\mathpiparallel{\cdots}{\mathpireplication{\tau_{n}}}})}{\mathpireplication{\mleft(\mathpiparallel{\tau_{0}}{\mathpiparallel{\cdots}{\tau_{n}}}\mright)}}$.
	 \label{item:rep-and-parallel-in-prop:property-of-subtyping}
   \item Let $\tau$ be a type.
	 Then $\mathissubtype{\mathlevelraise{t_{I}}{t_{O}}{\tau}}{\tau}$.
	 \label{item:levelraise-is-subtype-in-prop:property-of-subtyping}
   \item Let $\tau$ be a type.
	 Then $\mathissubtype{\mathlevelraiseuni{\tau}}{\tau}$.
	 \label{item:levelraiseuni-is-subtype-in-prop:property-of-subtyping}
  \end{enumerate}  
 \end{proposition}

 \begin{proof}
  We show each claim.

  \noindent \cref{item:comm-parallel-in-prop:property-of-subtyping}
  We show 
  $\mathissubtype{\mathpiparallel{\tau_{0}}{\tau_{1}}}{\mathpiparallel{\tau_{1}}{\tau_{0}}}$.
  
  Assume $\tau_{0}$ is a base type and ${\tau_{0}}={\tau_{1}}$.
  Then ${\mathpiparallel{\tau_{0}}{\tau_{1}}}={\tau_{0}}$.
  Since ${\tau_{0}}={\tau_{1}}$, we see that $\tau_{1}$ is a base type.
  Then ${\mathpiparallel{\tau_{1}}{\tau_{0}}}={\tau_{1}}$.
  Since ${\tau_{0}}={\tau_{1}}$, we have
  ${\mathpiparallel{\tau_{0}}{\tau_{1}}}={\mathpiparallel{\tau_{1}}{\tau_{0}}}$.
  Hence, we have 
  $\mathissubtype{\mathpiparallel{\tau_{0}}{\tau_{1}}}{\mathpiparallel{\tau_{1}}{\tau_{0}}}$.

  Assume ${\tau_{0}} = {\xi/U_{0}}$ and $\tau_{1} = \xi/U_{1}$.
  Then, we have
  ${\mathpiparallel{\tau_{0}}{\tau_{1}}}\equiv{{\xi}/{\mathpiparallel{U_{0}}{U_{1}}}}$
  and
  ${\mathpiparallel{\tau_{1}}{\tau_{0}}}\equiv{{\xi}/{\mathpiparallel{U_{1}}{U_{0}}}}$.
  Since ${\mathpiparallel{U_{0}}{U_{1}}}\mathusagestructuralpo{\mathpiparallel{U_{1}}{U_{0}}}$,
  \cref{prop:property-of-subusage} \cref{item:usagestructuralpo-in-prop:pos} implies
  $\mathissubusage{\mathpiparallel{U_{0}}{U_{1}}}{\mathpiparallel{U_{1}}{U_{0}}}$.
  Thus, $\mathissubtype{\mathpiparallel{\tau_{0}}{\tau_{1}}}{\mathpiparallel{\tau_{1}}{\tau_{0}}}$.

  \noindent \cref{item:assoc1-parallel-in-prop:property-of-subtyping}
  We show 
  $\mathissubtype{\mathpiparallel{(\mathpiparallel{\tau_{0}}{\tau_{1}})}{\tau_{2}}}{\mathpiparallel{\tau_{0}}{(\mathpiparallel{\tau_{1}}{\tau_{2}})}}$.
  
  Assume $\tau_{0}$ is a base type and ${\tau_{0}}={\tau_{1}}$.
  Then ${\mathpiparallel{\tau_{0}}{\tau_{1}}}={\tau_{0}}$.
  Because $\tau_{0}$ is a base type and ${\mathpiparallel{\tau_{0}}{\tau_{2}}}$ is defined,
  we have ${\tau_{0}}={\tau_{2}}$ and ${\mathpiparallel{\tau_{0}}{\tau_{2}}}={\tau_{0}}$.
  Hence,
  ${\mathpiparallel{(\mathpiparallel{\tau_{0}}{\tau_{1}})}{\tau_{2}}}={\tau_{0}}={\mathpiparallel{\tau_{0}}{\tau_{1}}}={\mathpiparallel{\tau_{0}}{(\mathpiparallel{\tau_{1}}{\tau_{2}})}}$.

  Assume ${\tau_{0}} = {{\xi}/{U_{0}}}$ and 
  ${\tau_{1}} = {\xi/U_{1}}$.
  Then ${\mathpiparallel{\tau_{0}}{\tau_{1}}}={{\xi}/{\mathpiparallel{U_{0}}{U_{1}}}}$.
  Because ${\mathpiparallel{(\mathpiparallel{\tau_{0}}{\tau_{1}})}{\tau_{2}}}$ is defined,
  there exists a usage $U_{2}$ such that ${\tau_{2}} = {\xi/U_{2}}$.
  Hence,
  ${\mathpiparallel{(\mathpiparallel{\tau_{0}}{\tau_{1}})}{\tau_{2}}}={{\xi}/{(\mathpiparallel{({\mathpiparallel{U_{0}}{U_{1}}})}{U_{2}})}}$.
  We also have 
  ${\mathpiparallel{\tau_{0}}{(\mathpiparallel{\tau_{1}}{\tau_{2}})}}={{\xi}/{(\mathpiparallel{U_{0}}{(\mathpiparallel{U_{1}}{U_{2}})})}}$.
  Since 
  ${(\mathpiparallel{{\mathpiparallel{U_{0}}{U_{1}}})}{U_{2}}}\mathusagestructuralpo{\mathpiparallel{U_{0}}{(\mathpiparallel{U_{1}}{U_{2}})}}$,
  we see 
  $\mathissubusage{\mathpiparallel{({\mathpiparallel{U_{0}}{U_{1}}})}{U_{2}}}{\mathpiparallel{U_{0}}{(\mathpiparallel{U_{1}}{U_{2}})}}$.
  Thus,
  $\mathissubtype{\mathpiparallel{(\mathpiparallel{\tau_{0}}{\tau_{1}})}{\tau_{2}}}{\mathpiparallel{\tau_{0}}{(\mathpiparallel{\tau_{1}}{\tau_{2}})}}$.

  \noindent \cref{item:parallel-in-prop:property-of-subtyping}
  Assume $\mathissubtype{\tau_{i}}{\tau'_{i}}$ for each $i=0, 1$.
  We show 
  $\mathissubtype{\mathpiparallel{\tau_{0}}{\tau_{1}}}{\mathpiparallel{\tau'_{0}}{\tau'_{1}}}$.

  Assume $\tau_{0}$ is a base type and ${\tau_{0}}={\tau_{1}}$.
  Then ${\mathpiparallel{\tau_{0}}{\tau_{1}}}={\tau_{0}}$.
  Since $\mathissubtype{\tau_{0}}{\tau'_{0}}$, we have ${\tau'_{0}}={\tau_{0}}$.
  Since ${\tau_{0}}={\tau_{1}}$, we see that $\tau_{1}$ is a base type.
  Since $\mathissubtype{\tau_{1}}{\tau'_{1}}$, we have ${\tau'_{1}}={\tau_{1}}$.
  Then 
  ${\mathpiparallel{\tau_{0}}{\tau_{1}}}\equiv{\mathpiparallel{\tau'_{0}}{\tau'_{1}}}$.
  Hence, we have
  $\mathissubtype{\mathpiparallel{\tau_{0}}{\tau_{1}}}{\mathpiparallel{\tau'_{0}}{\tau'_{1}}}$.

  Assume ${\tau_{0}} = {\xi/U_{0}}$ and $\tau_{1} = \xi/U_{1}$.
  Then 
  ${\mathpiparallel{\tau_{0}}{\tau_{1}}}\equiv{{\xi}/{\mathpiparallel{U_{0}}{U_{1}}}}$.
  For each $i=0, 1$, because of $\mathissubtype{\tau_{i}}{\tau'_{i}}$, we have
  ${\tau'_{i}} = {\xi/U'_{i}}$ and $\mathissubusage{U_{i}}{U'_{i}}$.
  By \cref{prop:property-of-subusage} \cref{item:cong-parallel-in-prop:pos},
  we have 
  $\mathissubusage{\mathpiparallel{U_{0}}{U_{1}}}{\mathpiparallel{U'_{0}}{U'_{1}}}$.
  Then
  \[
  {\mathpiparallel{\tau_{0}}{\tau_{1}}}
  \equiv{{\xi}/{\mathpiparallel{U_{0}}{U_{1}}}}
  \mathissubtypesy{{\xi}/{\mathpiparallel{U'_{0}}{U'_{1}}}}
  \equiv{\mathpiparallel{\tau'_{0}}{\tau'_{1}}}.
  \]

  \noindent \cref{item:assoc2-parallel-in-prop:property-of-subtyping}
  We show $\mathissubtype{\mathpiparallel{\tau_{0}}{(\mathpiparallel{\tau_{1}}{\tau_{2}})}}{\mathpiparallel{(\mathpiparallel{\tau_{0}}{\tau_{1}})}{\tau_{2}}}$.
  By \cref{item:comm-parallel-in-prop:property-of-subtyping},
  \cref{item:assoc1-parallel-in-prop:property-of-subtyping},
  \cref{item:parallel-in-prop:property-of-subtyping}, and
  transitivity of $\mathissubtypesy$, we have
  \begin{align*}
   {\mathpiparallel{\tau_{0}}{(\mathpiparallel{\tau_{1}}{\tau_{2}})}}
   &\mathissubtypesy{\mathpiparallel{(\mathpiparallel{\tau_{1}}{\tau_{2}})}{\tau_{0}}} \\
   &\mathissubtypesy{\mathpiparallel{(\mathpiparallel{\tau_{2}}{\tau_{1}})}{\tau_{0}}} \\
   &\mathissubtypesy{\mathpiparallel{\tau_{2}}{(\mathpiparallel{\tau_{1}}{\tau_{0}})}} \\
   &\mathissubtypesy{\mathpiparallel{(\mathpiparallel{\tau_{1}}{\tau_{0}})}{\tau_{2}}} \\
   &\mathissubtypesy{\mathpiparallel{(\mathpiparallel{\tau_{0}}{\tau_{1}})}{\tau_{2}}}.
  \end{align*}
  Thus, 
  $\mathissubtype{\mathpiparallel{\tau_{0}}{(\mathpiparallel{\tau_{1}}{\tau_{2}})}}{\mathpiparallel{(\mathpiparallel{\tau_{0}}{\tau_{1}})}{\tau_{2}}}$.

  \noindent \cref{item:ob-infty-and-parallel-in-prop:property-of-subtyping}
  Assume $\mathobligation{}{\tau}=\infty$.

  Assume that $\tau$ is a base type.
  Since ${\mathpiparallel{\tau}{\tau'}}$ is defined,
  we have $\tau'$ is a base type and ${\tau}={\tau'}$.
  Then, we have ${\mathpiparallel{\tau}{\tau'}}={\tau}={\tau'}$.
  Therefore, we have $\mathissubtype{\mathpiparallel{\tau}{\tau'}}{\tau'}$.

  Assume that ${\tau}\equiv{\xi/U}$ and $\mathobligation{}{U}=\infty$.
  Since ${\mathpiparallel{\tau}{\tau'}}$ is defined,
  there exists a usage $U'$ such that ${\tau'}\equiv{\xi/U'}$.
  Then, we have ${\mathpiparallel{\tau}{\tau'}}={{\xi}/{\mathpiparallel{U}{U'}}}$.
  By \cref{prop:property-of-subusage} \cref{item:ob-infty-and-parallel-in-prop:pos},
  we have $\mathissubusage{\mathpiparallel{U}{U'}}{U'}$.
  By \cref{def:subtyping}, $\mathissubtype{\mathpiparallel{\tau}{\tau'}}{\tau'}$.

  \noindent \cref{item:replication-cong-in-prop:property-of-subtyping}
  Assume $\mathissubtype{\tau}{\tau'}$.
  
  Assume that $\tau$ is a base type.
  Then, we have ${\tau}={\tau'}$.
  We also have ${\mathpireplication{\tau}}={\tau}$.
  Because $\tau'$ is a base type, ${\mathpireplication{\tau'}}={\tau'}$.
  Hence, ${\mathpireplication{\tau}}={\mathpireplication{\tau'}}$.
  Therefore, $\mathissubtype{\mathpireplication{\tau}}{\mathpireplication{\tau'}}$.
  
  Assume that ${\tau}\equiv{\xi/U}$.
  Then ${\mathpireplication{\tau}}\equiv{\xi/\mathpireplication{U}}$.
  Since $\mathissubtype{\tau}{\tau'}$,
  there exists $U'$ such that $\mathissubusage{U}{U'}$ and ${\tau'}\equiv{\xi/U'}$.
  Then ${\mathpireplication{\tau'}}\equiv{\xi/\mathpireplication{U'}}$.
  By \cref{def:subusage} \cref{item:sub-in-def:subusage},
  we have $\mathissubusage{\mathpireplication{U}}{\mathpireplication{U'}}$.
  Thus, $\mathissubtype{\mathpireplication{\tau}}{\mathpireplication{\tau'}}$.

  \noindent \cref{item:replication-in-prop:property-of-subtyping}
  We show $\mathissubtype{\mathpireplication{\tau}}{\mathpiparallel{\mathpireplication{\tau}}{\tau}}$.

  Assume that $\tau$ is a base type.
  Then, we have ${\mathpireplication{\tau}}={\tau}$.
  Since $\tau$ is a base type,
  we have
  ${\mathpiparallel{\mathpireplication{\tau}}{\tau}}={\mathpiparallel{\tau}{\tau}}={\tau}$.
  Therefore,
  $\mathissubtype{\mathpireplication{\tau}}{\mathpiparallel{\mathpireplication{\tau}}{\tau}}$.

  Assume that ${\tau}\equiv{\xi/U}$.
  Then ${\mathpireplication{\tau}}\equiv{\xi/\mathpireplication{U}}$ and
  ${\mathpiparallel{\mathpireplication{\tau}}{\tau}}\equiv{{\xi}/{{\mathpiparallel{\mathpireplication{U}}{U}}}}$.
  Since 
  ${\mathpireplication{U}}\mathusagestructuralpo{\mathpiparallel{\mathpireplication{U}}{U}}$,
  \cref{prop:property-of-subusage} \cref{item:usagestructuralpo-in-prop:pos} implies
  $\mathissubusage{\mathpireplication{U}}{\mathpiparallel{\mathpireplication{U}}{U}}$.
  Therefore,
  $\mathissubtype{\mathpireplication{\tau}}{\mathpiparallel{\mathpireplication{\tau}}{\tau}}$.

  \noindent \cref{item:rep-and-parallel-in-prop:property-of-subtyping}
  We show $\mathissubtype{(\mathpiparallel{\mathpireplication{\tau_{0}}}{\mathpiparallel{\cdots}{\mathpireplication{\tau_{n}}}})}{\mathpireplication{\mleft(\mathpiparallel{\tau_{0}}{\mathpiparallel{\cdots}{\tau_{n}}}\mright)}}$.

  Assume that $\tau_{0}$ is a base type.
  Then, we have ${\mathpireplication{\tau_{0}}}={\tau_{0}}$.
  Since $(\mathpiparallel{\mathpireplication{\tau_{0}}}{\mathpiparallel{\cdots}{\mathpireplication{\tau_{n}}}})$ is defined,
  ${\mathpireplication{\tau_{1}}}, \dots, {\mathpireplication{\tau_{n}}}$ are base types.
  Then, we see that
  ${\mathpireplication{\tau_{i}}}={\tau_{i}}$ and
  ${\tau_{i}}$ is a base type for each ${i}={0, \dots, n}$.
  Hence,
  $(\mathpiparallel{\mathpireplication{\tau_{0}}}{\mathpiparallel{\cdots}{\mathpireplication{\tau_{n}}}})=(\mathpiparallel{\tau_{0}}{\mathpiparallel{\cdots}{\tau_{n}}})$.
  Since $\mathpiparallel{\tau_{0}}{\mathpiparallel{\cdots}{\tau_{n}}}$ is defined and
  $\tau_{0}$ is a base type,
  we have ${\mathpiparallel{\tau_{0}}{\mathpiparallel{\cdots}{\tau_{n}}}}={\tau_{0}}$.
  Then ${\mathpireplication{\mathpiparallel{\tau_{0}}{\mathpiparallel{\cdots}{\tau_{n}}}}}={\mathpireplication{\tau_{0}}}={\tau_{0}}$.
  Therefore,
  $\mathissubtype{(\mathpiparallel{\mathpireplication{\tau_{0}}}{\mathpiparallel{\cdots}{\mathpireplication{\tau_{n}}}})}{\mathpireplication{(\mathpiparallel{\tau_{0}}{\mathpiparallel{\cdots}{\tau_{n}}})}}$.

  Assume that ${\tau_{0}}\equiv{\xi/U_{0}}$.
  Then ${\mathpireplication{\tau_{0}}}\equiv{\xi/\mathpireplication{U_{0}}}$.
  Since 
  ${\mathpireplication{(\mathpiparallel{\tau_{0}}{\mathpiparallel{\cdots}{\tau_{n}}})}}$ 
  is defined,
  ${\mathpiparallel{\tau_{0}}{\mathpiparallel{\cdots}{\tau_{n}}}}$ is defined.
  Hence, $\tau_{i}$ is the form of ${\xi/U_{i}}$ for each ${i}={1, \dots, n}$.
  Then, we have
  \begin{align*}
   {(\mathpiparallel{\mathpireplication{\tau_{0}}}{\mathpiparallel{\cdots}{\mathpireplication{\tau_{n}}}})} 
   &={{\xi}/{{(\mathpiparallel{\mathpireplication{U_{0}}}{\mathpiparallel{\cdots}{\mathpireplication{U_{n}}}})}}}
   \text{  and } \\
   {\mathpireplication{(\mathpiparallel{\tau_{0}}{\mathpiparallel{\cdots}{\tau_{n}}})}}
   &={{\xi}/{\mathpireplication{(\mathpiparallel{U_{0}}{\mathpiparallel{\cdots}{U_{n}}})}}}.
  \end{align*}
  By \cref{prop:property-of-subusage} \cref{item:rep-and-parallel-in-prop:pos},
  we have
  $\mathissubtype{(\mathpiparallel{\mathpireplication{\tau_{0}}}{\mathpiparallel{\cdots}{\mathpireplication{\tau_{n}}}})}{\mathpireplication{(\mathpiparallel{\tau_{0}}{\mathpiparallel{\cdots}{\tau_{n}}})}}$.

  \noindent \cref{item:levelraise-is-subtype-in-prop:property-of-subtyping}
  We show $\mathissubtype{\mathlevelraise{t_{I}}{t_{O}}{\tau}}{\tau}$.

  Assume $\tau$ is a base type. 
  Then ${\mathlevelraise{t_{I}}{t_{O}}{\tau}}={\tau}$.
  Hence, $\mathissubtype{\mathlevelraise{t_{I}}{t_{O}}{\tau}}{\tau}$

  Assume $\tau = {{\xi}/{U}}$.
  Then ${\mathlevelraise{t_{I}}{t_{O}}{\tau}}={{\xi}/{\mathlevelraise{t_{I}}{t_{O}}{U}}}$.
  By \cref{prop:property-of-subusage} \cref{item:gen-raise-op-in-prop:pos},
  we have $\mathissubusage{\mathlevelraise{t_{I}}{t_{O}}{U}}{U}$.
  Hence, $\mathissubtype{\mathlevelraise{t_{I}}{t_{O}}{\tau}}{\tau}$.

  \noindent \cref{item:levelraiseuni-is-subtype-in-prop:property-of-subtyping}
  We show $\mathissubtype{\mathlevelraiseuni{\tau}}{\tau}$.
  
  Assume $\tau$ is a base type. 
  Then ${\mathlevelraiseuni{\tau}}={\tau}$.
  Hence, $\mathissubtype{\mathlevelraiseuni{\tau}}{\tau}$

  Assume $\tau = {{\xi}/{U}}$.
  Then ${\mathlevelraiseuni{\tau}}={{\xi}/{\mathlevelraiseuni{U}}}$.
  By \cref{prop:property-of-subusage} \cref{item:uni-raise-op-in-prop:pos},
  we have $\mathissubusage{\mathlevelraiseuni{U}}{U}$.
  Hence, $\mathissubtype{\mathlevelraiseuni{\tau}}{\tau}$.

 \end{proof}

\subsection{A basic property of type environments}

  \begin{proposition}
   \label[proposition]{prop:property-subtyping-on-env}
   \begin{enumerate}
    \item For type environments $\Gamma_{0}$ and $\Gamma_{1}$,
	  $\mathissubtype{\mathpiparallel{\Gamma_{0}}{\Gamma_{1}}}{\mathpiparallel{\Gamma_{1}}{\Gamma_{0}}}$.
	  \label{item:commut-prop-property-subtyping-on-env}
    \item For type environments $\Gamma_{0}$, $\Gamma_{1}$, and $\Gamma_{2}$,
	  $\mathissubtype{\mathpiparallel{\Gamma_{0}}{\mleft(\mathpiparallel{\Gamma_{1}}{\Gamma_{2}}\mright)}}{\mathpiparallel{\mleft(\mathpiparallel{\Gamma_{0}}{\Gamma_{1}}\mright)}{\Gamma_{2}}}$.
	  \label{item:assoc1-prop-property-subtyping-on-env}
    \item For type environments $\Gamma_{0}$, $\Gamma_{1}$, and $\Gamma'_{0}$,
	  if $\mathissubtype{\Gamma_{0}}{\Gamma'_{0}}$,
	  then $\mathissubtype{\mathpiparallel{\Gamma_{0}}{\Gamma_{1}}}{\mathpiparallel{\Gamma'_{0}}{\Gamma_{1}}}$.
	  \label{item:cong-prop-property-subtyping-on-env}
    \item For type environments $\Gamma_{0}$, $\Gamma_{1}$, 
	  $\Gamma'_{0}$, and $\Gamma'_{1}$,
	  if $\mathissubtype{\Gamma_{0}}{\Gamma'_{0}}$ and 
	  $\mathissubtype{\Gamma_{1}}{\Gamma'_{1}}$,
	  then $\mathissubtype{\mathpiparallel{\Gamma_{0}}{\Gamma_{1}}}{\mathpiparallel{\Gamma'_{0}}{\Gamma'_{1}}}$.
	  \label{item:cong-full-prop-property-subtyping-on-env}
    \item For type environments $\Gamma_{0}$, $\Gamma_{1}$, and $\Gamma_{2}$,
	  $\mathissubtype{\mathpiparallel{\mleft(\mathpiparallel{\Gamma_{0}}{\Gamma_{1}}\mright)}{\Gamma_{2}}}{\mathpiparallel{\Gamma_{0}}{\mleft(\mathpiparallel{\Gamma_{1}}{\Gamma_{2}}\mright)}}$.
	  \label{item:assoc2-prop-property-subtyping-on-env}
    \item For type environments $\Gamma$ and $\Gamma'$,
	  if $\mathissubtype{\Gamma}{\Gamma'}$,
	  then $\mathissubtype{\mathpireplication{\Gamma}}{\mathpireplication{\Gamma'}}$.
	  \label{item:replication-cong-prop-property-subtyping-on-env}
    \item For type environments $\Gamma$,
	  $\mathissubtype{\mathpireplication{\Gamma}}{\mathpiparallel{\mathpireplication{\Gamma}}{\Gamma}}$.
	  \label{item:replication-prop-property-subtyping-on-env}
    \item For type environments $\Gamma$ and $\Gamma'$,
	  if $\mathissubtype{\Gamma}{\Gamma'}$, then 
	  $\mathissubtype{\Gamma, \mathistype{x}{\tau}}{\Gamma', \mathistype{x}{\tau}}$.
	  \label{item:addition-prop-property-subtyping-on-env}
    \item For type environments $\Gamma$ and $\Gamma'$, if 
	  $\mathissubtype{(\Gamma, \mathistype{x}{\tau})}{(\Gamma', \mathistype{x}{\tau})}$,
	 then $\mathissubtype{\Gamma}{\Gamma'}$.
	  \label{item:minus-prop-property-subtyping-on-env}
    \item For a type environment $\Gamma$, 
	  $\mathissubtype{\mathlevelraise{t_{I}}{t_{O}}{\Gamma}}{\Gamma}$.
	  \label{item:levelraise-prop-property-subtyping-on-env}
    \item For a type environment $\Gamma$, 
	  $\mathissubtype{\mathlevelraiseuni{\Gamma}}{\Gamma}$.
	  \label{item:levelraiseuni-prop-property-subtyping-on-env}
   \end{enumerate}
  \end{proposition}

   \begin{proof}
    We show each claim.
    
    \noindent \cref{item:commut-prop-property-subtyping-on-env}
    We show 
    $\mathissubtype{\mathpiparallel{\Gamma_{0}}{\Gamma_{1}}}{\mathpiparallel{\Gamma_{1}}{\Gamma_{0}}}$.
    
    \noindent \cref{item:dom-def-subtyping-on-env}
    Since ${\mathof{\mathdom}{\mathpiparallel{\Gamma_{0}}{\Gamma_{1}}}}={{\mathof{\mathdom}{\Gamma_{0}}}\cup{\mathof{\mathdom}{\Gamma_{1}}}}={\mathof{\mathdom}{\mathpiparallel{\Gamma_{1}}{\Gamma_{0}}}}$,
    we have ${\mathof{\mathdom}{\mathpiparallel{\Gamma_{0}}{\Gamma_{1}}}}\supseteq{\mathof{\mathdom}{\mathpiparallel{\Gamma_{1}}{\Gamma_{0}}}}$.

    \noindent \cref{item:subtyping-def-subtyping-on-env}
    Let ${x}\in{\mathof{\mathdom}{\mathpiparallel{\Gamma_{1}}{\Gamma_{0}}}}$.

    Assume $x\in{{\mathof{\mathdom}{\Gamma_{1}}}\cap{\mathof{\mathdom}{\Gamma_{0}}}}$.
    Then, we have 
    ${\mathof{\mathpiparallel{\Gamma_{0}}{\Gamma_{1}}}{x}}={\mathpiparallel{\mathof{\Gamma_{0}}{x}}{\mathof{\Gamma_{1}}{x}}}$ and 
    ${\mathof{\mathpiparallel{\Gamma_{1}}{\Gamma_{0}}}{x}}={\mathpiparallel{\mathof{\Gamma_{1}}{x}}{\mathof{\Gamma_{0}}{x}}}$.
    By \cref{prop:property-of-subtyping} \cref{item:comm-parallel-in-prop:property-of-subtyping}, 
    we have
    $\mathissubtype{\mathpiparallel{\mathof{\Gamma_{0}}{x}}{\mathof{\Gamma_{1}}{x}}}{\mathpiparallel{\mathof{\Gamma_{1}}{x}}{\mathof{\Gamma_{0}}{x}}}$.

    Assume $x\in{{\mathof{\mathdom}{\Gamma_{1}}}\setminus{\mathof{\mathdom}{\Gamma_{0}}}}$.
    Then, we have 
    ${\mathof{\mathpiparallel{\Gamma_{0}}{\Gamma_{1}}}{x}}={\mathof{\Gamma_{1}}{x}}$ and 
    ${\mathof{\mathpiparallel{\Gamma_{1}}{\Gamma_{0}}}{x}}={\mathof{\Gamma_{1}}{x}}$.
    Hence,
    $\mathissubtype{\mathpiparallel{\mathof{\Gamma_{0}}{x}}{\mathof{\Gamma_{1}}{x}}}{\mathpiparallel{\mathof{\Gamma_{1}}{x}}{\mathof{\Gamma_{0}}{x}}}$.

    Assume $x\in{{\mathof{\mathdom}{\Gamma_{0}}}\setminus{\mathof{\mathdom}{\Gamma_{1}}}}$.
    Then, we have 
    ${\mathof{\mathpiparallel{\Gamma_{0}}{\Gamma_{1}}}{x}}={\mathof{\Gamma_{0}}{x}}$ and 
    ${\mathof{\mathpiparallel{\Gamma_{1}}{\Gamma_{0}}}{x}}={\mathof{\Gamma_{0}}{x}}$.
    Hence,
    $\mathissubtype{\mathpiparallel{\mathof{\Gamma_{0}}{x}}{\mathof{\Gamma_{1}}{x}}}{\mathpiparallel{\mathof{\Gamma_{1}}{x}}{\mathof{\Gamma_{0}}{x}}}$.

    \noindent \cref{item:obligation-def-subtyping-on-env}
    ${\mathof{\mathdom}{\mathpiparallel{\Gamma_{0}}{\Gamma_{1}}}}={\mathof{\mathdom}{\mathpiparallel{\Gamma_{1}}{\Gamma_{0}}}}$.
    Then 
    ${{\mathof{\mathdom}{\mathpiparallel{\Gamma_{0}}{\Gamma_{1}}}}\setminus{\mathof{\mathdom}{\mathpiparallel{\Gamma_{1}}{\Gamma_{0}}}}}={\emptyset}$.
    Thus, \cref{def:subtyping-on-env} \cref{item:obligation-def-subtyping-on-env} holds
    obviously.
    
    \noindent \cref{item:assoc1-prop-property-subtyping-on-env}
    We show $\mathissubtype{\mathpiparallel{\Gamma_{0}}{\mleft(\mathpiparallel{\Gamma_{1}}{\Gamma_{2}}\mright)}}{\mathpiparallel{\mleft(\mathpiparallel{\Gamma_{0}}{\Gamma_{1}}\mright)}{\Gamma_{2}}}$.

    \noindent \cref{item:dom-def-subtyping-on-env}
    We see
    \[
    {\mathof{\mathdom}{\mathpiparallel{\Gamma_{0}}{\mleft(\mathpiparallel{\Gamma_{1}}{\Gamma_{2}}\mright)}}} 
    =
    {{\mathof{\mathdom}{\Gamma_{0}}} \cup {\mathof{\mathdom}{\Gamma_{1}}} \cup {\mathof{\mathdom}{\Gamma_{2}}}}
    =
    {\mathof{\mathdom}{\mathpiparallel{\mleft(\mathpiparallel{\Gamma_{0}}{\Gamma_{1}}\mright)}{\Gamma_{2}}}}.
    \]
    Then, we have
    \[
    {\mathof{\mathdom}{\mathpiparallel{\Gamma_{0}}{\mleft(\mathpiparallel{\Gamma_{1}}{\Gamma_{2}}\mright)}}} 
    \supseteq
    {\mathof{\mathdom}{\mathpiparallel{\mleft(\mathpiparallel{\Gamma_{0}}{\Gamma_{1}}\mright)}{\Gamma_{2}}}}.
    \]

    \noindent \cref{item:subtyping-def-subtyping-on-env}
    Let ${x}\in{\mathof{\mathdom}{\mathpiparallel{\mleft(\mathpiparallel{\Gamma_{0}}{\Gamma_{1}}\mright)}{\Gamma_{2}}}}$.
    
    Assume 
    ${x}\in{{\mathof{\mathdom}{\mathpiparallel{\Gamma_{0}}{\Gamma_{1}}}}\cap{\mathof{\mathdom}{\Gamma_{2}}}}$.
    Then 
    ${\mathof{\mathpiparallel{\mleft(\mathpiparallel{\Gamma_{0}}{\Gamma_{1}}\mright)}{\Gamma_{2}}}{x}}={\mathpiparallel{\mathof{\mathpiparallel{\Gamma_{0}}{\Gamma_{1}}}{x}}{\mathof{\Gamma_{2}}{x}}}$.
    
    Assume ${x}\in{{\mathof{\mathdom}{\Gamma_{0}}}\cap{\mathof{\mathdom}{\Gamma_{1}}}}$.
    Then
    ${\mathof{\mathpiparallel{\mleft(\mathpiparallel{\Gamma_{0}}{\Gamma_{1}}\mright)}{\Gamma_{2}}}{x}}={{\mathpiparallel{\mleft(\mathpiparallel{\mathof{\Gamma_{0}}{x}}{\mathof{\Gamma_{1}}{x}}\mright)}{\mathof{\Gamma_{2}}{x}}}}$.
    By \cref{prop:property-of-subtyping} \cref{item:assoc1-parallel-in-prop:property-of-subtyping},
    ${\mathof{\mathpiparallel{\mleft(\mathpiparallel{\Gamma_{0}}{\Gamma_{1}}\mright)}{\Gamma_{2}}}{x}}={\mathpiparallel{\mathof{\Gamma_{0}}{x}}{\mleft(\mathpiparallel{\mathof{\Gamma_{1}}{x}}{\mathof{\Gamma_{2}}{x}}\mright)}}$.
    Since
    $\mathof{\mathpiparallel{\Gamma_{0}}{\mleft(\mathpiparallel{\Gamma_{1}}{\Gamma_{2}}\mright)}}{x}=\mathpiparallel{\mathof{\Gamma_{0}}{x}}{\mleft(\mathpiparallel{\mathof{\Gamma_{1}}{x}}{\mathof{\Gamma_{2}}{x}}\mright)}$,
    we have
    ${\mathof{\mathpiparallel{\mleft(\mathpiparallel{\Gamma_{0}}{\Gamma_{1}}\mright)}{\Gamma_{2}}}{x}}={\mathof{\mathpiparallel{\Gamma_{0}}{\mleft(\mathpiparallel{\Gamma_{1}}{\Gamma_{2}}\mright)}}{x}}$.

    Assume 
    ${x}\in{{\mathof{\mathdom}{\Gamma_{0}}}\setminus{\mathof{\mathdom}{\Gamma_{1}}}}$.
    Then 
    ${\mathof{\mathpiparallel{\mleft(\mathpiparallel{\Gamma_{0}}{\Gamma_{1}}\mright)}{\Gamma_{2}}}{x}}={\mathpiparallel{\mathof{\Gamma_{0}}{x}}{\mathof{\Gamma_{2}}{x}}}$.
    Since
    ${\mathof{\mathpiparallel{\Gamma_{0}}{\mleft(\mathpiparallel{\Gamma_{1}}{\Gamma_{2}}\mright)}}{x}}={\mathof{\mathpiparallel{\Gamma_{0}}{\Gamma_{2}}}{x}}$,
    we have
    ${\mathof{\mathpiparallel{\mleft(\mathpiparallel{\Gamma_{0}}{\Gamma_{1}}\mright)}{\Gamma_{2}}}{x}}={\mathof{\mathpiparallel{\Gamma_{0}}{\mleft(\mathpiparallel{\Gamma_{1}}{\Gamma_{2}}\mright)}}{x}}$.

    Assume 
    ${x}\in{{\mathof{\mathdom}{\Gamma_{1}}}\setminus{\mathof{\mathdom}{\Gamma_{0}}}}$.
    Then 
    ${\mathof{\mathpiparallel{\mleft(\mathpiparallel{\Gamma_{0}}{\Gamma_{1}}\mright)}{\Gamma_{2}}}{x}}={\mathpiparallel{\mathof{\Gamma_{1}}{x}}{\mathof{\Gamma_{2}}{x}}}$.
    Since
    ${\mathof{\mathpiparallel{\Gamma_{0}}{\mleft(\mathpiparallel{\Gamma_{1}}{\Gamma_{2}}\mright)}}{x}}={\mathof{\mathpiparallel{\Gamma_{1}}{\Gamma_{2}}}{x}}$,
    we have
    ${\mathof{\mathpiparallel{\mleft(\mathpiparallel{\Gamma_{0}}{\Gamma_{1}}\mright)}{\Gamma_{2}}}{x}}={\mathof{\mathpiparallel{\Gamma_{0}}{\mleft(\mathpiparallel{\Gamma_{1}}{\Gamma_{2}}\mright)}}{x}}$.

    Assume 
    ${x}\in{{\mathof{\mathdom}{\mathpiparallel{\Gamma_{0}}{\Gamma_{1}}}}\setminus{\mathof{\mathdom}{\Gamma_{2}}}}$.
    Then
    ${\mathof{\mathpiparallel{\mleft(\mathpiparallel{\Gamma_{0}}{\Gamma_{1}}\mright)}{\Gamma_{2}}}{x}}={\mathof{\mathpiparallel{\Gamma_{0}}{\Gamma_{1}}}{x}}$.
    Since
    ${\mathof{\mathpiparallel{\Gamma_{0}}{\mleft(\mathpiparallel{\Gamma_{1}}{\Gamma_{2}}\mright)}}{x}}={\mathof{\mathpiparallel{\Gamma_{0}}{\Gamma_{1}}}{x}}$,
    we have
    ${\mathof{\mathpiparallel{\mleft(\mathpiparallel{\Gamma_{0}}{\Gamma_{1}}\mright)}{\Gamma_{2}}}{x}}={\mathof{\mathpiparallel{\Gamma_{0}}{\mleft(\mathpiparallel{\Gamma_{1}}{\Gamma_{2}}\mright)}}{x}}$.

    Assume 
    ${x}\in{{\mathof{\mathdom}{\Gamma_{2}}}\setminus{\mathof{\mathdom}{\mathpiparallel{\Gamma_{0}}{\Gamma_{1}}}}}$.
    Then
    ${\mathof{\mathpiparallel{\mleft(\mathpiparallel{\Gamma_{0}}{\Gamma_{1}}\mright)}{\Gamma_{2}}}{x}}={\mathof{\Gamma_{2}}{x}}$.
    Since
    ${\mathof{\mathpiparallel{\Gamma_{0}}{\mleft(\mathpiparallel{\Gamma_{1}}{\Gamma_{2}}\mright)}}{x}}={\mathof{\Gamma_{2}}{x}}$,
    we have
    ${\mathof{\mathpiparallel{\mleft(\mathpiparallel{\Gamma_{0}}{\Gamma_{1}}\mright)}{\Gamma_{2}}}{x}}={\mathof{\mathpiparallel{\Gamma_{0}}{\mleft(\mathpiparallel{\Gamma_{1}}{\Gamma_{2}}\mright)}}{x}}$.

    \noindent \cref{item:obligation-def-subtyping-on-env}
    ${\mathof{\mathdom}{\mathpiparallel{\Gamma_{0}}{\mleft(\mathpiparallel{\Gamma_{1}}{\Gamma_{2}}\mright)}}}={\mathof{\mathdom}{\mathpiparallel{\mleft(\mathpiparallel{\Gamma_{0}}{\Gamma_{1}}\mright)}{\Gamma_{2}}}}$.
    Then
    ${{\mathof{\mathdom}{\mathpiparallel{\Gamma_{0}}{\mleft(\mathpiparallel{\Gamma_{1}}{\Gamma_{2}}\mright)}}}\setminus{\mathof{\mathdom}{\mathpiparallel{\mleft(\mathpiparallel{\Gamma_{0}}{\Gamma_{1}}\mright)}{\Gamma_{2}}}}}={\emptyset}$.
    Thus, \cref{def:subtyping-on-env} \cref{item:obligation-def-subtyping-on-env} holds
    obviously.

    \noindent \cref{item:cong-prop-property-subtyping-on-env}
    Assume $\mathissubtype{\Gamma_{0}}{\Gamma'_{0}}$.
    We show 
    $\mathissubtype{\mathpiparallel{\Gamma_{0}}{\Gamma_{1}}}{\mathpiparallel{\Gamma'_{0}}{\Gamma_{1}}}$.
    
    \noindent \cref{item:dom-def-subtyping-on-env}
    Since $\mathissubtype{\Gamma_{0}}{\Gamma'_{0}}$,
    we have ${\mathof{\mathdom}{\Gamma_{0}}}\supseteq{\mathof{\mathdom}{\Gamma'_{0}}}$.
    Thus, 
    \[
    {\mathof{\mathdom}{\mathpiparallel{\Gamma_{0}}{\Gamma_{1}}}}={{\mathof{\mathdom}{\Gamma_{0}}}\cup{\mathof{\mathdom}{\Gamma_{1}}}}\supseteq{{\mathof{\mathdom}{\Gamma'_{0}}}\cup{\mathof{\mathdom}{\Gamma_{1}}}}={\mathof{\mathdom}{\mathpiparallel{\Gamma'_{0}}{\Gamma_{1}}}}. 
    \]

    \noindent \cref{item:subtyping-def-subtyping-on-env}
    Let ${x}\in{\mathof{\mathdom}{\mathpiparallel{\Gamma'_{0}}{\Gamma_{1}}}}$.
    Since ${\mathof{\mathdom}{\mathpiparallel{\Gamma_{0}}{\Gamma_{1}}}}\supseteq{\mathof{\mathdom}{\mathpiparallel{\Gamma'_{0}}{\Gamma_{1}}}}$,
    we have ${x}\in{\mathof{\mathdom}{\mathpiparallel{\Gamma_{0}}{\Gamma_{1}}}}$.

    Assume ${x}\in{{\mathof{\mathdom}{\Gamma'_{0}}}\cap{\mathof{\mathdom}{\Gamma_{1}}}}$.
    Then, we have
    ${\mathof{\mathpiparallel{\Gamma'_{0}}{\Gamma_{1}}}{x}}={\mathpiparallel{\mathof{\Gamma'_{0}}{x}}{\mathof{\Gamma_{1}}{x}}}$.
    Since ${\mathof{\mathdom}{\Gamma_{0}}}\supseteq{\mathof{\mathdom}{\Gamma'_{0}}}$,
    we have ${x}\in{{\mathof{\mathdom}{\Gamma_{0}}}\cap{\mathof{\mathdom}{\Gamma_{1}}}}$.
    Then, we see
    ${\mathof{\mathpiparallel{\Gamma_{0}}{\Gamma_{1}}}{x}}={\mathpiparallel{\mathof{\Gamma_{0}}{x}}{\mathof{\Gamma_{1}}{x}}}$.
    Since $\mathissubtype{\Gamma_{0}}{\Gamma'_{0}}$,
    we have $\mathissubtype{\mathof{\Gamma_{0}}{x}}{\mathof{\Gamma'_{0}}{x}}$.
    By \cref{prop:property-of-subtyping} \cref{item:parallel-in-prop:property-of-subtyping},
    $\mathissubtype{\mathpiparallel{\mathof{\Gamma_{0}}{x}}{\mathof{\Gamma_{1}}{x}}}{\mathpiparallel{\mathof{\Gamma'_{0}}{x}}{\mathof{\Gamma_{1}}{x}}}$.
    Thus, 
    $\mathissubtype{\mathof{\mathpiparallel{\Gamma_{0}}{\Gamma_{1}}}{x}}{\mathof{\mathpiparallel{\Gamma'_{0}}{\Gamma_{1}}}{x}}$.

    Assume ${x}\in{{\mathof{\mathdom}{\Gamma'_{0}}}\setminus{\mathof{\mathdom}{\Gamma_{1}}}}$.
    Then 
    ${\mathof{\mathpiparallel{\Gamma'_{0}}{\Gamma_{1}}}{x}}={\mathof{\Gamma'_{0}}{x}}$.
    Since ${\mathof{\mathdom}{\Gamma_{0}}}\supseteq{\mathof{\mathdom}{\Gamma'_{0}}}$,
    we have ${x}\in{{\mathof{\mathdom}{\Gamma'_{0}}}\setminus{\mathof{\mathdom}{\Gamma_{1}}}}$.
    Then 
    ${\mathof{\mathpiparallel{\Gamma_{0}}{\Gamma_{1}}}{x}}={\mathof{\Gamma_{0}}{x}}$.
    Since $\mathissubtype{\Gamma_{0}}{\Gamma'_{0}}$,
    we have $\mathissubtype{\mathof{\Gamma_{0}}{x}}{\mathof{\Gamma'_{0}}{x}}$.
    Thus, 
    $\mathissubtype{\mathof{\mathpiparallel{\Gamma_{0}}{\Gamma_{1}}}{x}}{\mathof{\mathpiparallel{\Gamma'_{0}}{\Gamma_{1}}}{x}}$.

    Assume ${x}\in{{\mathof{\mathdom}{\Gamma_{1}}}\setminus{\mathof{\mathdom}{\Gamma'_{0}}}}$.
    Then 
    ${\mathof{\mathpiparallel{\Gamma'_{0}}{\Gamma_{1}}}{x}}={\mathof{\Gamma_{1}}{x}}$.
 
    Assume ${x}\in{\mathof{\mathdom}{\Gamma_{0}}}$.
    Then ${\mathof{\mathpiparallel{\Gamma_{0}}{\Gamma_{1}}}{x}}={\mathpiparallel{\mathof{\Gamma_{0}}{x}}{\mathof{\Gamma_{1}}{x}}}$.
    Since ${x}\in{{\mathof{\mathdom}{\Gamma_{0}}}\setminus{\mathof{\mathdom}{\Gamma'_{0}}}}$, and $\mathissubtype{\Gamma_{0}}{\Gamma'_{0}}$,
    we have $\mathobligation{}{\mathof{\Gamma_{0}}{x}}={\infty}$.
    By \cref{prop:property-of-subtyping} \cref{item:ob-infty-and-parallel-in-prop:property-of-subtyping},
    we see $\mathissubtype{\mathpiparallel{\mathof{\Gamma_{0}}{x}}{\mathof{\Gamma_{1}}{x}}}{\mathof{\Gamma_{1}}{x}}$.
    Thus, 
    $\mathissubtype{\mathof{\mathpiparallel{\Gamma_{0}}{\Gamma_{1}}}{x}}{\mathof{\mathpiparallel{\Gamma'_{0}}{\Gamma_{1}}}{x}}$.

    Assume ${x}\notin{\mathof{\mathdom}{\Gamma_{0}}}$.
    Then 
    ${\mathof{\mathpiparallel{\Gamma_{0}}{\Gamma_{1}}}{x}}={\mathof{\Gamma_{1}}{x}}$.
    Thus, 
    $\mathissubtype{\mathof{\mathpiparallel{\Gamma_{0}}{\Gamma_{1}}}{x}}{\mathof{\mathpiparallel{\Gamma'_{0}}{\Gamma_{1}}}{x}}$.

    \noindent \cref{item:obligation-def-subtyping-on-env}
    Let ${x}\in{{\mathof{\mathdom}{\mathpiparallel{\Gamma_{0}}{\Gamma_{1}}}}\setminus{\mathof{\mathdom}{\mathpiparallel{\Gamma'_{0}}{\Gamma_{1}}}}}$.
    Then
    ${x}\in{{\mleft({\mathof{\mathdom}{\Gamma_{0}}}\cup{\mathof{\mathdom}{\Gamma_{1}}}\mright)}\setminus{\mleft({\mathof{\mathdom}{\Gamma'_{0}}}\cup{\mathof{\mathdom}{\Gamma_{1}}}\mright)}}$.
    Hence, 
    ${x}\in{{\mathof{\mathdom}{\Gamma_{0}}}\setminus{\mathof{\mathdom}{\Gamma'_{0}}}}$ and
    ${x}\notin{\mathof{\mathdom}{\Gamma_{1}}}$.
    Therefore,
    ${\mathof{\mathpiparallel{\Gamma_{0}}{\Gamma_{1}}}{x}}={\mathof{\Gamma_{0}}{x}}$.
    Since $\mathissubtype{\Gamma_{0}}{\Gamma'_{0}}$,
    we have $\mathobligation{}{\mathof{\Gamma_{0}}{x}}=\infty$.
    Thus,
    $\mathobligation{}{\mathof{\mathpiparallel{\Gamma_{0}}{\Gamma_{1}}}{x}}={\infty}$.

    \noindent \cref{item:cong-full-prop-property-subtyping-on-env}
    Assume $\mathissubtype{\Gamma_{0}}{\Gamma'_{0}}$ and 
    $\mathissubtype{\Gamma_{1}}{\Gamma'_{1}}$.
    By this proposition \cref{item:cong-prop-property-subtyping-on-env},
    $\mathissubtype{\mathpiparallel{\Gamma_{0}}{\Gamma_{1}}}{\mathpiparallel{\Gamma'_{0}}{\Gamma_{1}}}$.
    From this proposition \cref{item:commut-prop-property-subtyping-on-env},
    $\mathissubtype{\mathpiparallel{\Gamma'_{0}}{\Gamma_{1}}}{\mathpiparallel{\Gamma_{1}}{\Gamma'_{0}}}$.
    This proposition \cref{item:cong-prop-property-subtyping-on-env} implies
    $\mathissubtype{\mathpiparallel{\Gamma_{1}}{\Gamma'_{0}}}{\mathpiparallel{\Gamma'_{1}}{\Gamma'_{0}}}$.
    From this proposition \cref{item:commut-prop-property-subtyping-on-env},
    $\mathissubtype{\mathpiparallel{\Gamma'_{1}}{\Gamma'_{0}}}{\mathpiparallel{\Gamma'_{0}}{\Gamma'_{1}}}$.
    By transitivity,
    $\mathissubtype{\mathpiparallel{\Gamma_{0}}{\Gamma_{1}}}{\mathpiparallel{\Gamma'_{0}}{\Gamma'_{1}}}$.

    \noindent \cref{item:assoc2-prop-property-subtyping-on-env}
    By \cref{item:commut-prop-property-subtyping-on-env},
    \cref{item:assoc1-prop-property-subtyping-on-env},
    \cref{item:cong-full-prop-property-subtyping-on-env}, and
    transitivity of $\mathissubtypesy$, we have
    \begin{align*}
     {\mathpiparallel{\mleft(\mathpiparallel{\Gamma_{0}}{\Gamma_{1}}\mright)}{\Gamma_{2}}}
     &\mathissubtypesy{\mathpiparallel{\Gamma_{2}}{\mleft(\mathpiparallel{\Gamma_{0}}{\Gamma_{1}}\mright)}} \\
     &\mathissubtypesy{\mathpiparallel{\Gamma_{2}}{\mleft(\mathpiparallel{\Gamma_{1}}{\Gamma_{0}}\mright)}} \\
     &\mathissubtypesy{\mathpiparallel{\mleft(\mathpiparallel{\Gamma_{2}}{\Gamma_{1}}\mright)}{\Gamma_{0}}} \\
     &\mathissubtypesy{\mathpiparallel{\Gamma_{0}}{\mleft(\mathpiparallel{\Gamma_{2}}{\Gamma_{1}}\mright)}} \\
     &\mathissubtypesy{\mathpiparallel{\Gamma_{0}}{\mleft(\mathpiparallel{\Gamma_{1}}{\Gamma_{2}}\mright)}}.
    \end{align*}
    Thus, 
    $\mathissubtype{\mathpiparallel{\mleft(\mathpiparallel{\Gamma_{0}}{\Gamma_{1}}\mright)}{\Gamma_{2}}}{\mathpiparallel{\Gamma_{0}}{\mleft(\mathpiparallel{\Gamma_{1}}{\Gamma_{2}}\mright)}}$.

    \noindent \cref{item:replication-cong-prop-property-subtyping-on-env}
    Assume $\mathissubtype{\Gamma}{\Gamma'}$.
    We show $\mathissubtype{\mathpireplication{\Gamma}}{\mathpireplication{\Gamma'}}$.
    
    \noindent \cref{item:dom-def-subtyping-on-env}
    Since $\mathissubtype{\Gamma}{\Gamma'}$,
    we have ${\mathof{\mathdom}{\Gamma}}\supseteq{\mathof{\mathdom}{\Gamma'}}$.
    Since ${\mathof{\mathdom}{\mathpireplication{\Gamma}}}={\mathof{\mathdom}{\Gamma}}$ and
    ${\mathof{\mathdom}{\mathpireplication{\Gamma'}}}={\mathof{\mathdom}{\Gamma'}}$,
    we have
    ${\mathof{\mathdom}{\mathpireplication{\Gamma}}}\supseteq{\mathof{\mathdom}{\mathpireplication{\Gamma'}}}$.

    \noindent \cref{item:subtyping-def-subtyping-on-env}
    Let ${x}\in{\mathof{\mathdom}{\mathpireplication{\Gamma'}}}$.
    Since 
    ${\mathof{\mathdom}{\mathpireplication{\Gamma'}}}={\mathof{\mathdom}{\Gamma'}}$,
    we have ${x}\in{\mathof{\mathdom}{\Gamma'}}$.
    Since $\mathissubtype{\Gamma}{\Gamma'}$,
    we see $\mathissubtype{\mathof{\Gamma}{x}}{\mathof{\Gamma'}{x}}$.
    By \cref{prop:property-of-subtyping} 
    \cref{item:replication-in-prop:property-of-subtyping},
    $\mathissubtype{\mathpireplication{\mathof{\Gamma}{x}}}{\mathpireplication{\mathof{\Gamma'}{x}}}$.
    Then, we have
    $\mathissubtype{\mathof{\mathpireplication{\Gamma}}{x}}{\mathof{\mathpireplication{\Gamma'}}{x}}$.

    \noindent \cref{item:obligation-def-subtyping-on-env} 
    Let ${x}\in{{\mathof{\mathdom}{\mathpireplication{\Gamma}}}\setminus{\mathof{\mathdom}{\mathpireplication{\Gamma'}}}}$.
    Since ${\mathof{\mathdom}{\mathpireplication{\Gamma}}}={\mathof{\mathdom}{\Gamma}}$ and
    ${\mathof{\mathdom}{\mathpireplication{\Gamma'}}}={\mathof{\mathdom}{\Gamma'}}$,
    we have
    ${x}\in{{\mathof{\mathdom}{\Gamma}}\setminus{\mathof{\mathdom}{\Gamma'}}}$.
    Since $\mathissubtype{\Gamma}{\Gamma'}$, we see
    ${\mathobligation{}{\mathof{\Gamma}{x}}}={\infty}$.
    Then
    \[
    {\mathobligation{}{\mathof{\mathpireplication{\Gamma}}{x}}}={\mathobligation{}{\mathpireplication{\mathof{\Gamma}{x}}}}={\mathobligation{}{\mathof{\Gamma}{x}}}={\infty}. 
    \]

    \noindent \cref{item:replication-prop-property-subtyping-on-env}
    We show $\mathissubtype{\mathpireplication{\Gamma}}{\mathpiparallel{\mathpireplication{\Gamma}}{\Gamma}}$.

    \noindent \cref{item:dom-def-subtyping-on-env}
    Since 
    ${\mathof{\mathdom}{\mathpireplication{\Gamma}}}={\mathof{\mathdom}{\Gamma}}$ and
    ${\mathof{\mathdom}{\mathpiparallel{\mathpireplication{\Gamma}}{\Gamma}}}={{\mathof{\mathdom}{\mathpireplication{\Gamma}}}\cup{\mathof{\mathdom}{\Gamma}}}={{\mathof{\mathdom}{\Gamma}}\cup{\mathof{\mathdom}{\Gamma}}}={\mathof{\mathdom}{\Gamma}}$,
    we have 
    ${\mathof{\mathdom}{\mathpireplication{\Gamma}}}\supseteq{\mathof{\mathdom}{\mathpiparallel{\mathpireplication{\Gamma}}{\Gamma}}}$.

    \noindent \cref{item:subtyping-def-subtyping-on-env}
    Let ${x}\in{\mathof{\mathdom}{\mathpiparallel{\mathpireplication{\Gamma}}{\Gamma}}}$.
    By \cref{prop:property-of-subtyping} \cref{item:replication-in-prop:property-of-subtyping},
    we have
    $\mathissubtype{\mathof{\mathpireplication{\Gamma}}{x}}{\mathof{\mleft(\mathpiparallel{\mathpireplication{\Gamma}}{\Gamma}\mright)}{x}}$.

    \noindent \cref{item:obligation-def-subtyping-on-env}
    ${\mathof{\mathdom}{\mathpireplication{\Gamma}}}={\mathof{\mathdom}{\mathpiparallel{\mathpireplication{\Gamma}}{\Gamma}}}$.
    Then 
    ${{\mathof{\mathdom}{\mathpireplication{\Gamma}}}\setminus{\mathof{\mathdom}{\mathpiparallel{\mathpireplication{\Gamma}}{\Gamma}}}}={\emptyset}$.
    Thus, \cref{def:subtyping-on-env} \cref{item:obligation-def-subtyping-on-env} holds
    obviously.

    \noindent \cref{item:addition-prop-property-subtyping-on-env}
    Assume $\mathissubtype{\Gamma}{\Gamma'}$.
    We show $\mathissubtype{(\Gamma, \mathistype{x}{\tau})}{(\Gamma', \mathistype{x}{\tau})}$.

    \noindent \cref{item:dom-def-subtyping-on-env}
    Since $\mathissubtype{\Gamma}{\Gamma'}$,
    we have ${\mathof{\mathdom}{\Gamma}}\supseteq{\mathof{\mathdom}{\Gamma'}}$.
    Hence, ${\mathof{\mathdom}{\Gamma, \mathistype{x}{\tau}}}\supseteq{\mathof{\mathdom}{\Gamma', \mathistype{x}{\tau}}}$.

    \noindent \cref{item:subtyping-def-subtyping-on-env}
    Let ${y}\in{\mathof{\mathdom}{\Gamma', \mathistype{x}{\tau}}}$.
    If ${y}={x}$, then
    ${\mathof{({\Gamma', \mathistype{x}{\tau}})}{y}}={\tau}={\mathof{({\Gamma, \mathistype{x}{\tau}})}{y}}$.
    Assume ${y}\not={x}$. Then ${y}\in{\mathof{\mathdom}{\Gamma'}}$.
    Since $\mathissubtype{\Gamma}{\Gamma'}$,
    $\mathissubtype{\mathof{\Gamma}{y}}{\mathof{\Gamma'}{y}}$.
    Thus,
    $\mathissubtype{\mathof{({\Gamma, \mathistype{x}{\tau}})}{y}}{\mathof{({\Gamma', \mathistype{x}{\tau}})}{y}}$.

    \noindent \cref{item:obligation-def-subtyping-on-env}
    Let ${y}\in{{\mathof{\mathdom}{\Gamma, \mathistype{x}{\tau}}}\setminus{\mathof{\mathdom}{\Gamma', \mathistype{x}{\tau}}}}$.
    Then, we have
    ${y}\in{{\mathof{\mathdom}{\Gamma}}\setminus{\mathof{\mathdom}{\Gamma'}}}$.
    Since $\mathissubtype{\Gamma}{\Gamma'}$,
    we see
    ${\mathobligation{}{\mathof{\Gamma}{y}}}={\infty}$.
    Thus, ${\mathobligation{}{\mathof{(\Gamma, \mathistype{x}{\tau})}{y}}}={\infty}$.

    \noindent \cref{item:minus-prop-property-subtyping-on-env}
    Assume 
    $\mathissubtype{(\Gamma, \mathistype{x}{\tau})}{(\Gamma', \mathistype{x}{\tau})}$.
    Then
    ${x}\notin{\mathof{\mathdom}{\Gamma}}$ and ${x}\notin{\mathof{\mathdom}{\Gamma'}}$.
    We show $\mathissubtype{\Gamma}{\Gamma'}$.

    \noindent \cref{item:dom-def-subtyping-on-env}
    Since 
    $\mathissubtype{(\Gamma, \mathistype{x}{\tau})}{(\Gamma', \mathistype{x}{\tau})}$,
    ${x}\notin{\mathof{\mathdom}{\Gamma}}$, and ${x}\notin{\mathof{\mathdom}{\Gamma'}}$,
    we have ${\mathof{\mathdom}{\Gamma}}\supseteq{\mathof{\mathdom}{\Gamma'}}$.

    \noindent \cref{item:subtyping-def-subtyping-on-env}
    Let ${y}\in{\mathof{\mathdom}{\Gamma'}}$.
    Since ${x}\notin{\mathof{\mathdom}{\Gamma'}}$, we have ${y}\not={x}$. 
    We also see ${y}\in{\mathof{\mathdom}{\Gamma', \mathistype{x}{\tau}}}$.
    Since
    $\mathissubtype{(\Gamma, \mathistype{x}{\tau})}{(\Gamma', \mathistype{x}{\tau})}$,
    we have 
    $\mathissubtype{\mathof{(\Gamma', \mathistype{x}{\tau})}{y}}{\mathof{(\Gamma, \mathistype{x}{\tau})}{y}}$.
    Thus, $\mathissubtype{\mathof{\Gamma'}{y}}{\mathof{\Gamma}{y}}$.

    \noindent \cref{item:obligation-def-subtyping-on-env}
    Let ${y}\in{{\mathof{\mathdom}{\Gamma}}\setminus{\mathof{\mathdom}{\Gamma'}}}$.
    Then, we have
    ${y}\in{{\mathof{\mathdom}{(\Gamma, \mathistype{x}{\tau})}}\setminus{\mathof{\mathdom}{(\Gamma', \mathistype{x}{\tau})}}}$.
    Since $\mathissubtype{(\Gamma, \mathistype{x}{\tau})}{(\Gamma', \mathistype{x}{\tau})}$,
    we see
    ${\mathobligation{}{\mathof{(\Gamma, \mathistype{x}{\tau})}{y}}}={\infty}$.
    Thus, ${\mathobligation{}{\mathof{\Gamma}{y}}}={\infty}$.

    \noindent \cref{item:levelraise-prop-property-subtyping-on-env}
    We show $\mathissubtype{\mathlevelraise{t_{I}}{t_{O}}{\Gamma}}{\Gamma}$.
    
    \noindent \cref{item:dom-def-subtyping-on-env}
    Since
    ${\mathof{\mathdom}{\mathlevelraise{t_{I}}{t_{O}}{\Gamma}}}={\mathof{\mathdom}{\Gamma}}$,
    we have ${\mathof{\mathdom}{\mathlevelraise{t_{I}}{t_{O}}{\Gamma}}}\supseteq{\mathof{\mathdom}{\Gamma}}$. 

    \noindent \cref{item:subtyping-def-subtyping-on-env}
    Let ${x}\in{\mathof{\mathdom}{\Gamma}}$.
    By \cref{prop:property-of-subtyping} \cref{item:levelraise-is-subtype-in-prop:property-of-subtyping},
    we have
    ${\mathof{\mathlevelraise{t_{I}}{t_{O}}{\Gamma}}{x}}={\mathof{\Gamma}{x}}$.

    \noindent \cref{item:obligation-def-subtyping-on-env}
    ${\mathof{\mathdom}{\mathlevelraise{t_{I}}{t_{O}}{\Gamma}}}={\mathof{\mathdom}{\Gamma}}$.
    Then ${{\mathof{\mathdom}{\mathlevelraise{t_{I}}{t_{O}}{\Gamma}}}\setminus{\mathof{\mathdom}{\Gamma}}} = {\emptyset}$.
    Thus, \cref{def:subtyping-on-env} \cref{item:obligation-def-subtyping-on-env} holds
    obviously.

    \noindent \cref{item:levelraiseuni-prop-property-subtyping-on-env}
    We show $\mathissubtype{\mathlevelraiseuni{\Gamma}}{\Gamma}$.

    \noindent \cref{item:dom-def-subtyping-on-env}
    Since
    ${\mathof{\mathdom}{\mathlevelraiseuni{\Gamma}}}={\mathof{\mathdom}{\Gamma}}$,
    we have ${\mathof{\mathdom}{\mathlevelraiseuni{\Gamma}}}\supseteq{\mathof{\mathdom}{\Gamma}}$. 

    \noindent \cref{item:subtyping-def-subtyping-on-env}
    Let ${x}\in{\mathof{\mathdom}{\Gamma}}$.
    By \cref{prop:property-of-subtyping} \cref{item:levelraiseuni-is-subtype-in-prop:property-of-subtyping},
    we have ${\mathof{\mathlevelraiseuni{\Gamma}}{x}}={\mathof{\Gamma}{x}}$.

    \noindent \cref{item:obligation-def-subtyping-on-env}
    ${\mathof{\mathdom}{\mathlevelraiseuni{\Gamma}}}={\mathof{\mathdom}{\Gamma}}$.
    Then ${{\mathof{\mathdom}{\mathlevelraiseuni{\Gamma}}}\setminus{\mathof{\mathdom}{\Gamma}}} = {\emptyset}$.
    Thus, \cref{def:subtyping-on-env} \cref{item:obligation-def-subtyping-on-env} holds
    obviously.

   \end{proof}

   \begin{proposition}
    \label[proposition]{prop:property-subtyping-on-env-rel}
    For type environments $\Gamma_{0}$ and $\Gamma_{1}$,
    if $\mathissubusage{\Gamma_{0}}{\Gamma_{1}}$ and $\mathreliable{\Gamma_{0}}$,
    then $\mathreliable{\Gamma_{1}}$.
   \end{proposition}

   \begin{proof}
    Straightforward.
   \end{proof}

   \begin{lemma}
    Let $\Gamma$, $\Delta$ be type environments and $L$ be a lattice of secrecy levels.
    Assume that 
    both $\mathtypeenvandsecrecylatice{\Gamma}{L}$ and 
    $\mathtypeenvandsecrecylatice{\Delta}{L}$ 
    are $l$-secure.
    Then:
    \begin{enumerate}
     \item $\mathtypeenvandsecrecylatice{\mathpiparallel{\Gamma}{\Delta}}{L}$ is $l$-secure.
     \item $\mathtypeenvandsecrecylatice{\mathpireplication{\Gamma}}{L}$ is $l$-secure.
     \item $\mathtypeenvandsecrecylatice{\mathlevelraise{t_{I}}{t_{O}}{\Gamma}}{L}$ is
	   $l$-secure.
     \item $\mathtypeenvandsecrecylatice{\mathlevelraiseuni{\Gamma}}{L}$ is $l$-secure.
    \end{enumerate}
   \end{lemma}

   \begin{proof}
    Straightforward.
   \end{proof}

\section{The details of proof of subject reduction}
 \label[appendix]{sec:proof-of-subject-reduction}

\subsection{Inversion lemma}

\begin{lemma}[Inversion]
 \label[lemma]{lemma:inversion}
 Assume that
 $\mathtypejudgementwithsecrecy{\Gamma}{L}{m}{P}$ is $l$-securely derivable.
 \begin{enumerate}
  \item If ${P}\equiv{\mathusagenil}$,
	then $\mathissubtype{\Gamma}{\emptyset}$. 
	 \label{item:usagenil-lemma-inversion}
  \item If ${P}\equiv{\mathpiparallel{P_{0}}{P_{1}}}$,
	then there exist two type environments $\Gamma'_{0}$, $\Gamma'_{1}$,
	and ${m'}\in{L}$
	such that $\mathissubtype{\Gamma}{\mathpiparallel{\Gamma'_{0}}{\Gamma'_{1}}}$, 
	$\mathissublattice{L'}{L}$, and $m\leq_{L}m'$ and
	$\mathtypejudgementwithsecrecy{\Gamma'_{i}}{L'}{m'}{P_{i}}$ is 
	$l$-securely derivable for each $i=0, 1$.
	\label{item:parallel-lemma-inversion}
  \item If ${P}\equiv{\mathpioutput{x}{\mathtuple{\mathvect{v}}}.P_{0}}$,
	then there exist a type environments $\Gamma'$,
	secrecy levels ${l_{0}}, {m_{0}}\in{L}$,
	types ${\mathvect{\tau}}$, a usage $U$ and
	${t_{c}}\in{\mathnat \cup \mathsetextension{\infty}}$
	such that
	$\mathissubtype{\Gamma}{\Gamma''}$,
	${m}\leq_{L}{l_{0}}$, and 
	${m}\leq_{L}{m_{0}}$,
	$\mathtypejudgementwithsecrecy{\Gamma', {\mathistype{x}{\mathprogramtypetuple{\mathvect{\tau}}^{l_{0}}/U}}}{L}{m_{0}}{P_{0}}$
	is $l$-securely derivable,
	$\mathtypeenvandsecrecylatice{\Gamma''}{L}$ is $l$-secure, and
	$t_{c} = \infty$ implies ${l_{0}}\leq_{L}{m_{0}}$,
	where
	${\Gamma''}\equiv\mleft(\mathpiparallel{\mathlevelraise{t_{c}+1}{t_{c}+1}{\mathpiparallel{\Gamma'}{\mathistype{\mathvect{v}}{\mathlevelraiseuni{\mathvect{\tau}}}}}}{\mathistype{x}{\mathprogramtypetuple{\mathvect{\tau}}^{l_{0}}/O^{0}_{t_{c}}U}}\mright)$.
	 \label{item:output-lemma-inversion}
  \item If ${P}\equiv{\mathpiinput{x}{\mathtuple{\mathvect{y}}}.P_{0}}$,
	then there exist a type environments $\Gamma'$,
	secrecy levels ${l_{0}}, {m_{0}}\in{L}$, 
	types ${\mathvect{\tau}}$, a usage $U$ and
	${t_{c}}\in{\mathnat \cup \mathsetextension{\infty}}$
	such that
	$\mathissubtype{\Gamma}{\Gamma''}$,
	${m}\leq_{L}{l_{0}}$, and ${m}\leq_{L}{m_{0}}$ hold,
	$\mathtypejudgementwithsecrecy{\Gamma', {\mathistype{x}{\mathprogramtypetuple{\mathvect{\tau}}^{l_{0}}/U}}, {\mathistype{\mathvect{y}}{\mathvect{\tau}}}}{L}{m_{0}}{P_{0}}$
	is $l$-securely derivable, 
	$\mathtypeenvandsecrecylatice{\Gamma''}{L}$ is $l$-secure, and
	$t_{c} = \infty$ implies ${l_{0}}\leq_{L}{m_{0}}$, 
	where
	$\Gamma''\equiv\mleft(\mathlevelraise{t_{c}+1}{t_{c}+1}{\Gamma'}, \mathistype{x}{\mathprogramtypetuple{\mathvect{\tau}}^{l_{0}}/I^{0}_{t_{c}}U}\mright)$.
	 \label{item:input-lemma-inversion}
  \item If ${P}\equiv{\mathpireplication{P_{0}}}$,
	then there exist a type environments $\Gamma'$,
	and ${m'}\in{L}$ such that
	$m\leq_{L}m'$, and
	$\mathissubtype{\Gamma}{\mathpireplication{\Gamma'}}$, and
	$\mathtypejudgementwithsecrecy{\Gamma'}{L}{m'}{P_{0}}$
	is $l$-securely derivable.
	\label{item:replication-lemma-inversion}
  \item If ${P}\equiv{\mathpinew{\mathistype{x}{\xi}}{P_{0}}}$,
	then there exist a type environments $\Gamma'$, 
	a usage $U$, and ${m'}\in{L}$ such that
	$m\leq_{L}m'$,
	$\mathreliable{U}$ and $\mathissubtype{\Gamma}{\Gamma'}$, and
	$\mathtypejudgementwithsecrecy{\Gamma', {\mathistype{x}{\xi/U}}}{L}{m'}{P_{0}}$ 
	is $l$-securely derivable.
	 \label{item:pinew-lemma-inversion}
  \item If ${P}\equiv{\mathpinewsecrecylevel{l_{0}}{\mathvect{l_{1}}}{\mathvect{l_{2}}}{P_{0}}}$,
	then there exist a type environments $\Gamma'$, 
	and ${m'}\in{L}$ such that
	$m\leq_{L}m'$ and $\mathissubtype{\Gamma}{\Gamma'}$,
	$m' \leq_{L} l'$ for any $l'\in\mathvect{l_{1}}, \mathvect{l_{2}}$, and
	$\mathtypejudgementwithsecrecy{\Gamma'}{\mathpinewsecrecylevel{l_{0}}{\mathvect{l_{1}}}{\mathvect{l_{2}}}{L}}{m'}{P_{0}}$ 
	is $l$-securely derivable.
	\label{item:newlevel-lemma-inversion}
  \item If ${P}\equiv{\mathobif{v}{Q_{0}}{Q_{1}}}$,
	then there exist a type environments $\Gamma'$, 
	and ${m'}\in{L}$ such that
	$m\leq_{L}m'$, and $\mathissubtype{\Gamma}{\mathpiparallel{\Gamma'}{\mathistype{v}{\mathbooltypewithsec{m'}}}}$, and
	$\mathtypejudgementwithsecrecy{\Gamma'}{L}{m'}{Q_{0}}$ and
	$\mathtypejudgementwithsecrecy{\Gamma'}{L}{m'}{Q_{1}}$ are 
	$l$-securely derivable.
	\label{item:if-then-lemma-inversion}
 \end{enumerate}
\end{lemma}

\begin{proof}
 By induction on 
 the size of derivation tree of $\mathtypejudgementwithsecrecy{\Gamma}{L}{m}{P}$.
\end{proof}

\begin{lemma}
 \label[lemma]{lemma:minus-derivable}
 If $\mathtypejudgementwithsecrecy{\Gamma}{L}{m}{P}$
 is $l$-securely derivable and 
 ${x}\notin{\mathFVof{P}}$,
 then $\mathtypejudgementwithsecrecy{\Gamma'}{L}{m}{P}$
 is $l$-securely derivable and
 $\mathissubtype{\Gamma}{\Gamma'}$,
 where 
 $\Gamma'$ is the restriction of $\Gamma$
 to $(\mathof{\mathdom}{\Gamma}\setminus{\mathsetextension{x}})$.
\end{lemma}

\begin{proof}
 By induction on 
 the size of derivation tree of $\mathtypejudgementwithsecrecy{\Gamma}{L}{m}{P}$.
\end{proof}

\begin{lemma}
 \label[lemma]{lemma:FV-derivable}
 If $\mathtypejudgementwithsecrecy{\Gamma}{L}{m}{P}$
 is $l$-securely derivable and 
 ${x}\in{\mathFVof{P}}$,
 then $x\in\mathof{\mathdom}{\Gamma}$.
\end{lemma}

\begin{proof}
 By induction on 
 a derivation tree of $\mathtypejudgementwithsecrecy{\Gamma}{L}{m}{P}$.
\end{proof}

\subsection{Proof of \cref{lemma:spo-preservation}} 
\label[appendix]{sec:proof-of-lemma-spo-preservation}

We show \cref{lemma:spo-preservation}.

 Let $P$ and $P'$ be processes, $\Gamma$ be a type environment, 
 $L$ be a lattice of secrecy levels. Let ${m}\in {L}$.
 Assume that $\mathtypejudgementwithsecrecy{\Gamma}{L}{m}{P}$ 
 is $l$-securely derivable and 
 ${P}\mathpistructuralpo{P'}$.
 We show that $\mathtypejudgementwithsecrecy{\Gamma}{L}{m}{P'}$
 is $l$-securely derivable.
 The proof is by induction on the construction of ${P}\mathpistructuralpo{P'}$.
 We consider cases according to the last rule of the construction of
 ${P}\mathpistructuralpo{P'}$.
 
 \noindent Case 1.
 If ${P'}\equiv{P}$, the assumptions immediately imply that
 $\mathtypejudgementwithsecrecy{\Gamma}{L}{m}{P'}$
 is $l$-securely derivable.

 \noindent Case 2.
 Assume that there exists a process $P''$ such that
 ${P}\mathpistructuralpo{P''}$ and ${P''}\mathpistructuralpo{P'}$. 
 By the induction hypothesis,
 $\mathtypejudgementwithsecrecy{\Gamma}{L}{m}{P''}$ 
 is $l$-securely derivable.
 Then ${P''}\mathpistructuralpo{P'}$ and the induction hypothesis imply that
 $\mathtypejudgementwithsecrecy{\Gamma}{L}{m}{P'}$ 
 is $l$-securely derivable.

 \noindent Case 3. \rulename{SP-Zero1}.
 Assume ${P'}\equiv{\mathpiparallel{P}{\mathnil}}$.
 By assumption, there exists an $\bar{l}$-secure derivation tree $\pi$ of 
 $\mathtypejudgementwithsecrecy{\Gamma}{L}{m}{P}$.
 Then, we have an $l$-secure derivation tree as follows:
 \begin{center}
  \begin{inlineprooftree}
   \AxiomC{}
   \RightLabel{$\pi$}
   \DeduceC{$\mathtypejudgementwithsecrecy{\Gamma}{L}{m}{P}$}

   \AxiomC{}
   \RightLabel{\rulename{T-Zero}}
   \UnaryInfC{$\mathtypejudgementwithsecrecy{\emptyset}{L}{m}{\mathnil}$}
   
   \BinaryInfC{$\mathtypejudgementwithsecrecy{\mathpiparallel{\Gamma}{\emptyset}}{L}{m}{\mathpiparallel{P}{\mathnil}}$}
  \end{inlineprooftree}.
 \end{center}
 Since ${\mathpiparallel{\Gamma}{\emptyset}}\equiv{\Gamma}$,
 we see that $\mathtypejudgementwithsecrecy{\Gamma}{L}{m}{P'}$ 
 is $l$-securely derivable.

 Assume ${P}\equiv{\mathpiparallel{P'}{\mathnil}}$.
 By \cref{lemma:inversion} \cref{item:parallel-lemma-inversion},
 there exist two type environments $\Gamma'_{0}$, $\Gamma'_{1}$,
 and ${m'}\in{L}$
 such that $\mathissubtype{\Gamma}{\mathpiparallel{\Gamma'_{0}}{\Gamma'_{1}}}$, 
 and $m\leq_{L}m'$ and both
 $\mathtypejudgementwithsecrecy{\Gamma'_{0}}{L}{m'}{P'}$ and
 $\mathtypejudgementwithsecrecy{\Gamma'_{1}}{L}{m'}{\mathusagenil}$
 are an $l$-securely derivable.
 By \cref{lemma:inversion} \cref{item:usagenil-lemma-inversion}, we have
 $\mathissubtype{\Gamma'_{1}}{\emptyset}$.
 By \cref{prop:property-subtyping-on-env} \cref{item:cong-full-prop-property-subtyping-on-env},
  $\mathissubtype{\mathpiparallel{\Gamma'_{0}}{\Gamma'_{1}}}{\mathpiparallel{\Gamma'_{0}}{\emptyset}}$.
 Since ${\mathpiparallel{\Gamma'_{0}}{\emptyset}}\equiv{\Gamma'_{0}}$,
 we have $\mathissubtype{\mathpiparallel{\Gamma'_{0}}{\Gamma'_{1}}}{\Gamma'_{0}}$.
 By $\mathissubtype{\Gamma}{\mathpiparallel{\Gamma'_{0}}{\Gamma'_{1}}}$,
 we have $\mathissubtype{\Gamma}{\Gamma'_{0}}$.
 Let $\pi'$ be an $l$-secure derivation tree of 
 $\mathtypejudgementwithsecrecy{\Gamma'_{0}}{L}{l'}{P'}$.
 Then, we have an $l$-secure derivation tree as follows:
 \begin{center}
  \begin{inlineprooftree}
   \AxiomC{}
   \RightLabel{$\pi'$}
   \DeduceC{$\mathtypejudgementwithsecrecy{\Gamma'_{0}}{L}{m'}{P'}$}
   \AxiomC{$\mathissubtype{\Gamma}{\Gamma'_{0}}$}
   \AxiomC{$m\leq_{L}m'$}
   \RightLabel{\rulename{T-Weak}}
   \TrinaryInfC{$\mathtypejudgementwithsecrecy{\Gamma}{L}{m}{P'}$}
  \end{inlineprooftree}.
 \end{center}
 Thus, we see that $\mathtypejudgementwithsecrecy{\Gamma}{L}{m}{P'}$ is derivable.
 
 \noindent Case 4. \rulename{SP-Zero2}.
 Assume ${P}\equiv{\mathnil}$ and ${P'}\equiv{\mathpinew{\mathistype{x}{\xi}}{\mathnil}}$.
 By \cref{lemma:inversion} \cref{item:usagenil-lemma-inversion},
 $\mathissubtype{\Gamma}{\emptyset}$.
 We have an $l$-secure derivation tree as follows:
 \begin{center}
  \begin{inlineprooftree}
   \AxiomC{}
   \RightLabel{\rulename{T-Zero}}
   \UnaryInfC{$\mathtypejudgementwithsecrecy{\emptyset}{L}{m}{\mathnil}$}
   \AxiomC{$\mathissubtype{\mathistype{x}{\xi/\mathusagenil}}{\emptyset}$}

   \RightLabel{\rulename{T-Weak}}
   \BinaryInfC{$\mathtypejudgementwithsecrecy{\emptyset}{L}{m}{P'}$}
   \AxiomC{$\mathissubtype{\Gamma}{\emptyset}$}

   \RightLabel{\rulename{T-Weak}}
   \BinaryInfC{$\mathtypejudgementwithsecrecy{\Gamma}{L}{m}{P'}$}
  \end{inlineprooftree}.
 \end{center}
 Thus, we see that $\mathtypejudgementwithsecrecy{\Gamma}{L}{m}{P'}$ 
 is $l$-securely derivable.

 Assume ${P}\equiv{\mathpinew{\mathistype{x}{\xi}}{\mathnil}}$ and ${P'}\equiv{\mathnil}$.
 By \cref{lemma:inversion} \cref{item:pinew-lemma-inversion},
 there exist a type environments $\Gamma'$, 
 a usage $U$, and ${m'}\in{L}$ such that
 $m\leq_{L}m'$,
 $\mathreliable{U}$ and $\mathissubtype{\Gamma}{\Gamma'}$, and
 $\mathtypejudgementwithsecrecy{\Gamma', {\mathistype{x}{\xi/U}}}{L}{m'}{\mathnil}$
 is $l$-securely derivable.
 By \cref{lemma:inversion} \cref{item:usagenil-lemma-inversion},
 $\mathissubtype{\Gamma', {\mathistype{x}{\xi/U}}}{\emptyset}$.
 By \cref{def:subtyping-on-env} \cref{item:obligation-def-subtyping-on-env},
 we have ${\mathobligation{}{\xi/U}}={\infty}$.
 Then $\mathissubtype{\Gamma'}{\Gamma', {\mathistype{x}{\xi/U}}}$.
 Hence, we have $\mathissubtype{\Gamma'}{\emptyset}$.
 Therefore, $\mathissubtype{\Gamma}{\emptyset}$.
 We have an $l$-secure derivation tree as follows:
 \begin{center}
  \begin{inlineprooftree}
   \AxiomC{}
   \RightLabel{\rulename{T-Zero}}
   \UnaryInfC{$\mathtypejudgementwithsecrecy{\emptyset}{L}{m}{\mathnil}$}
   \AxiomC{$\mathissubtype{\Gamma}{\emptyset}$}
   \AxiomC{$m\leq_{L}m$}
   \RightLabel{\rulename{T-Weak}}
   \TrinaryInfC{$\mathtypejudgementwithsecrecy{\Gamma}{L}{m}{P'}$}
  \end{inlineprooftree}.
 \end{center}
 Thus, we see that $\mathtypejudgementwithsecrecy{\Gamma}{L}{m}{P'}$
 is $l$-securely derivable.

 \noindent Case 5. \rulename{SP-Commut}.
 Assume ${P}\equiv{\mathpiparallel{P_{0}}{P_{1}}}$ and 
 ${P'}\equiv{\mathpiparallel{P_{1}}{P_{0}}}$.
 By \cref{lemma:inversion} \cref{item:parallel-lemma-inversion},
 there exist two type environments $\Gamma'_{0}$, $\Gamma'_{1}$,
 and ${m'}\in{L}$
 such that $\mathissubtype{\Gamma}{\mathpiparallel{\Gamma'_{0}}{\Gamma'_{1}}}$
 and $m\leq_{L}m'$, and
 $\mathtypejudgementwithsecrecy{\Gamma'_{i}}{L}{m'}{P_{i}}$
 is $l$-securely derivable for each $i=0, 1$.
 By \cref{prop:property-subtyping-on-env} \cref{item:commut-prop-property-subtyping-on-env},
 $\mathissubtype{\mathpiparallel{\Gamma'_{0}}{\Gamma'_{1}}}{\mathpiparallel{\Gamma'_{1}}{\Gamma'_{0}}}$.
 Hence, $\mathissubtype{\Gamma}{\mathpiparallel{\Gamma'_{1}}{\Gamma'_{0}}}$.
 Let $\pi_{i}$ be an $l$-secure derivation tree of 
 $\mathtypejudgementwithsecrecy{\Gamma'_{i}}{L}{m'}{P_{i}}$ for each $i=0, 1$.
 Then, we have an $l$-secure derivation tree as follows:
 \begin{center}
  \begin{inlineprooftree}
   \AxiomC{}
   \RightLabel{$\pi_{1}$}
   \DeduceC{$\mathtypejudgementwithsecrecy{\Gamma'_{1}}{L}{m'}{P_{1}}$}

   \AxiomC{}
   \RightLabel{$\pi_{0}$}
   \DeduceC{$\mathtypejudgementwithsecrecy{\Gamma'_{0}}{L}{m'}{P_{0}}$}
   
   \RightLabel{\rulename{T-Par}}
   \BinaryInfC{$\mathtypejudgementwithsecrecy{\mathpiparallel{\Gamma'_{1}}{\Gamma'_{0}}}{L}{m'}{\mathpiparallel{P_{1}}{P_{0}}}$}

   \AxiomC{$\mathissubtype{\Gamma}{\mathpiparallel{\Gamma'_{1}}{\Gamma'_{0}}}$}
   \AxiomC{$m\leq_{L}m'$}
   \RightLabel{\rulename{T-Weak}}
   \TrinaryInfC{$\mathtypejudgementwithsecrecy{\Gamma}{L}{m}{P'}$}
  \end{inlineprooftree}.
 \end{center}
 Thus, we see that $\mathtypejudgementwithsecrecy{\Gamma}{L}{m}{P'}$ 
 is $l$-securely derivable.

 In case ${P}\equiv{\mathpiparallel{P_{1}}{P_{0}}}$ and 
 ${P'}\equiv{\mathpiparallel{P_{0}}{P_{1}}}$,
 we have an $l$-secure derivation tree of
 $\mathtypejudgementwithsecrecy{\Gamma}{L}{m}{P'}$
 in the same way.

 \noindent Case 6. \rulename{SP-Assoc}. 
 Assume ${P}\equiv{\mathpiparallel{\mleft(\mathpiparallel{P_{0}}{P_{1}}\mright)}{P_{2}}}$
 and
 ${P'}\equiv{\mathpiparallel{P_{0}}{\mleft(\mathpiparallel{P_{1}}{P_{2}}\mright)}}$.
 By \cref{lemma:inversion} \cref{item:parallel-lemma-inversion},
 there exist two type environments $\Gamma'_{01}$, $\Gamma'_{2}$,
 and ${m'}\in{L}$
 such that $\mathissubtype{\Gamma}{\mathpiparallel{\Gamma'_{01}}{\Gamma'_{2}}}$, 
 and $m\leq_{L}m'$ and
 $\mathtypejudgementwithsecrecy{\Gamma'_{01}}{L}{m'}{\mathpiparallel{P_{0}}{P_{1}}}$
 and 
 $\mathtypejudgementwithsecrecy{\Gamma'_{2}}{L}{m'}{P_{2}}$ are 
 $l$-securely derivable.
 Because
 $\mathtypejudgementwithsecrecy{\Gamma'_{01}}{L}{m'}{\mathpiparallel{P_{0}}{P_{1}}}$
 is $l$-securely derivable,
 \cref{lemma:inversion} \cref{item:parallel-lemma-inversion} implies that
 there exist two type environments $\Gamma''_{0}$, $\Gamma''_{1}$,
 and ${m''}\in{L}$
 such that $\mathissubtype{\Gamma'_{01}}{\mathpiparallel{\Gamma'_{0}}{\Gamma'_{1}}}$, 
 and $m'\leq_{L}m''$ and
 $\mathtypejudgementwithsecrecy{\Gamma''_{i}}{L}{m''}{P_{i}}$
 is $l$-securely derivable 
 for each $i=0, 1$.
 By \cref{prop:property-subtyping-on-env}
 \cref{item:cong-full-prop-property-subtyping-on-env} and 
 transitivity of $\mathissubtypesy$, we have
 $\mathissubtype{\Gamma}{\mathpiparallel{({\mathpiparallel{\Gamma''_{0}}{\Gamma''_{1}}})}{\Gamma'_{2}}}$.
 By \cref{prop:property-subtyping-on-env} 
 \cref{item:assoc2-prop-property-subtyping-on-env} and 
 transitivity of $\mathissubtypesy$, we see
 $\mathissubtype{\Gamma}{\mathpiparallel{\Gamma''_{0}}{({\mathpiparallel{\Gamma''_{1}}{\Gamma'_{2}}})}}$. 
 Let $\pi_{i}$ be an $l$-secure derivation tree of
 $\mathtypejudgementwithsecrecy{\Gamma''_{i}}{L}{m''}{P_{i}}$ for each $i=0, 1$.
 Let $\pi_{2}$ be an $l$-secure derivation tree of
 $\mathtypejudgementwithsecrecy{\Gamma'_{2}}{L}{m'}{P_{2}}$.
 Then, we have an $l$-secure derivation tree as follows:
 \begin{center}
  \begin{inlineprooftree}
   \AxiomC{}
   \RightLabel{$\pi_{0}$}
   \DeduceC{$\mathtypejudgementwithsecrecy{\Gamma''_{0}}{L}{m''}{P_{0}}$}
   \RightLabel{\rulename{T-Weak}}
   \UnaryInfC{$\mathtypejudgementwithsecrecy{\Gamma''_{0}}{L}{m'}{P_{0}}$}

   \AxiomC{}
   \RightLabel{$\pi_{1}$}
   \DeduceC{$\mathtypejudgementwithsecrecy{\Gamma''_{1}}{L}{m''}{P_{1}}$}
  
   \RightLabel{\rulename{T-Weak}}
   \UnaryInfC{$\mathtypejudgementwithsecrecy{\Gamma''_{1}}{L}{m'}{P_{1}}$}

   \AxiomC{}
   \RightLabel{$\pi_{2}$}
   \DeduceC{$\mathtypejudgementwithsecrecy{\Gamma'_{2}}{L}{m'}{P_{2}}$}

   \RightLabel{\rulename{T-Par}}
   \BinaryInfC{$\mathtypejudgementwithsecrecy{\mathpiparallel{\Gamma''_{1}}{\Gamma'_{2}}}{L}{m'}{\mathpiparallel{P_{1}}{P_{2}}}$}

   \RightLabel{\rulename{T-Par}}
   \BinaryInfC{$\mathtypejudgementwithsecrecy{\mathpiparallel{\Gamma''_{0}}{(\mathpiparallel{\Gamma''_{1}}{\Gamma'_{2}})}}{L}{m'}{\mathpiparallel{P_{0}}{(\mathpiparallel{P_{1}}{P_{2}})}}$}

   \RightLabel{\rulename{T-Weak}}
   \UnaryInfC{$\mathtypejudgementwithsecrecy{\Gamma}{L}{m}{P'}$}
  \end{inlineprooftree}.
 \end{center}
 Thus, we see that $\mathtypejudgementwithsecrecy{\Gamma}{L}{m}{P'}$ 
 is $l$-securely derivable.

 \noindent Case 7. \rulename{SP-New}.
 Assume ${P}\equiv{\mathpiparallel{\mathpinew{\mathistype{x}{\xi}}{(P_{0})}}{P_{1}}}$,
 ${P'}\equiv{\mathpinew{\mathistype{x}{\xi}}{(\mathpiparallel{P_{0}}{P_{1}})}}$, and
 $x\notin\mathFNof{P_{1}}$.
 By \cref{lemma:inversion} \cref{item:parallel-lemma-inversion},
 there exist two type environments $\Gamma'_{0}$, $\Gamma'_{1}$,
 and ${m'}\in{L}$
 such that $\mathissubtype{\Gamma}{\mathpiparallel{\Gamma'_{0}}{\Gamma'_{1}}}$, 
 and $m\leq_{L}m'$ and
 $\mathtypejudgementwithsecrecy{\Gamma'_{0}}{L}{m'}{\mathpinew{\mathistype{x}{\xi}}{P_{0}}}$ and
 $\mathtypejudgementwithsecrecy{\Gamma'_{1}}{L}{m'}{P_{1}}$ are 
 $l$-securely derivable.
 Because
 $\mathtypejudgementwithsecrecy{\Gamma'_{0}}{L}{m'}{\mathpinew{\mathistype{x}{\xi}}{P_{0}}}$
 is $l$-securely derivable,
 \cref{lemma:inversion} \cref{item:pinew-lemma-inversion} implies that
 there exist a type environments $\Gamma''_{0}$, 
 a usage $U$, and ${m''}\in{L}$ such that
 $m'\leq_{L}m''$,
 $\mathreliable{U}$ and $\mathissubtype{\Gamma'_{0}}{\Gamma''_{0}}$, and
 $\mathtypejudgementwithsecrecy{\Gamma''_{0}, {\mathistype{x}{\xi/U}}}{L}{m''}{P_{0}}$ 
 is $l$-securely derivable.
 Since $\mathtypejudgementwithsecrecy{\Gamma'_{1}}{L}{m'}{P_{1}}$ is
 $l$-securely derivable, and
 $x\notin\mathFNof{P_{1}}$,
 \cref{lemma:minus-derivable} implies that
 $\mathtypejudgementwithsecrecy{\Gamma''_{1}}{L}{m'}{P_{1}}$ is
 $l$-securely derivable and
 $\mathissubtype{\Gamma'_{1}}{\Gamma''_{1}}$, where 
 $\Gamma''_{1}$ is the restriction of $\Gamma'_{1}$
 to $(\mathof{\mathdom}{\Gamma'_{1}}\setminus{\mathsetextension{x}})$.
 Then 
 ${\mathpiparallel{(\Gamma''_{0}, {\mathistype{x}{\xi/U}})}{\Gamma''_{1}}}\equiv{(\mathpiparallel{\Gamma''_{0}}{\Gamma''_{1}}), {\mathistype{x}{\xi/U}}}$.
 Since $\mathissubtype{\Gamma'_{0}}{\Gamma''_{0}}$ and
 $\mathissubtype{\Gamma'_{1}}{\Gamma''_{1}}$,
 \cref{prop:property-subtyping-on-env} \cref{item:cong-full-prop-property-subtyping-on-env}
 implies that
 $\mathissubtype{\mathpiparallel{\Gamma'_{0}}{\Gamma'_{1}}}{\mathpiparallel{\Gamma''_{0}}{\Gamma''_{1}}}$.
 By transitivity of $\mathissubtypesy$, we have
 $\mathissubtype{\Gamma}{\mathpiparallel{\Gamma''_{0}}{\Gamma''_{1}}}$.
 Let $\pi_{0}$ be an $l$-secure derivation tree of 
 $\mathtypejudgementwithsecrecy{\Gamma''_{0}, {\mathistype{x}{\xi/U}}}{L}{m''}{P_{0}}$,
 and $\pi_{1}$ be an $l$-secure derivation tree of 
 $\mathtypejudgementwithsecrecy{\Gamma''_{1}}{L}{m'}{P_{1}}$.
 Then, we have an $l$-secure derivation tree as follows:
 \begin{center}
  \begin{inlineprooftree}
   \AxiomC{}
   \RightLabel{$\pi_{0}$}
   \DeduceC{$\mathtypejudgementwithsecrecy{\Gamma''_{0}, {\mathistype{x}{\xi/U}}}{L}{m''}{P_{0}}$}

   \AxiomC{}
   \RightLabel{$\pi_{1}$}
   \DeduceC{$\mathtypejudgementwithsecrecy{\Gamma''_{1}}{L}{m''}{P_{1}}$}

   \RightLabel{\rulename{T-Par}}
   \BinaryInfC{$\mathtypejudgementwithsecrecy{(\mathpiparallel{\Gamma''_{0}}{\Gamma''_{1}}), {\mathistype{x}{\xi/U}}}{L}{m'}{\mathpiparallel{P_{0}}{P_{1}}}$}

   \RightLabel{\rulename{T-New}}
   \UnaryInfC{$\mathtypejudgementwithsecrecy{\mathpiparallel{\Gamma''_{0}}{\Gamma''_{1}}}{L}{m'}{\mathpinew{\mathistype{x}{\xi}}{(\mathpiparallel{P_{0}}{P_{1}})}}$}
   \RightLabel{\rulename{T-Weak}}
   \UnaryInfC{$\mathtypejudgementwithsecrecy{\Gamma}{L}{m}{P'}$}
  \end{inlineprooftree}.
 \end{center}
 Thus, we see that $\mathtypejudgementwithsecrecy{\Gamma}{L}{m}{P'}$ is
 $l$-securely derivable. 

 The case
 ${P}={\mathpinew{\mathistype{x}{\xi}}{(\mathpiparallel{P_{0}}{P_{1}})}}$, 
 ${P'}={\mathpiparallel{\mathpinew{\mathistype{x}{\xi}}{(P_{0})}}{P_{1}}}$, and
 $x\notin\mathFNof{P_{1}}$
 is straightforward.

 \noindent Case 8. \rulename{SP-IfT}. 
 Assume ${P}={\mathobif{\mathobtrue^{m''}}{P_{0}}{P_{1}}}$ and ${P'}={P_{0}}$.
 By \cref{lemma:inversion} \cref{item:if-then-lemma-inversion},
 there exist a type environments $\Gamma'$
 and ${m'}\in{L}$ such that
 $m\leq_{L}m'$, and $\mathissubtype{\Gamma}{\mathpiparallel{\Gamma'}{\mathistype{\mathobtrue^{m''}}{\mathbooltypewithsec{m'}}}}$, and
 $\mathtypejudgementwithsecrecy{\Gamma'}{L}{m'}{P_{0}}$ and
 $\mathtypejudgementwithsecrecy{\Gamma'}{L}{m'}{P_{1}}$ 
 are $l$-securely derivable.
 Since the type of ${\mathobtrue^{m''}}$ is $\mathbooltypewithsec{m''}$,
 we see ${m''}={m'}$ and
 ${\mathpiparallel{\Gamma'}{\mathistype{\mathobtrue^{m'}}{\mathbooltypewithsec{m'}}}}={\Gamma'}$.
 Hence, $\mathissubtype{\Gamma}{\Gamma'}$.
 Let $\pi$ be an $l$-secure derivation tree of 
 $\mathtypejudgementwithsecrecy{\Gamma'}{L}{m'}{P_{0}}$.
 Then, we have an $l$-secure derivation tree as follows:
 \begin{center}
  \begin{inlineprooftree}
   \AxiomC{}
   \RightLabel{$\pi$}
   \DeduceC{$\mathtypejudgementwithsecrecy{\Gamma'}{L}{m'}{P_{0}}$}
   \RightLabel{\rulename{T-Weak}}
   \UnaryInfC{$\mathtypejudgementwithsecrecy{\Gamma}{L}{m}{P'}$}
  \end{inlineprooftree}.
 \end{center}
 Thus, we see that $\mathtypejudgementwithsecrecy{\Gamma}{L}{m}{P'}$ is
 $l$-securely derivable. 

 \noindent Case 9. \rulename{SP-IfF}. 
 Assume ${P}={\mathobif{\mathobfalse^{m''}}{P_{0}}{P_{1}}}$ and ${P'}={P_{1}}$.
 By \cref{lemma:inversion} \cref{item:if-then-lemma-inversion},
 there exist a type environments $\Gamma'$, 
 and ${m'}\in{L}$ such that
 $m\leq_{L}m'$, and $\mathissubtype{\Gamma}{\mathpiparallel{\Gamma'}{\mathistype{\mathobfalse^{m''}}{\mathbooltypewithsec{m'}}}}$, and
 $\mathtypejudgementwithsecrecy{\Gamma'}{L}{m'}{P_{0}}$ and
 $\mathtypejudgementwithsecrecy{\Gamma'}{L}{m'}{P_{1}}$ are
 $l$-securely derivable.
 Since the type of ${\mathobfalse^{m''}}$ is $\mathbooltypewithsec{m''}$,
 we see ${m''}={m'}$ and
 ${\mathpiparallel{\Gamma'}{\mathistype{\mathobfalse^{m'}}{\mathbooltypewithsec{m'}}}}={\Gamma'}$.
 Hence, $\mathissubtype{\Gamma}{\Gamma'}$.
 Let $\pi$ be an $l$-secure derivation tree of 
 $\mathtypejudgementwithsecrecy{\Gamma'}{L}{m}{P_{1}}$.
 Then, we have an $l$-secure derivation tree as follows:
 \begin{center}
  \begin{inlineprooftree}
   \AxiomC{}
   \RightLabel{$\pi$}
   \DeduceC{$\mathtypejudgementwithsecrecy{\Gamma'}{L}{m'}{P_{1}}$}
   \RightLabel{\rulename{T-Weak}}
   \UnaryInfC{$\mathtypejudgementwithsecrecy{\Gamma}{L}{m}{P'}$}
  \end{inlineprooftree}.
 \end{center}
 Thus, we see that $\mathtypejudgementwithsecrecy{\Gamma}{L}{m}{P'}$ is
 $l$-securely derivable. 

 \noindent Case 10. \rulename{SP-Rep}. 
 Assume ${P}={{\mathpireplication{P_{0}}}}$ and ${P'}={\mathpiparallel{\mathpireplication{P_{0}}}{P_{0}}}$.
 By \cref{lemma:inversion} \cref{item:replication-lemma-inversion},
 there exist a type environments $\Gamma'$,
 and ${m'}\in{L}$ such that
 $m\leq_{L}m'$, and
 $\mathissubtype{\Gamma}{\mathpireplication{\Gamma'}}$, and
 $\mathtypejudgementwithsecrecy{\Gamma'}{L}{m'}{P_{0}}$ is
 $l$-securely derivable. 
 By \cref{prop:property-subtyping-on-env} \cref{item:replication-prop-property-subtyping-on-env},
 we have 
 $\mathissubtype{\mathpireplication{\Gamma'}}{\mathpiparallel{\mathpireplication{\Gamma'}}{\Gamma'}}$.
 Hence, we have $\mathissubtype{\Gamma}{\mathpiparallel{\mathpireplication{\Gamma'}}{\Gamma'}}$.
 Let $\pi$ be an $l$-secure derivation tree of 
 $\mathtypejudgementwithsecrecy{\Gamma'}{L}{m'}{P_{0}}$.
 Then, we have an $l$-secure derivation tree as follows:
 \begin{center}
  \begin{inlineprooftree}
   \AxiomC{}
   \RightLabel{$\pi$}
   \DeduceC{$\mathtypejudgementwithsecrecy{\Gamma'}{L}{m'}{P_{0}}$}
   \RightLabel{\rulename{T-Rep}}
   \UnaryInfC{$\mathtypejudgementwithsecrecy{\mathpireplication{\Gamma'}}{L}{m'}{\mathpireplication{P_{0}}}$}

   \AxiomC{}
   \RightLabel{$\pi$}
   \DeduceC{$\mathtypejudgementwithsecrecy{\Gamma'}{L}{m'}{P_{0}}$}

   \RightLabel{\rulename{T-Par}}
   \BinaryInfC{$\mathtypejudgementwithsecrecy{\mathpiparallel{\mathpireplication{\Gamma'}}{\Gamma'}}{L}{m'}{\mathpiparallel{\mathpireplication{P_{0}}}{P_{0}}}$}

   \RightLabel{\rulename{T-Weak}}
   \UnaryInfC{$\mathtypejudgementwithsecrecy{\Gamma}{L}{m}{P'}$}
  \end{inlineprooftree}.
 \end{center}
 Thus, we see that $\mathtypejudgementwithsecrecy{\Gamma}{L}{m}{P'}$ is
 $l$-securely derivable. 

 \noindent Case 11. \rulename{SP-Par}. 
 Assume ${P}={\mathpiparallel{P_{0}}{Q}}$ and ${P'}={\mathpiparallel{P_{1}}{Q}}$
 with ${P_{0}}\mathpistructuralpo{P_{1}}$.
 By \cref{lemma:inversion} \cref{item:parallel-lemma-inversion},
 there exist two type environments $\Gamma'$, $\Gamma''$,
 and ${m'}\in{L}$
 such that $\mathissubtype{\Gamma}{\mathpiparallel{\Gamma'}{\Gamma''}}$, 
 and $m\leq_{L}m'$ and
 $\mathtypejudgementwithsecrecy{\Gamma'}{L}{m'}{P_{0}}$ and
 $\mathtypejudgementwithsecrecy{\Gamma''}{L}{m'}{Q}$ are
 $l$-securely derivable.
 Since ${P_{0}}\mathpistructuralpo{P_{1}}$,
 the induction hypothesis implies that 
 $\mathtypejudgementwithsecrecy{\Gamma'}{L}{m'}{P_{1}}$ is 
 $l$-securely derivable.
 Let $\pi_{1}$ be an $l$-secure derivation tree of 
 $\mathtypejudgementwithsecrecy{\Gamma'}{L}{m'}{P_{1}}$, 
 and $\pi$ be an $l$-secure derivation tree of 
 $\mathtypejudgementwithsecrecy{\Gamma''}{L}{m'}{Q}$.
 Then, we have an $l$-secure derivation tree as follows:
 \begin{center}
  \begin{inlineprooftree}
   \AxiomC{}
   \RightLabel{$\pi_{1}$}
   \DeduceC{$\mathtypejudgementwithsecrecy{\Gamma'}{L}{m'}{P_{1}}$}

   \AxiomC{}
   \RightLabel{$\pi$}
   \DeduceC{$\mathtypejudgementwithsecrecy{\Gamma''}{L}{m'}{Q}$}

   \RightLabel{\rulename{T-Par}}
   \BinaryInfC{$\mathtypejudgementwithsecrecy{\mathpiparallel{\Gamma'}{\Gamma''}}{L}{m'}{\mathpiparallel{P_{1}}{Q}}$}

   \RightLabel{\rulename{T-Weak}}
   \UnaryInfC{$\mathtypejudgementwithsecrecy{\Gamma}{L}{m}{P'}$}
  \end{inlineprooftree}.
 \end{center}
 Thus, we see that $\mathtypejudgementwithsecrecy{\Gamma}{L}{m}{P'}$ is
 $l$-securely derivable. 

 \noindent Case 12. \rulename{SP-CNew}. 
 Assume ${P}={\mathpinew{\mathistype{x}{\xi}}{P_{0}}}$ and
 ${P'}={\mathpinew{\mathistype{x}{\xi}}{P_{1}}}$
 with ${P_{0}}\mathpistructuralpo{P_{1}}$.
 By \cref{lemma:inversion} \cref{item:pinew-lemma-inversion},
 there exist a type environments $\Gamma'$, 
 a usage $U$, and ${m'}\in{L}$ such that
 $m\leq_{L}m'$,
 $\mathreliable{U}$ and $\mathissubtype{\Gamma}{\Gamma'}$, and
 $\mathtypejudgementwithsecrecy{\Gamma', {\mathistype{x}{\xi/U}}}{L}{m'}{P_{0}}$ 
 is $l$-securely derivable.
 Since ${P_{0}}\mathpistructuralpo{P_{1}}$,
 the induction hypothesis implies that
 $\mathtypejudgementwithsecrecy{\Gamma', {\mathistype{x}{\xi/U}}}{L}{m'}{P_{1}}$ 
 is $l$-securely derivable.
 Let $\pi_{1}$ be an $l$-secure derivation tree of 
 $\mathtypejudgementwithsecrecy{\Gamma', {\mathistype{x}{\xi/U}}}{L}{m'}{P_{1}}$.
 Then, we have an $l$-secure derivation tree as follows:
 \begin{center}
  \begin{inlineprooftree}
   \AxiomC{}
   \RightLabel{$\pi_{1}$}
   \DeduceC{$\mathtypejudgementwithsecrecy{\Gamma', {\mathistype{x}{\xi/U}}}{L}{m'}{P_{1}}$}

   \RightLabel{\rulename{T-New}}
   \UnaryInfC{$\mathtypejudgementwithsecrecy{\Gamma'}{L}{m'}{\mathpinew{\mathistype{x}{\xi}}{P_{1}}}$}

   \RightLabel{\rulename{T-Weak}}
   \UnaryInfC{$\mathtypejudgementwithsecrecy{\Gamma}{L}{m}{P'}$}
  \end{inlineprooftree}.
 \end{center}
 Thus, we see that $\mathtypejudgementwithsecrecy{\Gamma}{L}{m}{P'}$ is 
 $l$-securely derivable.

\subsection{Proof of substitution lemma}
\label[appendix]{sec:proof-of-substitution-lemma}

\begin{lemma}
 \label[lemma]{lemma:substitution-preserve-subtyping}
 For type environments $\Gamma_{0}$, $\Gamma_{1}$,
 a tuple of variables $\mathvect{x}=\mathtuple{x_{0}, \ldots, x_{n}}$,
 and values $\mathvect{v}=\mathtuple{v_{0}, \ldots, v_{n}}$,
 if ${\mathsubstbox{\Gamma_{0}}{\mathsubst{\mathvect{x}}{\mathvect{v}}}}$
 is well-defined and
 $\mathissubtype{\Gamma_{0}}{\Gamma_{1}}$, 
 then
 ${\mathsubstbox{\Gamma_{1}}{\mathsubst{\mathvect{x}}{\mathvect{v}}}}$
 is well-defined and
 $\mathissubtype{{\mathsubstbox{\Gamma_{0}}{\mathsubst{\mathvect{x}}{\mathvect{v}}}}}{{\mathsubstbox{\Gamma_{1}}{\mathsubst{\mathvect{x}}{\mathvect{v}}}}}$.
\end{lemma}

\begin{proof}
 For type environments $\Gamma_{0}$, $\Gamma_{1}$,
 a tuple of variables $\mathvect{x}=\mathtuple{x_{0}, \ldots, x_{n}}$,
 and values $\mathvect{v}=\mathtuple{v_{0}, \ldots, v_{n}}$,
 assume that ${\mathsubstbox{\Gamma_{0}}{\mathsubst{\mathvect{x}}{\mathvect{v}}}}$ 
 is well-defined and
 $\mathissubtype{\Gamma_{0}}{\Gamma_{1}}$.
 Let 
 \[
 {D_{j}}={{\mleft({\mathof{\mathdom}{\Gamma_{j}}}\setminus{\mathsetextension{x_{0}, \ldots, x_{n}}}\mright)}\cup{\mathsetintension{v_{i}}{{x_{i}}\in{\mathof{\mathdom}{\Gamma_{j}}}}}}
 \]
 for $j=0, 1$.
 We note that ${D_{0}}\supseteq{D_{1}}$ 
 because of ${\mathof{\mathdom}{\Gamma_{0}}}\supseteq{\mathof{\mathdom}{\Gamma_{1}}}$.

 We show that 
 ${\mathsubstbox{\Gamma_{1}}{\mathsubst{\mathvect{x}}{\mathvect{v}}}}$ is well-defined.
 By the assumption $\mathissubtype{\Gamma_{0}}{\Gamma_{1}}$,
 we have ${\mathof{\mathdom}{\Gamma_{0}}}\supseteq{\mathof{\mathdom}{\Gamma_{1}}}$
 and $\mathissubtype{\mathof{\Gamma_{0}}{w}}{\mathof{\Gamma_{1}}{w}}$ 
 for each ${w}\in{\mathof{\mathdom}{\Gamma_{0}}}$.
 For each ${w}\in{\mathof{\mathdom}{\Gamma_{1}}}$,
 by $\mathissubtype{\mathof{\Gamma_{0}}{w}}{\mathof{\Gamma_{1}}{w}}$,
 we have ${\mathof{\Gamma_{0}}{w}}\mathequivexceptusagessy{\mathof{\Gamma_{1}}{w}}$.

 Let ${w}\in{D_{1}}$.

 When ${w}\notin{\mathvect{v}}$ holds, 
 $\mathof{\mathsubstbox{\Gamma_{1}}{\mathsubst{\mathvect{x}}{\mathvect{v}}}}{w}$
 is defined as $\mathof{\Gamma_{1}}{w}$.
 
 Assume that ${w}\in{\mathvect{v}}$, ${w}\notin{\mathof{\mathdom}{\Gamma_{1}}}$, and
 ${\mathsetintension{{x_{i}}\in{\mathvect{x}}}{{w}={v_{i}}\text{ and } {x_{i}}\in{\mathof{\mathdom}{\Gamma_{1}}}}}={\mathsetextension{x_{j_{0}}, \dots, x_{j_{k}}}}$
 with ${0} \leq {j_{0}} < \dots < {j_{k}} \leq {n}$.
 We show 
 ${\mathof{\Gamma_{1}}{x_{i_{0}}}}\mathequivexceptusagessy{\mathof{\Gamma_{1}}{x_{i_{1}}}}$
 for any $i_{0}$, ${i_{1}}\in{\mathsetextension{j_{0}, \dots, j_{k}}}$.
 From ${D_{1}}\subseteq{D_{0}}$, we have ${w}\in{D_{0}}$.
 Let ${\mathsetintension{{x_{i}}\in{\mathvect{x}}}{{w}={v_{i}}\text{ and } {x_{i}}\in{\mathof{\mathdom}{\Gamma_{0}}}}}={\mathsetextension{x_{j'_{0}}, \dots, x_{j'_{l}}}}$
 with ${0} \leq {j'_{0}} < \dots < {j'_{l}} \leq {n}$.
 Since ${\mathsubstbox{\Gamma_{0}}{\mathsubst{\mathvect{x}}{\mathvect{v}}}}$ 
 is well-defined,
 either
 $\mathpiparallel{\mathpiparallel{\mathof{\Gamma_{0}}{x_{j'_{0}}}}{\cdots}}{\mathof{\Gamma_{0}}{x_{j'_{k}}}}$
 or
 $\mathpiparallel{\mathpiparallel{\mathpiparallel{\mathof{\Gamma_{0}}{x_{j'_{0}}}}{\cdots}}{\mathof{\Gamma_{0}}{x_{j'_{l}}}}}{\mathof{\Gamma}{w}}$ 
 is defined.
 Hence, 
 ${\mathof{\Gamma_{0}}{x_{i_{0}}}}\mathequivexceptusagessy{\mathof{\Gamma_{0}}{x_{i_{1}}}}$
 for any $i_{0}$, ${i_{1}}\in{\mathsetextension{0, \dots, l}}$.
 Since ${\mathsetextension{x_{j_{0}}, \dots, x_{j_{k}}}}\subseteq{\mathsetextension{x_{j'_{0}}, \dots, x_{j'_{l}}}}$,
 we have 
 ${\mathof{\Gamma_{0}}{x_{i_{0}}}}\mathequivexceptusagessy{\mathof{\Gamma_{0}}{x_{i_{1}}}}$
 for any $i_{0}$, ${i_{1}}\in{\mathsetextension{0, \dots, k}}$.
 Since ${\mathof{\Gamma_{0}}{x}}\mathequivexceptusagessy{\mathof{\Gamma_{1}}{x}}$
 for each $x\in{\mathof{\mathdom}{\Gamma_{1}}}$,
 we see that
 ${\mathof{\Gamma_{1}}{x_{i_{0}}}}\mathequivexceptusagessy{\mathof{\Gamma_{0}}{x_{i_{0}}}}\mathequivexceptusagessy{\mathof{\Gamma_{0}}{x_{i_{1}}}}\mathequivexceptusagessy{\mathof{\Gamma_{1}}{x_{i_{1}}}}$.
 Therefore, 
 $\mathpiparallel{\mathpiparallel{\mathof{\Gamma_{1}}{x_{j_{0}}}}{\cdots}}{\mathof{\Gamma_{1}}{x_{j_{k}}}}$ 
 is defined.
 Thus, 
 $\mathof{\mathsubstbox{\Gamma_{1}}{\mathsubst{\mathvect{x}}{\mathvect{v}}}}{w}$
 is defined.
 
 In a similar way, we can show that
 $\mathof{\mathsubstbox{\Gamma_{1}}{\mathsubst{\mathvect{x}}{\mathvect{v}}}}{w}$
 is defined in case ${w}\in{\mathvect{v}}$ and ${w}\in{\mathof{\mathdom}{\Gamma_{1}}}$.
 
 We show
 $\mathissubtype{{\mathsubstbox{\Gamma_{0}}{\mathsubst{\mathvect{x}}{\mathvect{v}}}}}{{\mathsubstbox{\Gamma_{1}}{\mathsubst{\mathvect{x}}{\mathvect{v}}}}}$.

 \noindent \cref{item:dom-def-subtyping-on-env}
 We see
 ${\mathof{\mathdom}{{\mathsubstbox{\Gamma_{0}}{\mathsubst{\mathvect{x}}{\mathvect{v}}}}}}\supseteq{\mathof{\mathdom}{{\mathsubstbox{\Gamma_{1}}{\mathsubst{\mathvect{x}}{\mathvect{v}}}}}}$
 because of 
 ${\mathof{\mathdom}{{\mathsubstbox{\Gamma_{j}}{\mathsubst{\mathvect{x}}{\mathvect{v}}}}}}={D_{j}}$ 
 for $j=0, 1$.

 \noindent \cref{item:subtyping-def-subtyping-on-env}
 We show
   $\mathissubtype{\mathof{{\mathsubstbox{\Gamma_{0}}{\mathsubst{\mathvect{x}}{\mathvect{v}}}}}{w}}{\mathof{{\mathsubstbox{\Gamma_{1}}{\mathsubst{\mathvect{x}}{\mathvect{v}}}}}{w}}$
 for each ${w}\in{\mathof{\mathdom}{\mathsubstbox{\Gamma_{1}}{\mathsubst{\mathvect{x}}{\mathvect{v}}}}}$.

 Let ${w}\in{\mathof{\mathdom}{\mathsubstbox{\Gamma_{1}}{\mathsubst{\mathvect{x}}{\mathvect{v}}}}}$.
   
 When ${w}\notin{\mathvect{v}}$, 
 we have
 $\mathissubtype{{\mathof{{\mathsubstbox{\Gamma_{0}}{\mathsubst{\mathvect{x}}{\mathvect{v}}}}}{w}}={\mathof{\Gamma_{0}}{w}}}{{\mathof{\Gamma_{1}}{w}}={\mathof{{\mathsubstbox{\Gamma_{1}}{\mathsubst{\mathvect{x}}{\mathvect{v}}}}}{w}}}$.
   
 Assume \linebreak[2] that \linebreak[2] ${w}\in{\mathvect{v}}$, ${w}\notin{\mathof{\mathdom}{\Gamma_{1}}}$, \linebreak[2] and \linebreak[2]
 ${\mathsetintension{{x_{i}}\in{\mathvect{x}}}{{w}={v_{i}}\text{ and } {x_{i}}\in{\mathof{\mathdom}{\Gamma_{1}}}}}={\mathsetextension{x_{j_{0}}, \dots, x_{j_{k}}}}$
 with ${0} \leq {j_{0}} < \dots < {j_{k}} \leq {n}$.
 Then, we have \linebreak[2]
 ${\mathof{\mathsubstbox{\Gamma_{1}}{\mathsubst{\mathvect{x}}{\mathvect{v}}}}{w}}={\mathpiparallel{\mathpiparallel{\mathof{\Gamma_{1}}{x_{j_{0}}}}{\cdots}}{\mathof{\Gamma_{1}}{x_{j_{k}}}}}$.
 Assume \linebreak[4]
 ${\mathsetintension{{x_{i}}\in{\mathvect{x}}}{{w}={v_{i}}\text{ and } {x_{i}}\in{\mathof{\mathdom}{\Gamma_{1}}}}}={{\mathsetextension{y_{0}, \dots, y_{k}}}\cup{\mathsetextension{z_{0}, \dots, z_{l}}}}$. \linebreak[2]
 By \cref{prop:property-of-subtyping} \cref{item:comm-parallel-in-prop:property-of-subtyping} and \cref{item:assoc1-parallel-in-prop:property-of-subtyping}, 
 \linebreak[2] we \linebreak[2] have \linebreak[2]
 $\mathissubtype{\mathof{\mathsubstbox{\Gamma_{0}}{\mathsubst{\mathvect{x}}{\mathvect{v}}}}{w}}{\mathpiparallel{\mathpiparallel{\mathpiparallel{\mathof{\Gamma_{0}}{y_{0}}}{\cdots}}{\mathof{\Gamma_{0}}{y_{k}}}}{\mathpiparallel{\mathpiparallel{\mathof{\Gamma_{0}}{z_{0}}}{\cdots}}{\mathof{\Gamma_{0}}{z_{l}}}}}$.
 Since 
 ${z_{i}}\in{{\mathof{\mathdom}{\Gamma_{0}}}\setminus{\mathof{\mathdom}{\Gamma_{1}}}}$
 for ${i}={0, \dots, l}$,
 we have
 $\mathobligation{}{\mathof{\mathsubstbox{\Gamma_{0}}{\mathsubst{\mathvect{x}}{\mathvect{v}}}}{z_{i}}}=\infty$.
 By \cref{prop:property-of-subtyping} \cref{item:ob-infty-and-parallel-in-prop:property-of-subtyping},
 we have
 $\mathissubtype{\mathpiparallel{\mathpiparallel{\mathpiparallel{\mathof{\Gamma_{0}}{y_{0}}}{\cdots}}{\mathof{\Gamma_{0}}{y_{k}}}}{\mathpiparallel{\mathpiparallel{\mathof{\Gamma_{0}}{z_{0}}}{\cdots}}{\mathof{\Gamma_{0}}{z_{l}}}}}{\mathpiparallel{\mathpiparallel{\mathof{\Gamma_{0}}{y_{0}}}{\cdots}}{\mathof{\Gamma_{0}}{y_{k}}}}$.
 Since $\mathissubtype{\mathof{\Gamma_{0}}{w}}{\mathof{\Gamma_{1}}{w}}$ 
 for any ${w}\in{\mathof{\mathdom}{\Gamma_{1}}}$,
 we have $\mathissubtype{\mathof{\Gamma_{0}}{y_{i}}}{\mathof{\Gamma_{1}}{y_{i}}}$
 for $i=0, \dots, k$.
 From \cref{prop:property-of-subtyping} \cref{item:parallel-in-prop:property-of-subtyping},
 we have
 $\mathissubtype{\mathpiparallel{\mathpiparallel{\mathof{\Gamma_{0}}{y_{0}}}{\cdots}}{\mathof{\Gamma_{0}}{y_{k}}}}{\mathpiparallel{\mathpiparallel{\mathof{\Gamma_{1}}{y_{0}}}{\cdots}}{\mathof{\Gamma_{1}}{y_{k}}}}$.
 Then, we see 
 $\mathissubtype{\mathof{{\mathsubstbox{\Gamma_{0}}{\mathsubst{\mathvect{x}}{\mathvect{v}}}}}{w}}{\mathof{{\mathsubstbox{\Gamma_{1}}{\mathsubst{\mathvect{x}}{\mathvect{v}}}}}{w}}$.

 In a similar way, we can show 
 $\mathissubtype{\mathof{{\mathsubstbox{\Gamma_{0}}{\mathsubst{\mathvect{x}}{\mathvect{v}}}}}{w}}{\mathof{{\mathsubstbox{\Gamma_{1}}{\mathsubst{\mathvect{x}}{\mathvect{v}}}}}{w}}$
 in case ${w}\in{\mathvect{v}}$ and ${w}\in{\mathof{\mathdom}{\Gamma_{1}}}$.

 \noindent \cref{item:obligation-def-subtyping-on-env}
 We show
 $\mathobligation{}{\mathof{\mathsubstbox{\Gamma_{0}}{\mathsubst{\mathvect{x}}{\mathvect{v}}}}{w}}=\infty$
 for each 
 ${w}\in{{\mathof{\mathdom}{\mathsubstbox{\Gamma_{0}}{\mathsubst{\mathvect{x}}{\mathvect{v}}}}}\setminus{\mathof{\mathdom}{\mathsubstbox{\Gamma_{1}}{\mathsubst{\mathvect{x}}{\mathvect{v}}}}}}$.

 Assume ${w}\in{{\mathof{\mathdom}{\mathsubstbox{\Gamma_{0}}{\mathsubst{\mathvect{x}}{\mathvect{v}}}}}\setminus{\mathof{\mathdom}{\mathsubstbox{\Gamma_{1}}{\mathsubst{\mathvect{x}}{\mathvect{v}}}}}}$.
   
 Assuming ${w}\notin{\mathvect{v}}$,
 we have ${w}\in{{\mathof{\mathdom}{\Gamma_{0}}}\setminus{\mathof{\mathdom}{\Gamma_{1}}}}$.
 In case, 
 we have ${\mathobligation{}{\mathof{\mathsubstbox{\Gamma_{0}}{\mathsubst{\mathvect{x}}{\mathvect{v}}}}{w}}}={\mathobligation{}{\mathof{\Gamma_{0}}{w}}}={\infty}$
 because of $\mathissubtype{\Gamma_{0}}{\Gamma_{1}}$.

 Now, assume ${w}\in{\mathvect{v}}$ and ${w}\notin{\mathof{\mathdom}{\Gamma_{0}}}$.
 By ${w}\in{\mathof{\mathdom}{\mathsubstbox{\Gamma_{0}}{\mathsubst{\mathvect{x}}{\mathvect{v}}}}}$,
 we see that
 \[
 X=\mathsetintension{{x_{i}}\in{\mathvect{x}}}{{w}={v_{i}}\text{ and } {x_{i}}\in{\mathof{\mathdom}{\Gamma_{0}}}}
 \]
 is not empty.
 Assume
 ${X}={\mathsetextension{x_{j_{0}}, \dots, x_{j_{k}}}}$ with ${0} \leq {j_{0}} < \dots < {j_{k}} \leq {n}$.
 Then, we have
 ${\mathof{\mathsubstbox{\Gamma_{0}}{\mathsubst{\mathvect{x}}{\mathvect{v}}}}{w}}={\mathpiparallel{\mathpiparallel{\mathof{\Gamma_{0}}{x_{j_{0}}}}{\cdots}}{\mathof{\Gamma_{0}}{x_{j_{k}}}}}$.
 By ${w}\notin{\mathof{\mathdom}{\mathsubstbox{\Gamma_{1}}{\mathsubst{\mathvect{x}}{\mathvect{v}}}}}$,
 we see that
 $\mathsetintension{{x_{i}}\in{\mathvect{x}}}{{w}={v_{i}}\text{ and } {x_{i}}\in{\mathof{\mathdom}{\Gamma_{1}}}}$ 
 is empty.
 Hence, ${x_{j_{i}}}\notin{\mathof{\mathdom}{\Gamma_{1}}}$ for $i=0, \dots, k$.
 Therefore, 
 ${\mathobligation{}{\mathof{\Gamma_{0}}{x_{j_{i}}}}}={\infty}$ for $i=0, \dots, k$.
 Thus, 
 ${\mathobligation{}{\mathof{\mathsubstbox{\Gamma_{0}}{\mathsubst{\mathvect{x}}{\mathvect{v}}}}{w}}}={\infty}$.
  
 Now, assume ${w}\in{\mathvect{v}}$ and ${w}\in{\mathof{\mathdom}{\Gamma_{0}}}$.
 Let
 ${\mathsetintension{{x_{i}}\in{\mathvect{x}}}{{w}={v_{i}}\text{ and } {x_{i}}\in{\mathof{\mathdom}{\Gamma_{0}}}}}={\mathsetextension{x_{j_{0}}, \dots, x_{j_{k}}}}$.
 Then, we have
 ${\mathof{\mathsubstbox{\Gamma_{0}}{\mathsubst{\mathvect{x}}{\mathvect{v}}}}{w}}={\mathpiparallel{\mathpiparallel{\mathpiparallel{\mathof{\Gamma_{0}}{x_{j_{0}}}}{\cdots}}{\mathof{\Gamma_{0}}{x_{j_{k}}}}}{\mathof{\Gamma_{0}}{w}}}$.
 In a similar way to the case ${w}\notin{\mathof{\mathdom}{\Gamma_{0}}}$,
 we have
 ${\mathobligation{}{\mathof{\Gamma_{0}}{x_{j_{i}}}}}={\infty}$ for $i=0, \dots, k$.
 Since ${w}\notin{\mathof{\mathdom}{\Gamma_{1}}}$, we have
 ${\mathobligation{}{\mathof{\Gamma_{0}}{w}}}={\infty}$.
 Thus, 
 ${\mathobligation{}{\mathof{\mathsubstbox{\Gamma_{0}}{\mathsubst{\mathvect{x}}{\mathvect{v}}}}{w}}}={\infty}$.
\end{proof}

  \begin{lemma}
   \label[lemma]{lemma:substitution-and-some-operators}
   \begin{enumerate}
    \item For a type environment $\Gamma$, 
	  a tuple of variables $\mathvect{x}=\mathtuple{x_{0}, \ldots, x_{n}}$,
	  and values $\mathvect{v}=\mathtuple{v_{0}, \ldots, v_{n}}$,
	  if
	  ${\mathsubstbox{(\mathpireplication{\Gamma})}{\mathsubst{\mathvect{x}}{\mathvect{v}}}}$ 
	  is well-defined, then
	  ${\mathsubstbox{\Gamma}{\mathsubst{\mathvect{x}}{\mathvect{v}}}}$
	  is well-defined and
	  $\mathissubtype{\mathsubstbox{(\mathpireplication{\Gamma})}{\mathsubst{\mathvect{x}}{\mathvect{v}}}}{\mathpireplication{\mathsubstbox{\Gamma}{\mathsubst{\mathvect{x}}{\mathvect{v}}}}}$.
	  \label{item:replication-lemma:substitution-and-some-operators}
    \item For type environments $\Gamma_{0}$ and $\Gamma_{1}$,
	  a tuple of variables $\mathvect{x}=\mathtuple{x_{0}, \ldots, x_{n}}$,
	  and values $\mathvect{v}=\mathtuple{v_{0}, \ldots, v_{n}}$,
	  if ${\mathsubstbox{(\mathpiparallel{\Gamma_{0}}{\Gamma_{1}})}{\mathsubst{\mathvect{x}}{\mathvect{v}}}}$ 
	  is well-defined, then
	  ${\mathsubstbox{\Gamma_{0}}{\mathsubst{\mathvect{x}}{\mathvect{v}}}}$ and
	  ${\mathsubstbox{\Gamma_{1}}{\mathsubst{\mathvect{x}}{\mathvect{v}}}}$
	  are well-defined, and 
	  ${\mathsubstbox{(\mathpiparallel{\Gamma_{0}}{\Gamma_{1}})}{\mathsubst{\mathvect{x}}{\mathvect{v}}}}={\mathpiparallel{(\mathsubstbox{\Gamma_{0}}{\mathsubst{\mathvect{x}}{\mathvect{v}}})}{(\mathsubstbox{\Gamma_{1}}{\mathsubst{\mathvect{x}}{\mathvect{v}}})}}$. 
	  \label{item:parallel-lemma:substitution-and-some-operators}
   \end{enumerate}
  \end{lemma}

  \begin{proof}
   We show each statement.

   \noindent \cref{item:replication-lemma:substitution-and-some-operators}
   For a type environment $\Gamma$, 
   a tuple of variables $\mathvect{x}=\mathtuple{x_{0}, \ldots, x_{n}}$,
   and values $\mathvect{v}=\mathtuple{v_{0}, \ldots, v_{n}}$,
   assume that 
   ${\mathsubstbox{(\mathpireplication{\Gamma})}{\mathsubst{\mathvect{x}}{\mathvect{v}}}}$ 
   is well-defined.

   We show that
   ${\mathsubstbox{\Gamma}{\mathsubst{\mathvect{x}}{\mathvect{v}}}}$ is well-defined.
   Let ${D}={{\mleft({\mathof{\mathdom}{\Gamma}}\setminus{\mathsetextension{x_{0}, \ldots, x_{n}}}\mright)}\cup{\mathsetintension{v_{i}}{{x_{i}}\in{\mathof{\mathdom}{\Gamma}}}}}$.
   Since 
   ${\mathof{\mathdom}{\mathpireplication{\Gamma}}}={\mathof{\mathdom}{\Gamma}}$
   and 
   ${\mathsubstbox{(\mathpireplication{\Gamma})}{\mathsubst{\mathvect{x}}{\mathvect{v}}}}$ 
   is well-defined,
   ${\mathof{\mathdom}{\mathsubstbox{(\mathpireplication{\Gamma})}{\mathsubst{\mathvect{x}}{\mathvect{v}}}}}={D}$.
   
   Let ${w}\in{D}$.
   
   When ${w}\notin{\mathvect{v}}$ holds,
   ${\mathof{\mathsubstbox{\Gamma}{\mathsubst{\mathvect{x}}{\mathvect{v}}}}{w}}={\mathof{\Gamma}{w}}$. 
   Hence, ${\mathof{\mathsubstbox{\Gamma}{\mathsubst{\mathvect{x}}{\mathvect{v}}}}{w}}$ 
   is defined.

   Assume that ${w}\in{\mathvect{v}}$, ${w}\notin{\mathof{\mathdom}{\Gamma}}$, and
   ${\mathsetintension{{x_{i}}\in{\mathvect{x}}}{{w}={v_{i}}\text{ and } {x_{i}}\in{\mathof{\mathdom}{\Gamma}}}}={\mathsetextension{x_{j_{0}}, \dots, x_{j_{k}}}}$
   with ${0} \leq {j_{0}} < \dots < {j_{k}} \leq {n}$.
   In this case, ${w}\notin{\mathof{\mathdom}{\mathpireplication{\Gamma}}}$.
   Then
   ${\mathof{\mathsubstbox{(\mathpireplication{\Gamma})}{\mathsubst{\mathvect{x}}{\mathvect{v}}}}{w}}={\mathpiparallel{\mathpiparallel{\mathof{(\mathpireplication{\Gamma})}{x_{j_{0}}}}{\cdots}}{\mathof{(\mathpireplication{\Gamma})}{x_{j_{k}}}}}$.
   Since 
   ${\mathsubstbox{(\mathpireplication{\Gamma})}{\mathsubst{\mathvect{x}}{\mathvect{v}}}}$ 
   is well-defined,
   ${\mathpiparallel{\mathpiparallel{\mathpireplication{\mathof{\Gamma}{x_{j_{0}}}}}{\cdots}}{\mathpireplication{\mathof{\Gamma}{x_{j_{k}}}}}}$ 
   is defined.
   Then, we can show that
   ${\mathpiparallel{\mathpiparallel{\mathof{\Gamma}{x_{j_{0}}}}{\cdots}}{\mathof{\Gamma}{x_{j_{k}}}}}$ 
   is defined.
   Therefore,
   ${\mathof{\mathsubstbox{\Gamma}{\mathsubst{\mathvect{x}}{\mathvect{v}}}}{w}}$ is defined
   and
   ${\mathof{\mathsubstbox{\Gamma}{\mathsubst{\mathvect{x}}{\mathvect{v}}}}{w}}={\mathpiparallel{\mathpiparallel{\mathof{\Gamma}{x_{j_{0}}}}{\cdots}}{\mathof{\Gamma}{x_{j_{k}}}}}$.

   In a similar way, we can show that
   $\mathof{\mathsubstbox{\Gamma}{\mathsubst{\mathvect{x}}{\mathvect{v}}}}{w}$
   is defined in case ${w}\in{\mathvect{v}}$ and ${w}\in{\mathof{\mathdom}{\Gamma}}$.

   We show
   $\mathissubtype{\mathsubstbox{(\mathpireplication{\Gamma})}{\mathsubst{\mathvect{x}}{\mathvect{v}}}}{\mathpireplication{\mathsubstbox{\Gamma}{\mathsubst{\mathvect{x}}{\mathvect{v}}}}}$.
   
   \noindent \cref{item:dom-def-subtyping-on-env}
   We have
   \begin{align*}
    {\mathof{\mathdom}{\mathsubstbox{(\mathpireplication{\Gamma})}{\mathsubst{\mathvect{x}}{\mathvect{v}}}}}&={D} \\
    &={\mathof{\mathdom}{\mathsubstbox{\Gamma}{\mathsubst{\mathvect{x}}{\mathvect{v}}}}} \\
    &={\mathof{\mathdom}{\mathpireplication{\mathsubstbox{\Gamma}{\mathsubst{\mathvect{x}}{\mathvect{v}}}}}}.
   \end{align*}
   Hence, 
   ${\mathof{\mathdom}{\mathsubstbox{(\mathpireplication{\Gamma})}{\mathsubst{\mathvect{x}}{\mathvect{v}}}}}\supseteq{\mathof{\mathdom}{\mathpireplication{\mathsubstbox{\Gamma}{\mathsubst{\mathvect{x}}{\mathvect{v}}}}}}$.

   \noindent \cref{item:subtyping-def-subtyping-on-env}
   Let ${w}\in{\mathof{\mathdom}{\mathpireplication{\mathsubstbox{\Gamma}{\mathsubst{\mathvect{x}}{\mathvect{v}}}}}}$. Then ${w}\in{D}$.
   
   In case ${w}\notin{\mathvect{v}}$, we have 
   \begin{align*}
    {\mathof{\mathsubstbox{(\mathpireplication{\Gamma})}{\mathsubst{\mathvect{x}}{\mathvect{v}}}}{w}}&={\mathof{\mathpireplication{\Gamma}}{w}} \\
    &={\mathpireplication{\mathof{\Gamma}{w}}} \\
    &={\mathpireplication{\mathof{\mathsubstbox{\Gamma}{\mathsubst{\mathvect{x}}{\mathvect{v}}}}{w}}}.
   \end{align*}
   Then, we have
   $\mathissubtype{\mathof{\mathsubstbox{(\mathpireplication{\Gamma})}{\mathsubst{\mathvect{x}}{\mathvect{v}}}}{w}}{\mathpireplication{\mathof{\mathsubstbox{\Gamma}{\mathsubst{\mathvect{x}}{\mathvect{v}}}}{w}}}$.

   Assume that ${w}\in{\mathvect{v}}$, ${w}\notin{\mathof{\mathdom}{\Gamma}}$, and
   ${\mathsetintension{{x_{i}}\in{\mathvect{x}}}{{w}={v_{i}}\text{ and } {x_{i}}\in{\mathof{\mathdom}{\Gamma_{1}}}}}={\mathsetextension{x_{j_{0}}, \dots, x_{j_{k}}}}$
   with ${0} \leq {j_{0}} < \dots < {j_{k}} \leq {n}$.
   Then 
   ${\mathof{\mathsubstbox{\Gamma}{\mathsubst{\mathvect{x}}{\mathvect{v}}}}{w}}={\mathpiparallel{\mathpiparallel{\mathof{\Gamma}{x_{j_{0}}}}{\cdots}}{\mathof{\Gamma}{x_{j_{k}}}}}$
   and
   ${\mathof{\mathsubstbox{(\mathpireplication{\Gamma})}{\mathsubst{\mathvect{x}}{\mathvect{v}}}}{w}}={\mathpiparallel{\mathpiparallel{\mathof{(\mathpireplication{\Gamma})}{x_{j_{0}}}}{\cdots}}{\mathof{(\mathpireplication{\Gamma})}{x_{j_{k}}}}}={\mathpiparallel{\mathpiparallel{\mathpireplication{\mathof{\Gamma}{x_{j_{0}}}}}{\cdots}}{\mathpireplication{\mathof{\Gamma}{x_{j_{k}}}}}}$.
   We also have
   ${\mathof{\mathpireplication{\mathsubstbox{\Gamma}{\mathsubst{\mathvect{x}}{\mathvect{v}}}}}{w}}={\mathpireplication{\mathpiparallel{\mathpiparallel{\mathof{\Gamma}{x_{j_{0}}}}{\cdots}}{\mathof{\Gamma}{x_{j_{k}}}}}}$.
   By \cref{prop:property-of-subtyping} \cref{item:rep-and-parallel-in-prop:property-of-subtyping},
   we have
   \[
   \mathissubtype{\mathpiparallel{\mathpiparallel{\mathpireplication{\mathof{\Gamma}{x_{j_{0}}}}}{\cdots}}{\mathpireplication{\mathof{\Gamma}{x_{j_{k}}}}}}{\mathpireplication{\mathpiparallel{\mathpiparallel{\mathof{\Gamma}{x_{j_{0}}}}{\cdots}}{\mathof{\Gamma}{x_{j_{k}}}}}}.
   \]
   Thus,
   $\mathissubtype{\mathof{\mathsubstbox{(\mathpireplication{\Gamma})}{\mathsubst{\mathvect{x}}{\mathvect{v}}}}{w}}{\mathpireplication{\mathof{\mathsubstbox{\Gamma}{\mathsubst{\mathvect{x}}{\mathvect{v}}}}{w}}}$.

   In a similar way, we can show 
   $\mathissubtype{\mathof{\mathsubstbox{(\mathpireplication{\Gamma})}{\mathsubst{\mathvect{x}}{\mathvect{v}}}}{w}}{\mathpireplication{\mathof{\mathsubstbox{\Gamma}{\mathsubst{\mathvect{x}}{\mathvect{v}}}}{w}}}$
   in case ${w}\in{\mathvect{v}}$ and ${w}\in{\mathof{\mathdom}{\Gamma}}$.

   \noindent \cref{item:obligation-def-subtyping-on-env}
   Because ${\mathof{\mathdom}{\mathsubstbox{(\mathpireplication{\Gamma})}{\mathsubst{\mathvect{x}}{\mathvect{v}}}}}={\mathof{\mathdom}{\mathpireplication{\mathsubstbox{\Gamma}{\mathsubst{\mathvect{x}}{\mathvect{v}}}}}}$,
   \cref{def:subtyping-on-env} \cref{item:obligation-def-subtyping-on-env} holds
   obviously.

   \noindent \cref{item:parallel-lemma:substitution-and-some-operators}
   For type environments $\Gamma_{0}$ and $\Gamma_{1}$,
   a tuple of variables $\mathvect{x}=\mathtuple{x_{0}, \ldots, x_{n}}$,
   and values $\mathvect{v}=\mathtuple{v_{0}, \ldots, v_{n}}$,
   assume that 
   ${\mathsubstbox{(\mathpiparallel{\Gamma_{0}}{\Gamma_{1}})}{\mathsubst{\mathvect{x}}{\mathvect{v}}}}$ 
   is well-defined.
   
   We show that ${\mathsubstbox{\Gamma_{0}}{\mathsubst{\mathvect{x}}{\mathvect{v}}}}$ 
   is well-defined.
   Let ${D_{0}}={{\mleft({\mathof{\mathdom}{\Gamma_{0}}}\setminus{\mathsetextension{x_{0}, \ldots, x_{n}}}\mright)}\cup{\mathsetintension{v_{i}}{{x_{i}}\in{\mathof{\mathdom}{\Gamma_{0}}}}}}$.
   
   Let ${w}\in{D_{0}}$.
   
   When ${w}\notin{\mathvect{v}}$ holds,
   ${\mathof{\mathsubstbox{\Gamma_{0}}{\mathsubst{\mathvect{x}}{\mathvect{v}}}}{w}}={\mathof{\Gamma_{0}}{w}}$. 

   Assume that ${w}\in{\mathvect{v}}$, ${w}\notin{\mathof{\mathdom}{\Gamma_{0}}}$, and
   ${\mathsetintension{{x_{i}}\in{\mathvect{x}}}{{w}={v_{i}}\text{ and } {x_{i}}\in{\mathof{\mathdom}{\Gamma_{0}}}}}={\mathsetextension{x_{j_{0}}, \dots, x_{j_{k}}}}$
   with ${0} \leq {j_{0}} < \dots < {j_{k}} \leq {n}$.
   Let ${\mathsetintension{{x_{i}}\in{\mathvect{x}}}{{w}={v_{i}}\text{ and } {x_{i}}\in{\mathof{\mathdom}{\mathpiparallel{\Gamma_{0}}{\Gamma_{1}}}}}}={\mathsetextension{x_{j'_{0}}, \dots, x_{j'_{l}}}}$
   with ${0} \leq {j'_{0}} < \dots < {j'_{l}} \leq {n}$.
   Since ${\mathsubstbox{(\mathpiparallel{\Gamma_{0}}{\Gamma_{1}})}{\mathsubst{\mathvect{x}}{\mathvect{v}}}}$ 
   is well-defined,
   we have
   ${\mathof{\mathsubstbox{(\mathpiparallel{\Gamma_{0}}{\Gamma_{1}})}{\mathsubst{\mathvect{x}}{\mathvect{v}}}}{w}}={\mathpiparallel{\mathpiparallel{\mathof{(\mathpiparallel{\Gamma_{0}}{\Gamma_{1}})}{x_{j'_{0}}}}{\cdots}}{\mathof{(\mathpiparallel{\Gamma_{0}}{\Gamma_{1}})}{x_{j'_{l}}}}}$.
   Since
   ${\mathsetextension{x_{j_{0}}, \dots, x_{j_{k}}}}\subseteq{\mathsetextension{x_{j'_{0}}, \dots, x_{j'_{l}}}}$,
   we see that ${\mathof{\Gamma_{0}}{x_{j_{i}}}}$ is defined for each ${i}={0, \dots, k}$.
   Then, we have
   ${\mathof{\mathsubstbox{\Gamma_{0}}{\mathsubst{\mathvect{x}}{\mathvect{v}}}}{w}}={\mathpiparallel{\mathpiparallel{\mathof{\Gamma_{0}}{x_{j_{0}}}}{\cdots}}{\mathof{\Gamma_{0}}{x_{j_{k}}}}}$.
   Thus, ${\mathof{\mathsubstbox{\Gamma_{0}}{\mathsubst{\mathvect{x}}{\mathvect{v}}}}{w}}$
   is defined.

   In a similar way, we can show that
   $\mathof{\mathsubstbox{\Gamma_{0}}{\mathsubst{\mathvect{x}}{\mathvect{v}}}}{w}$
   is defined in case ${w}\in{\mathvect{v}}$ and ${w}\in{\mathof{\mathdom}{\Gamma_{0}}}$.
   Thus, we see that
   ${\mathsubstbox{\Gamma_{0}}{\mathsubst{\mathvect{x}}{\mathvect{v}}}}$ is well-defined.

   In a similar way to the case 
   ${\mathsubstbox{\Gamma_{0}}{\mathsubst{\mathvect{x}}{\mathvect{v}}}}$, we can show that
   ${\mathsubstbox{\Gamma_{1}}{\mathsubst{\mathvect{x}}{\mathvect{v}}}}$
   is well-defined.

   We show
   ${\mathsubstbox{(\mathpiparallel{\Gamma_{0}}{\Gamma_{1}})}{\mathsubst{\mathvect{x}}{\mathvect{v}}}}={\mathpiparallel{(\mathsubstbox{\Gamma_{0}}{\mathsubst{\mathvect{x}}{\mathvect{v}}})}{(\mathsubstbox{\Gamma_{1}}{\mathsubst{\mathvect{x}}{\mathvect{v}}})}}$.
   Let 
   \[
   {D_{i}}={{\mleft({\mathof{\mathdom}{\Gamma_{i}}}\setminus{\mathsetextension{x_{0}, \ldots, x_{n}}}\mright)}\cup{\mathsetintension{v_{i}}{{x_{i}}\in{\mathof{\mathdom}{\Gamma_{i}}}}}}
   \]
   for $i=0, 1$.
   Then
   ${\mathof{\mathdom}{\mathpiparallel{(\mathsubstbox{\Gamma_{0}}{\mathsubst{\mathvect{x}}{\mathvect{v}}})}{(\mathsubstbox{\Gamma_{1}}{\mathsubst{\mathvect{x}}{\mathvect{v}}})}}}={{D_{0}}\cup{D_{1}}}$.
   Since ${\mathsubstbox{(\mathpiparallel{\Gamma_{0}}{\Gamma_{1}})}{\mathsubst{\mathvect{x}}{\mathvect{v}}}}$ 
   is well-defined, we have
   \begin{align*}
    {\mathof{\mathdom}{{\mathsubstbox{(\mathpiparallel{\Gamma_{0}}{\Gamma_{1}})}{\mathsubst{\mathvect{x}}{\mathvect{v}}}}}}
    &= {{\mleft({\mathof{\mathdom}{\mathpiparallel{\Gamma_{0}}{\Gamma_{1}}}}\setminus{\mathsetextension{x_{0}, \ldots, x_{n}}}\mright)}\cup{\mathsetextension{v_{0}, \ldots, v_{n}}}} \\
    &= {{\mleft(\mleft({\mathof{\mathdom}{\Gamma_{0}}} \cup {\mathof{\mathdom}{\Gamma_{1}}} \mright)\setminus{\mathsetextension{x_{0}, \ldots, x_{n}}}\mright)}\cup{\mathsetextension{v_{0}, \ldots, v_{n}}}}.
   \end{align*}
   Then
   \begin{align*}
    &\phantom{=}{{\mleft(\mleft({\mathof{\mathdom}{\Gamma_{0}}} \cup {\mathof{\mathdom}{\Gamma_{1}}} \mright)\setminus{\mathsetextension{x_{0}, \ldots, x_{n}}}\mright)}\cup{\mathsetextension{v_{0}, \ldots, v_{n}}}} \\
    &={\mleft({{\mleft({{\mathof{\mathdom}{\Gamma_{0}}}\setminus{\mathsetextension{x_{0}, \ldots, x_{n}}}}\mright)} \cup {\mleft({{\mathof{\mathdom}{\Gamma_{1}}}\setminus{\mathsetextension{x_{0}, \ldots, x_{n}}}} \mright)}}\mright)\cup{\mathsetextension{v_{0}, \ldots, v_{n}}}} \\
    &={\mleft({\mleft({{\mathof{\mathdom}{\Gamma_{0}}}\setminus{\mathsetextension{x_{0}, \ldots, x_{n}}}}\mright)} \cup{\mathsetextension{v_{0}, \ldots, v_{n}}} \mright) \cup \mleft({\mleft({{\mathof{\mathdom}{\Gamma_{1}}}\setminus{\mathsetextension{x_{0}, \ldots, x_{n}}}} \mright)} \cup{\mathsetextension{v_{0}, \ldots, v_{n}}} \mright) } \\
    &={{D_{0}}\cup{D_{1}}}.
   \end{align*}
   Hence, we have
   \[
    {\mathof{\mathdom}{\mathsubstbox{(\mathpiparallel{\Gamma_{0}}{\Gamma_{1}})}{\mathsubst{\mathvect{x}}{\mathvect{v}}}}}={\mathof{\mathdom}{\mathpiparallel{(\mathsubstbox{\Gamma_{0}}{\mathsubst{\mathvect{x}}{\mathvect{v}}})}{(\mathsubstbox{\Gamma_{1}}{\mathsubst{\mathvect{x}}{\mathvect{v}}})}}}.
   \]

   Let ${w}\in{\mathof{\mathdom}{\mathpiparallel{(\mathsubstbox{\Gamma_{0}}{\mathsubst{\mathvect{x}}{\mathvect{v}}})}{(\mathsubstbox{\Gamma_{1}}{\mathsubst{\mathvect{x}}{\mathvect{v}}})}}}$. Then ${w}\in{{D_{0}}\cup{D_{1}}}$.
   
   If ${w}\notin{\mathvect{v}}$, then we have 
   \begin{align*}
    {\mathof{\mathpiparallel{(\mathsubstbox{\Gamma_{0}}{\mathsubst{\mathvect{x}}{\mathvect{v}}})}{(\mathsubstbox{\Gamma_{1}}{\mathsubst{\mathvect{x}}{\mathvect{v}}})}}{w}}
    &={\mathpiparallel{\mathof{\Gamma_{0}}{w}}{\mathof{\Gamma_{1}}{w}}} \\
    &={\mathof{\mathpiparallel{\Gamma_{0}}{\Gamma_{1}}}{w}} \\
    &={\mathof{\mathsubstbox{(\mathpiparallel{\Gamma_{0}}{\Gamma_{1}})}{\mathsubst{\mathvect{x}}{\mathvect{v}}}}{w}}.
   \end{align*}
   Hence, $\mathissubtype{\mathof{\mathsubstbox{(\mathpiparallel{\Gamma_{0}}{\Gamma_{1}})}{\mathsubst{\mathvect{x}}{\mathvect{v}}}}{w}}{\mathof{\mathpiparallel{(\mathsubstbox{\Gamma_{0}}{\mathsubst{\mathvect{x}}{\mathvect{v}}})}{(\mathsubstbox{\Gamma_{1}}{\mathsubst{\mathvect{x}}{\mathvect{v}}})}}{w}}$.

   Assume that ${w}\in{\mathvect{v}}$, ${w}\notin{\mathof{\mathdom}{\mathpiparallel{(\mathsubstbox{\Gamma_{0}}{\mathsubst{\mathvect{x}}{\mathvect{v}}})}{(\mathsubstbox{\Gamma_{1}}{\mathsubst{\mathvect{x}}{\mathvect{v}}})}}}$, and
   \[
   {\mathsetintension{{x_{i}}\in{\mathvect{x}}}{{w}={v_{i}}\text{ and } {x_{i}}\in{\mathof{\mathdom}{\mathpiparallel{(\mathsubstbox{\Gamma_{0}}{\mathsubst{\mathvect{x}}{\mathvect{v}}})}{(\mathsubstbox{\Gamma_{1}}{\mathsubst{\mathvect{x}}{\mathvect{v}}})}}}}}={\mathsetextension{x_{j_{0}}, \dots, x_{j_{k}}}}
   \]
   with ${0} \leq {j_{0}} < \dots < {j_{k}} \leq {n}$.
   Then
   \begin{align*}
    {\mathof{\mathpiparallel{(\mathsubstbox{\Gamma_{0}}{\mathsubst{\mathvect{x}}{\mathvect{v}}})}{(\mathsubstbox{\Gamma_{1}}{\mathsubst{\mathvect{x}}{\mathvect{v}}})}}{w}}
    &={\mathpiparallel{\mathpiparallel{\mathof{(\mathpiparallel{\Gamma_{0}}{\Gamma_{1}})}{x_{j_{0}}}}{\cdots}}{\mathof{(\mathpiparallel{\Gamma_{0}}{\Gamma_{1}})}{x_{j_{k}}}}} 
    \intertext{and}
    {\mathof{\mathsubstbox{(\mathpiparallel{\Gamma_{0}}{\Gamma_{1}})}{\mathsubst{\mathvect{x}}{\mathvect{v}}}}{w}}
    &={\mathpiparallel{\mathpiparallel{\mathof{(\mathpiparallel{\Gamma_{0}}{\Gamma_{1}})}{x_{j_{0}}}}{\cdots}}{\mathof{(\mathpiparallel{\Gamma_{0}}{\Gamma_{1}})}{x_{j_{k}}}}}.
   \end{align*}
   Hence, ${\mathof{\mathsubstbox{(\mathpiparallel{\Gamma_{0}}{\Gamma_{1}})}{\mathsubst{\mathvect{x}}{\mathvect{v}}}}{w}}={\mathof{\mathpiparallel{(\mathsubstbox{\Gamma_{0}}{\mathsubst{\mathvect{x}}{\mathvect{v}}})}{(\mathsubstbox{\Gamma_{1}}{\mathsubst{\mathvect{x}}{\mathvect{v}}})}}{w}}$.

   In a similar way, we can show 
   ${\mathof{\mathsubstbox{(\mathpiparallel{\Gamma_{0}}{\Gamma_{1}})}{\mathsubst{\mathvect{x}}{\mathvect{v}}}}{w}}={\mathof{\mathpiparallel{(\mathsubstbox{\Gamma_{0}}{\mathsubst{\mathvect{x}}{\mathvect{v}}})}{(\mathsubstbox{\Gamma_{1}}{\mathsubst{\mathvect{x}}{\mathvect{v}}})}}{w}}$
   in case ${w}\in{\mathvect{v}}$ and ${w}\in{\mathof{\mathdom}{\mathpiparallel{(\mathsubstbox{\Gamma_{0}}{\mathsubst{\mathvect{x}}{\mathvect{v}}})}{(\mathsubstbox{\Gamma_{1}}{\mathsubst{\mathvect{x}}{\mathvect{v}}})}}}$.
  \end{proof}

  \begin{proof}[Proof of \cref{lemma:substitution-lemma}]
   Assume that
   $\mathtypejudgementwithsecrecy{\Gamma}{L}{m}{P}$ is $l$-securely derivable,
   and ${\mathsubstbox{\Gamma}{\mathsubst{\mathvect{x}}{\mathvect{v}}}}$ is well-defined.
   Let $\mathvect{x}=\mathtuple{x_{0}, \dots, x_{n}}$, 
   $\mathvect{v}=\mathtuple{v_{0}, \dots, v_{n}}$,
   ${\tau_{i}}={\mathof{\Gamma}{x_{i}}}$ for $i=0, \dots, n$, and
   $\mathvect{\tau}=\mathtuple{\tau_{0}, \dots, \tau_{n}}$.

   We show the statement by induction on an $l$-secure derivation tree of
   $\mathtypejudgementwithsecrecy{\Gamma}{L}{m}{P}$.
   We proceed by a case analysis of the rule used at the root.

   \noindent Case 1. Assume that the rule used at the root is \rulename{T-Zero}.
   Then
   ${\Gamma}={\mathsubstbox{\Gamma}{\mathsubst{\mathvect{x}}{\mathvect{v}}}}={\emptyset}$ and
   ${P}={\mathsubstbox{P}{\mathsubst{\mathvect{x}}{\mathvect{v}}}}={\mathnil}$.
   Thus,
   $\mathtypejudgementwithsecrecy{\mathsubstbox{\Gamma}{\mathsubst{\mathvect{x}}{\mathvect{v}}}}{L}{m}{\mathsubstbox{P}{\mathsubst{\mathvect{x}}{\mathvect{v}}}}$
   is $l$-securely derivable.

   \noindent Case 2. Assume that the rule used at the root is \rulename{T-New}.
   Assume that $\mathtypejudgementwithsecrecy{\Gamma, \mathistype{y}{\xi/U}}{L}{m}{P'}$
   is the assumption of the rule instance, where
   ${P}={\mathpinew{\mathistype{y}{\xi}}{P'}}$ and $\mathreliable{U}$
   with ${y}\notin{\mathvect{x}}$.
   Then, we see that 
   $\mathtypejudgementwithsecrecy{\Gamma, \mathistype{y}{\xi/U}}{L}{m}{P'}$
   is $l$-securely derivable. 
   By the induction hypothesis,
   $\mathtypejudgementwithsecrecy{{\mathsubstbox{\Gamma}{\mathsubst{\mathvect{x}}{\mathvect{v}}}}, \mathistype{y}{\xi/U}}{L}{m}{\mathsubstbox{P'}{\mathsubst{\mathvect{x}}{\mathvect{v}}}}$
   is $l$-securely derivable.
   Let $\pi$ be an $l$-secure derivation tree of 
   $\mathtypejudgementwithsecrecy{{\mathsubstbox{\Gamma}{\mathsubst{\mathvect{x}}{\mathvect{v}}}}, \mathistype{y}{\xi/U}}{L}{m}{\mathsubstbox{P'}{\mathsubst{\mathvect{x}}{\mathvect{v}}}}$.
   Then, we have an $l$-secure derivation tree as follows:
   \begin{center}
    \begin{inlineprooftree}
     \AxiomC{}
     \RightLabel{$\pi$}
     \DeduceC{$\mathtypejudgementwithsecrecy{{\mathsubstbox{\Gamma}{\mathsubst{\mathvect{x}}{\mathvect{v}}}}, \mathistype{y}{\xi/U}}{L}{m}{\mathsubstbox{P'}{\mathsubst{\mathvect{x}}{\mathvect{v}}}}$}
     \RightLabel{\rulename{T-New}}
     \UnaryInfC{$\mathtypejudgementwithsecrecy{\mathsubstbox{\Gamma}{\mathsubst{\mathvect{x}}{\mathvect{v}}}}{L}{m}{\mathsubstbox{P}{\mathsubst{\mathvect{x}}{\mathvect{v}}}}$}
    \end{inlineprooftree}.
   \end{center}
   Thus,
   $\mathtypejudgementwithsecrecy{\mathsubstbox{\Gamma}{\mathsubst{\mathvect{x}}{\mathvect{v}}}}{L}{m}{\mathsubstbox{P}{\mathsubst{\mathvect{x}}{\mathvect{v}}}}$
   is $l$-securely derivable.

   \noindent Case 3. Assume that the rule used at the root is \rulename{T-Rep}.
   Assume that $\mathtypejudgementwithsecrecy{\Gamma'}{L}{m}{P'}$ is 
   the assumption of the rule instance, where
   ${P}={\mathpireplication{P'}}$ and
   ${\Gamma}={\mathpireplication{\Gamma'}}$.
   Then, we see that
   $\mathtypejudgementwithsecrecy{\Gamma'}{L}{m}{P'}$ is
   $l$-securely derivable.
   By the induction hypothesis,
   $\mathtypejudgementwithsecrecy{\mathsubstbox{\Gamma'}{\mathsubst{\mathvect{x}}{\mathvect{v}}}}{L}{m}{\mathsubstbox{P'}{\mathsubst{\mathvect{x}}{\mathvect{v}}}}$
   is $l$-securely derivable.
   By \cref{lemma:substitution-and-some-operators} \cref{item:replication-lemma:substitution-and-some-operators},
   ${\mathsubstbox{\Gamma'}{\mathsubst{\mathvect{x}}{\mathvect{v}}}}$
   is well-defined and
   $\mathissubtype{\mathsubstbox{(\mathpireplication{\Gamma'})}{\mathsubst{\mathvect{x}}{\mathvect{v}}}}{\mathpireplication{\mathsubstbox{\Gamma'}{\mathsubst{\mathvect{x}}{\mathvect{v}}}}}$.
   Let $\pi$ be an $l$-secure derivation tree of 
   $\mathtypejudgementwithsecrecy{\mathsubstbox{\Gamma'}{\mathsubst{\mathvect{x}}{\mathvect{v}}}}{L}{m}{\mathsubstbox{P'}{\mathsubst{\mathvect{x}}{\mathvect{v}}}}$.
   Then, we have an $l$-secure derivation tree as follows:
   \begin{center}
    \begin{inlineprooftree}
     \AxiomC{}
     \RightLabel{$\pi$}
     \DeduceC{$\mathtypejudgementwithsecrecy{\mathsubstbox{\Gamma'}{\mathsubst{\mathvect{x}}{\mathvect{v}}}}{L}{m}{\mathsubstbox{P'}{\mathsubst{\mathvect{x}}{\mathvect{v}}}}$}
     \RightLabel{\rulename{T-Rep}}
     \UnaryInfC{$\mathtypejudgementwithsecrecy{\mathpireplication{\mathsubstbox{\Gamma'}{\mathsubst{\mathvect{x}}{\mathvect{v}}}}}{L}{m}{\mathpireplication{\mathsubstbox{P'}{\mathsubst{\mathvect{x}}{\mathvect{v}}}}}$}
     \RightLabel{\rulename{T-Weak}}
     \UnaryInfC{$\mathtypejudgementwithsecrecy{\mathsubstbox{(\mathpireplication{\Gamma'})}{\mathsubst{\mathvect{x}}{\mathvect{v}}}}{L}{m}{\mathsubstbox{\mathpireplication{P'}}{\mathsubst{\mathvect{x}}{\mathvect{v}}}}$}
    \end{inlineprooftree}.
   \end{center}
   Thus,
   $\mathtypejudgementwithsecrecy{\mathsubstbox{\Gamma}{\mathsubst{\mathvect{x}}{\mathvect{v}}}}{L}{m}{\mathsubstbox{P}{\mathsubst{\mathvect{x}}{\mathvect{v}}}}$
   is $l$-securely derivable.

   \noindent Case 4. Assume that the rule used at the root is \rulename{T-Par}.
   Assume that
   $\mathtypejudgementwithsecrecy{\Gamma_{0}}{L}{m}{P_{0}}$ and
   $\mathtypejudgementwithsecrecy{\Gamma_{1}}{L}{m}{P_{1}}$ are
   the assumptions of the rule instance, where
   ${P}={\mathpiparallel{P_{0}}{P_{1}}}$ and
   ${\Gamma}={\mathpiparallel{\Gamma_{0}}{\Gamma_{1}}}$.
   Then, we see that
   $\mathtypejudgementwithsecrecy{\Gamma_{0}}{L}{m}{P_{0}}$ and
   $\mathtypejudgementwithsecrecy{\Gamma_{1}}{L}{m}{P_{1}}$ are
   $l$-securely derivable.
   By the induction hypothesis, we see that
   $\mathtypejudgementwithsecrecy{\mathsubstbox{\Gamma_{0}}{\mathsubst{\mathvect{x}}{\mathvect{v}}}}{L}{m}{\mathsubstbox{P_{0}}{\mathsubst{\mathvect{x}}{\mathvect{v}}}}$
   and
   $\mathtypejudgementwithsecrecy{\mathsubstbox{\Gamma_{1}}{\mathsubst{\mathvect{x}}{\mathvect{v}}}}{L}{m}{\mathsubstbox{P_{1}}{\mathsubst{\mathvect{x}}{\mathvect{v}}}}$
   are $l$-securely derivable.
   By \cref{lemma:substitution-and-some-operators} \cref{item:parallel-lemma:substitution-and-some-operators},
   we have 
   ${\mathsubstbox{(\mathpiparallel{\Gamma_{0}}{\Gamma_{1}})}{\mathsubst{\mathvect{x}}{\mathvect{v}}}}={\mathpiparallel{(\mathsubstbox{\Gamma_{0}}{\mathsubst{\mathvect{x}}{\mathvect{v}}})}{(\mathsubstbox{\Gamma_{1}}{\mathsubst{\mathvect{x}}{\mathvect{v}}})}}$.
   Let $\pi_{i}$ be an $l$-secure derivation tree of
   $\mathtypejudgementwithsecrecy{\mathsubstbox{\Gamma_{0}}{\mathsubst{\mathvect{x}}{\mathvect{v}}}}{L}{m}{\mathsubstbox{P_{0}}{\mathsubst{\mathvect{x}}{\mathvect{v}}}}$
   for ${i}={0, 1}$.
   Then, we have an $l$-secure derivation tree as follows:
   \begin{center}
    \begin{inlineprooftree}
     \AxiomC{}
     \RightLabel{$\pi_{0}$}
     \DeduceC{$\mathtypejudgementwithsecrecy{\mathsubstbox{\Gamma_{0}}{\mathsubst{\mathvect{x}}{\mathvect{v}}}}{L}{m}{\mathsubstbox{P_{0}}{\mathsubst{\mathvect{x}}{\mathvect{v}}}}$}

     \AxiomC{}
     \RightLabel{$\pi_{1}$}
     \DeduceC{$\mathtypejudgementwithsecrecy{\mathsubstbox{\Gamma_{1}}{\mathsubst{\mathvect{x}}{\mathvect{v}}}}{L}{m}{\mathsubstbox{P_{1}}{\mathsubst{\mathvect{x}}{\mathvect{v}}}}$}

     \RightLabel{\rulename{T-Par}}
     \BinaryInfC{$\mathtypejudgementwithsecrecy{\mathsubstbox{\Gamma}{\mathsubst{\mathvect{x}}{\mathvect{v}}}}{L}{m}{\mathsubstbox{P}{\mathsubst{\mathvect{x}}{\mathvect{v}}}}$}
    \end{inlineprooftree}.
   \end{center}
   Thus,
   $\mathtypejudgementwithsecrecy{\mathsubstbox{\Gamma}{\mathsubst{\mathvect{x}}{\mathvect{v}}}}{L}{m}{\mathsubstbox{P}{\mathsubst{\mathvect{x}}{\mathvect{v}}}}$
   is $l$-securely derivable.

   \noindent Case 5. Assume that the rule used at the root is \rulename{T-If}.
   Assume that
   $\mathtypejudgementwithsecrecy{\Gamma'}{L}{m}{Q_{0}}$ and
   $\mathtypejudgementwithsecrecy{\Gamma'}{L}{m}{Q_{1}}$ are
   the assumptions of the rule instance, where
   ${P}={\mathobif{w}{Q_{0}}{Q_{1}}}$ and
   ${\Gamma}={\mathpiparallel{\Gamma'}{\mathistype{w}{\mathbooltypewithsec{l}}}}$.
   Then, we see that
   $\mathtypejudgementwithsecrecy{\Gamma'}{L}{m}{Q_{0}}$ and
   $\mathtypejudgementwithsecrecy{\Gamma'}{L}{m}{Q_{1}}$ are 
   $l$-securely derivable.
   By the induction hypothesis,
   $\mathtypejudgementwithsecrecy{\mathsubstbox{\Gamma'}{\mathsubst{\mathvect{x}}{\mathvect{v}}}}{L}{m}{\mathsubstbox{Q_{0}}{\mathsubst{\mathvect{x}}{\mathvect{v}}}}$ and
   $\mathtypejudgementwithsecrecy{\mathsubstbox{\Gamma'}{\mathsubst{\mathvect{x}}{\mathvect{v}}}}{L}{m}{\mathsubstbox{Q_{1}}{\mathsubst{\mathvect{x}}{\mathvect{v}}}}$
   are $l$-securely derivable.
   Let $\pi_{i}$ be an $l$-secure derivation tree of
   $\mathtypejudgementwithsecrecy{\mathsubstbox{\Gamma'}{\mathsubst{\mathvect{x}}{\mathvect{v}}}}{L}{m}{\mathsubstbox{Q_{i}}{\mathsubst{\mathvect{x}}{\mathvect{v}}}}$
   for each $i=0, 1$.
   Then, we have an $l$-secure derivation tree as follows:
   \begin{center}
    \begin{inlineprooftree}
     \AxiomC{}
     \RightLabel{$\pi_{0}$}
     \DeduceC{$\mathtypejudgementwithsecrecy{\mathsubstbox{\Gamma'}{\mathsubst{\mathvect{x}}{\mathvect{v}}}}{L}{m}{\mathsubstbox{Q_{0}}{\mathsubst{\mathvect{x}}{\mathvect{v}}}}$}

     \AxiomC{}
     \RightLabel{$\pi_{1}$}
     \DeduceC{$\mathtypejudgementwithsecrecy{\mathsubstbox{\Gamma'}{\mathsubst{\mathvect{x}}{\mathvect{v}}}}{L}{m}{\mathsubstbox{Q_{1}}{\mathsubst{\mathvect{x}}{\mathvect{v}}}}$}

     \RightLabel{\rulename{T-If}}
     \BinaryInfC{$\mathtypejudgementwithsecrecy{\mathsubstbox{\Gamma}{\mathsubst{\mathvect{x}}{\mathvect{v}}}}{L}{m}{\mathsubstbox{P}{\mathsubst{\mathvect{x}}{\mathvect{v}}}}$}
    \end{inlineprooftree}.
   \end{center}
   Thus,
   $\mathtypejudgementwithsecrecy{\mathsubstbox{\Gamma}{\mathsubst{\mathvect{x}}{\mathvect{v}}}}{L}{m}{\mathsubstbox{P}{\mathsubst{\mathvect{x}}{\mathvect{v}}}}$
   is $l$-securely derivable.

   \noindent Case 6. Assume that the rule used at the root is \rulename{T-Out}.
   Assume that 
   $\mathtypejudgementwithsecrecy{\Gamma_{0}, {\mathistype{y}{\mathprogramtypetuple{\mathvect{\tau}}^{l_{1}}/U}}}{L}{m_{1}}{P'}$
   is the assumption of the rule instance,
   where ${P}={\mathpioutput{y}{\mathvect{w}}.{P'}}$,
   ${\Gamma}={\mathpiparallel{\mathlevelraise{t_{c}+1}{t_{c}+1}{\mathpiparallel{\Gamma_{0}}{\mathistype{\mathvect{w}}{\mathlevelraiseuni{\mathvect{\tau}}}}}}{\mathistype{y}{\mathprogramtypetuple{\mathvect{\tau}}^{l_{1}}/O^{0}_{t_{c}}U}}}$,
   ${l}\leq_{L}{l_{1}}$ and ${l}\leq_{L}{m_{1}}$ and
   $t_{c} = \infty$ implies ${l_{1}}\leq_{L}{m_{1}}$.
   Then, we see that
   $\mathtypejudgementwithsecrecy{\Gamma_{0}, {\mathistype{y}{\mathprogramtypetuple{\mathvect{\tau}}^{l_{1}}/U}}}{L}{m_{1}}{P'}$
   is $l$-securely derivable.
   By the induction hypothesis,
   $\mathtypejudgementwithsecrecy{\mathsubstbox{\Gamma_{0}}{\mathsubst{\mathvect{x}}{\mathvect{v}}}, {\mathistype{\mathsubstbox{y}{\mathsubst{\mathvect{x}}{\mathvect{v}}}}{\mathprogramtypetuple{\mathvect{\tau}}^{l_{1}}/U}}}{L}{m_{1}}{\mathsubstbox{P'}{\mathsubst{\mathvect{x}}{\mathvect{v}}}}$
   is $l$-securely derivable.
   Let $\pi$ be an $l$-secure derivation tree of
   $\mathtypejudgementwithsecrecy{\mathsubstbox{\Gamma_{0}}{\mathsubst{\mathvect{x}}{\mathvect{v}}}, {\mathistype{\mathsubstbox{y}{\mathsubst{\mathvect{x}}{\mathvect{v}}}}{\mathprogramtypetuple{\mathvect{\tau}}^{l_{1}}/U}}}{L}{m_{1}}{\mathsubstbox{P'}{\mathsubst{\mathvect{x}}{\mathvect{v}}}}$.
   Then, we have an $l$-secure derivation tree as follows:
   \begin{center}
    \begin{inlineprooftree}
     \AxiomC{}
     \RightLabel{$\pi$}
     \DeduceC{$\mathtypejudgementwithsecrecy{\mathsubstbox{\Gamma_{0}}{\mathsubst{\mathvect{x}}{\mathvect{v}}}, {\mathistype{\mathsubstbox{y}{\mathsubst{\mathvect{x}}{\mathvect{v}}}}{\mathprogramtypetuple{\mathvect{\tau}}^{l_{1}}/U}}}{L}{m_{1}}{\mathsubstbox{P'}{\mathsubst{\mathvect{x}}{\mathvect{v}}}}$}
     \RightLabel{\rulename{T-Out}}
     \UnaryInfC{$\mathtypejudgementwithsecrecy{\mathsubstbox{\Gamma}{\mathsubst{\mathvect{x}}{\mathvect{v}}}}{L}{m}{\mathpioutput{\mathsubstbox{y}{\mathsubst{\mathvect{x}}{\mathvect{v}}}}{\mathsubstbox{\mathvect{w}}{\mathsubst{\mathvect{x}}{\mathvect{v}}}}.\mathsubstbox{P'}{\mathsubst{\mathvect{x}}{\mathvect{v}}}}$}
    \end{inlineprooftree}.
   \end{center}
   Thus,
   $\mathtypejudgementwithsecrecy{\mathsubstbox{\Gamma}{\mathsubst{\mathvect{x}}{\mathvect{v}}}}{L}{m}{\mathsubstbox{P}{\mathsubst{\mathvect{x}}{\mathvect{v}}}}$
   is $l$-securely derivable.

   \noindent Case 7. Assume that the rule used at the root is \rulename{T-In}.
   Assume that
   $\mathtypejudgementwithsecrecy{\Gamma_{0}, {\mathistype{y}{\mathprogramtypetuple{\mathvect{\tau}}^{l_{1}}/U}}, {\mathistype{\mathvect{z}}{\mathvect{\tau}}}}{L}{m_{1}}{P'}$
   is the assumption of the rule instance, where
   ${l}\leq_{L}{l_{1}}$, ${l}\leq_{L}{m_{1}}$,
   ${P}={\mathpiinput{y}{\mathvect{z}}.{P'}}$, and
   ${\Gamma}= \mathlevelraise{t_{c}+1}{t_{c}+1}{\Gamma_{0}}, \mathistype{y}{\mathprogramtypetuple{\mathvect{\tau}}^{l_{1}}/I^{0}_{t_{c}}U} $, and
   $t_{c} = \infty$ implies ${l_{1}}\leq_{L}{m_{1}}$.
   We can assume ${z'}\notin{\mathvect{x}}$ and ${z'}\notin{\mathvect{v}}$
   for any ${z'}\in{\mathvect{z}}$.
   Since the assumption of the rule instance is $l$-securely derivable,
   we see that
   $\mathtypejudgementwithsecrecy{\Gamma_{0}, {\mathistype{y}{\mathprogramtypetuple{\mathvect{\tau}}^{l_{1}}/U}}, {\mathistype{\mathvect{z}}{\mathvect{\tau}}}}{L}{m_{1}}{P'}$
   is $l$-securely derivable.
   Let ${\Gamma'}={\Gamma_{0}, {\mathistype{y}{\mathprogramtypetuple{\mathvect{\tau}}^{l_{1}}/U}}, {\mathistype{\mathvect{z}}{\mathvect{\tau}}}}$.
   Since we have ${z'}\notin{\mathvect{x}}$ and ${z'}\notin{\mathvect{v}}$
   for any ${z'}\in{\mathvect{z}}$,
   we see that
   ${\mathsubstbox{\Gamma'}{\mathsubst{\mathvect{x}}{\mathvect{v}}}}={\mathsubstbox{\Gamma_{0}}{\mathsubst{\mathvect{x}}{\mathvect{v}}}, {\mathistype{\mathsubstbox{y}{\mathsubst{\mathvect{x}}{\mathvect{v}}}}{\mathprogramtypetuple{\mathvect{\tau}}^{l_{1}}/U}}, {\mathistype{\mathvect{z}}{\mathvect{\tau}}}}$
   is well-defined.
   By the induction hypothesis, we see that
   $\mathtypejudgementwithsecrecy{\mathsubstbox{\Gamma'}{\mathsubst{\mathvect{x}}{\mathvect{v}}}}{L}{m_{1}}{\mathsubstbox{P'}{\mathsubst{\mathvect{x}}{\mathvect{v}}}}$
   is $l$-securely derivable.
   Let $\pi$ be an $l$-secure derivation tree of
   $\mathtypejudgementwithsecrecy{\mathsubstbox{\Gamma'}{\mathsubst{\mathvect{x}}{\mathvect{v}}}}{L}{m_{1}}{\mathsubstbox{P'}{\mathsubst{\mathvect{x}}{\mathvect{v}}}}$.
   Then, we have an $l$-secure derivation tree as follows:
   \begin{center}
    \begin{inlineprooftree}
     \AxiomC{}
     \RightLabel{$\pi$}
     \DeduceC{$\mathtypejudgementwithsecrecy{\mathsubstbox{\Gamma'}{\mathsubst{\mathvect{x}}{\mathvect{v}}}}{L}{m_{1}}{\mathsubstbox{P'}{\mathsubst{\mathvect{x}}{\mathvect{v}}}}$}
     \RightLabel{\rulename{T-In}}
     \UnaryInfC{$\mathtypejudgementwithsecrecy{\mathsubstbox{\Gamma}{\mathsubst{\mathvect{x}}{\mathvect{v}}}}{L}{m_{1}}{\mathpiinput{\mathsubstbox{y}{\mathsubst{\mathvect{x}}{\mathvect{v}}}}{\mathvect{z}}.\mathsubstbox{P'}{\mathsubst{\mathvect{x}}{\mathvect{v}}}}$}
    \end{inlineprooftree}.
   \end{center}
   Thus,
   $\mathtypejudgementwithsecrecy{\mathsubstbox{\Gamma}{\mathsubst{\mathvect{x}}{\mathvect{v}}}}{L}{l}{\mathsubstbox{P}{\mathsubst{\mathvect{x}}{\mathvect{v}}}}$
   is $l$-securely derivable.

   \noindent Case 8. Assume that the rule used at the root is \rulename{T-NewSec}.
   Assume that
   $\mathtypejudgementwithsecrecy{\Gamma}{\mathpinewsecrecylevel{l_{0}}{\mathvect{l_{1}}}{\mathvect{l_{2}}}{L}}{m}{P'}$
   is the assumption of the rule instance, where
   ${P}={\mathpinewsecrecylevel{l_{0}}{\mathvect{l_{1}}}{\mathvect{l_{2}}}{P'}}$
   and $l' \leq_{L} l_{s}$ for any $l'\in\mathvect{l_{1}}$, $\mathvect{l_{2}}$.
   Then, we see that
   $\mathtypejudgementwithsecrecy{\Gamma}{\mathpinewsecrecylevel{l_{0}}{\mathvect{l_{1}}}{\mathvect{l_{2}}}{L}}{m}{P'}$
   is $l$-securely derivable.
   By the induction hypothesis, we see that \linebreak[3]
   $\mathtypejudgementwithsecrecy{\mathsubstbox{\Gamma}{\mathsubst{\mathvect{x}}{\mathvect{v}}}}{\mathpinewsecrecylevel{l_{0}}{\mathvect{l_{1}}}{\mathvect{l_{2}}}{L}}{m}{\mathsubstbox{P'}{\mathsubst{\mathvect{x}}{\mathvect{v}}}}$
   is $l$-securely derivable.
   Let $\pi$ be an $l$-secure derivation tree of  \linebreak[3]
   $\mathtypejudgementwithsecrecy{\mathsubstbox{\Gamma}{\mathsubst{\mathvect{x}}{\mathvect{v}}}}{\mathpinewsecrecylevel{l_{0}}{\mathvect{l_{1}}}{\mathvect{l_{2}}}{L}}{m}{\mathsubstbox{P'}{\mathsubst{\mathvect{x}}{\mathvect{v}}}}$.
   Then, we have an $l$-secure derivation tree as follows:
   \begin{center}
    \begin{inlineprooftree}
     \AxiomC{}
     \RightLabel{$\pi$}
     \DeduceC{$\mathtypejudgementwithsecrecy{\mathsubstbox{\Gamma}{\mathsubst{\mathvect{x}}{\mathvect{v}}}}{\mathpinewsecrecylevel{l_{0}}{\mathvect{l_{1}}}{\mathvect{l_{2}}}{L}}{m}{\mathsubstbox{P'}{\mathsubst{\mathvect{x}}{\mathvect{v}}}}$}
     \RightLabel{\rulename{T-NewSec}}
     \UnaryInfC{$\mathtypejudgementwithsecrecy{\mathsubstbox{\Gamma}{\mathsubst{\mathvect{x}}{\mathvect{v}}}}{L}{m}{\mathpinewsecrecylevel{l_{0}}{\mathvect{l_{1}}}{\mathvect{l_{2}}}{\mathsubstbox{P'}{\mathsubst{\mathvect{x}}{\mathvect{v}}}}}$}
    \end{inlineprooftree}.
   \end{center}
   Thus,
   $\mathtypejudgementwithsecrecy{\mathsubstbox{\Gamma}{\mathsubst{\mathvect{x}}{\mathvect{v}}}}{L}{m}{\mathsubstbox{P}{\mathsubst{\mathvect{x}}{\mathvect{v}}}}$
   is $l$-securely derivable.

   \noindent Case 9. Assume that the rule used at the root is \rulename{T-Weak}.
   Assume that $\mathtypejudgementwithsecrecy{\Gamma'}{L}{m'}{P}$ is 
   the assumption of the rule instance,
   where $\mathissubtype{\Gamma}{\Gamma'}$ and $l\leq_{L}m'$.
   Then, we see that 
   $\mathtypejudgementwithsecrecy{\Gamma'}{L}{m'}{P}$ is
   $l$-securely derivable.
   By the induction hypothesis, we see that 
   $\mathtypejudgementwithsecrecy{\mathsubstbox{\Gamma'}{\mathsubst{\mathvect{x}}{\mathvect{v}}}}{L}{m'}{\mathsubstbox{P}{\mathsubst{\mathvect{x}}{\mathvect{v}}}}$
   is $l$-securely derivable.
   By \cref{lemma:substitution-preserve-subtyping},
   ${\mathsubstbox{\Gamma'}{\mathsubst{\mathvect{x}}{\mathvect{v}}}}$
   is well-defined and
   $\mathissubtype{{\mathsubstbox{\Gamma}{\mathsubst{\mathvect{x}}{\mathvect{v}}}}}{{\mathsubstbox{\Gamma'}{\mathsubst{\mathvect{x}}{\mathvect{v}}}}}$.
   Let $\pi$ be an $l$-secure derivation tree of
   $\mathtypejudgementwithsecrecy{\mathsubstbox{\Gamma'}{\mathsubst{\mathvect{x}}{\mathvect{v}}}}{L}{m'}{\mathsubstbox{P}{\mathsubst{\mathvect{x}}{\mathvect{v}}}}$.
   Then, we have an $l$-secure derivation tree as follows:
   \begin{center}
    \begin{inlineprooftree}
     \AxiomC{}
     \RightLabel{$\pi$}
     \DeduceC{$\mathtypejudgementwithsecrecy{\mathsubstbox{\Gamma'}{\mathsubst{\mathvect{x}}{\mathvect{v}}}}{L}{m'}{\mathsubstbox{P}{\mathsubst{\mathvect{x}}{\mathvect{v}}}}$}
     \RightLabel{\rulename{T-Weak}}
     \UnaryInfC{$\mathtypejudgementwithsecrecy{\mathsubstbox{\Gamma}{\mathsubst{\mathvect{x}}{\mathvect{v}}}}{L}{m}{\mathsubstbox{P}{\mathsubst{\mathvect{x}}{\mathvect{v}}}}$}
    \end{inlineprooftree}.
   \end{center}
   Thus,
   $\mathtypejudgementwithsecrecy{\mathsubstbox{\Gamma}{\mathsubst{\mathvect{x}}{\mathvect{v}}}}{L}{m}{\mathsubstbox{P}{\mathsubst{\mathvect{x}}{\mathvect{v}}}}$
   is $l$-securely derivable.
  \end{proof}

  \subsection{Proof of subject reduction}
   \label[appendix]{sec:proof-of-subject-reduction}

   \begin{lemma}
   \label[lemma]{lemma:reduction-and-subtyping}
   For type environments $\Gamma_{0}$, $\Gamma'_{0}$, and $\Gamma_{1}$,
   if ${\Gamma_{1}}\mathtypeenvreduction{\Gamma'_{1}}$ and 
   $\mathissubtype{\Gamma_{0}}{\Gamma_{1}}$,
   then there exists $\Gamma'_{0}$ such that
   ${\Gamma_{0}}\mathtypeenvreduction{\Gamma'_{0}}$ and 
   $\mathissubtype{\Gamma'_{0}}{\Gamma'_{1}}$.
   \end{lemma}

  \begin{proof}
   Since ${\Gamma_{1}}\mathtypeenvreduction{\Gamma'_{1}}$,
   there exist a name $x$, a core channel type $\xi$, and usages $U_{1}$, $U'_{1}$ 
   such that ${\mathof{\Gamma_{1}}{x}}={{\xi}/{U_{1}}}$, 
   ${\mathof{\Gamma'_{1}}{x}}={{\xi}/{U'_{1}}}$, and 
   ${U_{1}}\mathusagereduction{U'_{1}}$.
   Since $\mathissubtype{\Gamma_{0}}{\Gamma_{1}}$,
   there exists a core channel type $\xi$, and a usage $U_{0}$ such that
   ${\mathof{\Gamma_{0}}{x}}={{\xi}/{U_{0}}}$ and $\mathissubusage{U_{0}}{U_{1}}$.
   By \cref{def:subusage} \cref{item:reduction-in-def:subusage},
   there exists a usage $U'_{0}$ such that
   ${U_{0}}\mathusagereduction{U'_{0}}$ and $\mathissubusage{U'_{0}}{U'_{1}}$.
   Let ${\Gamma'_{0}}={\mathsubstbox{\Gamma_{0}}{\mathsubst{x}{{\xi}/{U'_{0}}}}}$.
   Then ${\Gamma_{0}}\mathtypeenvreduction{\Gamma'_{0}}$ and 
   $\mathissubtype{\Gamma'_{0}}{\Gamma'_{1}}$.
  \end{proof}

  \begin{lemma}
   \label[lemma]{lemma:reduction-and-subtyping-parallel}
   For type environments $\Gamma_{0}$, $\Gamma'_{0}$, and $\Gamma_{1}$,
   if ${\Gamma_{0}}\mathtypeenvreduction{\Gamma'_{0}}$,
   then
   ${\mathpiparallel{\Gamma_{0}}{\Gamma_{1}}}\mathtypeenvreduction{\mathpiparallel{\Gamma'_{0}}{\Gamma_{1}}}$.
  \end{lemma}

  \begin{proof}
   Since ${\Gamma_{0}}\mathtypeenvreduction{\Gamma'_{0}}$,
   there exist a name $x$, a core channel type $\xi$, and usages $U_{0}$, $U'_{0}$ 
   such that ${\mathof{\Gamma_{0}}{x}}={{\xi}/{U_{0}}}$, 
   ${\mathof{\Gamma'_{0}}{x}}={{\xi}/{U'_{0}}}$, and 
   ${U_{0}}\mathusagereduction{U'_{0}}$.

   Then
   ${\mathof{\mathpiparallel{\Gamma_{0}}{\Gamma_{1}}}{x}}={{\xi}/{U_{0}}}$
   or
   ${\mathof{\mathpiparallel{\Gamma_{0}}{\Gamma_{1}}}{x}}={{\xi}/{\mathpiparallel{U_{0}}{U_{1}}}}$ with some usage $U_{1}$.

   If ${\mathof{\mathpiparallel{\Gamma_{0}}{\Gamma_{1}}}{x}}={{\xi}/{U_{0}}}$,
   then we have ${\mathof{\mathpiparallel{\Gamma'_{0}}{\Gamma_{1}}}{x}}={{\xi}/{U'_{0}}}$.
   Hence,
   ${\mathpiparallel{\Gamma_{0}}{\Gamma_{1}}}\mathtypeenvreduction{\mathpiparallel{\Gamma'_{0}}{\Gamma_{1}}}$.

   If ${\mathof{\mathpiparallel{\Gamma_{0}}{\Gamma_{1}}}{x}}={{\xi}/{\mathpiparallel{U_{0}}{U_{1}}}}$ with some usage $U_{1}$,
   then we have ${\mathof{\mathpiparallel{\Gamma'_{0}}{\Gamma_{1}}}{x}}={{\xi}/{\mathpiparallel{U'_{0}}{U_{1}}}}$.
   Hence,
   ${\mathpiparallel{\Gamma_{0}}{\Gamma_{1}}}\mathtypeenvreduction{\mathpiparallel{\Gamma'_{0}}{\Gamma_{1}}}$.
  \end{proof}

  Now, we prove \cref{prop:subject-reduction}.

  \begin{proof}[Proof of \cref{prop:subject-reduction}] 
   Let $P$ and $P'$ be processes.
   Assume that $\mathtypejudgementwithsecrecy{\Gamma}{L}{m}{P}$ is 
   $l$-securely derivable and 
   ${\mathtuple{P, L}}\mathpireduction{\mathtuple{P', L'}}$.
   We show that there exist a type environment $\Gamma'$ such that
   either ${\Gamma'}={\Gamma}$ or ${\Gamma}\mathtypeenvreduction{\Gamma'}$ and
   $\mathtypejudgementwithsecrecy{\Gamma'}{L'}{m}{P'}$ is 
   $l$-securely derivable.

   We show the statement by induction on the construction of
   ${\mathtuple{P, L}}\mathpireduction{\mathtuple{P', L'}}$.
   We proceed by a case analysis of the last rule used 
   to construct ${P}\mathpireduction{P'}$.

   \noindent Case 1. We consider the case \rulename{R-Com}.
   In this case, 
   ${P}={\mathpiparallel{\mathpioutput{x}{(v_{0}, \dots, v_{n})}. P_{0}}{\mathpiinput{x}{(y_{0}, \dots, y_{n})}.P_{1}}}$,  
   ${P'}={\mathpiparallel{P_{0}}{\mathsubstbox{P_{1}}{\mathsubst{y_{0}}{v_{0}}, \dots, \mathsubst{y_{n}}{v_{n}}}}}$,
   and ${L'}={L}$.
   Let ${\mathvect{y}}={(y_{0}, \dots, y_{n})}$ and ${\mathvect{v}}={(v_{0}, \dots, v_{n})}$.

   By \cref{lemma:inversion} \cref{item:parallel-lemma-inversion},
   there exist two type environments $\Gamma'_{0}$, $\Gamma'_{1}$,
   and ${l'}\in{L}$
   such that $\mathissubtype{\Gamma}{\mathpiparallel{\Gamma'_{0}}{\Gamma'_{1}}}$, 
   and $l\leq_{L}l'$,
   $\mathtypejudgementwithsecrecy{\Gamma'_{0}}{L}{m'}{\mathpioutput{x}{(v_{0}, \dots, v_{n})}. P_{0}}$
   and
   $\mathtypejudgementwithsecrecy{\Gamma'_{1}}{L}{m'}{\mathpiinput{x}{(y_{0}, \dots, y_{n})}.P_{1}}$
   are $\check{l}$-securely derivable.
   By \cref{lemma:inversion} \cref{item:output-lemma-inversion},
   there exist a type environments $\Gamma''_{0}$,
   ${l''_{00}}\in{L}$,
   ${m''_{01}}\in{L}$, types ${\mathvect{\tau}}$, a usage $U$ and
   ${t_{c}}\in{\mathnat \cup \mathsetextension{\infty}}$
   such that
   $\mathissubtype{\Gamma'_{0}}{\mleft(\mathpiparallel{\mathlevelraise{t_{c}+1}{t_{c}+1}{\mathpiparallel{\Gamma''_{0}}{\mathistype{\mathvect{v}}{\mathlevelraiseuni{\mathvect{\tau}}}}}}{\mathistype{x}{\mathprogramtypetuple{\mathvect{\tau}}^{l''_{00}}/O^{0}_{t_{c}}U}}\mright)}$,
   ${l_{0}}\leq_{L}{l''_{00}}$, and 
   ${l_{0}}\leq_{L}{m''_{01}}$,
   $\mathtypeenvandsecrecylatice{\mleft(\mathpiparallel{\mathlevelraise{t_{c}+1}{t_{c}+1}{\mathpiparallel{\Gamma''_{0}}{\mathistype{\mathvect{v}}{\mathlevelraiseuni{\mathvect{\tau}}}}}}{\mathistype{x}{\mathprogramtypetuple{\mathvect{\tau}}^{l''_{00}}/O^{0}_{t_{c}}U}}\mright)}{L}$ is $l$-secure, and
   $\mathtypejudgementwithsecrecy{\Gamma''_{0}, {\mathistype{x}{\mathprogramtypetuple{\mathvect{\tau}}^{l''_{00}}/U}}}{L}{m''_{01}}{P_{0}}$
   is $l$-securely derivable, and
   $t_{c} = \infty$ implies ${l''_{00}}\leq_{L}{m''_{01}}$.
   By \cref{lemma:inversion} \cref{item:input-lemma-inversion},
   there exist a type environments $\Gamma''_{1}$
   ${l''_{10}}\in{L}$,
   ${m''_{11}}\in{L}$, types ${\mathvect{\tau'}}$, a usage $U'$ and 
   ${t'_{c}}\in{\mathnat \cup \mathsetextension{\infty}}$ such that
   $\mathissubtype{\Gamma''_{1}}{\mleft(\mathlevelraise{t'_{c}+1}{t'_{c}+1}{\Gamma''_{1}}, \mathistype{x}{\mathprogramtypetuple{\mathvect{\tau'}}^{l''_{10}}/I^{0}_{t'_{c}}U'}\mright)}$,
   ${l_{1}}\leq_{L}{l''_{10}}$, and 
   ${l_{1}}\leq_{L}{m''_{11}}$,
   $\mathtypeenvandsecrecylatice{\mleft(\mathlevelraise{t'_{c}+1}{t'_{c}+1}{\Gamma''_{1}}, \mathistype{x}{\mathprogramtypetuple{\mathvect{\tau'}}^{l''_{10}}/I^{0}_{t'_{c}}U'}\mright)}{L}$ is $\check{l}$-secure,
   $\mathtypejudgementwithsecrecy{\Gamma''_{1}, {\mathistype{x}{\mathprogramtypetuple{\mathvect{\tau'}}^{l''_{10}}/U}}, {\mathistype{\mathvect{y}}{\mathvect{\tau'}}}}{L}{m''_{11}}{P_{1}}$
   is $l$-securely derivable, and
   $t'_{c} = \infty$ implies ${l''_{10}}\leq_{L}{m''_{11}}$.

   Since $\mathissubtype{\Gamma'_{0}}{\mleft(\mathpiparallel{\mathlevelraise{t_{c}+1}{t_{c}+1}{\mathpiparallel{\Gamma''_{0}}{\mathistype{\mathvect{v}}{\mathlevelraiseuni{\mathvect{\tau}}}}}}{\mathistype{x}{\mathprogramtypetuple{\mathvect{\tau}}^{l''_{00}}/O^{0}_{t_{c}}U}}\mright)}$,
   we have 
   ${\mathof{\Gamma'_{0}}{x}}\mathequivexceptusagessy{\mathprogramtypetuple{\mathvect{\tau}}^{l'''_{00}}/O^{0}_{t_{c}}U}$.
   Since $\mathissubtype{\Gamma'_{1}}{\mleft(\mathlevelraise{t'_{c}+1}{t'_{c}+1}{\Gamma'''_{1}}, \mathistype{x}{\mathprogramtypetuple{\mathvect{\tau'}}^{l'''_{10}}/I^{0}_{t'_{c}}U'}\mright)}$,
   we have 
   ${\mathof{\Gamma'_{1}}{x}}\mathequivexceptusagessy{\mathprogramtypetuple{\mathvect{\tau'}}^{l''_{10}}/I^{0}_{t'_{c}}U'}$.
   Since ${\mathpiparallel{\Gamma'_{0}}{\Gamma'_{1}}}$ is defined,
   ${\mathof{\Gamma'_{0}}{x}}\mathequivexceptusagessy{\mathof{\Gamma'_{1}}{x}}$.
   Hence,
   ${\mathprogramtypetuple{\mathvect{\tau}}^{l''_{00}}/O^{0}_{t_{c}}U}\mathequivexceptusagessy{\mathprogramtypetuple{\mathvect{\tau'}}^{l''_{10}}/I^{0}_{t'_{c}}U'}$.
   Therefore, ${\mathvect{\tau}}={\mathvect{\tau'}}$ and ${l''_{00}}={l''_{10}}$.
   
   Let $\pi_{0}$ be an $\check{l}$-secure derivation tree of
   $\mathtypejudgementwithsecrecy{\Gamma''_{0}, {\mathistype{x}{\mathprogramtypetuple{\mathvect{\tau}}^{l''_{00}}/U}}}{L}{m''_{01}}{P_{0}}$,
   and
   $\pi_{1}$ be an $\check{l}$-secure derivation tree of \linebreak[3]
   $\mathtypejudgementwithsecrecy{\Gamma''_{1}, {\mathistype{x}{\mathprogramtypetuple{\mathvect{\tau}}^{l''_{00}}/U}}, {\mathistype{\mathvect{y}}{\mathvect{\tau}}}}{L}{m''_{11}}{P_{1}}$.
   We have an $l$-secure derivation tree as follows:
   \begin{center}
    \begin{inlineprooftree}
     \AxiomC{}
     \RightLabel{$\pi_{0}$}
     \DeduceC{$\mathtypejudgementwithsecrecy{\Gamma''_{0}, {\mathistype{x}{\mathprogramtypetuple{\mathvect{\tau}}^{l''_{00}}/U}}}{L}{m''_{01}}{P_{0}}$}
     \UnaryInfC{$\mathtypejudgementwithsecrecy{\check{\Gamma}_{0}}{L}{m_{0}}{\mathpioutput{x}{(v_{0}, \dots, v_{n})}. P_{0}}$}
     \UnaryInfC{$\mathtypejudgementwithsecrecy{\check{\Gamma}_{0}}{L}{m'}{\mathpioutput{x}{(v_{0}, \dots, v_{n})}. P_{0}}$}

     \AxiomC{}
     \RightLabel{$\pi_{1}$}
     \DeduceC{$\mathtypejudgementwithsecrecy{\Gamma''_{1}, {\mathistype{x}{\mathprogramtypetuple{\mathvect{\tau}}^{l''_{00}}/U}}, {\mathistype{\mathvect{y}}{\mathvect{\tau}}}}{L}{m''_{11}}{P_{1}}$}
     \UnaryInfC{$\mathtypejudgementwithsecrecy{\check{\Gamma}_{1}}{L}{m_{1}}{\mathpiinput{x}{(y_{0}, \dots, y_{n})}.P_{1}}$}
     \UnaryInfC{$\mathtypejudgementwithsecrecy{\check{\Gamma}_{1}}{L}{m'}{\mathpiinput{x}{(y_{0}, \dots, y_{n})}.P_{1}}$}

     \BinaryInfC{$\mathtypejudgementwithsecrecy{\mathpiparallel{\check{\Gamma}_{0}}{\check{\Gamma}_{1}}}{L}{m'}{\mathpiparallel{\mathpioutput{x}{(v_{0}, \dots, v_{n})}. P_{0}}{\mathpiinput{x}{(y_{0}, \dots, y_{n})}.P_{1}}}$}
    \end{inlineprooftree},
   \end{center}
   where
   ${\check{\Gamma}_{0}}={\mleft(\mathpiparallel{\mathlevelraise{t_{c}+1}{t_{c}+1}{\mathpiparallel{\Gamma''_{0}}{\mathistype{\mathvect{v}}{\mathlevelraiseuni{\mathvect{\tau}}}}}}{\mathistype{x}{\mathprogramtypetuple{\mathvect{\tau}}^{l''_{00}}/O^{0}_{t_{c}}U'}}\mright)}$
   and
   ${\check{\Gamma}_{1}}={\mleft(\mathlevelraise{t'_{c}+1}{t'_{c}+1}{\Gamma''_{1}}, \mathistype{x}{\mathprogramtypetuple{\mathvect{\tau}}^{l''_{00}}/I^{0}_{t'_{c}}U'}\mright)}$.
   Thus, we see that 
   \[
   \mathtypejudgementwithsecrecy{\mathpiparallel{\mleft(\mathpiparallel{\mathlevelraise{t_{c}+1}{t_{c}+1}{\mathpiparallel{\Gamma''_{0}}{\mathistype{\mathvect{v}}{\mathlevelraiseuni{\mathvect{\tau}}}}}}{\mathistype{x}{\mathprogramtypetuple{\mathvect{\tau}}^{l''_{00}}/O^{0}_{t_{c}}U}}\mright)}{\mleft(\mathlevelraise{t'_{c}+1}{t'_{c}+1}{\Gamma''_{1}}, \mathistype{x}{\mathprogramtypetuple{\mathvect{\tau}}^{l''_{00}}/I^{0}_{t'_{c}}U'}\mright)}}{L}{m'}{P}
   \]
   is $l$-securely derivable.
   Then
   \begin{multline*}
    {\mathpiparallel{\mleft(\mathpiparallel{\mathlevelraise{t_{c}+1}{t_{c}+1}{\mathpiparallel{\Gamma''_{0}}{\mathistype{\mathvect{v}}{\mathlevelraiseuni{\mathvect{\tau}}}}}}{\mathistype{x}{\mathprogramtypetuple{\mathvect{\tau}}^{l''_{00}}/O^{0}_{t_{c}}U}}\mright)}{\mleft(\mathlevelraise{t'_{c}+1}{t'_{c}+1}{\Gamma''_{1}}, \mathistype{x}{\mathprogramtypetuple{\mathvect{\tau}}^{l''_{00}}/I^{0}_{t'_{c}}U'}\mright)}}
     \\
    \mathissubtypesy 
   {\mathpiparallel{\mathpiparallel{\mleft(\mathlevelraise{t_{c}+1}{t_{c}+1}{\mathpiparallel{\Gamma''_{0}}{\mathistype{\mathvect{v}}{\mathlevelraiseuni{\mathvect{\tau}}}}}\mright)}{\mleft(\mathlevelraise{t'_{c}+1}{t'_{c}+1}{\Gamma''_{1}} \mright)}}{\mathistype{x}{\mathprogramtypetuple{\mathvect{\tau}}^{l''_{00}}/(\mathpiparallel{O^{0}_{t_{c}}U}{I^{0}_{t'_{c}}U'})}}}. 
   \end{multline*}
   Hence, we have
   \[
   {\Gamma}
   \mathissubtypesy 
   {\mathpiparallel{\mathpiparallel{\mleft(\mathlevelraise{t_{c}+1}{t_{c}+1}{\mathpiparallel{\Gamma''_{0}}{\mathistype{\mathvect{v}}{\mathlevelraiseuni{\mathvect{\tau}}}}}\mright)}{\mleft(\mathlevelraise{t'_{c}+1}{t'_{c}+1}{\Gamma''_{1}} \mright)}}{\mathistype{x}{\mathprogramtypetuple{\mathvect{\tau}}^{l''_{00}}/(\mathpiparallel{O^{0}_{t_{c}}U}{I^{0}_{t'_{c}}U'})}}}. 
   \]

   Since ${y_{i}}\notin{\mathof{\mathdom}{\Gamma''_{1}}}$,
   we have ${\mathsubstbox{\Gamma''_{1}}{\mathsubst{\mathvect{y}}{\mathvect{v}}}}={\Gamma''_{1}}$.
   By \cref{lemma:substitution-lemma}, we see that
   \[
   \mathtypejudgementwithsecrecy{{\Gamma''_{1}}, \mathpiparallel{\mathistype{x}{\mathprogramtypetuple{\mathvect{\tau}}^{l''_{00}}/U'}}{\mathistype{\mathvect{v}}{\mathvect{\tau}}}}{L}{m''_{11}}{\mathsubstbox{P_{1}}{\mathsubst{\mathvect{y}}{\mathvect{v}}}} 
   \]
   is $l$-securely derivable.
   Let $\hat{\pi}_{0}$ be an $l$-secure derivation tree of
   $\mathtypejudgementwithsecrecy{\Gamma''_{0}, {\mathistype{x}{\mathprogramtypetuple{\mathvect{\tau}}^{l''_{00}}/U}}}{L}{m''_{01}}{P_{0}}$
   and
   $\hat{\pi}_{1}$ be an $l$-secure derivation tree of
   $\mathtypejudgementwithsecrecy{{\Gamma''_{1}}, \mathpiparallel{\mathistype{x}{\mathprogramtypetuple{\mathvect{\tau}}^{l''_{00}}/U'}}{\mathistype{\mathvect{v}}{\mathvect{\tau}}}}{L}{m''_{11}}{\mathsubstbox{P_{1}}{\mathsubst{\mathvect{y}}{\mathvect{v}}}}$.
   We have an $l$-secure derivation tree as follows:
 \begin{center}
  \begin{inlineprooftree}
   \AxiomC{}
   \RightLabel{$\hat{\pi}_{0}$}
   \DeduceC{$\mathtypejudgementwithsecrecy{\Gamma''_{0}, {\mathistype{x}{\mathprogramtypetuple{\mathvect{\tau}}^{l''_{00}}/U}}}{L}{m''_{01}}{P_{0}}$}
   \UnaryInfC{$\mathtypejudgementwithsecrecy{\Gamma''_{0}, {\mathistype{x}{\mathprogramtypetuple{\mathvect{\tau}}^{l''_{00}}/U}}}{L}{m'}{P_{0}}$}

   \AxiomC{}
   \RightLabel{$\hat{\pi}_{1}$}
   \DeduceC{$\mathtypejudgementwithsecrecy{{\Gamma''_{1}}, \mathpiparallel{\mathistype{x}{\mathprogramtypetuple{\mathvect{\tau}}^{l''_{00}}/U'}}{\mathistype{\mathvect{v}}{\mathvect{\tau}}}}{L}{m''_{11}}{\mathsubstbox{P_{1}}{\mathsubst{\mathvect{y}}{\mathvect{v}}}}$}
   \UnaryInfC{$\mathtypejudgementwithsecrecy{{\Gamma''_{1}}, \mathpiparallel{\mathistype{x}{\mathprogramtypetuple{\mathvect{\tau'}}^{l''_{00}}/U'}}{\mathistype{\mathvect{v}}{\mathvect{\tau}}}}{L}{m'}{\mathsubstbox{P_{1}}{\mathsubst{\mathvect{y}}{\mathvect{v}}}}$}

   \BinaryInfC{$\mathtypejudgementwithsecrecy{\mathpiparallel{\mleft(\Gamma''_{0}, {\mathistype{x}{\mathprogramtypetuple{\mathvect{\tau}}^{l''_{00}}/U}}\mright)}{\mleft({\Gamma''_{1}}, \mathpiparallel{\mathistype{x}{\mathprogramtypetuple{\mathvect{\tau}}^{l''_{00}}/U'}}{\mathistype{\mathvect{v}}{\mathvect{\tau}}}\mright)}}{L}{m'}{\mathpiparallel{P_{0}}{\mathsubstbox{P_{1}}{\mathsubst{\mathvect{y}}{\mathvect{v}}}}}$}
  \end{inlineprooftree}.
 \end{center}
 Thus, we see that
   \[
   \mathtypejudgementwithsecrecy{\mathpiparallel{\mleft(\Gamma''_{0}, {\mathistype{x}{\mathprogramtypetuple{\mathvect{\tau}}^{l''_{00}}/U}}\mright)}{\mleft({\Gamma''_{1}}, \mathpiparallel{\mathistype{x}{\mathprogramtypetuple{\mathvect{\tau}}^{l''_{00}}/U'}}{\mathistype{\mathvect{v}}{\mathvect{\tau}}}\mright)}}{L}{m'}{P'}
   \]
   is $l$-securely derivable.
   Then 
   \[
   {
   \mathpiparallel{\mleft( {\mathpiparallel{\Gamma''_{0}}{\Gamma''_{1}}} \mright) }{\mathpiparallel{\mathistype{\mathvect{v}}{\mathvect{\tau}}}{\mathistype{x}{\mathprogramtypetuple{\mathvect{\tau}}^{l''_{00}}/{(\mathpiparallel{U}{U'})}}}}
   }
   \mathissubtypesy
   {\mathpiparallel{\mleft(\Gamma''_{0}, {\mathistype{x}{\mathprogramtypetuple{\mathvect{\tau}}^{l''_{00}}/U}}\mright)}{\mleft({\Gamma''_{1}}, \mathpiparallel{\mathistype{x}{\mathprogramtypetuple{\mathvect{\tau'}}^{l''_{10}}/U'}}{\mathistype{\mathvect{v}}{\mathvect{\tau}}}\mright)}}.
   \]

   By \cref{prop:property-subtyping-on-env} \cref{item:levelraise-prop-property-subtyping-on-env} and \cref{item:levelraiseuni-prop-property-subtyping-on-env},
   we have
   \begin{multline*}
    {\mathpiparallel{\mathpiparallel{\mleft(\mathlevelraise{t_{c}+1}{t_{c}+1}{\mathpiparallel{\Gamma''_{0}}{\mathistype{\mathvect{v}}{\mathlevelraiseuni{\mathvect{\tau}}}}}\mright)}{\mleft(\mathlevelraise{t'_{c}+1}{t'_{c}+1}{\Gamma''_{1}} \mright)}}{\mathistype{x}{\mathprogramtypetuple{\mathvect{\tau}}^{l''_{00}}/(\mathpiparallel{U}{U'})}}} \\
    \mathissubtypesy
    {
    \mathpiparallel{\mleft( {\mathpiparallel{\mleft(\Gamma''_{0}\mright)}{\mleft({\Gamma''_{1}} \mright)}} \mright) }{\mathpiparallel{\mathistype{\mathvect{v}}{\mathvect{\tau}}}{\mathistype{x}{\mathprogramtypetuple{\mathvect{\tau}}^{l''_{00}}/{(\mathpiparallel{U}{U'})}}}}
    }. 
   \end{multline*}
   Hence,
   \[
   \mathtypejudgementwithsecrecy{\mathpiparallel{\mathpiparallel{\mleft(\mathlevelraise{t_{c}+1}{t_{c}+1}{\mathpiparallel{\Gamma''_{0}}{\mathistype{\mathvect{v}}{\mathlevelraiseuni{\mathvect{\tau}}}}}\mright)}{\mleft(\mathlevelraise{t'_{c}+1}{t'_{c}+1}{\Gamma''_{1}} \mright)}}{\mathistype{x}{\mathprogramtypetuple{\mathvect{\tau}}^{l''_{00}}/(\mathpiparallel{U}{U'})}}}{L}{m'}{P'}
   \]
   is $l$-securely derivable.
   
   Now, 
   \begin{multline*}
    {\mathpiparallel{\mathpiparallel{\mleft(\mathlevelraise{t_{c}+1}{t_{c}+1}{\mathpiparallel{\Gamma''_{0}}{\mathistype{\mathvect{v}}{\mathlevelraiseuni{\mathvect{\tau}}}}}\mright)}{\mleft(\mathlevelraise{t'_{c}+1}{t'_{c}+1}{\Gamma''_{1}} \mright)}}{\mathistype{x}{\mathprogramtypetuple{\mathvect{\tau}}^{l''_{00}}/(\mathpiparallel{O^{0}_{t_{c}}U}{I^{0}_{t'_{c}}U'})}}} \\
   \mathtypeenvreduction
   {\mathpiparallel{\mathpiparallel{\mleft(\mathlevelraise{t_{c}+1}{t_{c}+1}{\mathpiparallel{\Gamma''_{0}}{\mathistype{\mathvect{v}}{\mathlevelraiseuni{\mathvect{\tau}}}}}\mright)}{\mleft(\mathlevelraise{t'_{c}+1}{t'_{c}+1}{\Gamma''_{1}} \mright)}}{\mathistype{x}{\mathprogramtypetuple{\mathvect{\tau}}^{l''_{00}}/(\mathpiparallel{U}{U'})}}}.    
   \end{multline*}
   By \cref{lemma:reduction-and-subtyping},
   there exists $\Gamma'$ such that
   ${\Gamma}\mathtypeenvreduction{\Gamma'}$ and 
   \[
   \mathissubtype{\Gamma'}{\mathpiparallel{\mathpiparallel{\mleft(\mathlevelraise{t_{c}+1}{t_{c}+1}{\mathpiparallel{\Gamma''_{0}}{\mathistype{\mathvect{v}}{\mathlevelraiseuni{\mathvect{\tau}}}}}\mright)}{\mleft(\mathlevelraise{t'_{c}+1}{t'_{c}+1}{\Gamma''_{1}} \mright)}}{\mathistype{x}{\mathprogramtypetuple{\mathvect{\tau}}^{l''_{00}}/(\mathpiparallel{U}{U'})}}}. 
   \]
   Hence, $\mathtypejudgementwithsecrecy{\Gamma'}{L}{m}{P'}$ is 
   $l$-securely derivable.

   \noindent Case 2. We consider the case \rulename{R-NewLev}.
   In this case,
   ${P}={\mathpinewsecrecylevel{l_{0}}{\mathvect{l_{1}}}{\mathvect{l_{2}}}{P_{0}}}$,
   ${P'}={P_{0}}$ and
   ${L'}={\mathpinewsecrecylevel{l_{0}}{\mathvect{l_{1}}}{\mathvect{l_{2}}}{L}}$,
   where ${\mathvect{l_{1}}}$, ${\mathvect{l_{2}}}\subseteq{L}$ and 
   ${\mathpinewsecrecylevel{l_{0}}{\mathvect{l_{1}}}{\mathvect{l_{2}}}{L}}$ is defined.

   By \cref{lemma:inversion} \cref{item:newlevel-lemma-inversion},
   there exist a type environments $\Gamma'$ 
   and ${m'}\in{L}$ such that
   $m\leq_{L}m'$, and $\mathissubtype{\Gamma}{\Gamma'}$,
   $m' \leq_{L} l''$ for any $l''\in\mathvect{l_{1}}, \mathvect{l_{2}}$, and
   $\mathtypejudgementwithsecrecy{\Gamma'}{\mathpinewsecrecylevel{l_{0}}{\mathvect{l_{1}}}{\mathvect{l_{2}}}{L}}{m'}{P_{0}}$ 
   is $l$-securely derivable.

   Let $\pi$ be an $l$-secure derivation tree of
   $\mathtypejudgementwithsecrecy{\Gamma'}{\mathpinewsecrecylevel{l_{0}}{\mathvect{l_{1}}}{\mathvect{l_{2}}}{L}}{m'}{P_{0}}$.
   We have an $l$-secure derivation tree as follows:
   \begin{center}
    \begin{inlineprooftree}
     \AxiomC{}
     \RightLabel{$\pi$}
     \DeduceC{$\mathtypejudgementwithsecrecy{\Gamma'}{\mathpinewsecrecylevel{l_{0}}{\mathvect{l_{1}}}{\mathvect{l_{2}}}{L}}{m'}{P_{0}}$}
     \RightLabel{\rulename{T-Weak}}
     \UnaryInfC{$\mathtypejudgementwithsecrecy{\Gamma}{\mathpinewsecrecylevel{l_{0}}{\mathvect{l_{1}}}{\mathvect{l_{2}}}{L}}{m}{P_{0}}$}
    \end{inlineprooftree}.
   \end{center}
   Hence, we have the statement.

   \noindent Case 3. We consider the case \rulename{R-Par}.
   In this case,
   ${P}={\mathpiparallel{P_{0}}{P_{1}}}$ and
   ${P'}={\mathpiparallel{P'_{0}}{P_{1}}}$ with
   ${\mathtuple{{P_{0}}, L}}\mathpireduction{\mathtuple{{P'_{0}}, L'}}$.
   By \cref{lemma:inversion} \cref{item:parallel-lemma-inversion},
   there exist two type environments $\Gamma'_{0}$, $\Gamma'_{1}$,
   and ${m'}\in{L}$
   such that $\mathissubtype{\Gamma}{\mathpiparallel{\Gamma'_{0}}{\Gamma'_{1}}}$, 
   and $m\leq_{L}m'$ and
   $\mathtypejudgementwithsecrecy{\Gamma'_{i}}{L}{m'}{P_{i}}$ is 
   $l$-securely derivable for each $i=0, 1$.
   Since 
   $m\leq_{L}m'$ and
   $\mathtypejudgementwithsecrecy{\Gamma'_{i}}{L}{m'}{P_{i}}$ is
   $l$-securely derivable,
   we see that
   $\mathtypejudgementwithsecrecy{\Gamma'_{i}}{L}{m}{P_{i}}$ is 
   $l$-securely derivable for each $i=0, 1$.
   By the induction hypothesis,
   then there exists a type environment $\Gamma'_{0}$ such that 
   either ${\Gamma'_{0}}={\Gamma'_{0}}$ or
   ${\Gamma'_{0}}\mathtypeenvreduction{\Gamma'_{0}}$ and
   $\mathtypejudgementwithsecrecy{\Gamma'_{0}}{L'}{m}{P'_{0}}$ is 
   $l$-securely derivable.
   By \cref{prop:reduction-basic},
   either $L'=L$ or 
   $L'=\mathpinewsecrecylevel{l}{\mathvect{l_{0}}}{\mathvect{l_{1}}}{L}$.
   By \cref{lemma:weakening-lattice},
   $\mathtypejudgementwithsecrecy{\Gamma'_{1}}{L'}{m}{P'_{1}}$ is 
   $l$-securely derivable.

   Let $\pi_{i}$ be a derivation tree of 
   $\mathtypejudgementwithsecrecy{\Gamma'_{i}}{L}{m}{P_{i}}$ for each $i=0, 1$.
   Then, we have a derivation tree as follows:
   \begin{center}
    \begin{inlineprooftree}
     \AxiomC{}
     \RightLabel{$\pi_{0}$}
     \DeduceC{$\mathtypejudgementwithsecrecy{\Gamma'_{0}}{L}{m}{P_{0}}$}

     \AxiomC{}
     \RightLabel{$\pi_{1}$}
     \DeduceC{$\mathtypejudgementwithsecrecy{\Gamma'_{1}}{L}{m}{P_{1}}$}
     \RightLabel{\rulename{T-Weak}}
     \BinaryInfC{$\mathtypejudgementwithsecrecy{{\mathpiparallel{\Gamma'_{0}}{\Gamma'_{1}}}}{L}{m}{\mathpiparallel{P_{0}}{P_{1}}}$}
    \end{inlineprooftree}.
   \end{center}

   Let $\hat{\pi}_{0}$ be an $l$-secure derivation tree of 
   $\mathtypejudgementwithsecrecy{\Gamma'_{0}}{L'}{m}{P'_{0}}$
   and $\hat{\pi}_{1}$ be an $\check{l}$-secure derivation tree of 
   $\mathtypejudgementwithsecrecy{\Gamma'_{1}}{L'}{m}{P'_{1}}$.
   \begin{center}
    \begin{inlineprooftree}
     \AxiomC{}
     \RightLabel{$\hat{\pi}_{0}$}
     \DeduceC{$\mathtypejudgementwithsecrecy{\Gamma'_{0}}{L'}{m}{P_{0}}$}

     \AxiomC{}
     \RightLabel{$\hat{\pi}_{1}$}
     \DeduceC{$\mathtypejudgementwithsecrecy{\Gamma'_{1}}{L'}{m}{P_{1}}$}
     \RightLabel{\rulename{T-Weak}}
     \BinaryInfC{$\mathtypejudgementwithsecrecy{\mathpiparallel{\Gamma'_{0}}{\Gamma'_{1}}}{L'}{m}{\mathpiparallel{P_{0}}{P_{1}}}$}
    \end{inlineprooftree}.
   \end{center}

   If ${\Gamma'_{0}}={\Gamma'_{0}}$, then we have
   ${\mathpiparallel{\Gamma'_{0}}{\Gamma'_{1}}}={\mathpiparallel{\Gamma'_{0}}{\Gamma'_{1}}}$. 
   Since $\mathissubtype{\Gamma}{\mathpiparallel{\Gamma'_{0}}{\Gamma'_{1}}}$,
   we see that
   $\mathtypejudgementwithsecrecy{\Gamma}{L'}{m}{\mathpiparallel{P_{0}}{P_{1}}}$
   is $l$-securely derivable.

   Assume ${\Gamma'_{0}}\mathtypeenvreduction{\Gamma'_{0}}$.
   By \cref{lemma:reduction-and-subtyping-parallel},
   we have 
   ${\mathpiparallel{\Gamma'_{0}}{\Gamma'_{1}}}\mathtypeenvreduction{\mathpiparallel{\Gamma'_{0}}{\Gamma'_{1}}}$.
   By \cref{lemma:reduction-and-subtyping},
   there exists $\Gamma'$ such that
   ${\Gamma}\mathtypeenvreduction{\Gamma'}$ and 
   $\mathissubtype{\Gamma'}{\mathpiparallel{\Gamma'_{0}}{\Gamma'_{1}}}$.
   Since 
   $\mathissubtype{\Gamma'}{\mathpiparallel{\Gamma'_{0}}{\Gamma'_{1}}}$,
   we see that
   $\mathtypejudgementwithsecrecy{\Gamma'}{L'}{m}{\mathpiparallel{P_{0}}{P_{1}}}$
   is $l$-securely derivable.

   \noindent Case 4. We consider the case \rulename{R-New}.
   In this case,
   ${P}={\mathtuple{{\mathpinew{\mathistype{x}{\xi}}{P_{0}}}, L}}$ 
   and
   ${P'}={\mathtuple{{\mathpinew{\mathistype{x}{\xi}}{P'_{0}}}, L'}}$, 
   where
   ${\mathtuple{{P_{0}}, L}}\mathpireduction{\mathtuple{{P'_{0}}, L'}}$. 
   By \cref{lemma:inversion} \cref{item:pinew-lemma-inversion},
   there exist a type environments $\Gamma'$, 
   a usage $U$, and ${m'}\in{L'}$ such that
   $m\leq_{L}m'$,
   $\mathreliable{U}$ and $\mathissubtype{\Gamma}{\Gamma'}$, and
   $\mathtypejudgementwithsecrecy{\Gamma', {\mathistype{x}{\xi/U}}}{L}{m'}{P_{0}}$ 
   is $l$-securely derivable.
   By the induction hypothesis,
   there exist a type environment $\Gamma''$ such that
   $\mathtypejudgementwithsecrecy{\Gamma''}{L'}{m'}{P'_{0}}$
   is $l$-securely derivable, where
   either
   ${\Gamma''}={\Gamma', {\mathistype{x}{\xi/U}}}$ or
   ${\Gamma', {\mathistype{x}{\xi/U}}}\mathtypeenvreduction{\Gamma''}$.

   Let $\pi$ be an $l$-secure derivation tree of 
   $\mathtypejudgementwithsecrecy{\Gamma''}{L'}{m'}{P'_{0}}$.

   If ${\Gamma''}={\Gamma', {\mathistype{x}{\xi/U}}}$,
   then we have an $l$-secure derivation tree as follows:
   \begin{center}
    \begin{inlineprooftree}
     \AxiomC{}
     \RightLabel{$\pi$}
      \DeduceC{$\mathtypejudgementwithsecrecy{\Gamma', {\mathistype{x}{\xi/U}}}{L'}{m'}{P'_{0}}$}
     \RightLabel{\rulename{T-New}}
     \UnaryInfC{$\mathtypejudgementwithsecrecy{\Gamma'}{L'}{m'}{\mathpinew{\mathistype{x}{\xi}}{P'_{0}}}$}
     \RightLabel{\rulename{T-Weak}}
     \UnaryInfC{$\mathtypejudgementwithsecrecy{\Gamma}{L'}{m}{\mathpinew{\mathistype{x}{\xi}}{P'_{0}}}$}
    \end{inlineprooftree}.
   \end{center}

   Assume ${\Gamma', {\mathistype{x}{\xi/U}}}\mathtypeenvreduction{\Gamma''}$.

   Assume ${\mathof{\Gamma'}{y}}\mathusagereduction{\mathof{\Gamma''}{y}}$ 
   with ${y}\in{\mathof{\mathdom}{\Gamma'}}$.
   Then, there exists a type environment ${\Gamma''}_{0}$ such that
   ${\Gamma''}={{\Gamma''}_{0}, \mathistype{x}{\xi/U}}$ and
   ${\Gamma'}\mathtypeenvreduction{{\Gamma''}_{0}}$.
   Since ${\Gamma''}={{{\Gamma''}_{0}}, {\mathistype{x}{\xi/U}}}$,
   we have an $l$-secure derivation tree as follows:
   \begin{center}
    \begin{inlineprooftree}
     \AxiomC{}
     \RightLabel{$\pi$}
     \DeduceC{$\mathtypejudgementwithsecrecy{{{\Gamma''}_{0}}, {\mathistype{x}{\xi/U}}}{L'}{m'}{P'_{0}}$}
     \RightLabel{\rulename{T-New}}
     \UnaryInfC{$\mathtypejudgementwithsecrecy{{\Gamma''}_{0}}{L'}{m}{\mathpinew{\mathistype{x}{\xi}}{P'_{0}}}$}
    \end{inlineprooftree}.
   \end{center}
   Since ${\Gamma'}\mathtypeenvreduction{{\Gamma''}_{0}}$,
   \cref{lemma:reduction-and-subtyping} implies that
   there exists $\Gamma'_{0}$ such that
   ${\Gamma}\mathtypeenvreduction{\Gamma'}$ and 
   $\mathissubtype{\Gamma'}{{\Gamma''}_{0}}$.
   Hence, 
   $\mathtypejudgementwithsecrecy{\Gamma'}{L'}{m}{\mathpinew{\mathistype{x}{\xi}}{P'_{0}}}$ 
   is $l$-securely derivable.

   Assume ${\xi/U}\mathusagereduction{\xi/U'}$ with some usage $U'$.
   Then ${\Gamma''}={\Gamma', {\mathistype{x}{\xi/U'}}}$.
   Hence, we have an $l$-secure derivation tree as follows:
   \begin{center}
    \begin{inlineprooftree}
     \AxiomC{}
     \RightLabel{$\pi$}
     \DeduceC{$\mathtypejudgementwithsecrecy{\Gamma', {\mathistype{x}{\xi/U'}}}{L'}{m'}{P'_{0}}$}
     \RightLabel{\rulename{T-New}}
     \UnaryInfC{$\mathtypejudgementwithsecrecy{\Gamma'}{L'}{m'}{\mathpinew{\mathistype{x}{\xi}}{P'_{0}}}$}
     \RightLabel{\rulename{T-Weak}}
     \UnaryInfC{$\mathtypejudgementwithsecrecy{\Gamma}{L'}{m}{\mathpinew{\mathistype{x}{\xi}}{P'_{0}}}$}
    \end{inlineprooftree}.
   \end{center}
   Therefore, we have the statement.

   Case 5. We consider the case \rulename{R-SP}.
   In this case,
   ${P}={P_{0}}$ and
   ${P'}={P_{1}}$, where
   ${\mathtuple{P_{0}, L}}\mathpistructuralpo{\mathtuple{P'_{0}, L}}$, ${\mathtuple{P'_{0}, L}}\mathpireduction{\mathtuple{P'_{1}, L'}}$, and ${\mathtuple{P'_{1}, L'}}\mathpistructuralpo{\mathtuple{P_{1}, L'}}$.
   By \cref{lemma:spo-preservation},
   $\mathtypejudgementwithsecrecy{\Gamma}{L}{m}{P'_{0}}$ is 
   $l$-securely derivable.
   By the induction hypothesis,
   there exist a type environment $\Gamma'$ such that
   either
   ${\Gamma'}={\Gamma}$ or
   ${\Gamma}\mathtypeenvreduction{\Gamma'}$ and
   $\mathtypejudgementwithsecrecy{\Gamma'}{L'}{m}{P'_{1}}$ is
   $l$-securely derivable.
   By \cref{lemma:spo-preservation},
   $\mathtypejudgementwithsecrecy{\Gamma'}{L'}{m}{P'}$ is
   $l$-securely derivable.
   Thus, we have the statement.
  \end{proof}

\section{Lemmata for lock-freedom and the details of its proof}

\subsection{Properties of $\mathpisecreduction{\Gamma}{l}$}

  \begin{lemma}
   \label[lemma]{lemma:reduction-under-Gamma-and-Delta-domain}
   For type environments $\Gamma$ and $\Gamma'$, a lattice of secrecy levels $L$,
   a secrecy level ${l}\in{L}$ and a process $P$,
   if ${\mathof{\Gamma}{x}}\mathequivexceptusagessy{\mathof{\Gamma'}{x}}$
   for any value $x$ belonging to the domain of $\Gamma$, 
   the domain of $\Gamma$ is a subset of the domain of $\Gamma'$, and
   $\mathtuple{P, L}\mathpisecreduction{\Gamma}{l}\mathtuple{P', L'}$,
   then $\mathtuple{P, L}\mathpisecreduction{\Gamma'}{l}\mathtuple{P', L'}$.
  \end{lemma}

  \begin{proof}
   Straightforward.
  \end{proof}

  \begin{lemma}
   \label[lemma]{lemma:reduction-under-Gamma-and-Delta-P}
   For type environments $\Gamma$ and $\Gamma'$, a lattice of secrecy levels $L$,
   a secrecy level ${l}\in{L}$ and a process $P$,
   if ${\mathof{\Gamma}{x}}\mathequivexceptusagessy{\mathof{\Gamma'}{x}}$
   for any value $x$ occurring in $P$, and
   $\mathtuple{P, L}\mathpisecreduction{\Gamma}{l}\mathtuple{P', L'}$,
   then $\mathtuple{P, L}\mathpisecreduction{\Gamma'}{l}\mathtuple{P', L'}$.
  \end{lemma}

  \begin{proof}
   Straightforward.
  \end{proof}

  \subsection{Proof of \cref{lemma:lock-freedom}} 
  \label{sec:proof-of-lock-freedom}

  We show \cref{lemma:lock-freedom}.

   Assume that $\mathtypejudgementwithsecrecy{\Gamma}{L}{l}{P}$ and
   $\mathtypejudgementwithsecrecy{\Delta}{L}{l}{Q}$ 
   are $\lk$-securely derivable,
   $\mathpiparallel{\Gamma}{\Delta}$ are reliable,
   $\mathtypeenvandsecrecylatice{\mathpiparallel{\Gamma}{\Delta}}{L}$ is $\lk$-secure, 
   and $\mathobligation{\alpha}{\mathof{\Gamma}{x}}$ is finite,
   where either $\alpha={I}$ or $\alpha={O}$.

   Let $n$ be $\mathobligation{\alpha}{\mathof{\Gamma}{x}}$, and
   $l_{P}$ be the length of $P$.
   By induction on $\mathtuple{n, l_{P}}$,
   we prove that there exists $R$
   such that $\mathtuple{\mathpiparallel{P}{Q}, L}\mathpisecreductionkc{\mathpiparallel{\Gamma}{\Delta}}{\lk}{\mathtuple{R, \hat{L}}}$ 
   and ${x}\in{\maththestrongbarbsof{\alpha}{R}}$.
   Let $\mathof{\Gamma}{x}={\xi/{U_{x}}}$ and $\mathof{\Delta}{x}={\xi/{U'_{x}}}$. 

   Assume $\mathsuitable{\mathcoaction{\alpha}}{U_{x}}$ does not hold.
   Then 
   ${\mathobligation{\alpha}{U_{x}}}>{\mathcapability{\mathcoaction{\alpha}}{U_{x}}}$.
   Because $\mathpiparallel{\Gamma}{\Delta}$ is reliable, we have
   ${\mathobligation{\alpha}{\mathpiparallel{U_{x}}{U'_{x}}}}\leq{\mathcapability{\mathcoaction{\alpha}}{\mathpiparallel{U_{x}}{U'_{x}}}}$.
   Hence, 
   $\mathof{\min}{\mathobligationsub{\alpha}{\emptyset}{U_{x}}, \mathobligationsub{\alpha}{\emptyset}{U'_{x}}}\leq\mathof{\min}{\mathcapability{\mathcoaction{\alpha}}{U_{x}}, \mathcapability{\mathcoaction{\alpha}}{U'_{x}}}$.
   Then
   ${\mathobligation{\alpha}{U_{x}}}>{\mathcapability{\mathcoaction{\alpha}}{U_{x}}}\geq\mathof{\min}{\mathcapability{\mathcoaction{\alpha}}{U_{x}}, \mathcapability{\mathcoaction{\alpha}}{U'_{x}}}\geq\mathof{\min}{\mathobligationsub{\alpha}{\emptyset}{U_{x}}, \mathobligationsub{\alpha}{\emptyset}{U'_{x}}}$.
   Hence, ${n}>{\mathobligation{\alpha}{U'_{x}}}$.
   By the induction hypothesis,
   there exists $R$ such that 
   $\mathtuple{\mathpiparallel{Q}{P}, L}\mathpisecreductionkc{\mathpiparallel{\Delta}{\Gamma}}{\lk}{\mathtuple{R, \hat{L}}}$ 
   and ${x}\in{\maththestrongbarbsof{\alpha}{R}}$.
   Since $\mathpiparallel{P}{Q} \mathpistructuralpo \mathpiparallel{Q}{P}$,
   we have 
   $\mathtuple{\mathpiparallel{P}{Q}, L}\mathpisecreductionkc{\mathpiparallel{\Delta}{\Gamma}}{\lk}{\mathtuple{R, \lk}}$.
   By \cref{lemma:reduction-under-Gamma-and-Delta-domain},
   $\mathtuple{\mathpiparallel{P}{Q}, L}\mathpisecreductionkc{\mathpiparallel{\Gamma}{\Delta}}{\lk}{\mathtuple{R, \hat{L}}}$.

   Assume $\mathsuitable{\mathcoaction{\alpha}}{U_{x}}$.
   We consider cases according to the form of $P$.

   Case 1. Since $\mathobligation{\alpha}{\mathof{\Gamma}{x}}$ is finite,
   we have $P\not=\mathnil$.

   Case 2. ${P}={\mleft(\mathpiparallel{P_{0}}{P_{1}}\mright)}$.
   By \cref{lemma:inversion} \cref{item:parallel-lemma-inversion},
   there exist two type environments $\Gamma'_{0}$, $\Gamma'_{1}$,
   and ${l'}\in{L'}$
   such that $\mathissubtype{\Gamma}{\mathpiparallel{\Gamma'_{0}}{\Gamma'_{1}}}$
   and $l\leq_{L}l'$, and
   $\mathtypejudgementwithsecrecy{\Gamma'_{i}}{L}{l'}{P_{i}}$ is 
   $\lk$-securely derivable for each $i=0, 1$.
   By \cref{prop:property-subtyping-on-env} \cref{item:cong-full-prop-property-subtyping-on-env},
   $\mathissubtype{\mathpiparallel{\Gamma}{\Delta}}{\mathpiparallel{\mathpiparallel{\Gamma'_{0}}{\Gamma'_{1}}}{\Delta}}$.
   By \cref{prop:property-subtyping-on-env-rel},
   $\mathpiparallel{\mathpiparallel{\Gamma'_{0}}{\Gamma'_{1}}}{\Delta}$ is reliable.
   Because $\mathissubtype{\Gamma}{\mathpiparallel{\Gamma'_{0}}{\Gamma'_{1}}}$ and
   $\mathobligation{\alpha}{\mathof{\Gamma}{x}}$ is finite,
   we have $x\in\mathof{\mathdom}{\mathpiparallel{\Gamma'_{0}}{\Gamma'_{1}}}$.
   By \cref{def:subusage} \cref{item:obligation-in-def:subusage},
   we have 
   $n\geq\mathobligation{\alpha}{\mathof{\mathpiparallel{\Gamma'_{0}}{\Gamma'_{1}}}{x}}$.
   Hence, either $n\geq\mathobligation{\alpha}{\mathof{\Gamma'_{0}}{x}}$
   or $n\geq\mathobligation{\alpha}{\mathof{\Gamma'_{1}}{x}}$.
   
   Assume $n\geq\mathobligation{\alpha}{\mathof{\Gamma'_{0}}{x}}$.
   The length of $P_{0}$ is less then $l_{P}$.
   By the induction hypothesis, we see that
   there exist $R_{0}$ and $\hat{L}_{0}$
   such that $\mathtuple{\mathpiparallel{P_{0}}{Q}, L}\mathpisecreductionkc{\mathpiparallel{\Gamma'_{0}}{\Delta}}{\lk}{\mathtuple{R_{0}, \hat{L}_{0}}}$ 
   and ${x}\in{\maththestrongbarbsof{\alpha}{R_{0}}}$.
   By \cref{lemma:reduction-under-Gamma-and-Delta-domain},
   $\mathtuple{\mathpiparallel{P_{0}}{Q}, L}\mathpisecreductionkc{\mathpiparallel{\Gamma}{\Delta}}{\lk}{\mathtuple{R_{0}, \hat{L}_{0}}}$.
   Then
   $\mathtuple{\mathpiparallel{\mathpiparallel{P_{0}}{Q}}{P_{1}}, L}\mathpisecreductionkc{\mathpiparallel{\Gamma}{\Delta}}{\lk}{\mathtuple{\mathpiparallel{R_{0}}{P_{1}}, \hat{L}_{0}}}$.
   Since 
   $\mathpiparallel{\mathpiparallel{P_{0}}{P_{1}}}{Q}\mathpistructuralpo\mathpiparallel{\mathpiparallel{P_{0}}{Q}}{P_{1}}$,
   we have
   $\mathtuple{\mathpiparallel{P}{Q}, L}\mathpisecreductionkc{\mathpiparallel{\Gamma}{\Delta}}{\lk}{\mathtuple{\mathpiparallel{R_{0}}{P_{1}}, \hat{L}_{0}}}$.
   By ${x}\in{\maththestrongbarbsof{\alpha}{R_{0}}}$, 
   we see ${x}\in{\maththestrongbarbsof{\alpha}{\mathpiparallel{R_{0}}{P_{1}}}}$.

   In the similar way to the case $n\geq\mathobligation{\alpha}{\mathof{\Gamma'_{0}}{x}}$,
   under the assumption that $n\geq\mathobligation{\alpha}{\mathof{\Gamma'_{1}}{x}}$,
   we can show that there exists $R$
   such that $\mathtuple{\mathpiparallel{P}{Q}, L}\mathpisecreductionkc{\mathpiparallel{\Gamma}{\Delta}}{\lk}{\mathtuple{R, \hat{L}}}$ 
   and ${x}\in{\maththestrongbarbsof{\alpha}{R}}$.

   Case 3. $P={\mathpioutput{y}{\mathvect{v}}.P_{0}}$.
   
   Assume $y= x$.
   Let ${R}=\mathpiparallel{P}{Q}$ and $\hat{L}=L$.
   Then $\mathtuple{\mathpiparallel{P}{Q}, L}\mathpisecreductionkc{\mathpiparallel{\Gamma}{\Delta}}{\lk}{\mathtuple{R, \hat{L}}}$ 
   and ${x}\in{\maththestrongbarbsof{\alpha}{R}}$.

   Assume $y\not= x$.
   By \cref{lemma:inversion} \cref{item:output-lemma-inversion},
   there exist a type environments $\Gamma'$,
   secrecy levels ${l_{0}}, {l_{1}}\in{L}$,
   types ${\mathvect{\tau}}$, a usage $U$ and
   ${t_{c}}\in{\mathnat \cup \mathsetextension{\infty}}$
   such that
   $\mathissubtype{\Gamma}{\Gamma''}$,
   ${l}\leq_{L}{l_{0}}$, and 
   ${l}\leq_{L}{l_{1}}$,
   $\mathtypejudgementwithsecrecy{\Gamma', {\mathistype{y}{\mathprogramtypetuple{\mathvect{\tau}}^{l_{0}}/U_{y}}}}{L}{\lk}{l_{1}}{P_{0}}$
   is $\lk$-securely derivable,
   $\mathtypeenvandsecrecylatice{\Gamma''}{L}$ is $\lk$-secure, and
   $t_{c} = \infty$ implies ${l_{0}}\leq_{L}{l_{1}}$,
   where \linebreak[4]
   ${\Gamma''}= \mathpiparallel{\mathlevelraise{t_{c}+1}{t_{c}+1}{\mathpiparallel{\Gamma'}{\mathistype{\mathvect{v}}{\mathlevelraiseuni{\mathvect{\tau}}}}}}{\mathistype{y}{\mathprogramtypetuple{\mathvect{\tau}}^{l_{0}}/O^{0}_{t_{c}}U_{y}}}$.
   Since
   $\mathissubtype{\Gamma}{\Gamma''}$,
   we have
   ${t_{c}}<{t_{c}+1}\leq\mathobligation{\alpha}{\mathof{\Gamma''}{x}}\leq n$.
   Since $\mathtypeenvandsecrecylatice{\Gamma''}{L}$ is $\lk$-secure,
   we have $l_{0}\not<_{L}\lk$.
   By \cref{prop:property-subtyping-on-env} \cref{item:cong-full-prop-property-subtyping-on-env},
   $\mathissubtype{\mathpiparallel{\Gamma}{\Delta}}{\mathpiparallel{\Gamma''}{\Delta}}$.
   By \cref{prop:property-subtyping-on-env-rel},
   $\mathpiparallel{\Gamma''}{\Delta}$ is reliable.
   Then, we see that
   $\mathpiparallel{\mathprogramtypetuple{\mathvect{\tau}}^{l_{0}}/O^{0}_{t_{c}}U_{y}}{\mathof{\Delta}{y}}$ is reliable.
   Let $\mathof{\Delta}{y}=\mathprogramtypetuple{\mathvect{\tau}}^{l_{0}}/{U'_{y}}$.
   Then, we have 
   $\mathobligation{I}{\mathpiparallel{O^{0}_{t_{c}}U_{y}}{U'_{y}}}\leq\mathcapability{O}{\mathpiparallel{O^{0}_{t_{c}}U_{y}}{U'_{y}}}$.
   Then 
   $\mathobligation{I}{U'_{y}}\leq\mathcapability{O}{\mathpiparallel{O^{0}_{t_{c}}U_{y}}{U'_{y}}}\leq{t_{c}}<n$.
   By the induction hypothesis,
   there exist $R_{0}$ and $\hat{L}_{0}$
   such that $\mathtuple{\mathpiparallel{Q}{\mathnil}, L}\mathpisecreductionkc{\mathpiparallel{\Delta}{\emptyset}}{\lk}{\mathtuple{R_{0}, \hat{L}_{0}}}$ 
   and ${y}\in{\maththestrongbarbsof{I}{R_{0}}}$.
   Hence, 
   $\mathtuple{Q, L}\mathpisecreductionkc{\mathpiparallel{\Delta}{\emptyset}}{\lk}{\mathtuple{\mathpinew{\mathvect{w}}{\mathpiparallel{\mathpiinput{y}{\mathvect{z}}. Q_{0}}{Q_{1}}}, \hat{L}_{0}}}$
   and ${{y}\notin{\mathvect{w}}}$ for some $Q_{0}$ and $Q_{1}$.
   By \cref{prop:subject-reduction},
   there exists a type environment $\hat{\Delta}$ such that 
   ${\Delta}\mathtypeenvreductionkc{\hat{\Delta}}$ and
   $\mathtypejudgementwithsecrecy{\hat{\Delta}}{\hat{L}_{0}}{l}{\mathpinew{\mathvect{w}}{\mathpiparallel{\mathpiinput{y}{\mathvect{z}}. Q_{0}}{Q_{1}}}}$
   is $\lk$-securely derivable.
   By \cref{lemma:inversion}
   there exist type environments $\Delta_{0}$ and $\Delta_{1}$
   and $l_{0}, l_{1}\in{\hat{L}_{0}}$
   such that
   $\mathtypejudgementwithsecrecy{\Delta_{0}, {\mathistype{y}{\mathprogramtypetuple{\mathvect{\tau}}^{l_{0}}/V}}, {\mathistype{\mathvect{z}}{\mathvect{\tau}}}}{\hat{L}_{0}}{\lk}{l_{1}}{Q_{0}}$
   and
   $\mathtypejudgementwithsecrecy{\Delta_{1}}{\hat{L}_{0}}{l'}{Q_{1}}$ 
   are $\lk$-securely derivable, and
   $\mathissubtype{\hat{\Delta}, \mathistype{\mathvect{w}}{\mathvect{\tau'}}}{\mathpiparallel{\mleft(\mathlevelraise{t'_{c}+1}{t'_{c}+1}{\Delta_{0}}, \mathistype{y}{\mathprogramtypetuple{\mathvect{\tau}}^{l_{0}}/I^{0}_{t'_{c}}V}\mright)}{\Delta_{1}}}$,
   ${l}\leq_{L}{l_{0}}$, ${l}\leq_{L}{l_{1}}$ and ${l}\leq_{L}{l'}$.
   By \cref{lemma:substitution-lemma},
   $\mathtypejudgementwithsecrecy{\mathsubstbox{\mleft(\Delta_{0}, {\mathistype{y}{\mathprogramtypetuple{\mathvect{\tau}}^{l_{0}}/V}}, {\mathistype{\mathvect{z}}{\mathvect{\tau}}}\mright)}{\mathsubst{\mathvect{z}}{\mathvect{v}}}}{\hat{L}_{0}}{l_{1}}{\mathsubstbox{Q_{0}}{\mathsubst{\mathvect{z}}{\mathvect{v}}}}$
   is $\lk$-securely derivable.
   By \cref{prop:reduction-basic} and \cref{lemma:weakening-lattice},
   $\mathtypejudgementwithsecrecy{\Gamma', {\mathistype{y}{\mathprogramtypetuple{\mathvect{\tau}}^{l_{0}}/U_{y}}}}{\hat{L}_{0}}{l_{1}}{P_{0}}$
   is $\lk$-securely derivable.   
   Then, we see that
   \[
    \mathtypejudgementwithsecrecy{\mathsubstbox{\mleft(\mathpiparallel{\mathpiparallel{\Gamma', {\mathistype{y}{\mathprogramtypetuple{\mathvect{\tau}}^{l_{0}}/U_{y}}}}{\mleft(\Delta_{0}, {\mathistype{y}{\mathprogramtypetuple{\mathvect{\tau}}^{l_{0}}/V}}, {\mathistype{\mathvect{z}}{\mathvect{\tau}}}\mright)}}{\Delta_{1}}\mright)}{\mathsubst{\mathvect{z}}{\mathvect{v}}}}{\hat{L}_{0}}{l_{1}}{\mathsubstbox{\mleft(\mathpiparallel{P_{0}}{\mathpiparallel{Q_{0}}{Q_{1}}}\mright)}{\mathsubst{\mathvect{z}}{\mathvect{v}}}}
   \]
   is $\lk$-securely derivable.
   By \cref{prop:property-subtyping-on-env-rel},
   $\mathpiparallel{\mathpiparallel{\Gamma', {\mathistype{y}{\mathprogramtypetuple{\mathvect{\tau}}^{l_{0}}/U_{y}}}}{\mathsubstbox{\mleft(\Delta_{0}, {\mathistype{y}{\mathprogramtypetuple{\mathvect{\tau}}^{l_{0}}/V}}, {\mathistype{\mathvect{z}}{\mathvect{\tau}}}\mright)}{\mathsubst{\mathvect{z}}{\mathvect{v}}}}}{\Delta_{1}}$
   is reliable.

   Because
   $\mathissubtype{\Gamma}{\Gamma''}$ and $\mathsuitable{\mathcoaction{\alpha}}{U_{x}}$,
   we have
   \[
   \mathobligation{\alpha}{\mathof{\Gamma', {\mathistype{y}{\mathprogramtypetuple{\mathvect{\tau}}^{l_{0}}/U_{y}}}}{x}}\leq\mathobligation{\alpha}{\mathof{\mleft(\mathpiparallel{\mathlevelraise{t_{c}+1}{t_{c}+1}{\mathpiparallel{\Gamma'}{\mathistype{\mathvect{v}}{\mathlevelraiseuni{\mathvect{\tau}}}}}}{\mathistype{y}{\mathprogramtypetuple{\mathvect{\tau}}^{l_{0}}/O^{0}_{t_{c}}U_{y}}}\mright)}{x}}\leq\mathobligation{\alpha}{\mathof{\Gamma}{x}}=n. 
   \]   
   Since the length of $P_{0}$ is less then $l_{P}$,
   we see that there exist $R$ and $\lk$
   such that \linebreak[3] 
   \[
   \mathtuple{\mathpiparallel{P_{0}}{\mathpiparallel{\mathsubstbox{Q_{0}}{\mathsubst{\mathvect{z}}{\mathvect{v}}}}{Q_{1}}}, L}\mathpisecreductionkc{\mathpiparallel{\mathpiparallel{\Gamma', {\mathistype{y}{\mathprogramtypetuple{\mathvect{\tau}}^{l_{0}}/U_{y}}}}{\mathsubstbox{\mleft(\Delta_{0}, {\mathistype{y}{\mathprogramtypetuple{\mathvect{\tau}}^{l_{0}}/V}}, {\mathistype{\mathvect{z}}{\mathvect{\tau}}}\mright)}{\mathsubst{\mathvect{z}}{\mathvect{v}}}}}{\Delta_{1}}}{\mathtuple{R, \hat{L}}}
   \]
   and ${x}\in{\maththestrongbarbsof{\alpha}{R}}$.
   Then, we have the claimed result.
   
   Case 4. $P={\mathpiinput{y}{\mathvect{z}}.P_{0}}$.
   Straightforward.

   Case 5. $P={\mathpireplication{P_{0}}}$.
   Straightforward.

   Case 6. $P={\mathpinew{\mathistype{y}{\xi}}{P_{0}}}$.
   Straightforward.

   Case 7. $P={\mathpinewsecrecylevel{l_{0}}{\mathvect{l_{1}}}{\mathvect{l_{2}}}{P_{0}}}$.
   By \cref{lemma:inversion} \cref{item:newlevel-lemma-inversion},
   $l'' \not<_{L} \lk$ for some $l''\in\mathvect{l_{1}}, \mathvect{l_{2}}$  and
   there exist a type environments $\Gamma'$, 
   and ${l'}\in{L}$ such that
   $l\leq_{L}l'$ and $\mathissubtype{\Gamma}{\Gamma'}$,
   $l' \leq_{L} l''$ for any $l''\in\mathvect{l_{1}}, \mathvect{l_{2}}$, and
   $\mathtypejudgementwithsecrecy{\Gamma'}{\mathpinewsecrecylevel{l_{0}}{\mathvect{l_{1}}}{\mathvect{l_{2}}}{L}}{l'}{P_{0}}$ 
   is $\lk$-securely derivable.
   By \cref{lemma:weakening-lattice},
   $\mathtypejudgementwithsecrecy{\Delta}{\mathpinewsecrecylevel{l_{0}}{\mathvect{l_{1}}}{\mathvect{l_{2}}}{L}}{l}{Q}$ 
   is $\lk$-securely derivable.
   Because
   $\mathissubtype{\Gamma}{\Gamma'}$ and $\mathsuitable{\mathcoaction{\alpha}}{U_{x}}$,
   we have $\mathobligation{\alpha}{\mathof{\Gamma'}{x}}\leq\mathobligation{\alpha}{\mathof{\Gamma}{x}}$.
   Since the length of $P_{0}$ is less then $l_{P}$,
   we see that there exist $R$ and $\lk$
   such that $\mathtuple{\mathpiparallel{P_{0}}{Q}, \mathpinewsecrecylevel{l_{0}}{\mathvect{l_{1}}}{\mathvect{l_{2}}}{L}}\mathpisecreductionkc{\mathpiparallel{\Gamma'}{\Delta}}{\lk}{\mathtuple{R, \hat{L}}}$ 
   and ${x}\in{\maththestrongbarbsof{\alpha}{R}}$.
   By \cref{lemma:reduction-under-Gamma-and-Delta-domain},
   $\mathtuple{\mathpiparallel{P_{0}}{Q}, \mathpinewsecrecylevel{l_{0}}{\mathvect{l_{1}}}{\mathvect{l_{2}}}{L}}\mathpisecreductionkc{\mathpiparallel{\Gamma}{\Delta}}{\lk}{\mathtuple{R, \hat{L}}}$.
   Hence,
   \[
   \mathtuple{\mathpiparallel{P}{Q}, L}\mathpisecreduction{\mathpiparallel{\Gamma}{\Delta}}{\lk} \mathtuple{\mathpiparallel{P_{0}}{Q}, \mathpinewsecrecylevel{l_{0}}{\mathvect{l_{1}}}{\mathvect{l_{2}}}{L}}\mathpisecreductionkc{\mathpiparallel{\Gamma}{\Delta}}{\lk}{\mathtuple{R, \hat{L}}}. 
   \]

   Case 8. $P={\mathobif{v}{P_{0}}{P_{1}}}$.
   Straightforward. \qed

\section{Proof of non-interference theorems}
 \label[appendix]{section:app-proof-of-non-interference}

 \subsection{Basic properties for bisimulation}

\begin{lemma}
 \label[lemma]{lemma:structualeq-barb-equiv}
 For processes $P$, $P'$ and a lattice of secrecy levels $L$,
 if ${P}\mathpistructuraleq{P'}$,
 then ${\maththebarbsof{P, L}}={\maththebarbsof{P', L}}$.
\end{lemma}

\begin{proof}
 Assume ${P}\mathpistructuralpo{P'}$.
 We show ${\maththebarbsof{P, L}}={\maththebarbsof{P', L}}$.

 We show ${\maththebarbsof{P, L}}\subseteq{\maththebarbsof{P', L}}$.
 Assume ${x}\in{\maththebarbsof{P, L}}$. We show ${x}\in{\maththebarbsof{P', L}}$.

 Assume ${\mathtuple{P, L}\mathpireductionkc}\mathtuple{P'', L'}$, 
 ${P''}={\mathpinew{\mathvect{y}}{\mathpiparallel{\mathpioutput{x}{\mathvect{v}}. P_{0}}{P_{1}}}}$ and ${x}\notin{\mathvect{y}}$.
 Since ${P'}\mathpistructuralpo{P}$,
 we have ${\mathtuple{P', L}\mathpireductionkc}\mathtuple{P'', L'}$.
 Hence, ${x}\in{\maththebarbsof{P', L}}$.

 In the similar way to the case ${P''}={\mathpinew{\mathvect{y}}{\mathpiparallel{\mathpioutput{x}{\mathvect{v}}. P_{0}}{P_{1}}}}$,
 we have ${x}\in{\maththebarbsof{P', L}}$ if
 ${\mathtuple{P, L}\mathpireductionkc}\mathtuple{P'', L'}$, 
 ${P''}\mathpistructuralpo{\mathpinew{\mathvect{y}}{\mathpiparallel{\mathpiinput{x}{\mathvect{z}}. P_{0}}{P_{1}}}}$ and ${x}\notin{\mathvect{y}}$.

 In the same way to the case ${\maththebarbsof{P, L}}\subseteq{ \maththebarbsof{P', L}}$,
 we have ${\maththebarbsof{P', L}}\subseteq{\maththebarbsof{P, L}}$.
\end{proof}

  \begin{lemma}
   \label[lemma]{lemma:barbed-bisimilar-transitive}
   $\mathbisimilarsy$ is transitive i.e.\ 
   if $\mathisbisimilar{\mathtuple{P_{0}, L_{0}}}{\mathtuple{P_{1}, L_{1}}}$ and 
   $\mathisbisimilar{\mathtuple{P_{1}, L_{1}}}{\mathtuple{P_{2}, L_{2}}}$,
   then $\mathisbisimilar{\mathtuple{P_{0}, L_{0}}}{\mathtuple{P_{2}, L_{2}}}$.
  \end{lemma}

  \begin{proof}
   Easy.
  \end{proof}

  \begin{lemma}
   \label[lemma]{lemma:barbed-bisimilar-structual-preorder}
   For processes $P$, $P'$, $Q$, $Q'$ and a lattice of secrecy level $L$,
   if $\mathisbisimilar{\mathtuple{P, L}}{\mathtuple{Q, L}}$,
   ${P'}\mathpistructuraleq{P}$, and ${Q}\mathpistructuraleq{Q'}$,
   then $\mathisbisimilar{\mathtuple{P', L}}{\mathtuple{Q', L}}$.
  \end{lemma}

  \begin{proof}
   Assume $\mathisbisimilar{\mathtuple{P, L}}{\mathtuple{Q, L}}$,
   ${P'}\mathpistructuraleq{P}$, and ${Q}\mathpistructuraleq{Q'}$.
   Since $\mathisbisimilar{\mathtuple{P, L}}{\mathtuple{Q, L}}$,
   there exists a barbed bisimulation ${\mathbarbedbisim{R}}$ such that
   ${\mathtuple{\mathtuple{P, L}, \mathtuple{Q, L}}}\in{\mathbarbedbisim{R}}$.
   Let
   \[
    {\mathbarbedbisim{R'}}={\mathsetintension{\mathtuple{\mathtuple{P', L}, \mathtuple{Q', L'}}}{ {\mathtuple{\mathtuple{P, L}, \mathtuple{Q, L'}}}\in{\mathbarbedbisim{R}}, {P'}\mathpistructuraleq{P}, \text{ and } {Q}\mathpistructuraleq{Q'},}}.
   \]

   We show that $\mathbarbedbisim{R'}$ is a barbed bisimulation.
   Let ${\mathtuple{\mathtuple{P'_{0}, L_{0}}, \mathtuple{P'_{1}, L_{1}}}}\in{\mathbarbedbisim{R'}}$.
   Then, there exist processes $P_{0}$ and $P_{1}$ such that
   ${\mathtuple{\mathtuple{P_{0}, L_{0}}, \mathtuple{P_{1}, L_{1}}}}\in{\mathbarbedbisim{R}}$,
   ${P'_{0}}\mathpistructuraleq{P_{0}}$, and ${P_{1}}\mathpistructuraleq{P'_{1}}$.
   We note ${\mathbarbedbisim{R}}\subseteq{\mathbarbedbisim{R'}}$.

   \noindent \cref{item:left-def-Barbed-bisimulation}
   Assume ${\mathtuple{P'_{0}, L_{0}}}\mathpireduction{\mathtuple{P''_{0}, L'_{0}}}$.
   We show that 
   there exists ${\mathtuple{P''_{1}, L'_{1}}}$ such that
   ${\mathtuple{P'_{1}, L_{1}}}\mathpireductionkc{\mathtuple{P''_{1}, L'_{1}}}$
   and
   ${\mathtuple{\mathtuple{P''_{0}, L'_{0}}, \mathtuple{P''_{1}, L'_{1}}}}\in{\mathbarbedbisim{R'}}$.
   Since ${P'_{0}}\mathpistructuraleq{P_{0}}$,
   we have ${\mathtuple{P_{0}, L_{0}}}\mathpireduction{\mathtuple{P''_{0}, L'_{0}}}$.
   Since 
   ${\mathtuple{\mathtuple{P_{0}, L_{0}}, \mathtuple{P_{1}, L_{1}}}}\in{\mathbarbedbisim{R}}$,
   there exists ${\mathtuple{P''_{1}, L'_{1}}}$ such that
   ${\mathtuple{P_{1}, L_{1}}}\mathpireductionkc{\mathtuple{P''_{1}, L'_{1}}}$
   and
   ${\mathtuple{\mathtuple{P''_{0}, L'_{0}}, \mathtuple{P''_{1}, L'_{1}}}}\in{\mathbarbedbisim{R}}$.
   By ${P_{1}}\mathpistructuraleq{P'_{1}}$, we have
   ${\mathtuple{P'_{1}, L_{1}}}\mathpireductionkc{\mathtuple{P''_{1}, L'_{1}}}$.
   By ${\mathbarbedbisim{R}}\subseteq{\mathbarbedbisim{R'}}$, we see
   ${\mathtuple{\mathtuple{P''_{0}, L'_{0}}, \mathtuple{P''_{1}, L'_{1}}}}\in{\mathbarbedbisim{R'}}$.

   \noindent \cref{item:right-def-Barbed-bisimulation}
   In the same way to \cref{item:left-def-Barbed-bisimulation}.

   \noindent \cref{item:barbs-def-Barbed-bisimulation}
   By \cref{lemma:structualeq-barb-equiv},
   we have ${\maththebarbsof{P'_{0}, L}}={\maththebarbsof{P_{0}, L}}$ and ${\maththebarbsof{P_{1}, L}}={\maththebarbsof{P'_{1}, L}}$.
   Since 
   ${\mathtuple{\mathtuple{P_{0}, L_{0}}, \mathtuple{P_{1}, L_{1}}}}\in{\mathbarbedbisim{R}}$,
   we have ${\maththebarbsof{P_{0}, L}}={\maththebarbsof{P_{1}, L}}$.
   Hence, we have ${\maththebarbsof{P'_{0}, L}}={\maththebarbsof{P'_{1}, L}}$.

   Now, we see that $\mathbarbedbisim{R'}$ is a barbed bisimulation.
   By definition of $\mathbarbedbisim{R'}$,
   we have ${\mathtuple{\mathtuple{P', L}, \mathtuple{Q', L}}}\in{\mathbarbedbisim{R'}}$.
   Thus, $\mathisbisimilar{\mathtuple{P', L}}{\mathtuple{Q', L}}$.
  \end{proof}

 \begin{lemma}
  \label[lemma]{lemma:context-structural-po-and-set-of-free-names}
  \begin{enumerate}
   \item If ${P_{0}}\mathpistructuralpo{P_{1}}$, 
	 then 
	 ${\mathFNof{\mathcontext{C}{P_{0}}}}\supseteq{\mathFNof{\mathcontext{C}{P_{1}}}}$ for any context $C$. 
	 \label{item:po-lemma-context-structural-po-and-set-of-free-names}
   \item If ${P_{0}}\mathpistructuraleq{P_{1}}$, 
	 then 
	 ${\mathFNof{\mathcontext{C}{P_{0}}}}={\mathFNof{\mathcontext{C}{P_{1}}}}$ 
	 for any context $C$. 
	 \label{item:eq-lemma-context-structural-po-and-set-of-free-names}
  \end{enumerate}
 \end{lemma}

 \begin{proof}
  It suffices to show \cref{item:po-lemma-context-structural-po-and-set-of-free-names}.
  We see \cref{item:po-lemma-context-structural-po-and-set-of-free-names} by induction on 
  the construction of ${C}$.
 \end{proof}

  \begin{lemma}
   \label[lemma]{lemma-of-lemma:barbed-bisimilar-structual-preorder-with-context}
   For processes $P_{0}$ and $P_{1}$, 
   if ${P_{0}}\mathpistructuraleq{P_{1}}$ and 
   ${\mathcontext{C}{P_{0}}}\mathpistructuralpo{P'_{0}}$,
   then there exists a context $C'$ such that
   ${P'_{0}}={\mathcontext{C'}{P_{0}}}$ and
   ${\mathcontext{C}{P_{1}}}\mathpistructuralpo{\mathcontext{C'}{P_{1}}}$.
  \end{lemma}

  \begin{proof}
   Assume ${P_{0}}\mathpistructuraleq{P_{1}}$ and
   ${\mathcontext{C}{P_{0}}}\mathpistructuralpo{P'_{0}}$.
   We show that there exists a context $C'$ such that
   ${P'_{0}}={\mathcontext{C'}{P_{0}}}$ and
   ${\mathcontext{C}{P_{1}}}\mathpistructuralpo{\mathcontext{C'}{P_{1}}}$.
   We proceed by induction on the construction of 
   ${\mathcontext{C}{P_{0}}}\mathpistructuralpo{P'_{0}}$.
   We consider cases according to the form of $C$.

   If ${\mathcontext{}{\;}}$ does not occur in $C$, 
   then the required condition holds obviously.

   Assume ${C}={\mathcontext{}{\;}}$. Let ${C'}={\mathcontext{}{\;}}$. 
   Then, the required condition holds.

   Assume ${C}\not={\mathcontext{}{\;}}$.
   We consider cases according to the last rule of the construction of 
   ${\mathcontext{C}{P_{0}}}\mathpistructuralpo{P'_{0}}$.
   
   Case 1. Assume ${P'_{0}}={\mathcontext{C}{P_{0}}}$.
   Let ${C'}={C}$. Then, the required condition holds.

   Case 2. Assume that there exists a process $Q$ such that
   ${\mathcontext{C}{P_{0}}}\mathpistructuralpo{Q}$ and
   ${Q}\mathpistructuralpo{P'_{0}}$.
   By the induction hypothesis, we see that there exists a context $C''$ such that
   ${Q}={\mathcontext{C''}{P_{0}}}$ and
   ${\mathcontext{C}{P_{1}}}\mathpistructuralpo{\mathcontext{C''}{P_{1}}}$.
   Then, we have ${\mathcontext{C''}{P_{0}}}\mathpistructuralpo{P'_{0}}$.
   By the induction hypothesis, we see that there exists a context $C'$ such that
   ${\mathcontext{C''}{P_{0}}}={\mathcontext{C'}{P_{0}}}$ and
   ${\mathcontext{C''}{P_{1}}}\mathpistructuralpo{\mathcontext{C'}{P_{1}}}$.
   Since ${\mathcontext{C}{P_{1}}}\mathpistructuralpo{\mathcontext{C''}{P_{1}}}$ and
   ${\mathcontext{C''}{P_{1}}}\mathpistructuralpo{\mathcontext{C'}{P_{1}}}$,
   we have 
   ${\mathcontext{C}{P_{1}}}\mathpistructuralpo{\mathcontext{C'}{P_{1}}}$.

   Case 3. \rulename{SP-Zero1}.
   Assume ${P'_{0}}={\mathpiparallel{\mathcontext{C}{P_{0}}}{\mathnil}}$.
   Let ${C'}={\mathpiparallel{C}{\mathnil}}$.
   Then ${\mathcontext{C}{P_{1}}}\mathpistructuralpo{\mathcontext{C'}{P_{1}}}$.
   
   Assume ${C}={\mathpiparallel{C_{0}}{\mathcontext{}{\;}}}$ and 
   ${P_{0}}={\mathnil}$ for some context $C_{0}$.
   Then ${P'_{0}}={\mathcontext{C_{0}}{P_{0}}}$.
   Let ${C'}={C_{0}}$.
   Then, we have ${P'_{0}}={\mathcontext{C'}{P_{0}}}$.
   By \cref{lemma:structural-po-extra-rules-cong},
   we have 
   ${\mathcontext{C}{P_{1}}}={\mathpiparallel{\mathcontext{C_{0}}{P_{1}}}{P_{1}}}\mathpistructuralpo{\mathpiparallel{\mathcontext{C_{0}}{P_{1}}}{\mathnil}}$.
   Hence, ${\mathcontext{C}{P_{1}}}\mathpistructuralpo{\mathcontext{C'}{P_{1}}}$.

   Assume ${C}={\mathpiparallel{C_{0}}{\mathnil}}$ for some context $C_{0}$.
   Then ${P'_{0}}={\mathcontext{C_{0}}{P_{0}}}$.
   Let ${C'}={C_{0}}$.
   Then, we have ${P'_{0}}={\mathcontext{C'}{P_{0}}}$.
   By \rulename{SP-Zero1},
   ${\mathcontext{C}{P_{1}}}={\mathpiparallel{\mathcontext{C'}{P_{1}}}{\mathnil}}\mathpistructuralpo{\mathcontext{C'}{P_{1}}}$.

   Case 4. \rulename{SP-Zero2}.
   Assume ${C}={\mathpinew{\mathistype{x}{\xi}}{\mathcontext{}{\;}}}$ and 
   ${P_{0}}={\mathnil}$.
   Then ${P'_{0}}={\mathnil}$.
   Let ${C'}={\mathnil}$.
   Then, we have ${P'_{0}}={\mathcontext{C'}{P_{0}}}$.
   By \rulename{SP-CNew},
   ${\mathcontext{C}{P_{1}}}={\mathpinew{\mathistype{x}{\xi}}{P_{1}}}\mathpistructuralpo{\mathpinew{\mathistype{x}{\xi}}{\mathnil}}$.
   Then
   ${\mathcontext{C}{P_{1}}}\mathpistructuralpo{\mathnil}={\mathcontext{C'}{P_{1}}}$.

   Case 5. \rulename{SP-Commut}.
   Assume ${C}={\mathpiparallel{C_{0}}{C_{1}}}$ for some contexts $C_{0}$ and $C_{1}$.
   Then 
   ${P'_{0}}={\mathpiparallel{\mathcontext{C_{1}}{P_{0}}}{\mathcontext{C_{0}}{P_{0}}}}$.
   Let ${C'}={\mathpiparallel{C_{1}}{C_{0}}}$.
   Then, we have ${P'_{0}}={\mathcontext{C'}{P_{0}}}$.
   By \rulename{SP-Commut},
   ${\mathcontext{C}{P_{1}}}={\mathpiparallel{\mathcontext{C_{0}}{P_{1}}}{\mathcontext{C_{1}}{P_{1}}}}\mathpistructuralpo{\mathpiparallel{\mathcontext{C_{1}}{P_{1}}}{\mathcontext{C_{0}}{P_{1}}}}$.

   Case 6. \rulename{SP-Assoc}.
   Assume ${C}={\mathpiparallel{\mathcontext{}{\;}}{C_{2}}}$ and
   ${P_{0}}={\mathpiparallel{Q_{0}}{Q_{1}}}$ 
   for some context $C_{2}$ and processes $Q_{0}$ and $Q_{1}$.
   Then
   ${P'_{0}}={\mathpiparallel{Q_{0}}{\mleft(\mathpiparallel{Q_{1}}{\mathcontext{C_{2}}{P_{0}}}\mright)}}$.
   Let ${C'}={\mathpiparallel{Q_{0}}{\mleft(\mathpiparallel{Q_{1}}{C_{2}}\mright)}}$.
   Then, we have ${P'_{0}}={\mathcontext{C'}{P_{0}}}$.
   By \rulename{SP-Par}, we have 
   ${\mathcontext{C}{P_{1}}}={\mathpiparallel{P_{1}}{\mathcontext{C_{2}}{P_{1}}}}\mathpistructuralpo{\mathpiparallel{P_{0}}{\mathcontext{C_{2}}{P_{1}}}}={\mathpiparallel{\mleft(\mathpiparallel{Q_{0}}{Q_{1}}\mright)}{\mathcontext{C_{2}}{P_{1}}}}$.
   By \rulename{SP-Assoc},
   ${\mathcontext{C}{P_{1}}}\mathpistructuralpo{\mathpiparallel{Q_{0}}{\mleft(\mathpiparallel{Q_{1}}{\mathcontext{C_{2}}{P_{1}}}\mright)}}={\mathcontext{C'}{P_{1}}}$.

   Assume ${C}={\mathpiparallel{\mleft(\mathpiparallel{C_{0}}{C_{1}}\mright)}{C_{2}}}$
   for some contexts $C_{0}$, $C_{1}$, and $C_{2}$.
   Then
   ${P'_{0}}={\mathpiparallel{\mathcontext{C_{0}}{P_{0}}}{\mleft(\mathpiparallel{\mathcontext{C_{1}}{P_{0}}}{\mathcontext{C_{2}}{P_{0}}}\mright)}}$.
   Let ${C'}={\mathpiparallel{C_{0}}{\mleft(\mathpiparallel{C_{1}}{C_{2}}\mright)}}$.
   Then, we have ${P'_{0}}={\mathcontext{C'}{P_{0}}}$.
   By \rulename{SP-Assoc},
   ${\mathcontext{C}{P_{1}}}\mathpistructuralpo\mathpiparallel{\mathcontext{C_{0}}{P_{1}}}{\mleft(\mathpiparallel{\mathcontext{C_{1}}{P_{1}}}{\mathcontext{C_{2}}{P_{1}}}\mright)}=\mathcontext{C'}{P_{1}}$.

   Case 7. \rulename{SP-New}.
   Assume ${C}={\mathpiparallel{\mathcontext{}{\;}}{C_{1}}}$,
   ${P_{0}}={\mathpinew{\mathistype{x}{\xi}}{Q}}$, and
   $x\notin\mathFNof{\mathcontext{C_{1}}{P_{0}}}$ for some context $C_{1}$.
   Then 
   ${P'_{0}}={\mathpinew{\mathistype{x}{\xi}}{\mathpiparallel{Q}{\mathcontext{C_{1}}{P_{0}}}}}$.
   Let ${C'}={\mathpinew{\mathistype{x}{\xi}}{\mathpiparallel{Q}{C_{1}}}}$.
   By \rulename{SP-Par}, we have 
   ${\mathcontext{C}{P_{1}}}={\mathpiparallel{P_{1}}{\mathcontext{C_{1}}{P_{1}}}}\mathpistructuralpo{\mathpiparallel{P_{0}}{\mathcontext{C_{1}}{P_{1}}}}={\mathpiparallel{\mathpinew{\mathistype{x}{\xi}}{Q}}{\mathcontext{C_{1}}{P_{1}}}}$.
   By \cref{lemma:context-structural-po-and-set-of-free-names} \cref{item:eq-lemma-context-structural-po-and-set-of-free-names}
   and \rulename{SP-New},
   ${\mathcontext{C}{P_{1}}}\mathpistructuralpo{\mathpinew{\mathistype{x}{\xi}}{\mathpiparallel{Q}{\mathcontext{C_{1}}{P_{1}}}}}={\mathcontext{C'}{P_{1}}}$.

   Assume ${C}={\mathpiparallel{\mathpinew{\mathistype{x}{\xi}}{C_{0}}}{C_{1}}}$ and 
   $x\notin\mathFNof{\mathcontext{C_{1}}{P_{0}}}$ for some contexts $C_{0}$ and $C_{1}$.
   Then 
   ${P'_{0}}={\mathpinew{\mathistype{x}{\xi}}{\mathpiparallel{\mathcontext{C_{0}}{P_{0}}}{\mathcontext{C_{1}}{P_{0}}}}}$.
   Let ${C'}={\mathpinew{\mathistype{x}{\xi}}{\mathpiparallel{C_{0}}{C_{1}}}}$.
   By \cref{lemma:context-structural-po-and-set-of-free-names} \cref{item:eq-lemma-context-structural-po-and-set-of-free-names}
   and \rulename{SP-New},
   ${\mathcontext{C}{P_{1}}}\mathpistructuralpo{\mathpinew{\mathistype{x}{\xi}}{\mathpiparallel{\mathcontext{C_{0}}{P_{1}}}{\mathcontext{C_{1}}{P_{1}}}}}={\mathcontext{C'}{P_{1}}}$.

   Case 8. \rulename{SP-IfT}.
   Assume ${C}={\mathobif{\mathcontext{}{\;}}{C_{0}}{C_{1}}}$ and
   ${P_{0}}={\mathobtrue^{l}}$ for some contexts $C_{0}$ and $C_{1}$.
   Then ${P'_{0}}={\mathcontext{C_{0}}{P_{0}}}$.
   Let ${C'}={C_{0}}$.
   Since ${P_{0}}\mathpistructuraleq{P_{1}}$, 
   we have ${P_{1}}={\mathobtrue^{l}}$.
   By \rulename{SP-IfT},
   ${\mathcontext{C}{P_{1}}}={\mathobif{\mathobtrue^{l}}{\mathcontext{C_{0}}{P_{1}}}{\mathcontext{C_{1}}{P_{1}}}}\mathpistructuralpo{\mathcontext{C_{0}}{P_{1}}}={\mathcontext{C'}{P_{1}}}$.

   Assume ${C}={{\mathobif{\mathobtrue^{l}}{C_{0}}{C_{1}}}}$ 
   for some contexts $C_{0}$ and $C_{1}$.
   Then ${P'_{0}}={\mathcontext{C_{0}}{P_{0}}}$.
   Let ${C'}={C_{0}}$.
   By \rulename{SP-IfT},
   ${\mathcontext{C}{P_{1}}}={\mathobif{\mathobtrue^{l}}{\mathcontext{C_{0}}{P_{1}}}{\mathcontext{C_{1}}{P_{1}}}}\mathpistructuralpo{\mathcontext{C_{0}}{P_{1}}}={\mathcontext{C'}{P_{1}}}$.

   Case 9. \rulename{SP-IfF}.
   In the similar way to \rulename{SP-IfT}.

   Case 10. \rulename{SP-Rep}.
   Assume ${C}={\mathpireplication{C_{0}}}$ for some context $C_{0}$.
   Then ${P'_{0}}={\mathpiparallel{\mathcontext{\mathpireplication{C_{0}}}{P_{0}}}{\mathcontext{C_{0}}{P_{0}}}}$.
   Let ${C'}={\mathpiparallel{\mathpireplication{C_{0}}}{C_{0}}}$.
   Then, we have ${P'_{0}}={\mathcontext{C'}{P_{0}}}$.
   By \rulename{SP-Rep},
   ${\mathcontext{C}{P_{1}}}={\mathcontext{\mathpireplication{C_{0}}}{P_{1}}}\mathpistructuralpo{\mathpiparallel{\mathcontext{\mathpireplication{C_{0}}}{P_{1}}}{\mathcontext{C_{0}}{P_{1}}}}={\mathcontext{C'}{P_{1}}}$.

   Case 11. \rulename{SP-Par}.
   Assume ${C}={\mathpiparallel{C_{0}}{C_{1}}}$ and
   ${\mathcontext{C_{0}}{P_{0}}}\mathpistructuralpo{Q_{0}}$
   for some contexts $C_{0}$ and $C_{1}$, and a process $Q_{0}$.
   Then ${P'_{0}}={\mathpiparallel{Q_{0}}{\mathcontext{C_{1}}{P_{0}}}}$.
   By the induction hypothesis, we see that there exists a context $C'_{0}$ such that
   ${Q_{0}}={\mathcontext{C'_{0}}{P_{0}}}$ and
   ${\mathcontext{C_{0}}{P_{1}}}\mathpistructuralpo{\mathcontext{C'_{0}}{P_{1}}}$.
   Let ${C'}={\mathpiparallel{C'_{0}}{C_{1}}}$.
   Then, we have ${P'_{0}}={\mathcontext{C'}{P_{0}}}$.
   By \rulename{SP-Par},
   ${\mathcontext{C}{P_{1}}}={\mathpiparallel{\mathcontext{C_{0}}{P_{1}}}{\mathcontext{C_{1}}{P_{1}}}}\mathpistructuralpo{\mathpiparallel{\mathcontext{C'_{0}}{P_{1}}}{\mathcontext{C_{1}}{P_{1}}}}={\mathcontext{C'}{P_{1}}}$.

   Case 12. \rulename{SP-CNew}.
   Assume ${C}={\mathpinew{\mathistype{x}{\xi}}{C_{0}}}$ and
   ${\mathcontext{C_{0}}{P_{0}}}\mathpistructuralpo{Q_{0}}$
   for some context $C_{0}$ and a process $Q_{0}$.
   Then ${P'_{0}}={\mathpinew{\mathistype{x}{\xi}}{Q_{0}}}$.
   By the induction hypothesis, we see that there exists a context $C'_{0}$ such that
   ${Q_{0}}={\mathcontext{C'_{0}}{P_{0}}}$ and
   ${\mathcontext{C_{0}}{P_{1}}}\mathpistructuralpo{\mathcontext{C'_{0}}{P_{1}}}$.
   Let ${C'}={\mathpinew{\mathistype{x}{\xi}}{C'_{0}}}$.
   Then, we have ${P'_{0}}={\mathcontext{C'}{P_{0}}}$.
   By \rulename{SP-CNew},
   ${\mathcontext{C}{P_{1}}}={\mathpinew{\mathistype{x}{\xi}}{\mathcontext{C_{0}}{P_{1}}}}\mathpistructuralpo{\mathpinew{\mathistype{x}{\xi}}{\mathcontext{C'_{0}}{P_{1}}}}={\mathcontext{C'}{P_{1}}}$.
  \end{proof}

  \begin{lemma}
   \label[lemma]{lemma-of-lemma:barbed-bisimilar-reduction-with-context}
   For processes $P_{0}$ and $P_{1}$, 
   if ${P_{0}}\mathpistructuraleq{P_{1}}$ and 
   ${\mathtuple{{\mathcontext{C}{P_{0}}}, L}}\mathpireduction{\mathtuple{P'_{0}, L'}}$,
   then there exists a context $C'$ such that
   ${P'_{0}}={\mathcontext{C'}{P_{0}}}$ and
   ${\mathtuple{{\mathcontext{C}{P_{1}}}, L}}\mathpireduction{\mathtuple{{\mathcontext{C'}{P_{1}}}, L'}}$.
  \end{lemma}

  \begin{proof}
   Assume ${P_{0}}\mathpistructuraleq{P_{1}}$ and 
   ${\mathtuple{{\mathcontext{C}{P_{0}}}, L}}\mathpireduction{\mathtuple{P'_{0}, L'}}$.

   Assume ${C}={\mathcontext{}{\;}}$.
   In this case, ${\mathcontext{C}{P_{i}}}={P_{i}}$ for ${i}={0, 1}$.
   Assume ${\mathtuple{P_{0}, L}}\mathpireduction{\mathtuple{P'_{0}, L'}}$.
   Since ${P_{1}}\mathpistructuralpo{P_{0}}$,
   we have ${\mathtuple{P_{1}, L}}\mathpireduction{\mathtuple{P'_{0}, L'}}$.
   Let ${C'}={P'_{0}}$.
   Then ${P'_{0}}={\mathcontext{C'}{P_{0}}}$ and
   ${\mathcontext{C}{P_{1}}}\mathpireduction{\mathcontext{C'}{P_{1}}}$.

   Assume ${C}\not={\mathcontext{}{\;}}$.
   The proof proceeds by induction on the construction of
   ${\mathtuple{{\mathcontext{C}{P_{0}}}, L}}\mathpireduction{\mathtuple{P'_{0}, L'}}$.
   We consider cases according to the last rule of the construction of ${\mathtuple{{\mathcontext{C}{P_{0}}}, L}}\mathpireduction{\mathtuple{P'_{0}, L'}}$.

   Case 1. \rulename{R-Com}.
   Assume ${C}={\mathpiparallel{C_{0}}{C_{1}}}$
   with contexts $C_{0}$ and $C_{1}$ and $L' = L$.

   Assume ${C_{0}}={\mathcontext{}{\;}}$, 
   ${C_{1}}={\mathpiinput{x}{(y_{0}, \dots, y_{n})}.C'_{1}}$, and
   ${P_{0}}={\mathpioutput{x}{(v_{0}, \dots, v_{n})}. P}$.
   Then 
   ${P'_{0}}={\mathpiparallel{P}{\mathcontext{C'_{1}}{P_{0}}}}$.
   By \cref{def:structural-preorder}, we see
   ${P_{1}}={\mathpioutput{x}{(v_{0}, \dots, v_{n})}. P}$.
   Then, we have
   ${\mathcontext{C}{P_{1}}}= \mathpiparallel{\mathpioutput{x}{(v_{0}, \dots, v_{n})}. P}{{\mathpiinput{x}{(y_{0}, \dots, y_{n})}.\mathcontext{C'_{1}}{P_{1}}}}$.
   Let ${C'}={\mathpiparallel{P}{C'_{1}}}$.
   Then ${P'_{0}}={\mathcontext{C'}{P_{0}}}$.
   By \rulename{R-Com},
   ${\mathtuple{{\mathcontext{C}{P_{1}}}, L}}\mathpireduction{\mathtuple{{\mathcontext{C'}{P_{1}}}, L}}$.

   In the similar way, we can show the case
   ${C_{0}}={\mathpioutput{x}{(v_{0}, \dots, v_{n})}. C'_{0}}$,
   ${C_{1}}={\mathcontext{}{\;}}$, and
   ${P_{0}}={\mathpiinput{x}{(y_{0}, \dots, y_{n})}.P}$.

   Assume ${C_{0}}={\mathpioutput{x}{(v_{0}, \dots, v_{n})}. C'_{0}}$ and
   ${C_{1}}={\mathpiinput{x}{(y_{0}, \dots, y_{n})}.C'_{1}}$.
   Then
   ${P'_{0}}={\mathpiparallel{\mathcontext{C'_{0}}{P_{0}}}{\mathcontext{C'_{1}}{P_{0}}}}$.
   Let ${C'}={\mathpiparallel{C'_{0}}{C'_{1}}}$.
   Then ${P'_{0}}={\mathcontext{C'}{P_{0}}}$.
   By \rulename{R-Com},
   ${\mathtuple{{\mathcontext{C}{P_{1}}}, L}}\mathpireduction{\mathtuple{{\mathcontext{C'}{P_{1}}}, L}}$.
   
   Case 2. \rulename{R-NewLev}.
   Assume
   ${C}={\mathpinewsecrecylevel{l'}{\mathvect{l_{0}}}{\mathvect{l_{1}}}{C'_{0}}}$,
   ${L'}={\mathpinewsecrecylevel{l}{\mathvect{l_{0}}}{\mathvect{l_{1}}}{L}}$,
   ${\mathvect{l_{0}}}$ and ${\mathvect{l_{1}}}\subseteq{L}$ 
   with a context $C_{0}$.
   Then ${P'_{0}}={\mathcontext{C'_{0}}{P_{0}}}$.
   Let ${C'}={C'_{0}}$.
   Then ${P'_{0}}={\mathcontext{C'}{P_{0}}}$.
   By \rulename{R-NewLev},
   ${\mathtuple{{\mathcontext{C}{P_{1}}}, L}}\mathpireduction{\mathtuple{{\mathcontext{C'}{P_{1}}}, L'}}$.

   Case 3. \rulename{R-Par}.
   Assume ${C}={\mathpiparallel{C_{0}}{C_{1}}}$ with contexts $C_{0}$ and $C_{1}$.
   
   Assume ${C_{0}}={\mathcontext{}{\;}}$.
   Then, there exists a process $Q$ such that
   ${P'_{0}}={\mathpiparallel{Q}{\mathcontext{C_{1}}{P_{0}}}}$ and 
   ${\mathtuple{P_{0}, L}}\mathpireduction{\mathtuple{Q, L'}}$.
   Since ${P_{1}}\mathpistructuralpo{P_{0}}$,
   we have ${\mathtuple{P_{1}, L}}\mathpireduction{\mathtuple{Q, L'}}$.
   Let ${C'}={\mathpiparallel{Q}{C_{1}}}$.
   Then ${P'_{0}}={\mathcontext{C'}{P_{0}}}$.
   By \rulename{R-Par},
   ${\mathtuple{{\mathcontext{C}{P_{1}}}, L}}\mathpireduction{\mathtuple{{\mathcontext{C'}{P_{1}}}, L}}$.

   Assume ${C_{0}}\not={\mathcontext{}{\;}}$.
   Then, there exists a process $Q_{0}$ such that
   ${P'_{0}}={\mathpiparallel{Q_{0}}{\mathcontext{C_{1}}{P_{0}}}}$ and 
   ${\mathtuple{\mathcontext{C_{0}}{P_{0}}, L}}\mathpireduction{\mathtuple{Q_{0}, L'}}$.
   By the induction hypothesis, there exists a context $C'_{0}$ such that
   ${Q_{0}}={\mathcontext{C'_{0}}{P_{0}}}$ and
   ${\mathcontext{C_{0}}{P_{1}}}\mathpireduction{\mathcontext{C'_{0}}{P_{1}}}$.
   Let ${C'}={\mathpiparallel{C'_{0}}{C_{1}}}$.
   Then ${P'_{0}}={\mathcontext{C'}{P_{0}}}$.
   Since 
   ${\mathtuple{\mathcontext{C_{0}}{P_{1}}, L}}\mathpireduction{\mathtuple{\mathcontext{C'_{0}}{P_{1}}, L'}}$
   and \rulename{R-Par},
   we have 
   ${\mathtuple{{\mathcontext{C}{P_{1}}}, L}}\mathpireduction{\mathtuple{{\mathcontext{C'}{P_{1}}}, L'}}$.

   Case 4. \rulename{R-New}.
   Assume that ${C}={\mathpinew{\mathistype{x}{\xi}}{C_{0}}}$
   with a context $C_{0}$, and there exists a process $Q_{0}$ such that
   ${\mathtuple{\mathcontext{C_{0}}{P_{0}}, L}}\mathpireduction{\mathtuple{Q_{0}, L'}}$ and
   ${P'_{0}}={\mathpinew{\mathistype{x}{\xi}}{Q_{0}}}$.
   By the induction hypothesis, there exists a context $C'_{0}$ such that
   ${Q_{0}}={\mathcontext{C'_{0}}{P_{0}}}$ and
   ${\mathcontext{C_{0}}{P_{1}}}\mathpireduction{\mathcontext{C'_{0}}{P_{1}}}$.
   Let ${C'}={\mathpinew{\mathistype{x}{\xi}}{C'_{0}}}$.
   Then ${P'_{0}}={\mathcontext{C'}{P_{0}}}$.
   Since 
   ${\mathtuple{\mathcontext{C_{0}}{P_{1}}, L}}\mathpireduction{\mathtuple{\mathcontext{C'_{0}}{P_{1}}, L'}}$ and
   \rulename{R-New},
   we have 
   ${\mathtuple{{\mathcontext{C}{P_{1}}}, L}}\mathpireduction{\mathtuple{{\mathcontext{C'}{P_{1}}}, L'}}$.

   Case 5. \rulename{R-SP}.
   Assume that there exist processes $Q_{0}$ and $Q'_{0}$ such that
   ${\mathcontext{C}{P_{0}}}\mathpistructuralpo{Q_{0}}$,
   ${\mathtuple{Q_{0}, L}}\mathpireduction{\mathtuple{Q'_{0}, L'}}$, and 
   ${Q'_{0}}\mathpistructuralpo{P'_{0}}$.
   By \cref{lemma-of-lemma:barbed-bisimilar-structual-preorder-with-context},
   there exists a context $C''$ such that
   ${Q_{0}}={\mathcontext{C''}{P_{0}}}$ and
   ${\mathcontext{C}{P_{1}}}\mathpistructuralpo{\mathcontext{C''}{P_{1}}}$.
   By the induction hypothesis, there exists a context $C'''$ such that
   ${Q'_{0}}={\mathcontext{C'''}{P_{0}}}$ and
   ${\mathtuple{\mathcontext{C''}{P_{1}}, L}}\mathpireduction{\mathtuple{\mathcontext{C'''}{P_{1}}, L'}}$.
   By \cref{lemma-of-lemma:barbed-bisimilar-structual-preorder-with-context},
   there exists a context $C'$ such that
   ${P'_{0}}={\mathcontext{C'}{P_{0}}}$ and
   ${\mathcontext{C'''}{P_{1}}}\mathpistructuralpo{\mathcontext{C'}{P_{1}}}$.   
   Since 
   ${\mathcontext{C}{P_{1}}}\mathpistructuralpo{\mathcontext{C''}{P_{1}}}$,
   ${\mathtuple{\mathcontext{C''}{P_{1}}, L}}\mathpireduction{\mathtuple{\mathcontext{C'''}{P_{1}}, L'}}$,
   ${\mathcontext{C'''}{P_{1}}}\mathpistructuralpo{\mathcontext{C'}{P_{1}}}$
   and \rulename{R-SP},
   we have 
   ${\mathtuple{{\mathcontext{C}{P_{1}}}, L}}\mathpireduction{\mathtuple{{\mathcontext{C'}{P_{1}}}, L'}}$.
  \end{proof}

  \begin{lemma}
   \label[lemma]{lemma:barbed-bisimilar-structual-preorder-with-context}
   For processes $P_{0}$ and $P_{1}$, and a lattices for secrecy levels $L$, 
   if ${P_{0}}\mathpistructuraleq{P_{1}}$,
   then 
   $\mathisbisimilar{\mathtuple{\mathcontext{C}{P_{0}}, L}}{\mathtuple{\mathcontext{C}{P_{1}}, L}}$ with any context $C$.
  \end{lemma}

  \begin{proof}
   Let
   \[
  {\mathbarbedbisim{R}}={\mathsetintension{\mathtuple{\mathtuple{\mathcontext{C}{P_{0}}, L}, \mathtuple{\mathcontext{C}{P_{1}}, L}}}{{P_{0}}\mathpistructuraleq{P_{1}}, C \text{ is a context}, L  \text{ is a lattice of secrecy levels}}}
   \]

   To show the claim, 
   it suffices to show that ${\mathbarbedbisim{R}}$ is a barbed bisimulation.

   Fix $C$ be a  context, and 
   processes $P_{0}$ and $P_{1}$  with ${P_{0}}\mathpistructuraleq{P_{1}}$. 
   Then
   $\mathtuple{\mathtuple{\mathcontext{C}{P_{0}}, L}, \mathtuple{\mathcontext{C}{P_{1}}, L}}\in{\mathbarbedbisim{R}}$.
   We show that the conditions in \cref{def:Barbed-bisimulation} hold.

   \noindent \cref{item:left-def-Barbed-bisimulation} and
   \cref{item:right-def-Barbed-bisimulation}
   By \cref{lemma-of-lemma:barbed-bisimilar-reduction-with-context}.

   \noindent \cref{item:barbs-def-Barbed-bisimulation}
   We show 
   ${\maththebarbsof{\mathcontext{C}{P_{0}}, L}}={\maththebarbsof{\mathcontext{C}{P_{1}}, L}}$.
   To show the claim, we show
   ${\maththebarbsof{\mathcontext{C}{P_{0}}, L}}\subseteq{\maththebarbsof{\mathcontext{C}{P_{1}}, L}}$.
   Let ${x}\in{\maththebarbsof{\mathcontext{C}{P_{0}}, L}}$.
   
   Assume 
   ${\mathtuple{\mathcontext{C}{P_{0}}, L}}\mathpireductionkc{\mathtuple{Q, L'}}, {Q}={\mathpinew{\mathvect{y}}{\mathpiparallel{\mathpioutput{x}{\mathvect{v}}. Q_{0}}{Q_{1}}}}$
   with ${x}\notin{\mathvect{y}}$.
   By  \cref{lemma-of-lemma:barbed-bisimilar-structual-preorder-with-context} and
   \cref{lemma-of-lemma:barbed-bisimilar-reduction-with-context},
   there exists a context $C'$ such that
   ${\mathpinew{\mathvect{y}}{\mathpiparallel{\mathpioutput{x}{\mathvect{v}}. Q_{0}}{Q_{1}}}}={\mathcontext{C'}{P_{0}}}$,
   $\mathtuple{\mathcontext{C}{P_{1}}, L}\mathpireductionkc{\mathtuple{Q', L'}}$, and
   ${Q'}\mathpistructuralpo{\mathcontext{C'}{P_{1}}}$.

   Assume ${C'}={\mathcontext{}{\;}}$ and 
   ${P_{0}}={\mathpinew{\mathvect{y}}{\mathpiparallel{\mathpioutput{x}{\mathvect{v}}. Q_{0}}{Q_{1}}}}$.
   In this case, $\mathtuple{\mathcontext{C}{P_{1}}, L}\mathpireductionkc{\mathtuple{P_{1}, L'}}$.
   Since ${P_{1}}\mathpistructuralpo{P_{0}}$,
   we have ${\mathtuple{\mathcontext{C}{P_{1}}, L}}\mathpireductionkc\mathtuple{{\mathpinew{\mathvect{y}}{\mathpiparallel{\mathpioutput{x}{\mathvect{v}}. Q_{0}}{Q_{1}}}}, L'}$.
   Hence, ${x}\in{\maththebarbsof{\mathcontext{C}{P_{1}}, L}}$.

   Assume ${C'}={\mathpinew{\mathvect{y}}{\mathcontext{}{\;}}}$ and 
   ${P_{0}}={\mathpiparallel{\mathpioutput{x}{\mathvect{v}}. Q_{0}}{Q_{1}}}$.
   In this case, ${\mathtuple{\mathcontext{C}{P_{1}}, L}}\mathpireductionkc{\mathtuple{\mathpinew{\mathvect{y}}{P_{1}}, L'}}$.
   Since ${P_{1}}\mathpistructuralpo{P_{0}}$,
   we have ${\mathpinew{\mathvect{y}}{P_{1}}}\mathpistructuralpo{\mathpinew{\mathvect{y}}{P_{1}}}$.
   Hence, we have ${\mathtuple{\mathcontext{C}{P_{1}}, L}}\mathpireductionkc{\mathtuple{{\mathpinew{\mathvect{y}}{\mathpioutput{x}{\mathvect{v}}. \mathpiparallel{Q_{0}}{Q_{1}}}}, L'}}$.
   Therefore, ${x}\in{\maththebarbsof{\mathcontext{C}{P_{1}}, L}}$.

   Assume
   ${C'}={\mathpinew{\mathvect{y}}{{\mathpiparallel{\mathcontext{}{\;}}{C_{1}}}}}$ 
   and 
   ${P_{0}}={\mathpioutput{x}{\mathvect{v}}. Q_{0}}$
   for some context $C_{1}$.
   In this case, ${\mathtuple{\mathcontext{C}{P_{1}}, L}}\mathpireductionkc{\mathtuple{\mathpinew{\mathvect{y}}{{\mathpiparallel{P_{1}}{\mathcontext{C_{1}}{P_{1}}}}}, L'}}$.
   Since ${P_{1}}\mathpistructuralpo{P_{0}}$,
   we have ${\mathpinew{\mathvect{y}}{{\mathpiparallel{P_{1}}{\mathcontext{C_{1}}{P_{1}}}}}}\mathpistructuralpo{\mathpinew{\mathvect{y}}{{\mathpiparallel{\mathpioutput{x}{\mathvect{v}}. Q_{0}}{\mathcontext{C_{1}}{P_{1}}}}}}$.
   Hence, we have ${\mathtuple{\mathcontext{C}{P_{1}}, L}}\mathpireductionkc{\mathtuple{{\mathpinew{\mathvect{y}}{{\mathpiparallel{\mathpioutput{x}{\mathvect{v}}. Q_{0}}{\mathcontext{C_{1}}{P_{1}}}}}}, L'}}$.
   Therefore, ${x}\in{\maththebarbsof{\mathcontext{C}{P_{1}}, L}}$.   

   Assume
   ${C'}={\mathpinew{\mathvect{y}}{\mathpiparallel{\mathpioutput{x}{\mathvect{v}}. C_{1}}{C_{1}}}}$ 
   for some contexts $C_{0}$ and $C_{1}$.
   In this case, \linebreak[4]
   $\mathtuple{\mathcontext{C}{P_{1}}, L}\mathpireductionkc \mathtuple{\mathpinew{\mathvect{y}}{\mathpiparallel{\mathpioutput{x}{\mathvect{v}}. \mathcontext{C_{0}}{P_{1}}}{\mathcontext{C_{1}}{P_{1}}}}, L'}$.
   Hence, ${x}\in{\maththebarbsof{\mathcontext{C}{P_{1}}, L}}$.   

   In the similar way to the case 
   ${\mathtuple{\mathcontext{C}{P_{0}}, L}}\mathpireductionkc{\mathtuple{Q, L'}}, {Q}={\mathpinew{\mathvect{y}}{\mathpioutput{x}{\mathvect{v}}. \mathpiparallel{Q_{0}}{Q_{1}}}}$,
   we have ${x}\in{\maththebarbsof{\mathcontext{C}{P_{1}}, L}}$ 
   if ${\mathtuple{\mathcontext{C}{P_{0}}, L}}\mathpireductionkc{\mathtuple{Q, L'}}, {Q}={\mathpinew{\mathvect{y}}{\mathpiinput{x}{\mathvect{z}}. \mathpiparallel{Q_{0}}{Q_{1}}}}$ with ${x}\notin{\mathvect{y}}$.
   Therefore, ${\maththebarbsof{\mathcontext{C}{P_{0}}, L}}\subseteq{\maththebarbsof{\mathcontext{C}{P_{1}}, L}}$.

   In the same way to the case ${\maththebarbsof{\mathcontext{C}{P_{0}}, L}}\subseteq{\maththebarbsof{\mathcontext{C}{P_{1}}, L}}$,
   we have ${\maththebarbsof{\mathcontext{C}{P_{1}}, L}}\subseteq{\maththebarbsof{\mathcontext{C}{P_{0}}, L}}$.
   Thus,
   ${\maththebarbsof{\mathcontext{C}{P_{0}}, L}}={\maththebarbsof{\mathcontext{C}{P_{1}}, L}}$.
  \end{proof}

  \begin{lemma}
   \label[lemma]{lemma:barbed-conguruence-structual-preorder}
   For processes $P$, $P'$, $Q$, $Q'$ and a lattice of secrecy level $L$,
   if $\mathiscongruentwithenvandlevel{\Gamma}{L}{l}{P}{Q}$,
   ${P'}\mathpistructuraleq{P}$, and ${Q}\mathpistructuraleq{Q'}$,
   then $\mathiscongruentwithenvandlevel{\Gamma}{L}{l}{P'}{Q'}$.
  \end{lemma}

  \begin{proof}
   Assume $\mathiscongruentwithenvandlevel{\Gamma}{L}{l}{P}{Q}$,
   ${P'}\mathpistructuraleq{P}$, and ${Q}\mathpistructuraleq{Q'}$.
   We show that the conditions in \cref{def:barbed-congruence} hold.

   \noindent \cref{item:type-in-def-barbed-congruence}
   By \cref{lemma:substitution-preserve-subtyping}.

   \noindent \cref{item:bisimilar-in-def-barbed-congruence}
   Fix $C$ be an 
   $\mathenvandlevel{\Gamma}{L}{l}$-$\mathenvandlevel{\Delta}{L'}{l'}$-context.
   Then, we have
   $\mathisbisimilar{\mathtuple{\mathcontext{C}{P}, L'}}{\mathtuple{\mathcontext{C}{Q}, L'}}$.
   Since ${P'}\mathpistructuraleq{P}$, and ${Q}\mathpistructuraleq{Q'}$,
   \cref{lemma:barbed-bisimilar-structual-preorder-with-context} implies
   $\mathisbisimilar{\mathtuple{\mathcontext{C}{P'}, L'}}{\mathtuple{\mathcontext{C}{P}, L'}}$ 
   and
   $\mathisbisimilar{\mathtuple{\mathcontext{C}{Q}, L'}}{\mathtuple{\mathcontext{C}{Q'}, L'}}$.
   By \cref{lemma:barbed-bisimilar-transitive},
   we have
   $\mathisbisimilar{\mathtuple{\mathcontext{C}{P'}, L'}}{\mathtuple{\mathcontext{C}{Q'}, L'}}$.
  \end{proof}

\subsection{Definition of $\mathErsy$}
   \label{sec:def-Er}

  \begin{definition}[$\mathErsy$]
   \label[definition]{def:Er}
   For a type environment $\Gamma$, a lattice of secrecy levels $L$, and 
   a secrecy level ${l}\in{L}$,
   we inductively define $\mathEr{\Gamma}{L, l}{P}$ as follows:
   \begin{align*}
    {\mathEr{\Gamma}{L, l}{\mathnil}}&={\mathnil} &&\text{if ${P}={\mathnil}$},  \\[1.0em]
    {\mathEr{\Gamma}{L, l}{v}}&={\mathobunit} &&
    \begin{aligned}
     &\text{if ${P}={v}$, $v$ is a value, and} \\
     &\text{ $\mathof{\Gamma}{v}$ is not $l$ and not lower than $l$ in $L$}, 
    \end{aligned} \\[1.0em] 
    {\mathEr{\Gamma}{L, l}{v}}&={v} &&
    \begin{aligned}
     &\text{if ${P}={v}$, $v$ is a value, and} \\
     &\text{ $\mathof{\Gamma}{v}$ is $l$ or lower than $l$ in $L$}, 
    \end{aligned} \\[1.0em]
    {\mathEr{\Gamma}{L, l}{\mathpiparallel{P_{0}}{P_{1}}}}&={\mathpiparallel{\mathEr{\Gamma}{L, l}{P_{0}}}{\mathEr{\Gamma}{L, l}{P_{1}}}} &&
 \begin{aligned}
 &\text{if ${P}={\mathpiparallel{P_{0}}{P_{1}}}$} \\
 &\text{ for processes $P_{0}$ and $P_{1}$}, 
 \end{aligned}
 \\[1.0em]
    {\mathEr{\Gamma}{L, l}{\mathpireplication{P'}}}&={\mathnil} &&
    \begin{aligned}
     &\text{if ${P}={\mathpireplication{P'}}$ and } \\
     &\text{ ${\mathEr{\Gamma}{L, l}{P'}}\mathpistructuraleq{\mathnil}$} \\
     &\text{ for a process $P'$}, 
    \end{aligned}
    \\[1.0em]
    {\mathEr{\Gamma}{L, l}{\mathpireplication{P'}}}&={\mathpireplication{\mathEr{\Gamma}{L, l}{P'}}} &&
    \begin{aligned}
     &\text{if ${P}={\mathpireplication{P'}}$ and } \\
     &\text{ ${\mathEr{\Gamma}{L, l}{P'}}\not\mathpistructuraleq{\mathnil}$} \\
     &\text{ for a process $P'$}, 
    \end{aligned}
    \\[1.0em]
    {\mathEr{\Gamma}{L, l}{\mathpioutput{x}{\mathtuple{v_{0}, \dots, v_{n}}}.P'}}&={\mathpioutput{x}{v'_{0}, \dots, v'_{n}}. \mathEr{\Gamma}{L, l}{P'}} &&
    \begin{aligned}
     &\text{if ${P}={\mathpioutput{x}{\mathtuple{v_{0}, \dots, v_{n}}}.P'}$} \\
     &\text{ for a process $P'$, } \\
     &\text{  $\mathof{\Gamma}{x}$ is the form $\mathprogramtypetuple{\tau_{0}, \dots, \tau_{n}}^{l'}/U$ } \\
     &\text{ with ${l'}\leq_{L}{l}$, and} \\
     &\text{ ${v'_{i}}={\mathEr{\Gamma}{L}{v_{i}}}$} \\
     &\text{ for all $i=0, \dots, n$}, 
    \end{aligned} \\[1.0em]
    {\mathEr{\Gamma}{L, l}{\mathpioutput{x}{\mathtuple{v_{0}, \dots, v_{n}}}.P'}}&={\mathEr{\Gamma}{L, l}{P'}} &&
    \begin{aligned}
     &\text{if ${P}={\mathpioutput{x}{\mathtuple{v_{0}, \dots, v_{n}}}.P'}$} \\
     &\text{ for a process $P'$, and} \\
     &\text{ $\mathof{\Gamma}{x}$ is \emph{not} the form } \\
     &\text{ $\mathprogramtypetuple{\tau_{0}, \dots, \tau_{n}}^{l'}/U$ with ${l'}\leq_{L}{l}$}, 
    \end{aligned} \\[1.0em]
    {\mathEr{\Gamma}{L, l}{\mathpiinput{x}{\mathtuple{y_{0}, \dots, y_{n}}}.P'}}&={\mathpiinput{x}{y_{0}, \dots, y_{n}}. \mathEr{\Gamma, \mathistype{y_{0}}{\tau_{0}}, \dots, \mathistype{y_{n}}{\tau_{n}} }{L, l}{P'}} &&
    \begin{aligned}
     &\text{if ${P}={\mathpiinput{x}{\mathtuple{y_{0}, \dots, y_{n}}}.P'}$ } \\
     &\text{ for a program $P'$, and} \\ 
     &\text{ $\mathof{\Gamma}{x}$ is the form $\mathprogramtypetuple{\tau_{0}, \dots, \tau_{n}}^{l'}/U$} \\
     &\text{ with ${l'}\leq_{L}{l}$},
    \end{aligned} \\[1.0em]
    {\mathEr{\Gamma}{L, l}{\mathpiinput{x}{\mathtuple{y_{0}, \dots, y_{n}}}.P'}}&={\mathEr{\Gamma, \mathistype{y_{0}}{\tau_{0}}, \dots, \mathistype{y_{n}}{\tau_{n}} }{L, l}{P'}} &&
    \begin{aligned}
     &\text{if ${P}={\mathpiinput{x}{\mathtuple{y_{0}, \dots, y_{n}}}.P'}$ } \\
     &\text{ for a process $P'$, and} \\ 
    &\text{ $\mathof{\Gamma}{x}$ is the form $\mathprogramtypetuple{\tau_{0}, \dots, \tau_{n}}^{l'}/U$} \\
    &\text{ with ${l'}\not \leq_{L}{l}$},
    \end{aligned} \\[1.0em]
    {\mathEr{\Gamma}{L, l}{\mathpiinput{x}{\mathtuple{y_{0}, \dots, y_{n}}}.P'}}&={\mathEr{\Gamma, \mathistype{y_{0}}{\mathunittype}, \dots, \mathistype{y_{n}}{\mathunittype}}{L, l}{P'}} &&
  \begin{aligned}
   &\text{if ${P}={\mathpiinput{x}{\mathtuple{y_{0}, \dots, y_{n}}}.P'}$ } \\
   &\text{ for a process $P'$, and} \\ 
   &\text{ $\mathof{\Gamma}{x}$ is \emph{not} } \\
   &\text{ the form $\mathprogramtypetuple{\tau_{0}, \dots, \tau_{n}}^{l'}/U$},
  \end{aligned}  \\[1.0em] 
    {\mathEr{\Gamma}{L, l}{\mathpinew{\mathistype{x}{\xi}}{P'}}}&={{\mathpinew{\mathistype{x}{\xi}}{\mathEr{\Gamma, \mathistype{x}{\xi/\mathusagenil}}{L, l}{P'}}}} &&
    \begin{aligned}
    &\text{if ${P}={\mathpinew{\mathistype{x}{\xi}}{P'}}$} \\
    &\text{ for a process $P'$, and} \\
    &\text{ $\xi$ is the form $\mathprogramtypetuple{\tau_{0}, \dots, \tau_{n}}^{l'}$ } \\
    &\text{ with ${l'}\leq_{L}{l}$},
    \end{aligned} \\[1.0em]
    {\mathEr{\Gamma}{L, l}{\mathpinew{\mathistype{x}{\xi}}{P'}}}&={\mathEr{\Gamma, \mathistype{x}{\xi/\mathusagenil}}{L, l}{P'}} &&
    \begin{aligned}
     &\text{if ${P}={\mathpinew{\mathistype{x}{\xi}}{P'}}$} \\
     &\text{ for a process $P'$, and } \\
     &\text{ $\xi$ is \emph{not} the form $\mathprogramtypetuple{\tau_{1}, \dots, \tau_{n}}^{l'}$} \\
     &\text{ with ${l'}\leq_{L}{l}$},
    \end{aligned}  \\[1.0em]
    {\mathEr{\Gamma}{L, l}{\mathpinewsecrecylevel{l_{0}}{\mathvect{l_{1}}}{\mathvect{l_{2}}}{P'}}}&= \mathpinewsecrecylevel{l_{0}}{\mathvect{l_{1}}}{\mathvect{l_{2}}}{\mathEr{\Gamma}{\mathpinewsecrecylevel{l_{0}}{\mathvect{l_{1}}}{\mathvect{l_{2}}}{L}, l}{P'}} &&
    \begin{aligned}
    &\text{if ${P}={\mathpinewsecrecylevel{l_{0}}{\mathvect{l_{1}}}{\mathvect{l_{2}}}{P'}}$} \\
    &\text{ for a process $P'$, and} \\
    &\text{ $l' \leq_{L} l$ for any  $l'\in\mathvect{l_{1}}$, $\mathvect{l_{2}}$}, \\
    \end{aligned} \\[1.0em]
    {\mathEr{\Gamma}{L, l}{\mathpinewsecrecylevel{l_{0}}{\mathvect{l_{1}}}{\mathvect{l_{2}}}{P'}}}&={\mathEr{\Gamma}{\mathpinewsecrecylevel{l_{0}}{\mathvect{l_{1}}}{\mathvect{l_{2}}}{L}, l}{P'}} &&
    \begin{aligned}
    &\text{if ${P}={\mathpinewsecrecylevel{l_{0}}{\mathvect{l_{1}}}{\mathvect{l_{2}}}{P'}}$} \\
    &\text{ for a process $P'$, and} \\
    &\text{ $l' \not\leq_{L} l$ for some  $l'\in\mathvect{l_{1}}$, $\mathvect{l_{2}}$}, 
    \end{aligned} \\[1.0em]
    {\mathEr{\Gamma}{L, l}{\mathobif{v}{P'}{P''}}}&={\mathobif{v}{\mathEr{\Gamma}{L, l}{P'}}{\mathEr{\Gamma}{L, l}{P''}}} &&
    \begin{aligned}
     &\text{if ${P}={\mathobif{v}{P'}{P''}}$ } \\
     &\text{ for processes $P'$, $P''$, and} \\
     &\text{ ${\mathof{\Gamma}{v}}={\mathbooltypewithsec{l'}}$ with ${l'}\leq_{L}{l}$},
    \end{aligned}  \\[1.0em]
    {\mathEr{\Gamma}{L, l}{\mathobif{v}{P'}{P''}}}&={\mathnil} &&
    \begin{aligned}
     &\text{if ${P}={\mathobif{v}{P'}{P''}}$ } \\
     &\text{ for processes $P'$,  $P''$, and}  \\
     &\text{ ${\mathof{\Gamma}{v}}={\mathbooltypewithsec{l'}}$ does \emph{not} hold } \\
     &\text{ with ${l'}\leq_{L}{l}$}.
    \end{aligned} 
   \end{align*}
  \end{definition}

  \begin{definition}[The order of occurrences of $\mathpinewsecrecylevelhead{\mathplaceholder}{\mathplaceholder}{\mathplaceholder}$'s in a process $P$]
   \label[definition]{def:order-of-occurences}
   For a process $P$,
   we define the order $\mathpinewsecrecylevelheadpo{P}$
   of occurrences of 
   $\mathpinewsecrecylevelhead{\mathplaceholder}{\mathplaceholder}{\mathplaceholder}$'s
   in $P$ as follows:
   
   $\mathpinewsecrecylevelhead{l_{0}}{\mathvect{l_{1}}}{\mathvect{l_{2}}}\mathpinewsecrecylevelheadpo{P}\mathpinewsecrecylevelhead{l'_{0}}{\mathvect{l'_{1}}}{\mathvect{l'_{2}}}$
   if and only if
   $\mathpinewsecrecylevel{l_{0}}{\mathvect{l_{1}}}{\mathvect{l_{2}}}{P'}$
   is a subexpression of $P$, and
   $\mathpinewsecrecylevelhead{l'_{0}}{\mathvect{l'_{1}}}{\mathvect{l'_{2}}}$
   occurs in $P'$.
  \end{definition}

  \begin{definition}[$\mathEr{\Gamma}{L, l}{P}$-sublattice]
   \label[definition]{def:Er-sublattice}
   We say that a sublattice $L'$ of $L$ is an $\mathEr{\Gamma}{L, l}{P}$-sublattice 
   if the following conditions hold: 
   \begin{enumerate}
    \item For occurrences of 
	  $\mathpinewsecrecylevelhead{l_{0_{0}}}{\mathvect{l_{1_{0}}}}{\mathvect{l_{2_{0}}}}, \dots, \mathpinewsecrecylevelhead{l_{0_{n}}}{\mathvect{l_{1_{n}}}}{\mathvect{l_{2_{n}}}}$
	  in $P$, where
	  $\mathpinewsecrecylevelhead{l_{0_{i}}}{\mathvect{l_{1_{i}}}}{\mathvect{l_{2_{i}}}} \not\mathpinewsecrecylevelheadop{P}\mathpinewsecrecylevelhead{l_{0_{j}}}{\mathvect{l_{1_{j}}}}{\mathvect{l_{2_{j}}}}$ 
	  for $i<j$, if
	  $\mathpinewsecrecylevel{l_{0_{0}}}{\mathvect{l_{1_{0}}}}{\mathvect{l_{2_{0}}}}{\dots \mleft(\mathpinewsecrecylevel{l_{0_{n}}}{\mathvect{l_{1_{n}}}}{\mathvect{l_{2_{n}}}}{L} \mright.}$
	  is defined, then there exist $j_{0}<\dots < j_{m}$ such that
	  $\mathpinewsecrecylevel{l_{0_{j_{0}}}}{\mathvect{l_{1_{j_{0}}}}}{\mathvect{l_{2_{j_{0}}}}}{\dots\mleft( \mathpinewsecrecylevel{l_{0_{j_{m}}}}{\mathvect{l_{1_{j_{m}}}}}{\mathvect{l_{2_{j_{m}}}}}{L'}\mright.}$
	  is defined, 
	  \begin{multline*}
	   \mathsetextension{\mathpinewsecrecylevelhead{l_{0_{j_{0}}}}{\mathvect{l_{1_{j_{0}}}}}{\mathvect{l_{2_{j_{0}0}}}}, \dots, \mathpinewsecrecylevelhead{l_{0_{j_{m}}}}{\mathvect{l_{1_{j_{m}}}}}{\mathvect{l_{2_{j_{m}}}}}} = \\
	   \mathsetintension{\mathpinewsecrecylevelhead{l_{0}}{\mathvect{l_{1}}}{\mathvect{l_{2}}} }{\mathpinewsecrecylevelhead{l_{0}}{\mathvect{l_{1}}}{\mathvect{l_{2}}} \text{ occurs in } \mathEr{\Gamma}{L, l}{P}} \cap \\
	  \mathsetextension{\mathpinewsecrecylevelhead{l_{0_{0}}}{\mathvect{l_{1_{0}}}}{\mathvect{l_{2_{0}}}}, \dots, \mathpinewsecrecylevelhead{l_{0_{n}}}{\mathvect{l_{1_{n}}}}{\mathvect{l_{2_{n}}}}}.
	  \end{multline*}
	  \label{item:P-def-Er-sublattice}
    \item For occurrences of \linebreak[3]
	  $\mathpinewsecrecylevelhead{l'_{0_{0}}}{\mathvect{l'_{1_{0}}}}{\mathvect{l'_{2_{0}}}}, \dots, \mathpinewsecrecylevelhead{l'_{0_{m}}}{\mathvect{l'_{1_{m}}}}{\mathvect{l'_{2_{m}}}}$
	  in $\mathEr{\Gamma}{L, l}{P}$, where \linebreak[4]
	  $\mathpinewsecrecylevelhead{l'_{0_{i}}}{\mathvect{l'_{1_{i}}}}{\mathvect{l'_{2_{i}}}} \not\mathpinewsecrecylevelheadop{\mathEr{\Gamma}{L, l}{P}} \mathpinewsecrecylevelhead{l'_{0_{j}}}{\mathvect{l'_{1_{j}}}}{\mathvect{l'_{2_{j}}}}$ for $i<j$,
	  if $\mathpinewsecrecylevel{l'_{0_{0}}}{\mathvect{l'_{1_{0}}}}{\mathvect{l'_{2_{0}}}}{\dots\mleft( \mathpinewsecrecylevel{l'_{0_{m}}}{\mathvect{l'_{1_{m}}}}{\mathvect{l'_{2_{m}}}}{L'}\mright.}$
	  is defined, then
	  there exist $\mathpinewsecrecylevelhead{l_{0_{0}}}{\mathvect{l_{1_{0}}}}{\mathvect{l_{2_{0}}}}, \dots, \mathpinewsecrecylevelhead{l_{0_{n}}}{\mathvect{l_{1_{n}}}}{\mathvect{l_{2_{n}}}}$
	  such that
	  $\mathpinewsecrecylevel{l_{0_{0}}}{\mathvect{l_{1_{0}}}}{\mathvect{l_{2_{0}}}}{\dots \mleft(\mathpinewsecrecylevel{l_{0_{n}}}{\mathvect{l_{1_{n}}}}{\mathvect{l_{2_{n}}}}{L} \mright.}$
	  is defined,
	  \begin{multline*}
	   \mathsetextension{\mathpinewsecrecylevelhead{l_{0_{0}}}{\mathvect{l_{1_{0}}}}{\mathvect{l_{2_{0}}}}, \dots, \mathpinewsecrecylevelhead{l_{0_{n}}}{\mathvect{l_{1_{n}}}}{\mathvect{l_{2_{n}}}}} = \\
	   \mathsetintension{\mathpinewsecrecylevelhead{l_{0}}{\mathvect{l_{1}}}{\mathvect{l_{2}}} }{ \mathpinewsecrecylevelhead{l_{0}}{\mathvect{l_{1}}}{\mathvect{l_{2}}} \mathpinewsecrecylevelheadpo{P} \mathpinewsecrecylevelhead{l'_{0_{i}}}{\mathvect{l'_{1_{i}}}}{\mathvect{l'_{2_{i}}}}  \text{ for some $i=0, \dots, m$ } },
	  \end{multline*}
	  and
	  $\mathpinewsecrecylevelhead{l_{0_{i}}}{\mathvect{l_{1_{i}}}}{\mathvect{l_{2_{i}}}} \not\mathpinewsecrecylevelheadop{P}\mathpinewsecrecylevelhead{l_{0_{j}}}{\mathvect{l_{1_{j}}}}{\mathvect{l_{2_{j}}}}$ 
	  for $i<j$.
	  \label{item:Er-def-Er-sublattice}
   \end{enumerate}

   We note that $\mathof{\Gamma}{v}$ is lower than $l$ in $L$ for any value $v$ 
   freely occurring in $\mathEr{\Gamma}{L, l}{P}$.
  \end{definition}

   \subsection{Basic properties of $\mathErsy$}

  \begin{lemma}
   \label[lemma]{lemma:Er-essential-property}
   For a type environments $\Gamma$, a lattice for secrecy levels $L$,
   a secrecy level ${l}\in{L}$ and a process $P$,
   if $\mathof{\Gamma}{x}$ is not lower than $l$ in $L$,
   then $x$ does not occur in $\mathEr{\Gamma}{L, l}{P}$.
  \end{lemma}

  \begin{proof}
   By induction on the construction of $P$.
  \end{proof}

  \begin{lemma}
   \label[lemma]{lemma:Er-property-core-type}
   For type environments $\Gamma$ and $\Gamma'$, a lattice for secrecy levels $L$,
   a secrecy level ${l}\in{L}$ and a process $P$,
   if ${\mathof{\Gamma}{x}}\mathequivexceptusagessy{\mathof{\Gamma'}{x}}$
   for any value $x$ occurring in $P$,
   then ${\mathEr{\Gamma}{L, l}{P}}\equiv{\mathEr{\Gamma'}{L, l}{P}}$.
  \end{lemma}

  \begin{proof}
   By induction on the construction of $P$.
  \end{proof}

  \begin{lemma}
   \label{lemma:Er-property-weakning-lattice}
   For a type environments $\Gamma$, lattices for secrecy levels $L, L'$,
   a secrecy level $l \in{L}$, and a process $P$,
   if $\mathissublattice{L'}{L}$ and 
   ${\mathEr{\Gamma}{L', l}{P}}\mathpistructuraleq{\mathnil}$,
   then ${\mathEr{\Gamma}{L, l}{P}}\mathpistructuraleq{\mathnil}$.
  \end{lemma}

  \begin{proof}
   By induction on the construction of $P$. 
  \end{proof}

  \begin{lemma}
   \label{lemma:Er-property-weakning-level} 
   For a type environment $\Gamma$,a lattice for secrecy levels $L$,
   secrecy levels $l, l' \in{L}$, and a process $P$,
   if ${l}\leq_{L} {l'}$ and 
   ${\mathEr{\Gamma}{L, l'}{P}}\mathpistructuraleq{\mathnil}$,
   then ${\mathEr{\Gamma}{L, l}{P}}\mathpistructuraleq{\mathnil}$.
  \end{lemma}

  \begin{proof}
   By induction on the construction of $P$. 
  \end{proof}

  \begin{lemma}
   \label{lemma:Er-derivable-process-to-nil}
   If $\mathtypejudgementwithsecrecy{\Gamma}{L}{m}{P}$
   is $k$-securely derivable, 
   then ${\mathEr{\Gamma}{L, l}{P}}\mathpistructuraleq{\mathnil}$
   for any secrecy level ${l}\not \geq_{L}{m}$.
  \end{lemma}

  \begin{proof}
   Assume that
   $\mathtypejudgementwithsecrecy{\Gamma}{L}{m}{P}$ 
   is $k$-securely derivable.
   We show ${\mathEr{\Gamma}{L, l}{P}}\equiv{\mathnil}$ by induction on 
   $k$-secure derivation tree of
   $\mathtypejudgementwithsecrecy{\Gamma}{L}{l}{P}$.
   We proceed by a case analysis of the rule used at the root.
   
   \noindent Case 1. In case the rule used at the root is \rulename{T-Zero}, 
   the claimed result holds obviously.

   \noindent Case 2. Assume that the rule used at the root is \rulename{T-New}.
   In this case, there exists a process $P'$ such that
   ${P}\equiv{\mathpinew{\mathistype{x}{\xi}}{P'}}$.
   Then 
   $\mathtypejudgementwithsecrecy{\Gamma, {\mathistype{x}{\xi/U}}}{L}{m}{P'}$ 
   is derivable.
   Let ${l}\not \geq_{L}{m}$.
   By the induction hypothesis,
   ${\mathEr{\Gamma, {\mathistype{x}{\xi/U}}}{L, l}{P'}}\mathpistructuraleq{\mathnil}$.

   Assume that
   $\xi$ is the form $\mathprogramtypetuple{\tau_{1}, \dots, \tau_{n}}^{l''}$
   for ${l''}\leq_{L}{l}$.
   Then ${\mathEr{\Gamma}{L, l}{\mathpinew{\mathistype{x}{\xi}}{P'}}}\equiv{{\mathpinew{\mathistype{x}{\xi}}{\mathEr{\Gamma, \mathistype{x}{\xi/\mathusagenil}}{L, l}{P'}}}}$.
   By \cref{lemma:Er-property-core-type}, we have
   ${\mathpinew{\mathistype{x}{\xi}}{\mathEr{\Gamma, \mathistype{x}{\xi/\mathusagenil}}{L, l}{P'}}}\mathpistructuraleq{\mathpinew{\mathistype{x}{\xi}}{\mathnil}}\mathpistructuraleq{\mathnil}$.
   Hence, we see ${\mathEr{\Gamma}{L, l}{P}}\mathpistructuraleq{\mathnil}$.

   Assume that
   $\xi$ is not the form $\mathprogramtypetuple{\tau_{1}, \dots, \tau_{n}}^{l''}$
   with ${l''}\leq_{L}{l}$.
   Then 
   ${\mathEr{\Gamma}{L, l}{\mathpinew{\mathistype{x}{\xi}}{P'}}}\equiv{\mathEr{\Gamma, \mathistype{x}{\xi/\mathusagenil}}{L, l}{P'}}$.
   By \cref{lemma:Er-property-core-type},
   ${\mathEr{\Gamma, {\mathistype{x}{\xi/U}}}{L, l}{P'}}\mathpistructuraleq{\mathnil}$.
   Hence, we have ${\mathEr{\Gamma}{L, l}{P}}\mathpistructuraleq{\mathnil}$.

   \noindent Case 3. Assume that the rule used at the root is \rulename{T-Rep}.
   In this case, there exist a process $P'$ and a type environment $\Gamma'$ such that
   ${P}\equiv{\mathpireplication{P'}}$ and ${\Gamma}={\mathpireplication{\Gamma'}}$.
   Then $\mathtypejudgementwithsecrecy{\Gamma'}{L}{m}{P'}$ is derivable.
   Let ${l}\not \geq_{L}{m}$.
   By the induction hypothesis,
   ${\mathEr{\Gamma'}{L, l}{P'}}\mathpistructuraleq{\mathnil}$.
   By \cref{lemma:Er-property-core-type},
   ${\mathEr{\Gamma}{L, l}{P'}}\mathpistructuraleq{\mathnil}$.
   Hence, ${\mathEr{\Gamma}{L, l}{P}}\equiv{\mathnil}$.

   \noindent Case 4. Assume that the rule used at the root is \rulename{T-Par}.
   In this case, there exist processes $P_{0}$ and $P_{1}$ such that
   ${P}\equiv{\mathpiparallel{P_{0}}{P_{1}}}$.
   Then, there exist type environments $\Gamma_{0}$ and $\Gamma_{1}$ such that
   ${\Gamma}\equiv{\mathpiparallel{\Gamma_{0}}{\Gamma_{1}}}$ and
   $\mathtypejudgementwithsecrecy{\Gamma_{i}}{L}{m}{P_{i}}$ is derivable
   for ${i}={0, 1}$.
   Let ${l}\not \geq_{L}{m}$.
   By the induction hypothesis,
   ${\mathEr{\Gamma_{i}}{L, l}{P_{i}}}\mathpistructuraleq{\mathnil}$ for ${i}={0, 1}$.
   By \cref{lemma:Er-property-core-type}, we have
   ${\mathEr{\Gamma_{i}}{L, l}{P_{i}}} \equiv {\mathEr{\Gamma}{L, l}{P_{i}}}$ for ${i}={0, 1}$.
   Then
   ${\mathEr{\Gamma}{L, l}{\mathpiparallel{P_{0}}{P_{1}}}}\mathpistructuraleq{\mathpiparallel{\mathnil}{\mathnil}}\mathpistructuraleq{\mathnil}$.

   \noindent Case 5. Assume that the rule used at the root is \rulename{T-If}.
   In this case,  there exist a type environment $\Gamma'$ and
   processes $P_{0}$ and $P_{1}$ such that
   ${\Gamma}\equiv{{\mathpiparallel{\Gamma'}{\mathistype{v}{\mathbooltypewithsec{l}}}}}$ 
   and ${P}\equiv{\mathobif{v}{P_{0}}{P_{1}}}$ hold and 
   $\mathtypejudgementwithsecrecy{\Gamma'}{L}{m}{P_{i}}$ is derivable for ${i}={0, 1}$.
   Let ${l}\not \geq_{L}{m}$. 
   Since ${\mathof{\Gamma}{v}}={\mathbooltypewithsec{l}}$,
   we see ${\mathEr{\Gamma}{L, l}{\mathobif{v}{P_{0}}{P_{1}}}}\equiv{\mathnil}$.

   \noindent Case 6. Assume that the rule used at the root is \rulename{T-Out}.
   In this case, 
   there exist a process $P'$, a type environments $\Gamma'$,
   secrecy levels ${l'_{0}}, {l'_{1}}\in{L}$, types ${\mathvect{\tau}}$, a usage $U$ and
   ${t_{c}}\in{\mathnat \cup \mathsetextension{\infty}}$ such that
   ${P}\equiv{\mathpioutput{x}{\mathvect{v}}.P'}$, ${m}\leq_{L}{l'_{1}}$ and
   ${\Gamma}\equiv{\mathpiparallel{\mathlevelraise{t_{c}+1}{t_{c}+1}{\mathpiparallel{\Gamma'}{\mathistype{\mathvect{v}}{\mathlevelraiseuni{\mathvect{\tau}}}}}}{\mathistype{x}{\mathprogramtypetuple{\mathvect{\tau}}^{l'_{0}}/O^{0}_{t_{c}}U}}}$ with ${m}\leq_{L}{l'_{0}}$,
   $t_{c} = \infty$ implies ${l'_{0}}\leq_{L}{l'_{1}}$, and
   $\mathtypejudgementwithsecrecy{\Gamma', {\mathistype{x}{\mathprogramtypetuple{\mathvect{\tau}}^{l'_{0}}/U}}}{L}{l'_{1}}{P'}$ is derivable.
   Let ${l}\not \geq_{L}{m}$. 
   Since ${m}\leq_{L}{l'_{1}}$, we have ${l}\not \geq_{L}{l'_{1}}$. 
   By the induction hypothesis,
   ${\mathEr{\Gamma', {\mathistype{x}{\mathprogramtypetuple{\mathvect{\tau}}^{l'_{0}}/U}}}{L, l}{P'}}\mathpistructuraleq{\mathnil}$.
   Since ${l}\not \geq_{L}{m}$ and ${m}\leq_{L}{l'_{0}}$,
   we have ${l'_{0}}\not \leq_{L}{l}$. 
   Since ${l'_{0}}\not \leq_{L}{l}$, we have
   \[
    {\mathEr{\mathpiparallel{\mathlevelraise{t_{c}+1}{t_{c}+1}{\mathpiparallel{\Gamma'}{\mathistype{\mathvect{v}}{\mathlevelraiseuni{\mathvect{\tau}}}}}}{\mathistype{x}{\mathprogramtypetuple{\mathvect{\tau}}^{l'_{0}}/O^{0}_{t_{c}}U}}}{L, l}{\mathpioutput{x}{\mathvect{v}}.P'}}\equiv{\mathEr{\mathpiparallel{\mathlevelraise{t_{c}+1}{t_{c}+1}{\mathpiparallel{\Gamma'}{\mathistype{\mathvect{v}}{\mathlevelraiseuni{\mathvect{\tau}}}}}}{\mathistype{x}{\mathprogramtypetuple{\mathvect{\tau}}^{l'_{0}}/O^{0}_{t_{c}}U}}}{L, l}{P'}}.
   \]
   By \cref{lemma:Er-property-core-type},
   we have
   \[
   {\mathEr{\mathpiparallel{\mathlevelraise{t_{c}+1}{t_{c}+1}{\mathpiparallel{\Gamma'}{\mathistype{\mathvect{v}}{\mathlevelraiseuni{\mathvect{\tau}}}}}}{\mathistype{x}{\mathprogramtypetuple{\mathvect{\tau}}^{l'_{0}}/O^{0}_{t_{c}}U}}}{L, l}{\mathpioutput{x}{\mathvect{v}}.P'}}\equiv{\mathEr{\Gamma', {\mathistype{x}{\mathprogramtypetuple{\mathvect{\tau}}^{l'_{0}}/U}}}{L, l}{P'}}\mathpistructuraleq{\mathnil}.
   \]

   \noindent Case 7. Assume that the rule used at the root is \rulename{T-In}.
   In this case, there exist a process $P'$, a type environments $\Gamma'$,
   secrecy levels ${l'_{0}}, {l'_{1}}\in{L}$, types ${\mathvect{\tau}}$, a usage $U$ and
   ${t_{c}}\in{\mathnat \cup \mathsetextension{\infty}}$
   such that
   ${P}\equiv{\mathpiinput{x}{\mathtuple{\mathvect{y}}}.P'}$,
   ${\Gamma}\equiv{\mleft(\mathlevelraise{t_{c}+1}{t_{c}+1}{\Gamma'}, \mathistype{x}{\mathprogramtypetuple{\mathvect{\tau}}^{l'_{0}}/I^{0}_{t_{c}}U}\mright)}$,
   ${m}\leq_{L}{l'_{0}}$, and ${m}\leq_{L}{l'_{1}}$ hold,
   $t_{c} = \infty$ implies ${l'_{0}}\leq_{L}{l'_{1}}$, and
   $\mathtypejudgementwithsecrecy{\Gamma', {\mathistype{x}{\mathprogramtypetuple{\mathvect{\tau}}^{l'_{0}}/U}}, {\mathistype{\mathvect{y}}{\mathvect{\tau}}}}{L}{l'_{1}}{P'}$
   is derivable.
   Let ${l}\not \geq_{L}{m}$.
   Since ${m}\leq_{L}{l'_{1}}$, we have ${l}\not \geq_{L}{l'_{1}}$. 
   By the induction hypothesis,
   we have ${\mathEr{\Gamma', {\mathistype{x}{\mathprogramtypetuple{\mathvect{\tau}}^{l'_{0}}/U}}, {\mathistype{\mathvect{y}}{\mathvect{\tau}}}}{L, l}{P'}}\mathpistructuraleq{\mathnil}$.
   Since ${l}\not \geq_{L}{m}$ and ${m}\leq_{L}{l'_{0}}$, 
   we have ${l'_{0}}\not \leq_{L}{l}$. 
   Since ${l'_{0}} \not \leq_{L} {l}$, we have
   \[
   {\mathEr{\mleft(\mathlevelraise{t_{c}+1}{t_{c}+1}{\Gamma'}, \mathistype{x}{\mathprogramtypetuple{\mathvect{\tau}}^{l'_{0}}/I^{0}_{t_{c}}U}\mright)}{L, l}{\mathpiinput{x}{\mathtuple{\mathvect{y}}}.P'}}\equiv{\mathEr{\mleft(\mathlevelraise{t_{c}+1}{t_{c}+1}{\Gamma'}, \mathistype{x}{\mathprogramtypetuple{\mathvect{\tau}}^{l'_{0}}/I^{0}_{t_{c}}U}\mright), {\mathistype{\mathvect{y}}{\mathvect{\tau}}}}{L, l}{P'}}.
   \]
   By \cref{lemma:Er-property-core-type},
   we have
   \[
   {\mathEr{\mleft(\mathlevelraise{t_{c}+1}{t_{c}+1}{\Gamma'}, \mathistype{x}{\mathprogramtypetuple{\mathvect{\tau}}^{l'_{0}}/I^{0}_{t_{c}}U}\mright)}{L, l}{\mathpiinput{x}{\mathtuple{\mathvect{y}}}.P'}}\equiv{\mathEr{\mleft(\mathlevelraise{t_{c}+1}{t_{c}+1}{\Gamma'}, \mathistype{x}{\mathprogramtypetuple{\mathvect{\tau}}^{l'_{0}}/I^{0}_{t_{c}}U}\mright), {\mathistype{\mathvect{y}}{\mathvect{\tau}}}}{L, l}{P'}}\mathpistructuraleq{\mathnil}.
   \]

   \noindent Case 8. Assume that the rule used at the root is \rulename{T-NewSec}.
   In this case, 
   $m \leq_{L} l'$ for any $l''\in\mathvect{l_{2}}, \mathvect{l_{3}}$, and
   there exist a process $P'$ such that
   ${P}\equiv{\mathpinewsecrecylevel{l_{1}}{\mathvect{l_{2}}}{\mathvect{l_{3}}}{P'}}$.
   Then, the assumption of the rule instance
   $\mathtypejudgementwithsecrecy{\Gamma}{\mathpinewsecrecylevel{l_{1}}{\mathvect{l_{2}}}{\mathvect{l_{3}}}{L}}{m}{P'}$ 
   is derivable.
   Let ${l}\not \geq_{L}{m}$.
   By the induction hypothesis, we have 
   ${\mathEr{\Gamma}{\mathpinewsecrecylevel{l_{1}}{\mathvect{l_{2}}}{\mathvect{l_{3}}}{L}, l}{P'}}\mathpistructuraleq{\mathnil}$.
   For any $l'\in\mathvect{l_{2}}$ and $\mathvect{l_{3}}$,
   because of  $m \leq_{L} l'$, we have $l' \not\leq_{L} l$.
   Then 
   ${\mathEr{\Gamma}{L, l}{P}}\equiv{\mathEr{\Gamma}{\mathpinewsecrecylevel{l_{1}}{\mathvect{l_{2}}}{\mathvect{l_{3}}}{L}, l}{P'}}\mathpistructuraleq{\mathnil}$.

   \noindent Case 9. Assume that the rule used at the root is \rulename{T-Weak}.
   In this case, there exist a type environments $\Gamma'$,
   a lattice for secrecy levels $L'$, a secrecy level ${l'}\in{L'}$ such that
   $\mathissubtype{\Gamma}{\Gamma'}$, $\mathissublattice{L'}{L}$, and 
   $l\leq_{L}l'$, and
   $\mathtypejudgementwithsecrecy{\Gamma'}{L'}{l'}{P}$ is derivable.
   Let ${l}\not \geq_{L}{m}$. 
   By the induction hypothesis, we have
   ${\mathEr{\Gamma'}{L', l'}{P}}\mathpistructuraleq{\mathnil}$.
   By \cref{lemma:Er-property-core-type}, \cref{lemma:Er-property-weakning-lattice}, and \cref{lemma:Er-property-weakning-level},
   we see ${\mathEr{\Gamma}{L, l}{P}}\mathpistructuraleq{\mathnil}$.
  \end{proof}

  \subsection{${\mathEr{\Gamma}{L, l}{P}}$ can simulate $P$}

  \begin{lemma}
   \label{lemma:Er-and-structual-po}
   If $\mathtypejudgementwithsecrecy{\Gamma}{L}{m}{P}$
   is $k$-securely derivable,
   the secrecy level of $\mathtypeenvandsecrecylatice{\Gamma}{L}$ is $l_{1}$, and
   ${P}\mathpistructuralpo{P'}$,
   then ${\mathEr{\Gamma}{L, l}{P}}\mathpistructuralpo{\mathEr{\Gamma}{L, l}{P'}}$ for any secrecy level ${l}\not \leq_{L}{l_{1}}$.
  \end{lemma}

  \begin{proof}
   Assume that $\mathtypejudgementwithsecrecy{\Gamma}{L}{m}{P}$ 
   is $k$-securely derivable and
   ${P}\mathpistructuralpo{P'}$.
   Let $l_{1}$ be the secrecy level of $\mathtypeenvandsecrecylatice{\Gamma}{L}$.
   Fix ${l}\not \leq_{L}{l_{1}}$.
   We show 
   ${\mathEr{\Gamma}{L, l}{P}}\mathpistructuralpo{\mathEr{\Gamma}{L, l}{P'}}$ by
   induction on the construction of ${P}\mathpistructuralpo{P'}$.
   We consider cases according to the last rule of the construction of 
   ${P}\mathpistructuralpo{P'}$.

   Case 1. If ${P'}\equiv{P}$, 
   then ${\mathEr{\Gamma}{L, l}{P}}\mathpistructuralpo{\mathEr{\Gamma}{L, l}{P'}}$
   obviously.

   Case 2. Assume that there exists a process $Q$ such that
   ${P}\mathpistructuralpo{Q}$ and
   ${Q}\mathpistructuralpo{P'}$.
   By the induction hypothesis, 
   we have ${\mathEr{\Gamma}{L, l}{P}}\mathpistructuralpo{\mathEr{\Gamma}{L, l}{Q}}$
   and ${\mathEr{\Gamma}{L, l}{Q}}\mathpistructuralpo{\mathEr{\Gamma}{L, l}{P'}}$.
   Hence, 
   ${\mathEr{\Gamma}{L, l}{P}}\mathpistructuralpo{\mathEr{\Gamma}{L, l}{P'}}$.

   Case 3. \rulename{SP-Zero1}.
   Assume ${P'}\equiv{\mathpiparallel{P}{\mathnil}}$.
   Then 
   ${\mathEr{\Gamma}{L, l}{P'}}\equiv{\mathpiparallel{\mathEr{\Gamma}{L, l}{P}}{\mathEr{\Gamma}{L, l}{\mathnil}}}\equiv{\mathpiparallel{\mathEr{\Gamma}{L, l}{P}}{\mathnil}}$.
   Hence,
   ${\mathEr{\Gamma}{L, l}{P}}\mathpistructuralpo{\mathpiparallel{\mathEr{\Gamma}{L, l}{P}}{\mathnil}}\equiv{\mathEr{\Gamma}{L, l}{P'}}$.

   In the same way, we can show 
   ${\mathEr{\Gamma}{L, l}{P}}\mathpistructuralpo{\mathEr{\Gamma}{L, l}{P'}}$
   in case ${P}\equiv{\mathpiparallel{P'}{\mathnil}}$.
   
   Case 4. \rulename{SP-Zero2}.
   Assume ${P}\equiv{\mathnil}$ and 
   ${P'}\equiv{\mathpinew{\mathistype{x}{\xi}}{\mathnil}}$.
   Then ${\mathEr{\Gamma}{L, l}{P}}\equiv{\mathnil}$.
   If $\xi$ is the form $\mathprogramtypetuple{\tau_{1}, \dots, \tau_{n}}^{l'}$
   with ${l'}\leq_{L}{l}$, then
   ${\mathEr{\Gamma}{L, l}{P'}}\equiv{\mathpinew{\mathistype{x}{\xi}}{\mathEr{\Gamma, \mathistype{x}{\xi/\mathusagenil}}{L, l}{\mathnil}}}\equiv{\mathpinew{\mathistype{x}{\xi}}{\mathnil}}$.
   If $\xi$ is not the form $\mathprogramtypetuple{\tau_{1}, \dots, \tau_{n}}^{l'}$
   with ${l'}\leq_{L}{l}$, then
   ${\mathEr{\Gamma}{L, l}{P'}}\equiv{\mathnil}$.
   In both cases, we have 
   ${\mathEr{\Gamma}{L, l}{P}}\mathpistructuralpo{\mathEr{\Gamma}{L, l}{P'}}$.

   In the same way, we can show 
   ${\mathEr{\Gamma}{L, l}{P}}\mathpistructuralpo{\mathEr{\Gamma}{L, l}{P'}}$
   in case ${P}\equiv{\mathpinew{\mathistype{x}{\xi}}{\mathnil}}$ and 
   ${P'}\equiv{\mathnil}$.

   Case 5. \rulename{SP-Commut}.
   Assume ${P}\equiv{\mathpiparallel{P_{0}}{P_{1}}}$ and 
   ${P'}\equiv{\mathpiparallel{P_{1}}{P_{0}}}$ with processes $P_{0}$ and $P_{1}$.
   Then 
   ${\mathEr{\Gamma}{L, l}{P}}\equiv{\mathpiparallel{\mathEr{\Gamma}{L, l}{P_{0}}}{\mathEr{\Gamma}{L, l}{P_{1}}}}$ 
   and
   ${\mathEr{\Gamma}{L, l}{P'}}\equiv{\mathpiparallel{\mathEr{\Gamma}{L, l}{P_{1}}}{\mathEr{\Gamma}{L, l}{P_{0}}}}$.
   We have ${\mathEr{\Gamma}{L, l}{P}}\mathpistructuralpo{\mathEr{\Gamma}{L, l}{P'}}$.

   Case 6. \rulename{SP-Assoc}.
   Assume ${P}\equiv{\mathpiparallel{\mleft(\mathpiparallel{P_{0}}{P_{1}}\mright)}{P_{2}}}$
   and ${P'}\equiv{\mathpiparallel{P_{0}}{\mleft(\mathpiparallel{P_{1}}{P_{2}}\mright)}}$ 
   with processes $P_{0}$, $P_{1}$, and $P_{2}$.
   Then 
   ${\mathEr{\Gamma}{L, l}{P}}\equiv{\mathpiparallel{\mleft(\mathpiparallel{\mathEr{\Gamma}{L, l}{P_{0}}}{\mathEr{\Gamma}{L, l}{P_{1}}}\mright)}{\mathEr{\Gamma}{L, l}{P_{2}}}}$ 
   and
   ${\mathEr{\Gamma}{L, l}{P'}}\equiv{\mathpiparallel{\mathEr{\Gamma}{L, l}{P_{0}}}{\mleft(\mathpiparallel{\mathEr{\Gamma}{L, l}{P_{1}}}{\mathEr{\Gamma}{L, l}{P_{2}}}\mright)}}$.
   We have ${\mathEr{\Gamma}{L, l}{P}}\mathpistructuralpo{\mathEr{\Gamma}{L, l}{P'}}$.

   Case 7. \rulename{SP-New}.
   Assume ${P}\equiv{\mathpiparallel{\mathpinew{\mathistype{x}{\xi}}{P_{0}}}{P_{1}}}$ and 
   ${P'}\equiv{\mathpinew{\mathistype{x}{\xi}}{\mathpiparallel{P_{0}}{P_{1}}}}$ 
   with processes $P_{0}$, $P_{1}$, and $x\notin\mathFNof{P_{1}}$.
   If $\xi$ is the form $\mathprogramtypetuple{\tau_{1}, \dots, \tau_{n}}^{l'}$
   with ${l'}\leq_{L}{l}$, then
   ${\mathEr{\Gamma}{L, l}{P}}\equiv{\mathpiparallel{\mathpinew{\mathistype{x}{\xi}}{\mathEr{\Gamma, \mathistype{x}{\xi/\mathusagenil}}{L, l}{P_{0}}}}{\mathEr{\Gamma, \mathistype{x}{\xi/\mathusagenil}}{L, l}{P_{1}}}}$ 
   and
   ${\mathEr{\Gamma}{L, l}{P'}}\equiv{\mathpinew{\mathistype{x}{\xi}}{\mathpiparallel{\mathEr{\Gamma, \mathistype{x}{\xi/\mathusagenil}}{L, l}{P_{0}}}{\mathEr{\Gamma, \mathistype{x}{\xi/\mathusagenil}}{L, l}{P_{1}}}}}$.
   If $\xi$ is not the form $\mathprogramtypetuple{\tau_{1}, \dots, \tau_{n}}^{l'}$
   with ${l'}\leq_{L}{l}$, then
   ${\mathEr{\Gamma}{L, l}{P}}\equiv{\mathpiparallel{\mathEr{\Gamma, \mathistype{x}{\xi/\mathusagenil}}{L, l}{P_{0}}}{\mathEr{\Gamma, \mathistype{x}{\xi/\mathusagenil}}{L, l}{P_{1}}}}$ 
   and
   ${\mathEr{\Gamma}{L, l}{P'}}\equiv{\mathpiparallel{\mathEr{\Gamma, \mathistype{x}{\xi/\mathusagenil}}{L, l}{P_{0}}}{\mathEr{\Gamma, \mathistype{x}{\xi/\mathusagenil}}{L, l}{P_{1}}}}$.
   In both cases, we have 
   ${\mathEr{\Gamma}{L, l}{P}}\mathpistructuralpo{\mathEr{\Gamma}{L, l}{P'}}$.

   Case 8. \rulename{SP-IfT}.
   Assume ${P}\equiv{\mathobif{\mathobtrue^{l'}}{P_{0}}{P_{1}}}$ and ${P'}\equiv{P_{0}}$ 
   with processes $P_{0}$ and $P_{1}$.
   By \cref{lemma:inversion} \cref{item:if-then-lemma-inversion},
   there exist a type environments $\Gamma'$, 
   a lattice for secrecy levels $L'$, and ${l''}\in{L'}$ such that
   $\mathissublattice{L'}{L}$, $m\leq_{L}l''$, and $\mathissubtype{\Gamma}{\mleft(\mathpiparallel{\Gamma'}{\mathistype{\mathobtrue^{l'}}{\mathbooltypewithsec{l''}}}\mright)}$, and both
   $\mathtypejudgementwithsecrecy{\Gamma'}{L'}{l''}{}{P_{0}}$ and
   $\mathtypejudgementwithsecrecy{\Gamma'}{L'}{l''}{}{P_{1}}$ are derivable.
   Then ${l'}={l''}$.
   We consider cases according to $l'$.

   Assume ${l'}\leq_{L}{l}$.
   Then 
   ${\mathEr{\Gamma}{L, l}{\mathobif{\mathobtrue^{l'}}{P_{0}}{P_{1}}}}\equiv{\mathobif{\mathobtrue^{l'}}{\mathEr{\Gamma}{L, l}{P_{0}}}{\mathEr{\Gamma}{L, l}{P_{1}}}}$.
   Hence,
   ${\mathEr{\Gamma}{L, l}{\mathobif{\mathobtrue^{l'}}{P_{0}}{P_{1}}}}\mathpistructuralpo{\mathEr{\Gamma}{L, l}{P_{0}}}$.

   Assume ${l'}\not \leq_{L}{l}$.
   Then
   ${\mathEr{\Gamma}{L, l}{\mathobif{\mathobtrue^{l'}}{P_{0}}{P_{1}}}}\equiv{\mathnil}$.
   By \cref{lemma:Er-derivable-process-to-nil}, we have
   ${\mathEr{\Gamma'}{L', l}{P_{0}}}\mathpistructuraleq{\mathnil}$.
   By \cref{lemma:Er-property-core-type} and \cref{lemma:Er-property-weakning-lattice},
   we have ${\mathEr{\Gamma}{L, l}{P_{0}}}\mathpistructuraleq{\mathnil}$.
   Hence, 
   ${\mathEr{\Gamma}{L, l}{\mathobif{\mathobtrue^{l'}}{P_{0}}{P_{1}}}}\mathpistructuralpo{\mathEr{\Gamma}{L, l}{P_{0}}}$.

   Case 9. \rulename{SP-IfF}.
   Assume ${P}\equiv{\mathobif{\mathobfalse^{l'}}{P_{0}}{P_{1}}}$ and ${P'}\equiv{P_{1}}$ 
   with processes $P_{0}$ and $P_{1}$.
   By \cref{lemma:inversion} \cref{item:if-then-lemma-inversion},
   there exist a type environments $\Gamma'$, 
   a lattice for secrecy levels $L'$, and ${l''}\in{L'}$ such that
   $\mathissublattice{L'}{L}$, $m\leq_{L}l''$, and $\mathissubtype{\Gamma}{\mleft(\mathpiparallel{\Gamma'}{\mathistype{\mathobfalse^{l'}}{\mathbooltypewithsec{l''}}}\mright)}$, and both
   $\mathtypejudgementwithsecrecy{\Gamma'}{L'}{l''}{}{P_{0}}$ and
   $\mathtypejudgementwithsecrecy{\Gamma'}{L'}{l''}{}{P_{1}}$ are derivable.
   Then ${l'}={l''}$.
   We consider cases according to $l'$.

   Assume ${l'}\leq_{L}{l}$.
   Then 
   ${\mathEr{\Gamma}{L, l}{\mathobif{\mathobfalse^{l'}}{P_{0}}{P_{1}}}}\equiv{\mathobif{\mathobfalse^{l'}}{\mathEr{\Gamma}{L, l}{P_{0}}}{\mathEr{\Gamma}{L, l}{P_{1}}}}$.
   Hence,
   ${\mathEr{\Gamma}{L, l}{\mathobif{\mathobfalse^{l'}}{P_{0}}{P_{1}}}}\mathpistructuralpo{\mathEr{\Gamma}{L, l}{P_{1}}}$.

   Assume ${l'}\not \leq_{L}{l}$.
   Then
   ${\mathEr{\Gamma}{L, l}{\mathobif{\mathobfalse^{l'}}{P_{0}}{P_{1}}}}\equiv{\mathnil}$.
   By \cref{lemma:Er-derivable-process-to-nil}, we have
   ${\mathEr{\Gamma'}{L', l}{P_{1}}}\mathpistructuraleq{\mathnil}$.
   By \cref{lemma:Er-property-core-type} and \cref{lemma:Er-property-weakning-lattice},
   we have ${\mathEr{\Gamma}{L, l}{P_{1}}}\mathpistructuraleq{\mathnil}$.
   Hence, 
   ${\mathEr{\Gamma}{L, l}{\mathobif{\mathobfalse^{l'}}{P_{0}}{P_{1}}}}\mathpistructuralpo{\mathEr{\Gamma}{L, l}{P_{1}}}$.

   Case 10. \rulename{SP-Rep}.
   Assume ${P}\equiv{\mathpireplication{P_{0}}}$ and 
   ${P'}\equiv{\mathpiparallel{\mathpireplication{P_{0}}}{P_{0}}}$ with a process $P_{0}$.
   We consider cases according to ${\mathEr{\Gamma}{L, l}{P_{0}}}$.
   
   Assume ${\mathEr{\Gamma}{L, l}{P_{0}}}\mathpistructuraleq{\mathnil}$.
   Then, we have ${\mathEr{\Gamma}{L, l}{\mathpireplication{P_{0}}}}\equiv{\mathnil}$.
   We also have 
   ${\mathEr{\Gamma}{L, l}{\mathpiparallel{\mathpireplication{P_{0}}}{P_{0}}}}\equiv{\mathpiparallel{\mathnil}{\mathEr{\Gamma}{L, l}{P_{0}}}}$.
   By \rulename{SP-Zero1} and
   \cref{lemma:structural-po-and-set-of-free-names} \cref{item:po-lemma-structural-po-and-set-of-free-names},
   we have 
   ${\mathEr{\Gamma}{L, l}{\mathpireplication{P_{0}}}}\equiv{\mathnil}\mathpistructuralpo{\mathpiparallel{\mathnil}{\mathnil}}\mathpistructuralpo{\mathpiparallel{\mathnil}{\mathEr{\Gamma}{L, l}{P_{0}}}}\equiv{\mathEr{\Gamma}{L, l}{\mathpiparallel{\mathpireplication{P_{0}}}{P_{0}}}}$.

   Assume ${\mathEr{\Gamma}{L, l}{P_{0}}}\not\mathpistructuraleq{\mathnil}$.
   Then, we have
   ${\mathEr{\Gamma}{L, l}{\mathpireplication{P_{0}}}}\equiv{\mathpireplication{\mathEr{\Gamma}{L, l}{P_{0}}}}$.
   Hence,  we have 
   ${\mathEr{\Gamma}{L, l}{\mathpireplication{P_{0}}}}\equiv{\mathpireplication{\mathEr{\Gamma}{L, l}{P_{0}}}}\mathpistructuralpo{\mathpiparallel{\mathpireplication{\mathEr{\Gamma}{L, l}{P_{0}}}}{\mathEr{\Gamma}{L, l}{P_{0}}}}\mathpistructuralpo{\mathEr{\Gamma}{L, l}{\mathpiparallel{\mathpireplication{P_{0}}}{P_{0}}}}$.

   Case 11. \rulename{SP-Par}.
   Assume ${P}\equiv{\mathpiparallel{P_{0}}{P_{1}}}$ and
   ${P'}\equiv{\mathpiparallel{P'_{0}}{P_{1}}}$
   with ${P_{0}}\mathpistructuralpo{P'_{0}}$ for process $P_{0}$, $P_{1}$, and $P'_{0}$.
   Then 
   ${\mathEr{\Gamma}{L, l}{P}}\equiv{\mathpiparallel{\mathEr{\Gamma}{L, l}{P_{0}}}{\mathEr{\Gamma}{L, l}{P_{1}}}}$ and
   ${\mathEr{\Gamma}{L, l}{P'}}\equiv{\mathpiparallel{\mathEr{\Gamma}{L, l}{P'_{0}}}{\mathEr{\Gamma}{L, l}{P_{1}}}}$.
   By the induction hypothesis, we have
   ${\mathEr{\Gamma}{L, l}{P_{0}}}\mathpistructuralpo{\mathEr{\Gamma}{L, l}{P'_{0}}}$.
   We have ${\mathEr{\Gamma}{L, l}{P}}\mathpistructuralpo{\mathEr{\Gamma}{L, l}{P'}}$.
   
   Case 12. \rulename{SP-CNew}.
   Assume ${P}\equiv{\mathpinew{\mathistype{x}{\xi}}{P_{0}}}$ and
   ${P'}\equiv{\mathpinew{\mathistype{x}{\xi}}{P'_{0}}}$
   with ${P_{0}}\mathpistructuralpo{P'_{0}}$ for process $P_{0}$ and $P'_{0}$.
   By the induction hypothesis, we have
   ${\mathEr{\Gamma, \mathistype{x}{\xi/\mathusagenil}}{L, l}{P_{0}}}\mathpistructuralpo{\mathEr{\Gamma, \mathistype{x}{\xi/\mathusagenil}}{L, l}{P'_{0}}}$.
   If $\xi$ is the form $\mathprogramtypetuple{\tau_{1}, \dots, \tau_{n}}^{l'}$
   with ${l'}\leq_{L}{l}$, then
   ${\mathEr{\Gamma}{L, l}{P}}\equiv{\mathpinew{\mathistype{x}{\xi}}{\mathEr{\Gamma, \mathistype{x}{\xi/\mathusagenil}}{L, l}{P_{0}}}}$ 
   and
   ${\mathEr{\Gamma}{L, l}{P}}\equiv{\mathpinew{\mathistype{x}{\xi}}{\mathEr{\Gamma, \mathistype{x}{\xi/\mathusagenil}}{L, l}{P'_{0}}}}$. 
   If $\xi$ is not the form $\mathprogramtypetuple{\tau_{1}, \dots, \tau_{n}}^{l'}$
   with ${l'}\leq_{L}{l}$, then
   ${\mathEr{\Gamma}{L, l}{P}}\equiv{\mathEr{\Gamma, \mathistype{x}{\xi/\mathusagenil}}{L, l}{P_{0}}}$ 
   and
   ${\mathEr{\Gamma}{L, l}{P'}}\equiv{\mathEr{\Gamma, \mathistype{x}{\xi/\mathusagenil}}{L, l}{P'_{0}}}$.
   In both cases, we have 
   ${\mathEr{\Gamma}{L, l}{P}}\mathpistructuralpo{\mathEr{\Gamma}{L, l}{P'}}$.
  \end{proof}

  \begin{lemma}
   \label{lemma:Er-and-substitution}
   Let $y_{0}, \dots, y_{n}$ be channel names,
   $v_{0}, \dots, v_{n}$ be values and ${v'_{i}}$ be ${\mathEr{\Gamma}{L, l}{v_{i}}}$.
   Let ${\mathvect{y}}=\mathtuple{y_{0}, \dots, y_{n}}$, 
   ${\mathvect{v}}=\mathtuple{v_{0}, \dots, v_{n}}$, and
   ${\mathvect{v'}}=\mathtuple{v'_{0}, \dots, v'_{n}}$.
   
   ${\mathEr{\Gamma}{L, l}{\mathsubstbox{P}{\mathsubst{\mathvect{y}}{\mathvect{v}}}}}\equiv{\mathsubstbox{\mleft(\mathEr{\Gamma, \mathistype{\mathvect{y}}{\mathvect{\tau}}}{L, l}{P}\mright)}{\mathsubst{\mathvect{y}}{\mathvect{v'}}}}$ with 
   ${\mathvect{\tau}}=\mathtuple{\mathof{\Gamma}{v_{0}}, \dots, \mathof{\Gamma}{v_{n}}}$
   for a value or process $P$.
  \end{lemma}

  \begin{proof}
   We show
   ${\mathEr{\Gamma}{L, l}{\mathsubstbox{P}{\mathsubst{\mathvect{y}}{\mathvect{v}}}}}\equiv{\mathsubstbox{\mleft(\mathEr{\Gamma, \mathistype{\mathvect{y}}{\mathvect{\tau}}}{L, l}{P}\mright)}{\mathsubst{\mathvect{y}}{\mathvect{v'}}}}$
   by induction on the construction of $P$.
   We consider cases according to the form of $P$.

   Case 1. Assume ${P}\equiv{w}$ for a value $w$.
   
   Assume ${w}\not\equiv{y_{i}}$ for ${i}=0, \dots, n$.
   Then
   ${\mathEr{\Gamma}{L, l}{\mathsubstbox{w}{\mathsubst{\mathvect{y}}{\mathvect{v}}}}}\equiv{\mathEr{\Gamma}{L, l}{w}}$.
   By \cref{lemma:Er-property-core-type},
   ${\mathEr{\Gamma}{L, l}{w}}\equiv{\mathsubstbox{\mleft(\mathEr{\Gamma, \mathistype{\mathvect{y}}{\mathvect{\tau}}}{L, l}{w}\mright)}{\mathsubst{\mathvect{y}}{\mathvect{v'}}}}$.

   Assume ${w}\equiv{y_{i}}$ for ${i}=0, \dots, n$.
   Then 
   ${\mathEr{\Gamma}{L, l}{\mathsubstbox{y_{i}}{\mathsubst{\mathvect{y}}{\mathvect{v}}}}}\equiv{\mathEr{\Gamma}{L, l}{v_{i}}}$.
   
   We consider the case where $\mathof{\Gamma}{v_{i}}$ is not lower than $l$ in $L$.
   In this case, ${\mathEr{\Gamma}{L, l}{v_{i}}}\equiv{\mathobunit}$ and
   ${\mathEr{\Gamma, \mathistype{\mathvect{y}}{\mathvect{\tau}}}{L, l}{y_{i}}}\equiv{\mathobunit}$.
   Hence, we have ${\mathEr{\Gamma}{L, l}{v_{i}}}\equiv{\mathsubstbox{\mleft(\mathEr{\Gamma, \mathistype{\mathvect{y}}{\mathvect{\tau}}}{L, l}{y_{i}}\mright)}{\mathsubst{\mathvect{y}}{\mathvect{v'}}}}$.
   
   We consider the case where $\mathof{\Gamma}{v_{i}}$ is lower than $l$ in $L$.
   In this case, ${\mathEr{\Gamma}{L, l}{v_{i}}}\equiv{v_{i}}\equiv{v'_{i}}$ and 
   ${\mathEr{\Gamma, \mathistype{\mathvect{y}}{\mathvect{\tau}}}{L, l}{y_{i}}}\equiv{y_{i}}$.
   Hence, we have
   ${\mathEr{\Gamma}{L, l}{v_{i}}}\equiv{\mathsubstbox{\mleft(\mathEr{\Gamma, \mathistype{\mathvect{y}}{\mathvect{\tau}}}{L, l}{y_{i}}\mright)}{\mathsubst{\mathvect{y}}{\mathvect{v'}}}}$.

   Case 2. Assume ${P}\equiv{\mathnil}$. 
   In this case,
   ${\mathEr{\Gamma}{L, l}{\mathsubstbox{P}{\mathsubst{\mathvect{y}}{\mathvect{v}}}}}\equiv{\mathnil}$ and 
   ${\mathEr{\Gamma, \mathistype{\mathvect{y}}{\mathvect{\tau}}}{L, l}{P}}\equiv{\mathnil}$.
   Hence, we have
   ${\mathEr{\Gamma}{L, l}{\mathsubstbox{P}{\mathsubst{\mathvect{y}}{\mathvect{v}}}}} \equiv {\mathsubstbox{\mleft(\mathEr{\Gamma, \mathistype{\mathvect{y}}{\mathvect{\tau}}}{L, l}{P}\mright)}{\mathsubst{\mathvect{y}}{\mathvect{v'}}}}$.
   
   Case 3. Assume ${P}\equiv{\mathpiparallel{P_{0}}{P_{1}}}$. In this case,
   ${\mathEr{\Gamma}{L, l}{\mathsubstbox{P}{\mathsubst{\mathvect{y}}{\mathvect{v}}}}}\equiv{\mathpiparallel{\mathEr{\Gamma}{L, l}{\mathsubstbox{P_{0}}{\mathsubst{\mathvect{y}}{\mathvect{v}}}}}{\mathEr{\Gamma}{L, l}{\mathsubstbox{P_{1}}{\mathsubst{\mathvect{y}}{\mathvect{v}}}}}}$ and 
   ${\mathEr{\Gamma, \mathistype{\mathvect{y}}{\mathvect{\tau}}}{L, l}{P}}\equiv{\mathpiparallel{\mathEr{\Gamma, \mathistype{\mathvect{y}}{\mathvect{\tau}}}{L, l}{P_{0}}}{\mathEr{\Gamma, \mathistype{\mathvect{y}}{\mathvect{\tau}}}{L, l}{P_{1}}}}$.
   By the induction hypothesis, we have
   ${\mathEr{\Gamma}{L, l}{\mathsubstbox{P_{i}}{\mathsubst{\mathvect{y}}{\mathvect{v}}}}} \equiv {\mathsubstbox{\mleft(\mathEr{\Gamma, \mathistype{\mathvect{y}}{\mathvect{\tau}}}{L, l}{P_{i}}\mright)}{\mathsubst{\mathvect{y}}{\mathvect{v'}}}}$
   for $i=0, 1$.
   Then, we have
   ${\mathEr{\Gamma}{L, l}{\mathsubstbox{P}{\mathsubst{\mathvect{y}}{\mathvect{v}}}}} \equiv {\mathsubstbox{\mleft(\mathEr{\Gamma, \mathistype{\mathvect{y}}{\mathvect{\tau}}}{L, l}{P}\mright)}{\mathsubst{\mathvect{y}}{\mathvect{v'}}}}$.

   Case 4. Assume ${P}\equiv{\mathpireplication{P'}}$. 
   By the induction hypothesis, we have 
   ${\mathEr{\Gamma}{L, l}{\mathsubstbox{P'}{\mathsubst{\mathvect{y}}{\mathvect{v}}}}} \equiv {\mathsubstbox{\mleft(\mathEr{\Gamma, \mathistype{\mathvect{y}}{\mathvect{\tau}}}{L, l}{P'}\mright)}{\mathsubst{\mathvect{y}}{\mathvect{v'}}}}$.
   We consider cases according to the form of ${\mathEr{\Gamma}{L, l}{\mathsubstbox{P'}{\mathsubst{\mathvect{y}}{\mathvect{v}}}}}$.

   Assume ${\mathEr{\Gamma}{L, l}{\mathsubstbox{P'}{\mathsubst{\mathvect{y}}{\mathvect{v}}}}}\mathpistructuraleq{\mathnil}$.
   Then 
   ${\mathEr{\Gamma}{L, l}{\mathpireplication{\mathsubstbox{P'}{\mathsubst{\mathvect{y}}{\mathvect{v}}}}}}\equiv{\mathnil}$.
   Since
   ${\mathsubstbox{\mleft(\mathEr{\Gamma, \mathistype{\mathvect{y}}{\mathvect{\tau}}}{L, l}{P'}\mright)}{\mathsubst{\mathvect{y}}{\mathvect{v'}}}}\mathpistructuraleq{\mathnil}$,
   we have
   ${\mathEr{\Gamma, \mathistype{\mathvect{y}}{\mathvect{\tau}}}{L, l}{P'}}\mathpistructuraleq{\mathnil}$.
   Hence,
   ${\mathEr{\Gamma, \mathistype{\mathvect{y}}{\mathvect{\tau}}}{L, l}{\mathpireplication{P'}}}\equiv{\mathnil}$.
   Then, we have
   ${\mathsubstbox{\mleft(\mathEr{\Gamma, \mathistype{\mathvect{y}}{\mathvect{\tau}}}{L, l}{\mathpireplication{P'}}\mright)}{\mathsubst{\mathvect{y}}{\mathvect{v'}}}}\equiv{\mathnil}$.
   Thus,
   ${\mathEr{\Gamma}{L, l}{\mathsubstbox{P}{\mathsubst{\mathvect{y}}{\mathvect{v}}}}} \equiv {\mathsubstbox{\mleft(\mathEr{\Gamma, \mathistype{\mathvect{y}}{\mathvect{\tau}}}{L, l}{P}\mright)}{\mathsubst{\mathvect{y}}{\mathvect{v'}}}}$.

   Assume ${\mathEr{\Gamma}{L, l}{\mathsubstbox{P'}{\mathsubst{\mathvect{y}}{\mathvect{v}}}}}\not\mathpistructuraleq{\mathnil}$.
   Then
   ${\mathEr{\Gamma}{L, l}{\mathsubstbox{P}{\mathsubst{\mathvect{y}}{\mathvect{v}}}}}\equiv{\mathpireplication{\mathEr{\Gamma}{L, l}{\mathsubstbox{P'}{\mathsubst{\mathvect{y}}{\mathvect{v}}}}}}$ and 
   ${\mathEr{\Gamma, \mathistype{\mathvect{y}}{\mathvect{\tau}}}{L, l}{P}}\equiv{\mathpireplication{\mathEr{\Gamma, \mathistype{\mathvect{y}}{\mathvect{\tau}}}{L, l}{P'}}}$.
   We have
   \begin{align*}
    {\mathEr{\Gamma}{L, l}{\mathsubstbox{P}{\mathsubst{\mathvect{y}}{\mathvect{v}}}}}
    &\equiv{\mathpireplication{\mathEr{\Gamma}{L, l}{\mathsubstbox{P'}{\mathsubst{\mathvect{y}}{\mathvect{v}}}}}} \\
    &\equiv {\mathsubstbox{\mleft(\mathpireplication{\mathEr{\Gamma, \mathistype{\mathvect{y}}{\mathvect{\tau}}}{L, l}{P'}}\mright)}{\mathsubst{\mathvect{y}}{\mathvect{v'}}}} \\
    &\equiv {\mathsubstbox{\mleft(\mathEr{\Gamma, \mathistype{\mathvect{y}}{\mathvect{\tau}}}{L, l}{P}\mright)}{\mathsubst{\mathvect{y}}{\mathvect{v'}}}}.
   \end{align*}
   Thus,
   ${\mathEr{\Gamma}{L, l}{\mathsubstbox{P}{\mathsubst{\mathvect{y}}{\mathvect{v}}}}} \equiv {\mathsubstbox{\mleft(\mathEr{\Gamma, \mathistype{\mathvect{y}}{\mathvect{\tau}}}{L, l}{P}\mright)}{\mathsubst{\mathvect{y}}{\mathvect{v'}}}}$.

   Case 5. Assume ${P}\equiv{\mathpioutput{x}{\mathvect{w}}.P'}$. 
   Let ${x'}\equiv{\mathsubstbox{x}{\mathsubst{\mathvect{y}}{\mathvect{v'}}}}$ and
   ${\mathvect{w'}}\equiv{\mathsubstbox{\mathvect{w}}{\mathsubst{\mathvect{y}}{\mathvect{v'}}}}$.
   We consider cases according to the form of $\mathof{\Gamma}{x}$.
   
   Assume that $\mathof{\Gamma}{x}$ is the form
   $\mathprogramtypetuple{\tau_{1}, \dots, \tau_{n}}^{l'}/U$ with ${l'}\leq_{L}{l}$.
   Then, we have 
   ${\mathEr{\Gamma}{L, l}{\mathsubstbox{P}{\mathsubst{\mathvect{y}}{\mathvect{v}}}}}\equiv{\mathpioutput{x'}{\mathvect{w'}}. \mathEr{\Gamma}{L, l}{\mathsubstbox{P'}{\mathsubst{\mathvect{y}}{\mathvect{v}}}}}$ and 
   ${\mathEr{\Gamma, \mathistype{\mathvect{y}}{\mathvect{\tau}}}{L, l}{P}}\equiv{\mathpioutput{x}{\mathvect{w}}. \mathEr{\Gamma, \mathistype{\mathvect{y}}{\mathvect{\tau}}}{L, l}{P'}}$.
   By the induction hypothesis, we have \linebreak[4]
   ${\mathEr{\Gamma}{L, l}{\mathsubstbox{P'}{\mathsubst{\mathvect{y}}{\mathvect{v}}}}} \equiv {\mathsubstbox{\mleft(\mathEr{\Gamma, \mathistype{\mathvect{y}}{\mathvect{\tau}}}{L, l}{P'}\mright)}{\mathsubst{\mathvect{y}}{\mathvect{v'}}}}$.
   Then, we have
   \begin{align*}
    {\mathEr{\Gamma}{L, l}{\mathsubstbox{P}{\mathsubst{\mathvect{y}}{\mathvect{v}}}}}
    &\equiv{\mathpioutput{x'}{\mathvect{w'}}. \mathEr{\Gamma}{L, l}{\mathsubstbox{P'}{\mathsubst{\mathvect{y}}{\mathvect{v}}}}} \\
    &\equiv {\mathpioutput{x'}{\mathvect{w'}}. \mathsubstbox{\mleft({\mathEr{\Gamma, \mathistype{\mathvect{y}}{\mathvect{\tau}}}{L, l}{P'}}\mright)}{\mathsubst{\mathvect{y}}{\mathvect{v'}}}} \\
    &\equiv {\mathsubstbox{\mleft({\mathpioutput{x}{\mathvect{w}}. \mathEr{\Gamma, \mathistype{\mathvect{y}}{\mathvect{\tau}}}{L, l}{P'}}\mright)}{\mathsubst{\mathvect{y}}{\mathvect{v'}}}} \\
    &\equiv {\mathsubstbox{\mleft(\mathEr{\Gamma, \mathistype{\mathvect{y}}{\mathvect{\tau}}}{L, l}{P}\mright)}{\mathsubst{\mathvect{y}}{\mathvect{v'}}}}.
   \end{align*}

   Assume that $\mathof{\Gamma}{x}$ is \emph{not} the form
   $\mathprogramtypetuple{\tau_{1}, \dots, \tau_{n}}^{l'}/U$ with ${l'}\leq_{L}{l}$.
   Then, we have 
   ${\mathEr{\Gamma}{L, l}{\mathsubstbox{P}{\mathsubst{\mathvect{y}}{\mathvect{v}}}}}\equiv{\mathEr{\Gamma}{L, l}{\mathsubstbox{P'}{\mathsubst{\mathvect{y}}{\mathvect{v}}}}}$ and 
   ${\mathEr{\Gamma, \mathistype{\mathvect{y}}{\mathvect{\tau}}}{L, l}{P}}\equiv{\mathEr{\Gamma, \mathistype{\mathvect{y}}{\mathvect{\tau}}}{L, l}{P'}}$.
   By the induction hypothesis, we have 
   ${\mathEr{\Gamma}{L, l}{\mathsubstbox{P'}{\mathsubst{\mathvect{y}}{\mathvect{v}}}}} \equiv {\mathsubstbox{\mleft(\mathEr{\Gamma, \mathistype{\mathvect{y}}{\mathvect{\tau}}}{L, l}{P'}\mright)}{\mathsubst{\mathvect{y}}{\mathvect{v'}}}}$.
   Then, we have 
   ${\mathEr{\Gamma}{L, l}{\mathsubstbox{P}{\mathsubst{\mathvect{y}}{\mathvect{v}}}}} \equiv {\mathsubstbox{\mleft(\mathEr{\Gamma, \mathistype{\mathvect{y}}{\mathvect{\tau}}}{L, l}{P}\mright)}{\mathsubst{\mathvect{y}}{\mathvect{v'}}}}$.

   Case 6. Assume ${P}\equiv{\mathpiinput{x}{\mathvect{z}}.P'}$. 
   Let ${x'}\equiv{\mathsubstbox{x}{\mathsubst{\mathvect{y}}{\mathvect{v'}}}}$.
   We consider cases according to the form of $\mathof{\Gamma}{x}$.

   Assume that $\mathof{\Gamma}{x}$ is the form 
   $\mathprogramtypetuple{\tau_{1}, \dots, \tau_{n}}^{l'}/U$
   with ${l'}\leq_{L}{l}$.
   Then, we have
   $\mathEr{\Gamma}{L, l}{\mathsubstbox{P}{\mathsubst{\mathvect{y}}{\mathvect{v}}}} \equiv \mathpiinput{x'}{\mathvect{z}}. \mathEr{\Gamma, \mathistype{\mathvect{z}}{\mathvect{\tau'}} }{L, l}{\mathsubstbox{P'}{\mathsubst{\mathvect{y}}{\mathvect{v}}}}$ and 
   ${\mathEr{\Gamma, \mathistype{\mathvect{y}}{\mathvect{\tau}}}{L, l}{P}}\equiv{\mathpiinput{x}{\mathvect{z}}. \mathEr{\Gamma, \mathistype{\mathvect{y}}{\mathvect{\tau}}, \mathistype{\mathvect{z}}{\mathvect{\tau'}}}{L, l}{P'}}$.
   By the induction hypothesis, we have 
   ${\mathEr{\Gamma, \mathistype{\mathvect{z}}{\mathvect{\tau'}}}{L, l}{\mathsubstbox{P'}{\mathsubst{\mathvect{y}}{\mathvect{v}}}}} \equiv {\mathsubstbox{\mleft(\mathEr{\Gamma, \mathistype{\mathvect{y}}{\mathvect{\tau}}, \mathistype{\mathvect{z}}{\mathvect{\tau'}}}{L, l}{P'}\mright)}{\mathsubst{\mathvect{y}}{\mathvect{v'}}}}$.
   Then, we have
   \begin{align*}
    {\mathEr{\Gamma}{L, l}{\mathsubstbox{P}{\mathsubst{\mathvect{y}}{\mathvect{v}}}}}
    &\equiv{\mathpiinput{x'}{\mathvect{z}}. \mathEr{\Gamma, \mathistype{\mathvect{z}}{\mathvect{\tau'}}}{L, l}{\mathsubstbox{P'}{\mathsubst{\mathvect{y}}{\mathvect{v}}}}} \\
    &\equiv {\mathpiinput{x'}{\mathvect{z}}. \mathsubstbox{\mleft({\mathEr{\Gamma, \mathistype{\mathvect{y}}{\mathvect{\tau}}, \mathistype{\mathvect{z}}{\mathvect{\tau'}}}{L, l}{P'}}\mright)}{\mathsubst{\mathvect{y}}{\mathvect{v'}}}} \\
    &\equiv {\mathsubstbox{\mleft({\mathpiinput{x}{\mathvect{z}}. \mathEr{\Gamma, \mathistype{\mathvect{y}}{\mathvect{\tau}}}{L, l}{P'}}\mright)}{\mathsubst{\mathvect{y}}{\mathvect{v'}}}} \\
    &\equiv {\mathsubstbox{\mleft(\mathEr{\Gamma, \mathistype{\mathvect{y}}{\mathvect{\tau}}}{L, l}{P}\mright)}{\mathsubst{\mathvect{y}}{\mathvect{v'}}}}.
   \end{align*}

   Assume that $\mathof{\Gamma}{x}$ is the form 
   $\mathprogramtypetuple{\tau_{1}, \dots, \tau_{n}}^{l'}/U$ with ${l'}\not \leq_{L}{l}$.
   Then 
   ${\mathEr{\Gamma}{L, l}{\mathsubstbox{P}{\mathsubst{\mathvect{y}}{\mathvect{v}}}}}\equiv{\mathEr{\Gamma, \mathistype{\mathvect{z}}{\mathvect{\tau'}}}{L, l}{\mathsubstbox{P'}{\mathsubst{\mathvect{y}}{\mathvect{v}}}}}$ and 
   ${\mathEr{\Gamma, \mathistype{\mathvect{y}}{\mathvect{\tau}}}{L, l}{P}}\equiv{\mathEr{\Gamma, \mathistype{\mathvect{y}}{\mathvect{\tau}}, \mathistype{\mathvect{z}}{\mathvect{\tau'}}}{L, l}{P'}}$.
   By the induction hypothesis, we have \linebreak[4]
   $\mathEr{\Gamma, \mathistype{\mathvect{z}}{\mathvect{\tau'}}}{L, l}{\mathsubstbox{P'}{\mathsubst{\mathvect{y}}{\mathvect{v}}}} \equiv \mathsubstbox{\mleft(\mathEr{\Gamma, \mathistype{\mathvect{y}}{\mathvect{\tau}}, \mathistype{\mathvect{z}}{\mathvect{\tau'}}}{L, l}{P'}\mright)}{\mathsubst{\mathvect{y}}{\mathvect{v'}}}$.
   Then, we have
   ${\mathEr{\Gamma}{L, l}{\mathsubstbox{P}{\mathsubst{\mathvect{y}}{\mathvect{v}}}}} \equiv {\mathsubstbox{\mleft(\mathEr{\Gamma, \mathistype{\mathvect{y}}{\mathvect{\tau}}}{L, l}{P}\mright)}{\mathsubst{\mathvect{y}}{\mathvect{v'}}}}$.

   In case that $\mathof{\Gamma}{x}$ is not the form 
   $\mathprogramtypetuple{\tau_{1}, \dots, \tau_{n}}^{l'}/U$ for any $l'$,
   we can show
   ${\mathEr{\Gamma}{L, l}{\mathsubstbox{P}{\mathsubst{\mathvect{y}}{\mathvect{v}}}}} \equiv {\mathsubstbox{\mleft(\mathEr{\Gamma, \mathistype{\mathvect{y}}{\mathvect{\tau}}}{L, l}{P}\mright)}{\mathsubst{\mathvect{y}}{\mathvect{v'}}}}$
   in  the similar way to the case that $\mathof{\Gamma}{x}$ is the form 
   $\mathprogramtypetuple{\tau_{1}, \dots, \tau_{n}}^{l'}/U$ with ${l'}\not \leq_{L}{l}$.

   Case 7. Assume ${P}\equiv{\mathpinew{\mathistype{x}{\xi}}{P'}}$.
   We consider cases according to the form of $\xi$.

   Assume that $\xi$ is the form 
   $\mathprogramtypetuple{\tau_{1}, \dots, \tau_{n}}^{l'}$ with ${l'}\leq_{L}{l}$.
   Then, we have \linebreak[4]
   $\mathEr{\Gamma}{L, l}{\mathsubstbox{P}{\mathsubst{\mathvect{y}}{\mathvect{v}}}} \equiv  \mathpinew{\mathistype{x}{\xi}}{\mathEr{\Gamma, \mathistype{x}{\xi/\mathusagenil} }{L, l}{\mathsubstbox{P'}{\mathsubst{\mathvect{y}}{\mathvect{v}}}}}$ and 
   ${\mathEr{\Gamma, \mathistype{\mathvect{y}}{\mathvect{\tau}}}{L, l}{P}}\equiv{\mathpinew{\mathistype{x}{\xi}}{\mathEr{\Gamma, \mathistype{\mathvect{y}}{\mathvect{\tau}}, \mathistype{x}{\xi/\mathusagenil}}{L, l}{P'}}}$.
   By the induction hypothesis, we have 
   ${\mathEr{\Gamma, \mathistype{x}{\xi/\mathusagenil}}{L, l}{\mathsubstbox{P'}{\mathsubst{\mathvect{y}}{\mathvect{v}}}}} \equiv {\mathsubstbox{\mleft(\mathEr{\Gamma, \mathistype{\mathvect{y}}{\mathvect{\tau}}, \mathistype{x}{\xi/\mathusagenil}}{L, l}{P'}\mright)}{\mathsubst{\mathvect{y}}{\mathvect{v'}}}}$.
   Then, we have
   \begin{align*}
    \mathEr{\Gamma}{L, l}{\mathsubstbox{P}{\mathsubst{\mathvect{y}}{\mathvect{v}}}} 
    &\equiv  \mathpinew{\mathistype{x}{\xi}}{\mathEr{\Gamma, \mathistype{x}{\xi/\mathusagenil} }{L, l}{\mathsubstbox{P'}{\mathsubst{\mathvect{y}}{\mathvect{v}}}}} \\
    &\equiv \mathpinew{\mathistype{x}{\xi}}{\mathsubstbox{\mleft(\mathEr{\Gamma, \mathistype{\mathvect{y}}{\mathvect{\tau}}, \mathistype{x}{\xi/\mathusagenil}}{L, l}{P'}\mright)}{\mathsubst{\mathvect{y}}{\mathvect{v'}}}} \\
    &\equiv {\mathsubstbox{\mleft(\mathEr{\Gamma, \mathistype{\mathvect{y}}{\mathvect{\tau}}}{L, l}{P}\mright)}{\mathsubst{\mathvect{y}}{\mathvect{v'}}}}.
   \end{align*}

   Assume that $\xi$ is not the form 
   $\mathprogramtypetuple{\tau_{1}, \dots, \tau_{n}}^{l'}$ with ${l'}\leq_{L}{l}$.
   Then 
   ${\mathEr{\Gamma}{L, l}{\mathsubstbox{P}{\mathsubst{\mathvect{y}}{\mathvect{v}}}}}\equiv{\mathEr{\Gamma, \mathistype{x}{\xi/\mathusagenil}}{L, l}{\mathsubstbox{P'}{\mathsubst{\mathvect{y}}{\mathvect{v}}}}}$ and 
   ${\mathEr{\Gamma, \mathistype{\mathvect{y}}{\mathvect{\tau}}}{L, l}{P}}\equiv{\mathEr{\Gamma, \mathistype{\mathvect{y}}{\mathvect{\tau}}, \mathistype{x}{\xi/\mathusagenil}}{L, l}{P'}}$.
   By the induction hypothesis, we have 
   \[
    \mathEr{\Gamma, \mathistype{x}{\xi/\mathusagenil}}{L, l}{\mathsubstbox{P'}{\mathsubst{\mathvect{y}}{\mathvect{v}}}} \equiv \mathsubstbox{\mleft(\mathEr{\Gamma, \mathistype{\mathvect{y}}{\mathvect{\tau}}, \mathistype{x}{\xi/\mathusagenil}}{L, l}{P'}\mright)}{\mathsubst{\mathvect{y}}{\mathvect{v'}}}.
   \]
   Then, we have
   ${\mathEr{\Gamma}{L, l}{\mathsubstbox{P}{\mathsubst{\mathvect{y}}{\mathvect{v}}}}} \equiv {\mathsubstbox{\mleft(\mathEr{\Gamma, \mathistype{\mathvect{y}}{\mathvect{\tau}}}{L, l}{P}\mright)}{\mathsubst{\mathvect{y}}{\mathvect{v'}}}}$.

   Case 8. Assume
   ${P}\equiv{\mathpinewsecrecylevel{l_{0}}{\mathvect{l_{1}}}{\mathvect{l_{2}}}{P'}}$. 
   Let $L'=\mathpinewsecrecylevel{l_{0}}{\mathvect{l_{1}}}{\mathvect{l_{2}}}{L}$
   We consider cases according to $l$.

   Assume $l' \leq_{L} l$ for any  $l'\in\mathvect{l_{1}}$, $\mathvect{l_{2}}$.
   In this case, 
   ${\mathEr{\Gamma}{L, l}{\mathsubstbox{P}{\mathsubst{\mathvect{y}}{\mathvect{v}}}}}\equiv\mathpinewsecrecylevel{l_{0}}{\mathvect{l_{1}}}{\mathvect{l_{2}}}{\mathEr{\Gamma}{L', l}{\mathsubstbox{P'}{\mathsubst{\mathvect{y}}{\mathvect{v}}}}}$
   and 
   ${\mathEr{\Gamma, \mathistype{\mathvect{y}}{\mathvect{\tau}}}{L, l}{P}}\equiv \mathpinewsecrecylevel{l_{0}}{\mathvect{l_{1}}}{\mathvect{l_{2}}}{\mathEr{\Gamma, \mathistype{\mathvect{y}}{\mathvect{\tau}} }{L', l}{P'}}$.
   By the induction hypothesis, we have 
   ${\mathEr{\Gamma}{L', l}{\mathsubstbox{P'}{\mathsubst{\mathvect{y}}{\mathvect{v}}}}} \equiv {\mathsubstbox{\mleft(\mathEr{\Gamma, \mathistype{\mathvect{y}}{\mathvect{\tau}}}{L', l}{P'}\mright)}{\mathsubst{\mathvect{y}}{\mathvect{v'}}}}$.
   Then, we have
   ${\mathEr{\Gamma}{L, l}{\mathsubstbox{P}{\mathsubst{\mathvect{y}}{\mathvect{v}}}}} \equiv {\mathsubstbox{\mleft(\mathEr{\Gamma, \mathistype{\mathvect{y}}{\mathvect{\tau}}}{L, l}{P}\mright)}{\mathsubst{\mathvect{y}}{\mathvect{v'}}}}$.

   Assume $l' \not\leq_{L} l$ for some  $l'\in\mathvect{l_{1}}$, $\mathvect{l_{2}}$.
   In this case,
   ${\mathEr{\Gamma}{L, l}{\mathsubstbox{P}{\mathsubst{\mathvect{y}}{\mathvect{v}}}}}\equiv{\mathEr{\Gamma}{L', l}{\mathsubstbox{P'}{\mathsubst{\mathvect{y}}{\mathvect{v}}}}}$ 
   and 
   ${\mathEr{\Gamma, \mathistype{\mathvect{y}}{\mathvect{\tau}}}{L, l}{P}}\equiv {\mathEr{\Gamma, \mathistype{\mathvect{y}}{\mathvect{\tau}} }{L', l}{P'}}$.
   By the induction hypothesis, we have 
   ${\mathEr{\Gamma}{L', l}{\mathsubstbox{P'}{\mathsubst{\mathvect{y}}{\mathvect{v}}}}} \equiv {\mathsubstbox{\mleft(\mathEr{\Gamma, \mathistype{\mathvect{y}}{\mathvect{\tau}}}{L', l}{P'}\mright)}{\mathsubst{\mathvect{y}}{\mathvect{v'}}}}$.
   Then, we have
   ${\mathEr{\Gamma}{L, l}{\mathsubstbox{P}{\mathsubst{\mathvect{y}}{\mathvect{v}}}}} \equiv {\mathsubstbox{\mleft(\mathEr{\Gamma, \mathistype{\mathvect{y}}{\mathvect{\tau}}}{L, l}{P}\mright)}{\mathsubst{\mathvect{y}}{\mathvect{v'}}}}$.

   Case 9. Assume ${P}\equiv{\mathobif{w}{P_{0}}{P_{1}}}$. 
   Let ${\mathvect{w'}}\equiv{\mathsubstbox{\mathvect{w}}{\mathsubst{\mathvect{y}}{\mathvect{v'}}}}$.
   We consider cases according to the form of $\mathof{\Gamma}{w}$.

   Assume that $\mathof{\Gamma}{w}$ is the form 
   $\mathprogramtypetuple{\tau_{1}, \dots, \tau_{n}}^{l'}/U$ with ${l'}\leq_{L}{l}$.
   Then, we have
   ${\mathEr{\Gamma}{L, l}{\mathsubstbox{P}{\mathsubst{\mathvect{y}}{\mathvect{v}}}}}\equiv{\mathobif{w'}{\mathEr{\Gamma}{L, l}{\mathsubstbox{P_{0}}{\mathsubst{\mathvect{y}}{\mathvect{v}}}}}{\mathEr{\Gamma}{L, l}{\mathsubstbox{P_{1}}{\mathsubst{\mathvect{y}}{\mathvect{v}}}}}}$ and 
   ${\mathEr{\Gamma, \mathistype{\mathvect{y}}{\mathvect{\tau}}}{L, l}{P}}\equiv{\mathobif{w}{\mathEr{\Gamma}{L, l}{P_{0}}}{\mathEr{\Gamma}{L, l}{P_{1}}}}$.
   By the induction hypothesis, we have 
   ${\mathEr{\Gamma}{\mathpinewsecrecylevel{l_{0}}{\mathvect{l_{1}}}{\mathvect{l_{2}}}{L}, l}{\mathsubstbox{P_{i}}{\mathsubst{\mathvect{y}}{\mathvect{v}}}}} \equiv {\mathsubstbox{\mleft(\mathEr{\Gamma, \mathistype{\mathvect{y}}{\mathvect{\tau}}}{L, l}{P_{i}}\mright)}{\mathsubst{\mathvect{y}}{\mathvect{v'}}}}$ 
   for $i=0, 1$.
   Then, we have
   ${\mathEr{\Gamma}{L, l}{\mathsubstbox{P}{\mathsubst{\mathvect{y}}{\mathvect{v}}}}} \equiv {\mathsubstbox{\mleft(\mathEr{\Gamma, \mathistype{\mathvect{y}}{\mathvect{\tau}}}{L, l}{P}\mright)}{\mathsubst{\mathvect{y}}{\mathvect{v'}}}}$.

   Assume that $\mathof{\Gamma}{w}$ is not the form 
   $\mathprogramtypetuple{\tau_{1}, \dots, \tau_{n}}^{l'}/U$ with ${l'}\leq_{L}{l}$.
   Then, we have
   ${\mathEr{\Gamma}{L, l}{\mathsubstbox{P}{\mathsubst{\mathvect{y}}{\mathvect{v}}}}}\equiv\mathnil$ and 
   ${\mathEr{\Gamma, \mathistype{\mathvect{y}}{\mathvect{\tau}}}{L, l}{P}}\equiv \mathnil$.
   Then, we have
   ${\mathEr{\Gamma}{L, l}{\mathsubstbox{P}{\mathsubst{\mathvect{y}}{\mathvect{v}}}}} \equiv {\mathsubstbox{\mleft(\mathEr{\Gamma, \mathistype{\mathvect{y}}{\mathvect{\tau}}}{L, l}{P}\mright)}{\mathsubst{\mathvect{y}}{\mathvect{v'}}}}$.
  \end{proof}

  \begin{lemma}
   \label[lemma]{lemma:Er-simulate-reduction}
   \begin{enumerate}
    \item If $\mathtypejudgementwithsecrecy{\Gamma}{L}{m}{P}$
   is $k$-securely derivable, and
   ${\mathtuple{P, L}}\mathpisecreduction{\Gamma}{\hat{l}}{\mathtuple{\hat{P}, \hat{L}}}$,
   then
   ${\mathEr{\Gamma}{L, \hat{l}}{P}}\mathpistructuralpo{\mathEr{\Gamma}{\hat{L}, \hat{l}}{\hat{P}}}$. 
	  \label{item:high-lemma-Er-simulate-reduction}
    \item If $\mathtypejudgementwithsecrecy{\Gamma}{L}{m}{P}$
	  is $k$-securely derivable, and
	  ${\mathtuple{P, L}}\not\mathpisecreduction{\Gamma}{\hat{l}}{\mathtuple{\hat{P}, \hat{L}}}$
	  but
	  ${\mathtuple{P, L}}\mathpireduction{\mathtuple{\hat{P}, \hat{L}}}$,
	  then, for any $\mathEr{\Gamma}{L, \hat{l}}{P}$-sublattice $L'$ of $L$,
	  there exists a lattice for secrecy levels $\hat{L}'$ such that
	  ${\mathtuple{\mathEr{\Gamma}{L, \hat{l}}{P}, L'}}\mathpireduction{\mathtuple{\mathEr{\Gamma}{\hat{L}, \hat{l}}{\hat{P}}, \hat{L}'}}$ and
	  $\hat{L}'$ is an $\mathEr{\Gamma}{\hat{L}, \hat{l}}{\hat{P}}$-sublattice of $\hat{L}$. 
	  \label{item:law-lemma-Er-simulate-reduction}
   \end{enumerate}
  \end{lemma}

  \begin{proof}
   We show each statements

   \noindent \cref{item:high-lemma-Er-simulate-reduction}
   Assume that $\mathtypejudgementwithsecrecy{\Gamma}{L}{m}{P}$
   is $k$-securely derivable, and
   ${\mathtuple{P, L}}\mathpisecreduction{\Gamma}{\hat{l}}{\mathtuple{\hat{P}, \hat{L}}}$.
   By induction on the construction of
   ${\mathtuple{P, L}}\mathpireduction{\mathtuple{\hat{P}, \hat{L}}}$,
   we prove 
   ${\mathEr{\Gamma}{L, \hat{l}}{P}}\mathpistructuralpo{\mathEr{\Gamma}{\hat{L}, \hat{l}}{\hat{P}}}$.
   We consider cases according to the last rule of the construction of 
   ${\mathtuple{P, L}}\mathpisecreduction{\Gamma}{\hat{l}}{\mathtuple{\hat{P}, \hat{L}}}$.

   Case \cref{item:r-com-def-reduction-under-Gamma}.
   In this case, ${P}\equiv \mathpiparallel{\mathpioutput{x}{\mathvect{v}}. P_{0}}{\mathpiinput{x}{\mathvect{y}}.P_{1}}$
   and
   ${\hat{P}}\equiv{\mathpiparallel{P_{0}}{\mathsubstbox{P_{1}}{\mathsubst{\mathvect{y}}{\mathvect{v}}}}}$
   with ${\mathvect{y}}=\mathtuple{y_{0}, \dots, y_{n}}$ and
   ${\mathvect{v}}=\mathtuple{v_{0}, \dots, v_{n}}$.
   We also have ${\hat{L}}={L}$.
   Then
   ${\mathEr{\Gamma}{L, \hat{l}}{P}}\equiv{\mathpiparallel{\mathEr{\Gamma}{L, \hat{l}}{\mathpioutput{x}{\mathvect{v}}. P_{0}}}{\mathEr{\Gamma}{L, \hat{l}}{\mathpiinput{x}{\mathvect{y}}. P_{1}}}}$ 
   and
   ${\mathEr{\Gamma}{L, \hat{l}}{\hat{P}}}\equiv{\mathpiparallel{{\mathEr{\Gamma}{L, \hat{l}}{P_{0}}}}{\mathEr{\Gamma}{L, \hat{l}}{\mathsubstbox{P_{1}}{\mathsubst{\mathvect{y}}{\mathvect{v}}}}}}$.
   By \cref{lemma:inversion}, $\mathof{\Gamma}{x}$ is the form 
   $\mathprogramtypetuple{\tau_{0}, \dots, \tau_{n}}^{l'}/U$, where
   ${\tau_{i}}\mathequivexceptusagessy{\mathof{\Gamma}{v_{i}}}$ for $i = 0, \dots, n$.
   Since 
   ${\mathtuple{P, L}}\mathpisecreduction{\Gamma}{\hat{l}}{\mathtuple{\hat{P}, \hat{L}}}$,
   we have ${l'}\not \leq_{L}{\hat{l}}$.
   Then, we see
   ${\mathEr{\Gamma}{L, \hat{l}}{P}}\equiv \mathpiparallel{\mathEr{\Gamma}{L, \hat{l}}{P_{0}}}{\mathEr{\Gamma, \mathistype{\mathvect{y}}{\mathvect{\tau}}}{L, \hat{l}}{P_{1}}}$ 
   and
   ${\mathEr{\Gamma}{L, \hat{l}}{\hat{P}}}\equiv \mathpiparallel{{\mathEr{\Gamma}{L, \hat{l}}{P_{0}}}}{\mathEr{\Gamma}{L, \hat{l}}{\mathsubstbox{P_{1}}{\mathsubst{\mathvect{y}}{\mathvect{v}}}}}$.
   By \cref{lemma:Er-and-substitution}, we have
   \[
   {\mathEr{\Gamma}{L, \hat{l}}{\mathsubstbox{P_{1}}{\mathsubst{\mathvect{y}}{\mathvect{v}}}}}\equiv{\mathsubstbox{\mleft(\mathEr{\Gamma, \mathistype{y_{0}}{\mathof{\Gamma}{v_{n}}}, \dots, \mathistype{y_{n}}{\mathof{\Gamma}{v_{n}}}}{L, \hat{l}}{P_{1}}\mright)}{\mathsubst{\mathvect{y}}{\mathvect{v'}}}}.
   \]
   Because $\mathtypeenvandsecrecylatice{\Gamma}{L}$ is secure,
   $l' \leq_{L} l''$ for any secrecy type $l''$ occurring in $\tau_{i}$ with $i=0, 1, \dots$.
   Hence, $\tau_{i}$ is not lower than $\hat{l}$ for any $i=0, \dots, n$.
   Therefore, $\mathof{\Gamma}{v_{i}}$ is not lower than $\hat{l}$ for any $i=0, \dots, n$.
   By \cref{lemma:Er-essential-property},
   $y_{i}$ does not occur in $\mathEr{\Gamma, \mathistype{y_{0}}{\mathof{\Gamma}{v_{n}}}, \dots, \mathistype{y_{n}}{\mathof{\Gamma}{v_{n}}}}{L, \hat{l}}{P_{1}}$ for any $i=0, \dots, n$.
   Hence,
   \[
    {\mathsubstbox{\mleft(\mathEr{\Gamma, \mathistype{y_{0}}{\mathof{\Gamma}{v_{n}}}, \dots, \mathistype{y_{n}}{\mathof{\Gamma}{v_{n}}}}{L, \hat{l}}{P_{1}}\mright)}{\mathsubst{\mathvect{y}}{\mathvect{v'}}}}\equiv{\mleft(\mathEr{\Gamma, \mathistype{y_{0}}{\mathof{\Gamma}{v_{n}}}, \dots, \mathistype{y_{n}}{\mathof{\Gamma}{v_{n}}}}{L, \hat{l}}{P_{1}}\mright)}.
   \]
   By \cref{lemma:Er-property-core-type}, we have
   ${\mleft(\mathEr{\Gamma, \mathistype{y_{0}}{\mathof{\Gamma}{v_{n}}}, \dots, \mathistype{y_{n}}{\mathof{\Gamma}{v_{n}}}}{L, \hat{l}}{P_{1}}\mright)}\equiv{\mathEr{\Gamma, \mathistype{\mathvect{y}}{\mathvect{\tau}}}{L, \hat{l}}{P_{1}}}$.
   Thus, 
   ${\mathEr{\Gamma}{L, \hat{l}}{P}}\mathpistructuralpo{\mathEr{\Gamma}{\hat{L}, \hat{l}}{\hat{P}}}$.
   Then $L'$ is an $\mathEr{\Gamma}{\hat{L}, \hat{l}}{\hat{P}}$-sublattice of $\hat{L}$.

   Case \cref{item:r-newlev-reduction-under-Gamma}.
   Let ${\mathvect{l_{0}}}$, ${\mathvect{l_{1}}}\subseteq{L}$.
   Let ${P}\equiv{\mathpinewsecrecylevel{l'}{\mathvect{l_{0}}}{\mathvect{l_{1}}}{P'}}$.
   Assume that
   ${\mathpinewsecrecylevel{l'}{\mathvect{l_{0}}}{\mathvect{l_{1}}}{L}}$ is defined.
   Then ${\hat{P}}\equiv{P'}$ and
   ${\hat{L}}={\mathpinewsecrecylevel{l'}{\mathvect{l_{0}}}{\mathvect{l_{1}}}{L}}$.
   Since 
   ${\mathtuple{P, L}}\mathpisecreduction{\Gamma}{\hat{l}}{\mathtuple{\hat{P}, \hat{L}}}$,
   we have $l' \not\leq_{L} \hat{l}$ for some  $l'\in\mathvect{l_{0}}$, $\mathvect{l_{1}}$.
   Then
   ${\mathEr{\Gamma}{L, \hat{l}}{P}}\equiv{\mathEr{\Gamma}{\hat{L}, \hat{l}}{\hat{P}}}$.
   Hence, 
   ${\mathEr{\Gamma}{L, \hat{l}}{P}}\mathpistructuralpo{\mathEr{\Gamma}{\hat{L}, \hat{l}}{\hat{P}}}$,
   and $L'$ is an $\mathEr{\Gamma}{\hat{L}, \hat{l}}{\hat{P}}$-sublattice of $\hat{L}$.

   Case \cref{item:r-par-reduction-under-Gamma}. Straightforward.

   Case \cref{item:r-new-reduction-under-Gamma}.
   Straightforward.

   Case \cref{item:r-sp-reduction-under-Gamma}.
   Straightforward.

   \noindent \cref{item:law-lemma-Er-simulate-reduction}
   Assume that $\mathtypejudgementwithsecrecy{\Gamma}{L}{m}{P}$
   is $k$-securely derivable, and
   ${\mathtuple{P, L}}\not\mathpisecreduction{\Gamma}{\hat{l}}{\mathtuple{\hat{P}, \hat{L}}}$
   but
   ${\mathtuple{P, L}}\mathpireduction{\mathtuple{\hat{P}, \hat{L}}}$.
   Let $L'$ be an $\mathEr{\Gamma}{L, \hat{l}}{P}$-sublattice of $L$.
   By induction on the construction of
   ${\mathtuple{P, L}}\mathpireduction{\mathtuple{\hat{P}, \hat{L}}}$,
   we prove that
   there exists a lattice for secrecy levels $\hat{L}'$ such that
   ${\mathtuple{\mathEr{\Gamma}{L, \hat{l}}{P}, L'}}\mathpireduction{\mathtuple{\mathEr{\Gamma}{\hat{L}, \hat{l}}{\hat{P}}, \hat{L}'}}$ and
   $\hat{L}'$ is an $\mathEr{\Gamma}{\hat{L}, \hat{l}}{\hat{P}}$-sublattice of $\hat{L}$. 
   We consider cases according to the last rule of the construction of 
   ${\mathtuple{P, L}}\mathpireduction{\mathtuple{\hat{P}, \hat{L}}}$.

   Case 1. \rulename{R-Com}.
   In this case, ${P}\equiv \mathpiparallel{\mathpioutput{x}{\mathvect{v}}. P_{0}}{\mathpiinput{x}{\mathvect{y}}.P_{1}}$
   and
   ${\hat{P}}\equiv{\mathpiparallel{P_{0}}{\mathsubstbox{P_{1}}{\mathsubst{\mathvect{y}}{\mathvect{v}}}}}$
   with ${\mathvect{y}}=\mathtuple{y_{0}, \dots, y_{n}}$ and
   ${\mathvect{v}}=\mathtuple{v_{0}, \dots, v_{n}}$.
   We also have ${\hat{L}}={L}$.
   Then
   ${\mathEr{\Gamma}{L, \hat{l}}{P}}\equiv{\mathpiparallel{\mathEr{\Gamma}{L, \hat{l}}{\mathpioutput{x}{\mathvect{v}}. P_{0}}}{\mathEr{\Gamma}{L, \hat{l}}{\mathpiinput{x}{\mathvect{y}}. P_{1}}}}$ 
   and
   ${\mathEr{\Gamma}{L, \hat{l}}{\hat{P}}}\equiv{\mathpiparallel{{\mathEr{\Gamma}{L, \hat{l}}{P_{0}}}}{\mathEr{\Gamma}{L, \hat{l}}{\mathsubstbox{P_{1}}{\mathsubst{\mathvect{y}}{\mathvect{v}}}}}}$.
   By \cref{lemma:inversion}, $\mathof{\Gamma}{x}$ is the form 
   $\mathprogramtypetuple{\tau_{0}, \dots, \tau_{n}}^{l'}/U$, where
   ${\tau_{i}}\mathequivexceptusagessy{\mathof{\Gamma}{v_{i}}}$ for $i = 0, \dots, n$. 
   We consider cases according to $l'$.
   Since 
   ${\mathtuple{P, L}}\not\mathpisecreduction{\Gamma}{\hat{l}}{\mathtuple{\hat{P}, \hat{L}}}$,
   we have ${l'}\leq_{L}{\hat{l}}$.
   Let ${v'_{i}}\equiv{\mathEr{\Gamma}{L, {\hat{l}}}{v_{i}}}$ for all $i=0, \dots, n$.
   Let 
   ${\mathvect{v'}}=\mathtuple{v'_{0}, \dots, v'_{n}}$, and
   ${\mathvect{\tau}}=\mathprogramtypetuple{\tau_{0}, \dots, \tau_{n}}$.
   Then, we have
   ${\mathEr{\Gamma}{L, \hat{l}}{P}}\equiv \mathpiparallel{\mathpioutput{x}{\mathvect{v'}}. \mathEr{\Gamma}{L, \hat{l}}{P_{0}}}{\mathpiinput{x}{\mathvect{y}}. \mathEr{\Gamma, \mathistype{\mathvect{y}}{\mathvect{\tau}}}{L, \hat{l}}{P_{1}}}$ 
   and
   ${\mathEr{\Gamma}{L, \hat{l}}{\hat{P}}}\equiv{\mathpiparallel{{\mathEr{\Gamma}{L, \hat{l}}{P_{0}}}}{\mathEr{\Gamma}{L, \hat{l}}{\mathsubstbox{P_{1}}{\mathsubst{\mathvect{y}}{\mathvect{v}}}}}}$.
   By \cref{lemma:Er-and-substitution}, we have
   \[
   {\mathEr{\Gamma}{L, \hat{l}}{\mathsubstbox{P_{1}}{\mathsubst{\mathvect{y}}{\mathvect{v}}}}}\equiv{\mathsubstbox{\mleft(\mathEr{\Gamma, \mathistype{y_{0}}{\mathof{\Gamma}{v_{n}}}, \dots, \mathistype{y_{n}}{\mathof{\Gamma}{v_{n}}}}{L, \hat{l}}{P_{1}}\mright)}{\mathsubst{\mathvect{y}}{\mathvect{v'}}}}.
   \]
   By \cref{lemma:Er-property-core-type},
   ${\mathEr{\Gamma}{L, \hat{l}}{\mathsubstbox{P_{1}}{\mathsubst{\mathvect{y}}{\mathvect{v}}}}}\equiv{\mathsubstbox{\mleft(\mathEr{\Gamma, \mathistype{\mathvect{y}}{\mathvect{\tau}}}{L, \hat{l}}{P_{1}}\mright)}{\mathsubst{\mathvect{y}}{\mathvect{v'}}}}$.
   Hence, we have
   ${\mathEr{\Gamma}{L, \hat{l}}{P}} \equiv \mathpiparallel{\mathpioutput{x}{\mathvect{v'}}. \mathEr{\Gamma}{L, \hat{l}}{P_{0}}}{\mathpiinput{x}{\mathvect{y}}. \mathEr{\Gamma, \mathistype{\mathvect{y}}{\mathvect{\tau}}}{L, \hat{l}}{P_{1}}}$ and
   \begin{align*}
    {\mathEr{\Gamma}{L, \hat{l}}{\hat{P}}} 
    &\equiv
    {\mathpiparallel{{\mathEr{\Gamma}{L, \hat{l}}{P_{0}}}}{\mathEr{\Gamma}{L, \hat{l}}{\mathsubstbox{P_{1}}{\mathsubst{\mathvect{y}}{\mathvect{v}}}}}} \\
    &\equiv
    {\mathpiparallel{{\mathEr{\Gamma}{L, \hat{l}}{P_{0}}}}{\mathsubstbox{\mleft(\mathEr{\Gamma, \mathistype{\mathvect{y}}{\mathvect{\tau}}}{L, \hat{l}}{P_{1}}\mright)}{\mathsubst{\mathvect{y}}{\mathvect{v'}}}}}.    
   \end{align*}
   Therefore, we have ${\mathtuple{\mathEr{\Gamma}{L, \hat{l}}{P}, L'}}\mathpireduction{\mathtuple{\mathEr{\Gamma}{\hat{L}, \hat{l}}{\hat{P}}, L'}}$
   for any lattice for secrecy levels $L'$, 
   where $L'$ is an $\mathEr{\Gamma}{\hat{L}, \hat{l}}{\hat{P}}$-sublattice of $L$.

   Case 2. \rulename{R-NewLev}.
   Let ${\mathvect{l_{0}}}$, ${\mathvect{l_{1}}}\subseteq{L}$.
   Let ${P}\equiv{\mathpinewsecrecylevel{l'}{\mathvect{l_{0}}}{\mathvect{l_{1}}}{P'}}$.
   Assume that
   ${\mathpinewsecrecylevel{l'}{\mathvect{l_{0}}}{\mathvect{l_{1}}}{L}}$ is defined.
   Then ${\hat{P}}\equiv{P'}$ and
   ${\hat{L}}={\mathpinewsecrecylevel{l'}{\mathvect{l_{0}}}{\mathvect{l_{1}}}{L}}$.
   Since
   ${\mathtuple{P, L}}\not\mathpisecreduction{\Gamma}{\hat{l}}{\mathtuple{\hat{P}, \hat{L}}}$,
   we have $l' \leq_{L} \hat{l}$ for any  $l'\in\mathvect{l_{0}}$, $\mathvect{l_{1}}$.
   Then
   ${\mathEr{\Gamma}{L, \hat{l}}{P}}\equiv \mathpinewsecrecylevel{l'}{\mathvect{l_{0}}}{\mathvect{l_{1}}}{\mathEr{\Gamma}{\hat{L}, \hat{l}}{\hat{P}}}$.
   Let $L'$ be an $\mathEr{\Gamma}{L, \hat{l}}{P}$-sublattice of $L$.
   Then 
   ${\mathtuple{\mathEr{\Gamma}{L, \hat{l}}{P}, L'}}\mathpireduction{\mathtuple{\mathEr{\Gamma}{\hat{L}, \hat{l}}{\hat{P}}, \mathpinewsecrecylevel{l'}{\mathvect{l_{0}}}{\mathvect{l_{1}}}{L'}}}$.
   Then $\mathpinewsecrecylevel{l'}{\mathvect{l_{0}}}{\mathvect{l_{1}}}{L'}$ is
   an $\mathEr{\Gamma}{\hat{L}, \hat{l}}{\hat{P}}$-sublattice of ${\hat{L}}$.

   Case 3. \rulename{R-Par}. Straightforward.

   Case 4. \rulename{R-New}. Straightforward.

   Case 5. \rulename{R-SP}. Straightforward.
  \end{proof}

  \subsection{$P$ can simulate ${\mathEr{\Gamma}{L, l}{P}}$}

  \begin{lemma}
   \label[lemma]{lemma:P-simulate-Er-structualpo}
   For a reliable type environment $\Gamma$,
   if $\mathtypejudgementwithsecrecy{\Gamma}{L}{m}{P}$ 
   is $k$-securely derivable,
   ${P'}\mathpistructuralpo{\mathEr{\Gamma}{L, k}{P}}$ and
   ${P'}\mathpistructuralpo{\bar{P}'}$, 
   then there exist $\bar{P}$, $\bar{L}$ and $\bar{\Gamma}$ such that
   ${\mathtuple{P, L}}\mathpireductionkc{\mathtuple{\bar{P}, \bar{L}}}$,
   ${\Gamma}\mathtypeenvreductionkc{\bar{\Gamma}}$
   and
   ${\bar{P}'}\mathpistructuralpo{\mathEr{\bar{\Gamma}}{\bar{L}, k}{\bar{P}}}$. 
  \end{lemma}

  \begin{proof}
   Assume that
   $\mathtypejudgementwithsecrecy{\Gamma}{L}{m}{P}$
   is $k$-securely derivable,
   ${P'}\mathpistructuralpo{\mathEr{\Gamma}{L, k}{P}}$ and
   ${P'}\mathpistructuralpo{\bar{P}'}$. 
   By induction on the construction of ${P'}\mathpistructuralpo{\bar{P}'}$,
   we show that there exist $\bar{P}$, $\bar{L}$, and $\bar{\Gamma}$ such that
   ${\mathtuple{P, L}}\mathpireductionkc{\mathtuple{\bar{P}, \bar{L}}}$,
   ${\Gamma}\mathtypeenvreductionkc{\bar{\Gamma}}$ and
   ${\bar{P}'}\mathpistructuralpo{\mathEr{\bar{\Gamma}}{\bar{L}, k}{\bar{P}}}$.
   We consider cases according to the last rule of the construction of
   ${P'}\mathpistructuralpo{\bar{P}'}$.

   Case 1. Assume ${\bar{P}'}\equiv{P'}$. 
   Let $\bar{P}\equiv{P}$, $\bar{L}\equiv{L}$ and ${\bar{\Gamma}}\equiv{\Gamma}$.
   Then, we have the claimed result.

   Case 2. Assume tbar there exists a process $P_{0}$ such that
   ${P'}\mathpistructuralpo{P_{0}}$ and ${P_{0}}\mathpistructuralpo{\bar{P}'}$.
   By the induction hypothesis,
   we see tbar there exist $\bar{P}_{0}$, $\bar{L}_{0}$ and $\bar{\Gamma}_{0}$ such tbar
   ${\mathtuple{P, L}}\mathpireductionkc{\mathtuple{\bar{P}_{0}, \bar{L}_{0}}}$,
   ${\Gamma}\mathtypeenvreductionkc{\bar{\Gamma}_{0}}$ and
   ${P_{0}}\mathpistructuralpo{\mathEr{\bar{\Gamma}_{0}}{\bar{L}_{0}, k}{\bar{P}_{0}}}$.
   By 
   ${P_{0}}\mathpistructuralpo{\mathEr{\bar{\Gamma}_{0}}{\bar{L}_{0}, k}{\bar{P}_{0}}}$ 
   and the induction hypothesis,
   we see tbar there exist $\bar{P}$, $\bar{L}$ and $\bar{\Gamma}$ such that
   ${\mathtuple{\bar{P}_{0}, \bar{L}_{0}}}\mathpireductionkc{\mathtuple{\bar{P}, \bar{L}}}$,
   ${\bar{\Gamma}_{0}}\mathtypeenvreductionkc{\bar{\Gamma}}$ and
   ${\bar{P}'}\mathpistructuralpo{\mathEr{\bar{\Gamma}}{\bar{L}, k}{\bar{P}}}$.
   Therefore,  we have 
   ${\mathtuple{P, L}}\mathpireductionkc{\mathtuple{\bar{P}, \bar{L}}}$,
   ${\Gamma}\mathtypeenvreductionkc{\bar{\Gamma}}$ and
   ${\bar{P}'}\mathpistructuralpo{\mathEr{\Gamma}{\bar{L}, k}{\bar{P}}}$.

   Case 3. \rulename{SP-Zero1}. Straightforward.

   Case 4. \rulename{SP-Zero2}. Straightforward.

   Case 5. \rulename{SP-Commut}. Straightforward.

   Case 6. \rulename{SP-Assoc}. Straightforward.

   Case 7. \rulename{SP-New}. Straightforward.
   
   Case 8. \rulename{SP-IfT}.
   Let ${P'}\equiv{\mathobif{\mathobtrue^{l'}}{Q_{0}}{Q_{1}}}$ and
   $\bar{P}'\equiv{Q_{0}}$. 
   Then, we have either ${\mathEr{\Gamma}{L, k}{P}}\equiv{P'}$
   or $Q_{0}\mathpistructuralpo{\mathEr{\Gamma}{L, k}{P}}$.
   
   Assume ${\mathEr{\Gamma}{L, k}{P}}\equiv{P'}$.
   Then $l'\leq_{L}k$.
   By \cref{def:Er}, $P\equiv\mathcontext{C}{P'}$ for some finite level context,
   where $\mathof{\Gamma}{x}=\mathprogramtypetuple{\mathvect{\tau}}^{l'}/U$ 
   implies $l'\not\leq_{L}k$ for all $x\in\mathFNof{C}$ and,
   for all occurrence $\mathpinew{\mathistype{x}{\xi}}{\mathplaceholder}$ in $C$,
   the type of $\xi$ is not less than $k$ in $L$.
   From \cref{lemma:lock-freedom-new-free-finite-level-to-eval},
   there exists an evaluation context $E$ and a lattice for secrecy levels $\hat{L}$
   such that
   $\mathtuple{C, L}\mathpisecreductionkc{\Gamma}{k}\mathtuple{E, \hat{L}}$.
   Hence, $\mathtuple{\mathcontext{C}{P'}, L}\mathpireductionkc\mathtuple{\mathcontext{E}{P'}, \hat{L}}$ and
   $\mathtuple{\mathcontext{C}{Q_{0}}, L}\mathpireductionkc\mathtuple{\mathcontext{E}{Q_{0}}, \hat{L}}$.
   Since $\mathcontext{E}{P'}\mathpistructuralpo\mathcontext{E}{Q_{0}}$, we have
   $\mathtuple{\mathcontext{C}{P'}, L}\mathpireductionkc\mathtuple{\mathcontext{E}{Q_{0}}, \hat{L}}$.
   We see ${\mathEr{\Gamma}{\hat{L}, k}{\mathcontext{E}{Q_{0}}}}\equiv{Q_{0}}$.
   Let $\bar{P}\equiv\mathcontext{E}{Q_{0}}$, $\bar{L}\equiv{\hat{L}}$, ${\bar{\Gamma}}\equiv{\Gamma}$.
    Then, we have the claimed result.
   
   Assume $Q_{0}\mathpistructuralpo{\mathEr{\Gamma}{L, k}{P}}$.
   Then $\bar{P}'\mathpistructuralpo{\mathEr{\Gamma}{L, k}{P}}$.
   Let $\bar{P}\equiv{P}$, $\bar{L}\equiv{L}$ and ${\bar{\Gamma}}\equiv{\Gamma}$.
   Then, we have the claimed result.

   Case 9. \rulename{SP-IfF}.
   In the similar way to the case \rulename{SP-IfT}.

   Case 10. \rulename{SP-Rep}.
   Straightforward.

   Case 11. \rulename{SP-Par}.
   Straightforward.

   Case 12. \rulename{SP-CNew}.
   Straightforward.
  \end{proof}  

  \begin{lemma}
   \label{lemma:Er-context}
   For type environments $\Gamma$, $\Delta$, lattices for secrecy levels $L$, $L'$ and
   a ${k}$-$\mathenvandlevel{\Gamma}{L}{m}$-$\mathenvandlevel{\Delta}{L'}{m'}$-context $C$,
   if $\mathtypejudgementwithsecrecy{\Gamma}{L}{m}{P}$ is 
   $k$-securely derivable, 
   then 
   ${\mathEr{\Delta}{L', m'}{\mathcontext{C}{P}}}\equiv{\mathcontext{\mathEr{\Delta}{L', m'}{C}}{\mathEr{\Gamma}{L, m}{P}}}$.
  \end{lemma}

  \begin{proof}
   By induction on $k$-secure derivation tree of $\mathtypejudgementwithsecrecy{\Delta}{L'}{m'}{C}$ from $\mathtypejudgementwithsecrecy{\Gamma}{L}{m}{\mathcontext{}{\;}}$.
  \end{proof}

  \begin{lemma}
   \label[lemma]{lemma:P-simulate-Er}
   For a reliable type environment $\Gamma$,
   if $\mathtypejudgementwithsecrecy{\Gamma}{L}{m}{P}$
   is $k$-securely derivable,
   ${P'}\mathpistructuralpo{\mathEr{\Gamma}{L, k}{P}}$,
   $L'$ is an $\mathEr{\Gamma}{L, k}{P}$-sublattice of $L$, and
   ${\mathtuple{P', L'}}\mathpireduction{\mathtuple{\bar{P}', \bar{L}'}}$,
   then there exist $\bar{P}$ and $\bar{L}$ such that
   ${\mathtuple{P, L}}\mathpireductiontc{\mathtuple{\bar{P}, \bar{L}}}$,
   and
   ${\bar{P}'}\mathpistructuralpo{\mathEr{\Gamma}{\bar{L}, k}{\bar{P}}}$,
   $\bar{L}'$ is an $\mathEr{\Gamma}{\bar{L}, k}{P}$-sublattice of $\bar{L}$.
  \end{lemma}

  \begin{proof}
   Assume that
   $\mathtypejudgementwithsecrecy{\Gamma}{L}{m}{P}$ is $k$-securely derivable,
   ${P'}\mathpistructuralpo{\mathEr{\Gamma}{L, \hat{l}}{P}}$,
   $L'$ is an $\mathEr{\Gamma}{\hat{L}, \hat{l}}{P}$-sublattice of $L$, and
   ${\mathtuple{P', L'}}\mathpireduction{\mathtuple{\bar{P}', \bar{L}'}}$.
   By \cref{prop:reduction-basic}, either 
   \begin{enumerate}
    \item ${P'}\mathpistructuralpo{\mathpinew{\mathistype{\mathvect{x}}{\mathvect{\xi}}}{\mathpiparallel{\mathpiparallel{\mathpioutput{z}{\mathvect{v}}. P'_{0}}{\mathpiinput{z}{\mathvect{y}}.P'_{1}}}{P'_{2}}}}$, 
       ${\mathpinew{\mathistype{\mathvect{x}}{\mathvect{\xi}}}{\mathpiparallel{\mathpiparallel{P'_{0}}{\mathsubstbox{P'_{1}}{\mathsubst{\mathvect{y}}{\mathvect{v}}}}}{P'_{2}}}}\mathpistructuralpo{\bar{P}'}$ and 
       $\bar{L}'=L'$, or
	  \label{item:com-lemma-P-simulate-Er}
    \item ${P'}\mathpistructuralpo{\mathpinew{\mathistype{\mathvect{x}}{\mathvect{\xi}}}{\mathpiparallel{\mathpinewsecrecylevel{l}{\mathvect{l_{0}}}{\mathvect{l_{1}}}{P'_{0}}}{P'_{1}}}}$,
       ${\mathpinew{\mathistype{\mathvect{x}}{\mathvect{\xi}}}{\mathpiparallel{P'_{0}}{P'_{1}}}}\mathpistructuralpo{\bar{P}'}$ and
       $\bar{L}'=\mathpinewsecrecylevel{l}{\mathvect{l_{0}}}{\mathvect{l_{1}}}{L'}$.
	  \label{item:newlev-lemma-P-simulate-Er}
   \end{enumerate}

   We consider the case \cref{item:com-lemma-P-simulate-Er}.
   By \cref{lemma:P-simulate-Er-structualpo},
   there exist $\bar{P}_{0}$, $\bar{L}_{0}$ and $\bar{\Gamma}_{0}$ such that
   ${\mathtuple{P, L}}\mathpireductionkc{\mathtuple{\bar{P}_{0}, \bar{L}_{0}}}$,
   ${\Gamma}\mathtypeenvreductionkc{\bar{\Gamma}_{0}}$
   and
   ${\mathpinew{\mathistype{\mathvect{x}}{\mathvect{\xi}}}{\mathpiparallel{\mathpiparallel{\mathpioutput{z}{\mathvect{v}}. P'_{0}}{\mathpiinput{z}{\mathvect{y}}.P'_{1}}}{P'_{2}}}}\mathpistructuralpo{\mathEr{\bar{\Gamma}_{0}}{\bar{L}_{0}, k}{\bar{P}_{0}}}$, and
   $\mathtypejudgementwithsecrecy{\bar{\Gamma}_{0}}{L}{m}{\bar{P}_{0}}$ is 
   $k$-securely derivable.
   Then, there exists an evaluation context with two holes $E$ such that
   ${\mathEr{\bar{\Gamma}_{0}}{\bar{L}_{0}, k}{\bar{P}_{0}}}\equiv{\mathcontexttwoholes{E}{\mathpioutput{z}{\mathvect{v}}. P'_{0}}{\mathpiinput{z}{\mathvect{y}}.P'_{1}}}$
   and
   ${\mathpinew{\mathistype{\mathvect{x}}{\mathvect{\xi}}}{\mathpiparallel{\mathpiparallel{\firstholemathcontexttwoholes{\;}}{\secondholemathcontexttwoholes{\;}}}{P'_{2}}}}\mathpistructuralpo{E}$.
   Since 
   $\mathtypejudgementwithsecrecy{\bar{\Gamma}_{0}}{L}{m}{\bar{P}_{0}}$ is 
   $k$-securely derivable,
   there exists a finite level context with two holes $C$, process $P_{0}$, $P_{1}$ 
   and type environments $\Delta_{0}$, $\Delta_{1}$ such that
   $\bar{P}_{0}\equiv{\mathcontexttwoholes{C}{\mathpioutput{z}{\mathvect{v}}. P_{0}}{\mathpiinput{z}{\mathvect{y}}.P_{1}}}$,
   ${\mathEr{\bar{\Gamma}_{0}}{\bar{L}_{0}, k}{C}}\equiv{E}$,
   ${\mathEr{\Delta_{0}}{\bar{L}_{0}, k}{P_{0}}}\equiv{\mathpioutput{z}{\mathvect{v}}. P'_{0}}$,
   ${\mathEr{\Delta_{1}}{\bar{L}_{0}, k}{P_{0}}}\equiv{\mathpiinput{z}{\mathvect{y}}.P'_{1}}$,
   and
   $\mathtypejudgementwithsecrecy{\bar{\Gamma}_{0}}{L}{m}{C}$ is 
   $k$-securely derivable
   from 
   $\mathtypejudgementwithsecrecy{\Delta_{0}}{L'_{1}}{l'_{0}}{\firstholemathcontexttwoholes{\;}}$
   and
   $\mathtypejudgementwithsecrecy{\Delta_{1}}{L'_{1}}{l'_{1}}{\secondholemathcontexttwoholes{\;}}$,
   where $l'_{i}\geq m$ for $i=0, 1$.
   By \cref{lemma:lock-freedom-new-free-finite-level-to-eval},
   there exists an evaluation context $\bar{E}$ and a lattice for secrecy levels $\bar{L}$
   such that
   $\mathtuple{C, L}\mathpisecreductionkc{\Gamma}{k}\mathtuple{\bar{E}, \bar{L}}$.
   Hence,
   ${\mathtuple{\bar{P}_{0}, \bar{L}_{0}}}\mathpireductiontc \mathtuple{\mathcontexttwoholes{\bar{E}}{P_{0}}{\mathsubstbox{P_{1}}{\mathsubst{\mathvect{y}}{\mathvect{v}}}}, \bar{L}}$.
   By \cref{lemma:Er-simulate-reduction},
   we have
   ${\mathpinew{\mathistype{\mathvect{x}}{\mathvect{\xi}}}{\mathpiparallel{\mathpiparallel{\firstholemathcontexttwoholes{P'_{0}}}{\secondholemathcontexttwoholes{\mathsubstbox{P'_{1}}{\mathsubst{\mathvect{y}}{\mathvect{v}}}}}}{P'_{2}}}}\mathpistructuralpo{\mathcontexttwoholes{E}{P'_{0}}{\mathsubstbox{P'_{1}}{\mathsubst{\mathvect{y}}{\mathvect{v}}}}}\equiv{\mathcontexttwoholes{\mathEr{\bar{\Gamma}_{0}}{\bar{L}_{0}, k}{C}}{P'_{0}}{\mathsubstbox{P'_{1}}{\mathsubst{\mathvect{y}}{\mathvect{v}}}}}\mathpistructuralpo{\mathcontexttwoholes{\mathEr{\bar{\Gamma}_{0}}{\bar{L}_{0}, k}{\bar{E}}}{P'_{0}}{\mathsubstbox{P'_{1}}{\mathsubst{\mathvect{y}}{\mathvect{v}}}}}$.
   By \cref{lemma:P-simulate-Er-structualpo},
   there exist $\bar{P}$, $\bar{L}$,  such that
   ${\mathtuple{\bar{P}_{0}, \bar{L}_{0}}}\mathpireductionkc{\mathtuple{\bar{P}, \bar{L}}}$,
   and
   ${\mathpinew{\mathistype{\mathvect{x}}{\mathvect{\xi}}}{\mathpiparallel{\mathpiparallel{P'_{0}}{\mathsubstbox{P'_{1}}{\mathsubst{\mathvect{y}}{\mathvect{v}}}}}{P'_{2}}}}\mathpistructuralpo{\mathEr{\bar{\Gamma}_{0}}{\bar{L}_{0}, k}{\bar{P}}}$.
   Therefore,
   ${\mathtuple{P, L}}\mathpireductiontc{\mathtuple{\bar{P}, \bar{L}}}$.
  
   The case \cref{item:newlev-lemma-P-simulate-Er} is obvious.
  \end{proof}

  \begin{lemma}
   \label[lemma]{lemma:key-barbed-bisimulation}
   Define
   \[
   {\mathbarbedbisim{R}}={\mathsetintension{\mathtuple{\mathtuple{P, L}, \mathtuple{P', L'}}}{ 
   \begin{gathered}
    \mathtypejudgementwithsecrecy{\Gamma}{L}{m}{P} \text{ is $k$-securely derivable} \\ 
    \text{for a reliable type environment $\Gamma$}, \\
    \text{${P'}\mathpistructuralpo{\mathEr{\Gamma}{L, k}{P}}$, and} \\
    \text{$L'$ is an $\mathEr{\Gamma}{L, \hat{l}}{P}$-sublattice of $L$.}
   \end{gathered}
   }
   }.
   \]
   ${\mathbarbedbisim{R}}$ is a barbed bisimulation.
  \end{lemma}
  
  \begin{proof}
   It suffices to show that ${\mathbarbedbisim{R}}$ satisfies all the conditions of 
   \cref{def:Barbed-bisimulation}.

   Assume ${\mathtuple{\mathtuple{P, L}, \mathtuple{P', L'}}}\in{R}$.
   Then, we see that 
   $\mathtypejudgementwithsecrecy{\Gamma}{L}{m}{P}$ is derivable
   for a reliable type environment $\Gamma$,
   the secrecy level of $\mathtypeenvandsecrecylatice{\Gamma}{L}$ is $l_{1}$, and
   ${P'}\mathpistructuralpo{\mathEr{\Gamma}{L, \hat{l}}{P}}$.

   \noindent \cref{item:left-def-Barbed-bisimulation}
   Assume ${\mathtuple{P, L}}\mathpireduction{\mathtuple{\hat{P}, \hat{L}}}$.
   By \cref{prop:subject-reduction},
   there exists a type environment $\hat{\Gamma}$ such that 
   either ${\hat{\Gamma}}\equiv{\Gamma}$ or ${\Gamma}\mathtypeenvreduction{\hat{\Gamma}}$ and
   $\mathtypejudgementwithsecrecy{\hat{\Gamma}}{\hat{L}}{m}{\hat{P}}$
   is derivable.
   By \cref{lemma:Er-simulate-reduction},
   either 
   ${\mathEr{\Gamma}{L, \hat{l}}{P}}\mathpistructuralpo{\mathEr{\Gamma}{\hat{L}, \hat{l}}{\hat{P}}}$ and
   $L'$ is an $\mathEr{\Gamma}{\hat{L}, \hat{l}}{\hat{P}}$-sublattice of $\hat{L}$,
   or
   there exists a lattice for secrecy levels $\hat{L}'$ such that
   ${\mathtuple{\mathEr{\Gamma}{L, \hat{l}}{P}, L'}}\mathpireduction{\mathtuple{\mathEr{\Gamma}{\hat{L}, \hat{l}}{\hat{P}}, \hat{L}'}}$ and
   $\hat{L}'$ is an $\mathEr{\Gamma}{\hat{L}, \hat{l}}{P}$-sublattice of $\hat{L}$. 
   
   Assume
   ${\mathEr{\Gamma}{L, \hat{l}}{P}}\mathpistructuralpo{\mathEr{\Gamma}{\hat{L}, \hat{l}}{\hat{P}}}$.
   Since ${P'}\mathpistructuralpo{\mathEr{\Gamma}{L, \hat{l}}{P}}$,
   we have ${P'}\mathpistructuralpo{\mathEr{\Gamma}{\hat{L}, \hat{l}}{\hat{P}}}$.
   Hence, ${\mathtuple{P', L'}}\mathpireductionkc{\mathtuple{\mathEr{\Gamma}{\hat{L}, \hat{l}}{\hat{P}}, L'}}$.
   By \cref{lemma:Er-property-core-type}, we have 
   $\mathEr{\Gamma}{\hat{L}, \hat{l}}{\hat{P}}\equiv\mathEr{\hat{\Gamma}}{\hat{L}, \hat{l}}{\hat{P}}$.
   Since $L'$ is an $\mathEr{\hat{\Gamma}}{\hat{L}, \hat{l}}{\hat{P}}$-sublattice of $\hat{L}$,
   we have
   ${\mathtuple{\mathtuple{\hat{P}, \hat{L}}, \mathtuple{\mathEr{\hat{\Gamma}}{\hat{L}, \hat{l}}{\hat{P}}, L'}}}\in{R}$.

   Assume
   ${\mathEr{\Gamma}{L, \hat{l}}{P}} \not\mathpistructuralpo{\mathEr{\Gamma}{\hat{L}, \hat{l}}{\hat{P}}}$.
   Then, there exists a lattice for secrecy levels $\hat{L}'$ such that \linebreak[4]
   $\mathtuple{\mathEr{\Gamma}{L, \hat{l}}{P}, L'}\mathpireduction\mathtuple{\mathEr{\Gamma}{\hat{L}, \hat{l}}{\hat{P}}, \hat{L}'}$ and
   $\hat{L}'$ is an $\mathEr{\Gamma}{\hat{L}, \hat{l}}{P}$-sublattice of $\hat{L}$. 
   Since ${P'}\mathpistructuralpo{\mathEr{\Gamma}{L, \hat{l}}{P}}$,
   we have
   ${\mathtuple{P', L'}}\mathpireduction{\mathtuple{\mathEr{\Gamma}{\hat{L}, \hat{l}}{\hat{P}}, \hat{L}'}}$.
   By \cref{lemma:Er-property-core-type}, we have 
   $\mathEr{\Gamma}{\hat{L}, \hat{l}}{\hat{P}}\equiv\mathEr{\hat{\Gamma}}{\hat{L}, \hat{l}}{\hat{P}}$.
   Hence, ${\mathtuple{P', L'}}\mathpireductionkc{\mathtuple{\mathEr{\hat{\Gamma}}{\hat{L}, \hat{l}}{\hat{P}}, L'}}$.
   We also have
   ${\mathtuple{\mathtuple{\hat{P}, \hat{L}}, \mathtuple{\mathEr{\hat{\Gamma}}{\hat{L}, \hat{l}}{\hat{P}}, \hat{L}'}}}\in{R}$.

   \noindent \cref{item:right-def-Barbed-bisimulation}
   By \cref{lemma:P-simulate-Er}.

   \noindent \cref{item:barbs-def-Barbed-bisimulation}
   Straightforward.
  \end{proof}

  \begin{lemma}
   \label{lemma:barbed-bisimilar-Er}
   For a reliable type environment $\Gamma$, a lattice for secrecy levels $L$ and
   a process $P$,
   if $\mathtypejudgementwithsecrecy{\Gamma}{L}{m}{P}$
   is $k$-securely derivable,
   then $\mathisbisimilar{\mathtuple{P, L}}{\mathtuple{\mathEr{\Gamma}{L, l}{P}, L}}$.
  \end{lemma}

  \begin{proof}
   Assume that 
   $\mathtypejudgementwithsecrecy{\Gamma}{L}{m}{P}$
   is $k$-securely derivable, 
   where ${m}\not \leq_{L}{l}$.
   By \cref{lemma:key-barbed-bisimulation},
   we have $\mathisbisimilar{\mathtuple{P, L}}{\mathtuple{\mathEr{\Gamma}{L, l}{P}, L}}$.
  \end{proof}

  \begin{definition}
   \label{def:lower-env-context}
   We write $\mathlower{L}{l}{\Gamma}$ for the type environment obtained 
   from a type environment $\Gamma$ 
   by replacing all the secrecy annotations ${l'}\not \leq_{L}{l}$ with 
   the infimum of $\mathsetextension{l, l'}$ and 
   all the capability level annotations with $\infty$.

   We also write $\mathlower{L}{l}{C}$ for the context obtained from a context $C$ 
   by replacing every secrecy annotation ${l'}\not \leq_{L}{l}$ in a type or a constant value
   with the infimum of $\mathsetextension{l, l'}$.
  \end{definition}
  
  We note that, for a closed type environment $\Gamma$,
  $\mathlower{L}{l}{\Gamma}$ is a reliable type environment whose secrecy level is $l$.

  \begin{lemma}
   \label{lemma:lower-context}
   For type environments $\Gamma$, $\Delta$, lattices for secrecy levels $L$, $L'$ and
   a 
   $\mathenvandlevel{\Gamma}{L}{m}$-$\mathenvandlevel{\Delta}{L'}{m'}$-context 
   $C$,
   if the secrecy level of $\mathtypeenvandsecrecylatice{\Gamma}{L}$ is $l_{0}$,
   then $\mathlower{L'}{l_{0}}{C}$ is a 
   ${k}$-$\mathenvandlevel{\Gamma}{L}{m}$-$\mathenvandlevel{\mathlower{L'}{l_{0}}{\Delta}}{L'}{\hat{m}}$-context, 
   where $\hat{m}$ is the infimum of $\mathsetextension{l_{0}, l'}$ in $L'$.
  \end{lemma}

  \begin{proof}
   Straightforward.
  \end{proof}

  \begin{lemma}
   \label{lemma:derivability-exchanged}
   For a type environment $\Gamma$, a lattice for secrecy levels $L$,
   a process $P$, and secrecy levels $m$, $l_{1}$,
   if $\mathtypejudgementwithsecrecy{\Gamma}{L}{m}{\mathsubstbox{P}{\mathsubst{x}{\mathobtrue^{l_{1}}}}}$ is $k$-securely  derivable,
   then $\mathtypejudgementwithsecrecy{\Gamma}{L}{m}{\mathsubstbox{P}{\mathsubst{x}{\mathobfalse^{l_{1}}}}}$ is $k$-securely derivable.
  \end{lemma}
  
  \begin{proof}
   By induction on derivation tree of $\mathtypejudgementwithsecrecy{\Gamma}{L}{m}{\mathsubstbox{P}{\mathsubst{x}{\mathobtrue^{l_{1}}}}}$.
  \end{proof}

  \begin{lemma}
   \label{lemma:equiv-Er-exchanged}
   For a type environment $\Gamma$, a lattice for secrecy levels $L$,
   a process $P$, and secrecy levels $l_{0}$, $l_{1}$,
   if $\mathtypejudgementwithsecrecy{\Gamma}{L}{m}{\mathsubstbox{P}{\mathsubst{x}{\mathobtrue^{l_{1}}}}}$ is $k$-securely derivable,
   then ${\mathEr{\Gamma}{L, m}{\mathsubstbox{P}{\mathsubst{x}{\mathobtrue^{l_{1}}}}}}\equiv{\mathEr{\Gamma}{L, m}{\mathsubstbox{P}{\mathsubst{x}{\mathobfalse^{l_{1}}}}}}$.
  \end{lemma}
  
  \begin{proof}
   By induction on the construction of $P$.
  \end{proof}
  
  \begin{lemma}
   \label{lemma:lower-context-bisimilar}
   For type environments $\Gamma$ and $\Delta$,
   lattices for secrecy levels $L$,
   processes $P$, $Q$,
   $\mathenvandlevel{\Gamma}{L}{m}$-$\mathenvandlevel{\Delta}{L'}{m'}$-context  $C$,
   if $\Delta$ is closed and
   $\mathisbisimilar{\mathtuple{\mathcontext{{\mathlower{L'}{l''}{C}}}{P}, L }}{\mathtuple{\mathcontext{{\mathlower{L'}{l''}{C}}}{Q}, L }}$,
   then 
   $\mathisbisimilar{\mathtuple{\mathcontext{C}{P}, L }}{\mathtuple{\mathcontext{C}{Q}, L }}$.
   
  \end{lemma}

  \begin{proof}
   Straightforward.
  \end{proof}

\subsection{Proof of \cref{thm:non-interference-value}}
\label{subsec:proof-thm-non-interference-value}

   For a type environment $\Gamma$, a lattice for secrecy levels $L$,
   a process $P$, and secrecy levels $l$, $l'$,
   Assume that ${l'}\not \leq_{L}{l}$, 
   the secrecy level of $\mathtypeenvandsecrecylatice{\Gamma}{L}$ is $l$ 
   and $\mathtypejudgementwithsecrecy{\Gamma}{L}{m}{\mathsubstbox{P}{\mathsubst{x}{\mathobtrue^{l'}}}}$ is $k$-securely derivable.
   We show 
   $\mathiscongruentwithenvandlevel{\Gamma}{L}{m}{{\mathsubstbox{P}{\mathsubst{x}{\mathobtrue^{l'}}}}}{{\mathsubstbox{P}{\mathsubst{x}{\mathobfalse^{l'}}}}}$.
   By \cref{lemma:derivability-exchanged},
   we see that
   $\mathtypejudgementwithsecrecy{\Gamma}{L}{m}{\mathsubstbox{P}{\mathsubst{x}{\mathobfalse^{l'}}}}$ is $k$-securely derivable.
   Then, it suffices to show that, 
   for any closed $\Delta$, a lattice for secrecy levels $L'$,
   and a secrecy level $m'$, 
   $\mathisbisimilar{\mathtuple{\mathcontext{C}{\mathsubstbox{P}{\mathsubst{x}{\mathobtrue^{l'}}}}, L'}}{\mathtuple{\mathcontext{C}{{\mathsubstbox{P}{\mathsubst{x}{\mathobfalse^{l'}}}}, L'}}}$ with any 
   $\mathenvandlevel{\Gamma}{L}{l}$-$\mathenvandlevel{\Delta}{L'}{m'}$-context $C$.

   Let $\Delta$ be a closed type environment, and $C$ be 
   a $\mathenvandlevel{\Gamma}{L}{m}$-$\mathenvandlevel{\Delta}{L'}{m'}$-context.
   Then 
   $\mathlower{L}{l}{\Delta}$ is a reliable type environment whose secrecy level is $l$.
   Let ${C'}\equiv{\mathlower{L}{l}{C}}$.

   By \cref{lemma:lower-context},
   $C'$ is a $\mathenvandlevel{\Gamma}{L}{m}$-$\mathenvandlevel{\mathlower{L}{l}{\Delta}}{L}{l}$-context.
   From \cref{lemma:barbed-bisimilar-Er},
   we have
   \begin{align*}
    {\mathtuple{\mathcontext{C'}{\mathsubstbox{P}{\mathsubst{x}{\mathobtrue^{l'}}}}, L }}
    &\mathbisimilarsy{\mathtuple{\mathEr{\mathlower{L}{l}{\Delta}}{L, l}{\mathcontext{C'}{\mathsubstbox{P}{\mathsubst{x}{\mathobtrue^{l'}}}}}, L}} \text{ and } \\
{\mathtuple{\mathcontext{C'}{\mathsubstbox{P}{\mathsubst{x}{\mathobfalse^{l'}}}}, L }}
    &\mathbisimilarsy{\mathtuple{\mathEr{\mathlower{L}{l}{\Delta}}{L, l}{\mathcontext{C'}{\mathsubstbox{P}{\mathsubst{x}{\mathobfalse^{l'}}}}}, L}}.  
   \end{align*}
   \cref{lemma:Er-context} implies 
   \[
    {\mathEr{\mathlower{L'}{l}{\Delta}}{L', l}{\mathcontext{C'}{\mathsubstbox{P}{\mathsubst{x}{\mathobtrue^{l'}}}}}}
   \equiv
   {\mathcontext{\mathEr{\mathlower{L'}{l}{\Delta}}{L', l}{C'}}{\mathEr{\Gamma}{L, m}{\mathsubstbox{P}{\mathsubst{x}{\mathobtrue^{l'}}}}}}.
   \]
   Since 
   $\mathtypejudgementwithsecrecy{\Gamma}{L}{m}{\mathsubstbox{P}{\mathsubst{x}{\mathobtrue^{l'}}}}$ and
   $\mathtypejudgementwithsecrecy{\Gamma}{L}{m}{\mathsubstbox{P}{\mathsubst{x}{\mathobfalse^{l'}}}}$ are derivable,
   \cref{lemma:equiv-Er-exchanged} implies
   \[
   {\mathEr{\Gamma}{L, m}{\mathsubstbox{P}{\mathsubst{x}{\mathobtrue^{l'}}}}}
   \equiv
   {\mathEr{\Gamma}{L, m}{\mathsubstbox{P}{\mathsubst{x}{\mathobfalse^{l'}}}}}.
   \]
   Then, we have
   \begin{align*}
    {\mathEr{\mathlower{L'}{l}{\Delta}}{L', l}{\mathcontext{C'}{\mathsubstbox{P}{\mathsubst{x}{\mathobtrue^{l'}}}}}}
   &\equiv
   {\mathcontext{\mathEr{\mathlower{L'}{l}{\Delta}}{L', l}{C'}}{\mathEr{\Gamma}{L, m}{\mathsubstbox{P}{\mathsubst{x}{\mathobtrue^{l'}}}}}} \\
   &\equiv
   {\mathcontext{\mathEr{\mathlower{L'}{l}{\Delta}}{L', l}{C'}}{\mathEr{\Gamma}{L, m}{\mathsubstbox{P}{\mathsubst{x}{\mathobfalse^{l'}}}}}} \\
   &\equiv
    {\mathEr{\mathlower{L'}{l}{\Delta}}{L', l}{\mathcontext{C'}{\mathsubstbox{P}{\mathsubst{x}{\mathobfalse^{l'}}}}}}.
   \end{align*}
   By \cref{lemma:barbed-bisimilar-transitive}, we have
   \[
   {\mathtuple{\mathcontext{C'}{\mathsubstbox{P}{\mathsubst{x}{\mathobtrue^{l'}}}}, L }}\mathbisimilarsy{\mathtuple{\mathcontext{C'}{\mathsubstbox{P}{\mathsubst{x}{\mathobfalse^{l'}}}}, L }}. 
   \]
   \cref{lemma:lower-context-bisimilar} implies
   \[
   {\mathtuple{\mathcontext{C}{\mathsubstbox{P}{\mathsubst{x}{\mathobtrue^{l'}}}}, L }}\mathbisimilarsy{\mathtuple{\mathcontext{C}{\mathsubstbox{P}{\mathsubst{x}{\mathobfalse^{l'}}}}, L }}. 
   \]
   \qed

  \subsection{Proof of \cref{thm:non-interference-process}}
  \label{subsec:proof-thm-non-interference-process}

   For type environments $\Gamma$,  $\Delta$,
   lattices for secrecy levels $L$, $L'$,
   processes $P_{0}$, $P_{1}$, and \linebreak[4]
   a $\mathenvandlevel{\Delta}{L'}{m'}$-$\mathenvandlevel{\Gamma}{L}{m}$-context $\hat{C}$,
   assume that ${m''}\in{L}$ and ${m'}\not \leq_{L}{m''}$, 
   the secrecy level of $\mathtypeenvandsecrecylatice{\Gamma}{L}$ is $m''$, and
   $\mathtypejudgementwithsecrecy{\Delta}{L'}{m'}{}{P_{i}}$ is derivable
   for ${i}={0, 1}$.
   We show
   $\mathiscongruentwithenvandlevel{\Gamma}{L}{m}{\mathcontext{\hat{C}}{P_{0}}}{\mathcontext{\hat{C}}{P_{1}}}$.
   
   Since $\mathtypejudgementwithsecrecy{\Delta}{L'}{m'}{P_{i}}$ is $k$-securely derivable
   for ${i}={0, 1}$,
   we see that
   $\mathtypejudgementwithsecrecy{\Delta}{L'}{m'}{\mathcontext{\hat{C}}{P_{i}}}$
   is $k$-securely derivable for ${i}={0, 1}$.

   Let $\Pi$ be a closed type environment, and
   $C$ be $\mathenvandlevel{\Gamma}{L}{m}$-$\mathenvandlevel{\Pi}{L'}{m'''}$-context.
   Then 
   $\mathlower{L'}{m''}{\Pi}$ is a reliable type environment whose secrecy level is $m''$.
   Let ${C'}\equiv{\mathlower{L'}{m''}{C}}$.

   By \cref{lemma:lower-context},
   $C'$ is a \linebreak[3] $\mathenvandlevel{\Gamma}{L}{m}$-$\mathenvandlevel{\mathlower{L'}{m''}{\Pi}}{L'}{m''}$-context.
   From \cref{lemma:barbed-bisimilar-Er}, we have
   $\mathisbisimilar{\mathtuple{\mathcontext{C'}{\mathcontext{\hat{C}}{P_{i}}}, L}}{\mathtuple{\mathEr{\mathlower{L}{m''}{\Pi}}{L, m''}{\mathcontext{C'}{\mathcontext{\hat{C}}{P_{i}}}}, L'}}$
   for ${i}={0, 1}$.
   Since $\mathcontext{C'}{\hat{C}}$ is \linebreak[4] a $\mathenvandlevel{\Delta}{L'}{m'}$-$\mathenvandlevel{\mathlower{L}{m''}{\Pi}}{L'}{m''}$-context,
   \cref{lemma:Er-context} implies
   \[
    {\mathEr{\mathlower{L'}{m''}{\Pi}}{L, m''}{\mathcontext{\mathcontext{C'}{\hat{C}}}{P_{i}}}}
   \equiv
   {\mathcontext{\mathEr{\mathlower{L'}{m''}{\Pi}}{L, m''}{\mathcontext{C'}{\hat{C}}}}{\mathEr{\Delta}{L', m'}{P_{i}}}}.
   \]
   for ${i}={0, 1}$.
   By \cref{lemma:barbed-bisimilar-structual-preorder-with-context} and \cref{lemma:Er-derivable-process-to-nil}, we have
   \[
    {\mathcontext{\mathEr{\mathlower{L'}{m''}{\Pi}}{L, m''}{\mathcontext{C'}{\hat{C}}}}{\mathEr{\Delta}{L', m'}{P_{0}}}}
   \mathbisimilarsy
   {\mathcontext{\mathEr{\mathlower{L'}{m''}{\Pi}}{L, m''}{\mathcontext{C'}{\hat{C}}}}{\mathnil}}
   \mathbisimilarsy
   {\mathcontext{\mathEr{\mathlower{L'}{m''}{\Pi}}{L, m''}{\mathcontext{C'}{\hat{C}}}}{\mathEr{\Delta}{L', m'}{P_{1}}}}.
   \]
   Hence, we have
   $\mathisbisimilar{\mathtuple{\mathcontext{C'}{\mathcontext{\hat{C}}{P_{0}}}, L}}{\mathtuple{\mathcontext{C'}{\mathcontext{\hat{C}}{P_{1}}}, L}}$.
   Thus, we see
   $\mathiscongruentwithenvandlevel{\Gamma}{L}{m}{\mathcontext{\hat{C}}{P_{0}}}{\mathcontext{\hat{C}}{P_{1}}}$.

\qed

\end{document}